\documentclass[printer,longauth,usenatbib]{aa}

\usepackage{natbib}
\usepackage{times}
\usepackage[T1]{fontenc}
\usepackage{graphicx}
\usepackage{tabularx}
\usepackage[skip=1ex]{caption}
\usepackage{amsmath}

\usepackage{pdflscape}

\usepackage[utf8]{inputenc}
\usepackage{tabularx}

\def\deg{\hbox{$^\circ$}}
\usepackage[normalem]{ulem}

\begin{document}

\title{Properties of slowly rotating asteroids \break 
       from the Convex Inversion Thermophysical Model}

\authorrunning{Marciniak et al.}
\titlerunning{Properties of slowly rotating asteroids from CITPM}

\author{A. Marciniak \inst{1} 
  \and J. \v{D}urech \inst{2} 
  \and V. Al{\'i}-Lagoa \inst{3}
  \and W. Og{\l}oza \inst{4}
  \and R. Szak{\'a}ts \inst{5}
  \and T. G. M{\"u}ller \inst{3}
  \and L. Moln{\'a}r \inst{5,6,7}
  \and A. P{\'a}l \inst{5,8} % apal@szofi.net
  \and F.~Monteiro \inst{9} % filipeastro@on.br
  \and P.~Arcoverde \inst{9}
  \and R.~Behrend \inst{10}  % raoul.behrend@unige.ch
  \and Z. Benkhaldoun \inst{11}
  \and L. Bernasconi \inst{12} % laurent.bernasconi.51@wanadoo.fr
  \and J. Bosch \inst{13} % jean-gabriel.bosch@unige.ch
  \and S. Brincat \inst{14} % stephenbrincat@gmail.com
  \and L. Brunetto \inst{15} % l.brunetto@laposte.net
  \and M. Butkiewicz - B\k{a}k \inst{1}
  \and F.~Del~Freo \inst{16} %mail?
  \and R.~Duffard \inst{17}
  \and M. Evangelista-Santana \inst{9}
  \and G. Farroni \inst{18} %gino.farroni@sfr.fr 
  \and S. Fauvaud \inst{19,20}
  \and M. Fauvaud \inst{19,20} % mail? 
   \and M.~Ferrais \inst{21}
  \and S.~Geier \inst{22,23}
   \and J. Golonka \inst{24}
  \and J. Grice \inst{25}
  \and R.~Hirsch \inst{1}
  \and J. Horbowicz \inst{1}
   \and E. Jehin \inst{26}
  \and P. Julien \inst{14}
  \and Cs. Kalup \inst{5} % mail?
  \and K.~Kami{\'n}ski \inst{1}
  \and M.~K.~Kami{\'n}ska \inst{1}
  \and P. Kankiewicz \inst{27}
  \and V.~Kecskem{\'e}thy \inst{5}
  \and D.-H. Kim \inst{28,29} 
  \and M.-J. Kim \inst{29}
  \and I. Konstanciak \inst{1}
  \and J.~Krajewski \inst{1}
  \and V.~Kudak \inst{30,31}
  \and P. Kulczak \inst{1}
  \and T.~Kundera \inst{4} % mail?
  \and D. Lazzaro \inst{9}
  \and F. Manzini \inst{15} % manzini.ff@aruba.it
  \and H. Medeiros \inst{9,22}
  \and J.~Michimani-Garcia \inst{9}
  \and N. Morales \inst{17} % mail?
  \and J. Nadolny \inst{22,32}
  \and D.~Oszkiewicz \inst{1}
  \and E. Pak\v{s}tien{\.e} \inst{33}
  \and M. Paw{\l}owski \inst{1}
  \and V. Perig \inst{31} % mail?
  \and F.~Pilcher \inst{34}
  \and P.~Pinel$^\dagger$ \inst{18}
  \and E. Podlewska-Gaca \inst{1}
  \and T. Polakis \inst{35}
  \and F. Richard \inst{20} % mail?
  \and T. Rodrigues \inst{9}
  \and E. Rond{\'o}n \inst{9}
  \and R.~Roy \inst{36}  % rene.roy@wanadoo.fr
  \and J.~J.~Sanabria \inst{22}
  \and T.~Santana-Ros \inst{37,38}
  \and B. Skiff \inst{39}
  \and J.~Skrzypek \inst{1}
  \and K. Sobkowiak \inst{1}
  \and E. Sonbas \inst{40}
  \and G. Stachowski \inst{4} % mail?
  \and J.~Strajnic \inst{16}
  \and P. Trela \inst{1}
   \and {\L}.~Tychoniec \inst{41}
  \and S.~Urakawa \inst{42}
  \and E. Verebelyi \inst{5}
  \and K. Wagrez \inst{16} % mail?
  \and M. {\.Z}ejmo \inst{43}
  \and K. {\.Z}ukowski \inst{1}
}

  \institute{Astronomical Observatory Institute, Faculty of Physics, A. Mickiewicz University,
  S{\l}oneczna 36, 60-286 Pozna{\'n}, Poland. E-mail: am@amu.edu.pl %1
  \and Astronomical Institute, Faculty of Mathematics and Physics, Charles University, V Hole\v{s}ovi\v{c}k{\'a}ch 2, 
  180 00 Prague 8, Czech Republic %2
  \and Max-Planck-Institut f{\"u}r Extraterrestrische Physik (MPE), Giessenbachstrasse 1, 85748 Garching, Germany %3
  \and Mt. Suhora Observatory, Pedagogical University, Podchor\k{a}{\.z}ych 2, 30-084, Cracow, Poland %4
  \and Konkoly Observatory, Research Centre for Astronomy and Earth Sciences, E\H{o}tv\H{o}s Lor{\'a}nd Research Network (ELKH), 
       H-1121 Budapest, Konkoly Thege Mikl{\'o}s {\'u}t 15-17, Hungary %5
  \and MTA CSFK Lend{\"u}let Near-Field Cosmology Research Group %6
  \and ELTE E{\"o}tv{\"o}s Lor{\'a}nd University, Institute of Physics, 1117, P\'azm\'any P\'eter s\'et\'any 1/A, Budapest, Hungary %7
  \and Astronomy Department, E\"otv\"os Lor\'and University, P\'azm\'any P. s. 1/A, H-1171 Budapest, Hungary %8
  \and Observat{\'o}rio Nacional, R. Gen. Jos{\'e} Cristino, 77 - S{\~a}o Crist{\'o}v{\~a}o, 20921-400, Rio de Janeiro - RJ, Brazil %9
  \and Geneva Observatory, CH-1290 Sauverny, Switzerland %10
  \and Oukaimeden Observatory, High Energy Physics and Astrophysics Laboratory, Cadi Ayyad University, Marrakech, Morocco %11
  \and Les Engarouines Observatory, F-84570 Mallemort-du-Comtat, France %12
  \and Collonges Observatory, F-74160 Collonges, France %13
  \and Flarestar Observatory Fl.5/B, George Tayar Street, San Gwann SGN 3160, Malta %14
  \and Stazione Astronomica, 28060 Sozzago (Novara), Italy %15
  \and Haute-Provence Observatory, St-Michel l'Observatoire, France %16
  \and Departamento de Sistema Solar, Instituto de Astrof{\'i}sica de Andaluc{\'i}a (CSIC),
  Glorieta de la Astronom{\'i}a s/n, 18008 Granada, Spain %17
  \and 11 rue du Puits Coellier, F-37550 Saint-Avertin, France %18
  \and Observato{\'i}re du Bois de Bardon, 16110 Taponnat, France %19
  \and Association T60, Observato{\'i}re Midi-Pyr{\'e}n{\'e}es, 14, avenue Edouard Belin, 31400 Toulouse, France %20
  \and Aix Marseille Universit{\'e}, CNRS, CNES, Laboratoire d'Astrophysique de Marseille, Marseille, France %21
  \and Instituto de Astrof{\'i}sica de Canarias, C/ V{\'i}a Lactea, s/n, 38205 La Laguna, Tenerife, Spain %22
  \and Gran Telescopio Canarias (GRANTECAN), Cuesta de San Jos{\'e} s/n, E-38712, Bre{\~n}a Baja, La Palma, Spain %23
  \and Faculty of Physics, Astronomy and Informatics, Nicolaus Copernicus University in Toru{\'n} %24
  \and School of Physical Sciences, The Open University, MK7 6AA, UK %25
  \and Space sciences, Technologies and Astrophysics Research Institute, Universit{\'e} de Li{\`e}ge, All{\'e}e du 6 Ao{\^u}t 17, 
  4000 Li{\`e}ge, Belgium %26
  \and Institute of Physics, Jan Kochanowski University, ul. Uniwersytecka 7, 25-406 Kielce %27
  \and Chungbuk National University, 1, Chungdae-ro, Seowon-gu, Cheongju-si, Chungcheongbuk-do, Republic of Korea %28
  \and Korea Astronomy and Space Science Institute, 776 Daedeok-daero, Yuseong-gu, Daejeon 34055, Korea %29
  \and Institute of Physics, Faculty of Natural Sciences, University of P. J. \v{S}af{\'a}rik, Park Angelinum 9, 
  040 01 Ko\v{s}ice, Slovakia %30
  \and Laboratory of Space Researches, Uzhhorod National University, Daleka st. 2a, 88000, Uzhhorod, Ukraine %31
  \and Universidad de La Laguna, Dept. Astrofisica, E.38206 La Laguna, Tenerife, Spain %32
  \and Institute of Theoretical Physics and Astronomy, Vilnius University, Saul{\.e}tekio al. 3, 10257 Vilnius, Lithuania %33
  \and Organ Mesa Observatory, 4438 Organ Mesa Loop, Las Cruces, New Mexico 88011 USA %34
  \and Command Module Observatory, 121 W. Alameda Dr., Tempe, AZ 85282 USA %35
  \and Observatoire de Blauvac, 293 chemin de St Guillaume, F-84570 St-Est{\`e}ve, France %36
  \and Departamento de F{\'i}sica, Ingenier{\'i}a de Sistemas y Teor{\'i}a de la Se{\~n}al, Universidad de Alicante, 
  Alicante, Spain %37 
  \and Institut de Ci{\`e}ncies del Cosmos, Universitat de Barcelona (IEEC-UB), Barcelona, Spain %38
  \and Lowell Observatory, 1400 West Mars Hill Road, Flagstaff, Arizona, 86001 USA %39
  \and Department of Physics, Adiyaman University, 02040 Adiyaman, Turkey %40
  \and European Southern Observatory, Karl-Schwarzschild-Strasse 2, 85748 Garching bei M{\"u}nchen, Germany %41
  \and Japan Spaceguard Association, Bisei Spaceguard Center, 1716-3, Okura, Bisei, Ibara, Okayama 714-1411 Japan %42
  \and Kepler Institute of Astronomy, University of Zielona G{\'o}ra, Lubuska 2, 65-265 Zielona G{\'o}ra, Poland %43
}
\date{Received 02 Apr 2021 / Accepted 20 June 2021}

%\clearpage
%\newpage

\abstract
{%Context 
    Recent results for asteroid rotation periods from the TESS mission showed how strongly previous studies have underestimated 
    the number of slow rotators, revealing the importance of studying those targets. 
    For most slowly rotating asteroids (those with P > 12 hours), no spin and shape model is available 
    because of observation selection effects. This hampers determination of their thermal parameters 
    and accurate sizes. Also, it is still unclear whether signatures of different surface material properties 
    can be seen in thermal inertia determined from mid-infrared thermal flux fitting.
}
{%Aims
    We continue our campaign in minimising selection effects among main belt asteroids. 
    Our targets are slow rotators with low light-curve amplitudes. Our goal is to provide their 
    scaled spin and shape models together with thermal inertia, albedo, and surface roughness 
    to complete the statistics.   
}
{%Methods
  Rich multi-apparition datasets of dense light curves are supplemented with data from Kepler and TESS spacecrafts.
  In addition to data in the visible range, we also use thermal data from infrared space observatories 
  (mainly IRAS, Akari and WISE) in a combined optimisation process using the Convex Inversion Thermophysical 
  Model (CITPM). This novel method has so far been applied to only a few targets, and therefore in this work we further 
  validate the method itself.
}
{%Results
  We present the models of 16 slow rotators, including two updated models. All provide good fits to both 
  thermal and visible data. The obtained sizes are on average accurate at the 5\% precision level, 
  with diameters found to be in the range from 25 to 145 km. 
  The rotation periods of our targets range from 11 to 59 hours, and
  the thermal inertia covers a wide range of values, from 
  2 to <400 J\,m$^{-2}$\,s$^{-1/2}$\,K$^{-1}$, not showing any correlation with the period.
} 
{%Conclusions
  With this work we increase the sample of slow rotators with reliable spin and 
  shape models and known thermal inertia  by 40\%.  
  The thermal inertia values of our sample do not display a previously suggested increasing trend with rotation period, 
  which might be due to their small skin depth. 
}


\maketitle

\section{Introduction} 
 Physical parameters of asteroids, such as spin, shape, size, albedo, macroscopic roughness, and thermal inertia, 
 form the basis for a significant number of Solar System studies. In particular, these parameters are of great interest for 
 large asteroids as these are considered remnants of early phases of planetary formation \citep{Morbidelli2009}. 
 Studying the way in which asteroid surfaces react to heating by the Sun 
 (which, among others, depends on the spin axis inclination and spin rate), can reveal material properties 
 of these layers \citep{Murdoch2015, Keihm2012}. Slowly rotating asteroids, with periods longer than 12 hours, are 
 especially interesting in this respect; they experience long periods of irradiation of the same surface parts, 
 and the diurnal heat wave from solar irradiation can penetrate to larger thermal skin depths \citep{Delbo2015, CapekVokrouhlicky2010}. 
 Furthermore, the most recent results from the TESS mission \citep[Transiting Exoplanet Survey Satellite;][]{Ricker2015}
 reveal that slow rotators actually dominate the population of main-belt asteroids \citep[see fig. 7 in][]{Pal2020}. 
 So far, however, they have been largely omitted by most ground-based studies mainly because of telescope time limitations 
 and the small number of targeted campaigns \citep{WarnerHarris2011}.

 As a consequence of the scarcity of multi-apparition light curves which are needed for spin and shape reconstruction via 
 light-curve inversion, the statistics of available spin- and shape-modelled asteroids are strongly biased towards  
 faster rotators \citep{M2015}. This might have implications on our interpretation of the statistical properties of the 
 asteroid population, such as for example the role of the YORP effect \citep{Vokrouhlicky2015} on the 
 spatial distribution of spin axes \citep{Hanus2013}, or the estimated contribution of tumblers and binaries in various 
 asteroid populations \citep{ATLAS}. 

 Another hidden problem is that most of the well-studied asteroids, especially among  
 slow rotators, are those with large-amplitude light curves \citep{WarnerHarris2011}, 
 caused by an elongated shape, high spin axis inclination, or both.
 In our survey, described in detail in \cite{M2015}, we addressed two of these biases 
 at the same time, focusing on slow rotators (P>12 hours) with maximum amplitudes no larger than $0.25$ mag, 
 at least at the target-selection stage. During our study, we found that several targets have somewhat larger 
 amplitudes or shorter periods, but nevertheless we kept these in the final sample of this latter work. 

 The statistics of asteroids with reliably determined thermal inertia is even more biased. 
 Recompiling data from previous works, as well as new values from \cite{Hanus2018}, \cite{M2018}, and \cite{M2019}, 
 there are currently 36 main-belt slow rotators, compared to 120 fast rotators studied using 
 detailed thermophysical modelling (TPM). 
 This shows that, in terms of studying slow rotators in the infrared, we have only touched the tip of the iceberg. 
 
 Thermal inertia ($\Gamma=\sqrt{\kappa \rho c}$) depends on the density of surface regolith $\rho$, 
 thermal conductivity $\kappa$, and heat capacity c. 
 Larger thermal inertia implies coarser regolith composed of grain sizes of the order of millimetres to centimetres, 
 typical for young surfaces of small near-Earth asteroids (NEAs) \citep{Gundlach2013}, while much finer, lunar-like regolith 
 with grain sizes of between 10 and 100 microns is expected at large (D>100 km) main-belt asteroids 
 \citep[see e.g.][and references therein]{Delbo2009}. This picture 
 might however be complicated by various family formation ages, recent catastrophic events refreshing the surface, 
 or by the presence of surface cohesion forces \citep{Marchi2012, Rozitis2014}. 
 Also, as more asteroids become thermally characterised we can also understand how thermal processes like thermal 
 cracking \citep{Delbo2014, Ravaji2019} have shaped or are still shaping asteroid surfaces. 

 However, in light of recent results for two targets studied   in situ, Ryugu and Bennu \citep{Okada2020, Walsh2019}, 
 this standard interpretation of thermal inertia versus surface properties fails; there are boulders on the surface with
 relatively low thermal inertia, while one would expect regolith.
 Thermal conductivity, and thus thermal inertia dependance on 
 temperature at various subsurface depths, is another factor to be considered \citep{Hayne2017}. It has been shown  
 that submillimetre flux probes deeper layers, carrying information on the conditions in these layers \citep{Keihm2012}. 

 \cite{HarrisDrube2016} estimated thermal inertias based on beaming parameters derived from WISE data 
 \citep[][and references therein]{Masiero2011} and found that thermal inertia increases with rotation period. 
 This motivated us to add the thermophysical analysis to our study of slow rotators. At first, our results seemed to confirm 
 this hypothesis \citep{M2018}, as we found large and medium thermal inertia values for the first sample of five targets. 
 Later, with a  sample of twice the size, we found a rather wide range of thermal inertia \citep{M2019}, from very small to medium, 
 similarly to \cite{Hanus2018}, generally not showing any trend with the rotation period.
 Still, the size of the slow rotators sample with known thermal inertia remains small. In this work we continue our effort 
 to expand this sample employing a different approach, namely the Convex Inversion Thermophysical Model (CITPM, see Section 3).

 The light-curve inversion method \citep{Kaas2001} can robustly reproduce asteroid spin and shape, provided the visible data 
 cover a wide range of viewing geometries. However, for targets orbiting close to the ecliptic plane (i.e. most of the main-belt asteroids), the result usually 
 consists of two mirror pole solutions \citep{KaasalainenLamberg2006, Mikko_and_Josef2020}.  
 These are similar in spin axis ecliptic latitude, but differ in ecliptic longitude:
 both solutions are roughly 180$\deg$ apart, and have different associated shape models. One such mirror pole solution 
 sometimes happens to fit thermal data better than the other \citep[see e.g.][]{Delbo2009}. 
 However, this can stem from the high sensitivity of thermal flux to small-scale shape details, and might not point to a truly better spin solution \citep{Hanus2015, Mikko_and_Josef2020}. We therefore decided to switch from independent 
 light curve inversion followed by thermophysical modelling of a fixed shape to simultaneous optimisation of both types of data. 
 The method enabling this approach is the CITPM introduced in \cite{CITPM}. 
 This method also enables the user to weight two types of data relative to each other to avoid the dominance of one data type 
 over the other.
 \cite{Muller2017} applied this method for asteroid Ryugu and the
  derived size, albedo, and thermal inertia are very close to the in situ
  properties; however, the spin pole was not well determined by this method
  (probably because of the very low light-curve amplitude and the lack
  of high-quality measurements).

 In Section 2, we describe the visible and infrared data used for modelling. Section 3 presents the main features 
 of the method for combined optical and mid-infrared photometric inversion, which is followed in Section 4 by a description 
 of the method used to scale the models by multi-chord stellar occultations. 
 The resulting models, with their spin, shape, and thermal parameters with the occultation scaling are presented in Section 5.
 In Section 6 we summarise the results and discuss our ideas for future work. 
 All the plots and figures asssociated with the models can be found in the Appendix.

\section{Visible and infrared data}

 Data for traditional, dense light curves in the visible range have been gathered in the framework of our 
 long-term photometric campaign conducted since the year 2013, and are described in \cite{M2015}, including target-selection criteria. 
 In short, the aim of the project is to observe a few tens of slowly rotating main-belt asteroids with small brightness variation amplitudes. 
 It involves over 20 observing stations with telescopes of up to 1 m in diameter, including for example TRAPPIST telescopes \citep{Jehin2011}. 
 To compliment these data, we also use data from the \textit{Kepler Space Telescope} in the extended K2 mission \citep{Howell2014} downlinked 
 within our proposals accepted by Kepler and K2 Science Center, 
 as well as publicly available data from TESS \citep{Pal2020}\footnote{https://archive.konkoly.hu/pub/tssys/dr1/}, 
 and Super WASP sky survey \citep{Grice2017}\footnote{http://asteroids.neilparley.com/asteroids/lc.html}.
 From the latter archive, we only used the best-quality  subsets, choosing from targets with Super WASP datapoints already folded 
 into light curves. Trimming those vast datasets was necessary because of their abundance and in order to avoid dominance 
 of one apparition over others, but also because of their intrinsic noise. Noisy light curves can sometimes prevent the identification 
 of a unique model solution over the whole dataset. The selection criteria for the best Super WASP light-curve fragments were 
 the lowest photometric scatter and the widest possible range of observing dates. 
 
 The great majority of the dense light curve data from our photometric campaign were provided in the form of relative photometry, 
 and the rest were treated as such to ascertain light-curve inversion convergence.
 Separate light-curve fragments obtained during our observing campaign in the {\it R} filter or unfiltered were combined 
 to create composite light curves (Figs. \ref{108composit2015} - \ref{995composit2017} in the Appendix E) 
 using the criterion of minimum scatter between data points for initial period determinations. 
 We present the light curves that cover most of the rotation period and show clear brightness 
 variations. For modelling, however, we used all the data described in Table \ref{obs}. Determined synodic 
 periods are in agreement in all apparitions, with differences of only a few thousandths
 due to changes in relative velocity of the observer and the source. The synodic period range from various apparitions, 
 extended at least three times, is a range on which the precise, sidereal period is later searched for in the light-curve inversion procedure. 

 Composite light curves from various apparitions depict the general character of the asteroid shape (if regular and symmetric, 
 or quite the opposite). Light-curve differences are due to phase-angle effects caused by shadowing on topographic features, and different 
 viewing geometries (aspect angles). Apart from ensuring a full period coverage, sometimes tens of hours long, we paid special 
 attention to covering the widest possible range of ecliptic longitudes and phase angles (see Table \ref{obs} 
 in the Appendix), which is a necessary prerequisite for shape reconstruction \citep{Mikko_and_Josef2020}.  
 The small point-to-point scatter of our light curves (see Appendix A), of the order of 0.01 mag down to a few millimagnitudes, 
 captures brightness variations in great detail, even in those cases with very small amplitudes. 

 Relative photometric data described above were supplemented with the calibrated {\it V}-band sparse data from the USNO (US Naval Observatory)
 archive\footnote{downloaded from AstDys\break https://newton.spacedys.com/astdys2/index.php?pc=3.0}.   
 These are necessary for size and albedo determination in the application of the full CITPM. 
 We decided to exclusively use the USNO archive due to its relatively high quality among the available options. 
 As has been shown by \cite{Hanus2011} the median accuracy of USNO data is at the level of 0.15 mag.

 Thermal infrared data were downloaded from the SBNAF Infrared Database\footnote{https://ird.konkoly.hu/} 
 \citep{Szakats2020}. This database provides expert-reduced data products from major infrared  
 space missions (Akari,  Infrared Astronomical Satellite (IRAS), Wide-field Infrared Survey Explorer (WISE), Herschel, 
 Midcourse Space Experiment (MSX), and  Infrared Space Observatory (ISO)) 
 as well as all the necessary auxiliary information, 
 such as the observing geometry, colour correction, or overall measurement uncertainties. 
 SBNAF Infrared Database was developed within the `Small Bodies: Near And Far' Horizon 2020 project \citep{SBNAF}. 
 This database stores calibrated flux densities obtained via careful consideration of instrument-specific calibration 
 and processing procedures. 
 All the measurement uncertainty values have been reanalysed for the sake of database consistency, and include contributions 
 from in-band flux density uncertainty, absolute calibration errors, and colour correction uncertainties.
 The infrared data for our targets came mostly from three missions: WISE \citep{Wright2010, Mainzer2011a} 
 at 11.1 $\mu$m and 22.64 $\mu$m, Akari \citep{Usui2011} at 9 and 18 $\mu$m, and IRAS \citep{Neugebauer1984} 
 at 12, 25, 60 and 100 $\mu$m, occasionally supplemented with data from 
 MSX \citep{Egan2003} at 8.28, 12.13, 14.65, and 21.34 $\mu$m, where available. 
 All the infrared datapoints were used, except in specific single cases where clear outliers were detected that were unable to be fitted 
 by any of the models. Also, because of the large size of some targets resulting in large infrared flux, sometimes a subset or all WISE data 
 at 11$\mu$m were partially saturated, and could not be used in our analysis.

\section{Convex Inversion Thermophysical Model}
  
   To fit optical light curves and thermal infrared data, we used a
   combined inversion of both data types developed by \cite{CITPM}
   called the Convex Inversion Thermophysical Model. The method
   combines convex inversion of light curves \citep{Kaas2001} with a
   thermophysical model \citep{Lagerros1996, Lagerros1997, Lagerros1998}. The 
   shape of an asteroid is parametrized by coefficients of spherical functions that
   describe a convex polyhedron of size $D$ with typically hundreds of
   surface facets. For each facet, a 1D heat diffusion equation is
   solved to compute its temperature and infrared flux at the time of
   observation. The  response of the surface to solar radiation is
   parametrized by the thermal inertia $\Gamma$, surface roughness
   (described by spherical craters 
   of varying both the fraction of surface coverage $f$, and the opening angle $\gamma_c$), 
   and light-scattering properties. For emissivity, a fixed value of 0.9 is used, 
   following a standard approach \citep[e.g.][]{Lim2005}.
   Instead of using absolute magnitude, Bond albedo, and geometric
   albedo ---which are only unambiguously defined for a
   sphere--- we use Hapke's light-scattering model \citep{Hapke1981,
   Hapke1984, Hapke1986}, from which any albedos can be directly computed. To
   tie the reflectance of the surface with the size of the asteroid,
   absolutely calibrated photometry is needed. Because most of the
   light curves we collected are provided as the relative photometry, we
   also use the calibrated V-band photometry from the USNO 
   that covers a sufficiently wide range of solar phase
   angles. Parameters of Hapke's model can be optimised to fit the phase
   curve. The merit function that we minimise is a sum
   $\chi^2_\mathrm{VIS} + w \chi^2_\mathrm{IR}$ of $\chi^2$ values for
   optical and thermal data. The relative weight $w$ is iteratively set
   such that (in an ideal case) the fit to light curves is as good as
   without thermal data, and the fit to thermal data is good, that is, the
   normalised $\chi^2_\mathrm{IR}$ is $\sim 1$. The advantage here is that 
   the spin and shape model optimised against visible light curves only 
   in most cases would not be optimal in the thermal radiation, as shown by 
   \cite{Hanus2015} and \cite{Hanus2018}; here it is optimised to fit both 
   types of data.

 The visual part of $\chi^2$ is computed as
  $$
    \chi^2_\mathrm{VIS} = \sum_{j=1}^{N-1} \sum_i \left( \frac{L_{i,j}^\text{obs}}{\bar{L}_j^\text{obs}} - \frac{L_{i,j}^\text{model}}{\bar{L}_j^\text{model}} \right)^2 + 0.2\,
    \sum_i \left( \frac{{L_{i,N}^\text{obs}} - L_{i,N}^\text{model}}{\bar{L}_N^\text{obs}} \right)^2 \,,
  $$
  where $N$ is the total number of light curves, and $L_{i,j}$ is the brightness (in arbitrary intensity units, not magnitudes) 
  of the $i$-th point of the $j$-th light curve. The normalisation by the mean brightness of the $j$-th light curve $\bar{L}_j$ means 
  that we treat all $N-1$ light curves as relative and that we neglect differences in photometric accuracy between them. 
  The only exception is calibrated photometry in V filter from USNO (the $N$-th light curve), for which we directly  compare 
  the observed flux with that predicted by our model without normalising by $\bar{L}_j^\text{obs}$ and $\bar{L}_j^\text{model}$ 
  separately. The empirical factor of 0.2 gives less weight to USNO data which is intentional because these have larger errors.

  For thermal data, errors of individual measurements are known, and so the thermal part of the $\chi^2$ is computed classically as
  $$
    \chi^2_\mathrm{IR} = \sum_i \left( \frac{F_i^\text{obs} - F_i^\text{model}}{\sigma_i} \right)^2\,,
  $$
  where $F_i$ is observed or modelled flux and $\sigma_i$ is the error of the measurement. By dividing $\chi^2_\mathrm{IR}$ by the number 
  of degrees of freedom, we get reduced $\chi^2_\text{red}$, which we use in Sect.~5 when presenting our results.

\begin{table*}
\begin{small}
\begin{center}
%\footnotesize\addtolength{\tabcolsep}{-1pt}
%  \flushleft
%\captionsetup{width=1.15\linewidth}
\caption{Ancillary information on the data and physical properties of our targets. 
 The first two columns contain asteroid name and
 the number of apparitions $N_{app}$ during which the $N_{lc}$ of visible light curves were obtained.
 The next part of the table details the infrared dataset: the number of points provided by space observatories IRAS $N_I$, Akari $N_{A}$, and WISE 
 in W3 and W4 bands: $N_{W3}$, and $N_{W4}$ respectively.
 For comparison of the diameters and albedos obtained in this work (see Table \ref{results}), the diameters 
 $D_{\rm WISE}$ from WISE spacecraft \citep{Mainzer2011b, Masiero2011} and taxonomic types are added  
 \citep{Bus2002a,Bus2002b}, and \citep{Tholen1989}.
 Diameter in parentheses, due to a lack of size determination from WISE, comes from IRAS survey results \citep{Tedesco2004}.
 }
\label{ancillary_info}
\begin{tabular}{ccllllllc}
\hline
    Asteroid    &$N_{app}$&$N_{lc}$&$N_I$&$N_{A}$& $N_{W3}$&$N_{W4}$& D$_{WISE}$ &Taxonomic \\
                &         &        &     &       &         &        &   [km]     &   type   \\
\hline
(108) Hecuba    &  8      &  59    & 15  &  7    & 13 & 13          & 75.498     &  S  \\    
(202) Chryseis  &  7      &  70    &  7  &  8    &    & 12          & 97.948     &  S  \\    
(219) Thusnelda &  6      & 116    & 18  &  6    &    & 19          & 38.078     &  S  \\    
(223) Rosa      &  7      &  58    & 20  &  5    &    & 12          & 83.394     &  X  \\    
(362) Havnia    &  7      &  38    &     &  9    &    & 13          & 89.202     &  XC \\    
(478) Tergeste  &  6      &  48    & 27  &  8    &  9 &  9          & 84.975     &  S  \\    
(483) Seppina   &  8      &  56    & 34  & 12    & 12 & 12          & 84.975     &  S  \\    
(501) Urhixidur &  7      &  61    & 11  &  8    & 11 & 11          & 85.404     &  C  \\    
(537) Pauly     &  7      &  50    &  8  &  9    &  6 &  6          & 52.330     &  DU \\    
(552) Sigelinde &  6      &  65    &  8  &  6    &  4 &             & (77.56)    &  C  \\    
(618) Elfriede  &  9      &  68    & 17  &  5    &    & 12          & 131.165    &  C  \\    
(666) Desdemona & 7       &  60    & 21  &  5    &  13 & 13         &  31.485    &  S  \\   
(667) Denise    & 5       & 40     & 21  &  5    &  15 & 13         &  88.630    &  C  \\    
(780) Armenia   &  8      &  95    & 24  &  7    &     & 12         & 102.257    &  C  \\    
(923) Herluga   &  7      &  51    & 12  &  8    &  16 & 16         &  37.638    &  C  \\    
(995) Sternberga&  7      &  81    & 22  &  6    &  11 & 11         &  22.350    &  S  \\       
\hline
\end{tabular}
\end{center}
\end{small}
\end{table*}
%\end{landscape}

\section{Occultation fitting}

 For three targets of our current sample there were good quality, multi-chord stellar occultations available 
 in the PDS archive\footnote{http://sbn.psi.edu/pds/resource/occ.html} \citep{Herald2019, Herald2020}. 
 More recent occultation results were downloaded from the archive of the Occult programme\footnote{http://www.lunar-occultations.com/iota/occult4.htm}. 
 We used them to independently scale the shape models obtained here, using the method described in \cite{Durech2011}, in order to: compare 
 obtained sizes with those from thermal fitting; confirm the shape silhouette; and if possible, identify the preferred pole solution
 (see Figures \ref{Havnia_occult} - \ref{Denise_occult} in the Appendix C).

  When scaling the models with occultations, we computed
  the orientation of the model for the time of occultation and projected
  the model on the fundamental plane (sky-plane projection). Because all
  models are convex, their silhouettes are also convex. We then
  iteratively searched for a scale of the silhouette that would
  provide the best match with chords. The mutual shift between the
  silhouette and the chords was described by two free parameters that
  were also optimised. We used the $\chi^2$ minimisation, where the
  difference between the silhouette and the chords was measured as a
  distance in the fundamental plane between the ends of the chords and
  the silhouette (measured along the direction of the chord). We
  rejected the solutions in which a negative chord (no occultation
  was observed) intersected the silhouette.

\section{Results}

 Table \ref{ancillary_info} provides the ancillary information on 
 the visible and thermal datasets: number of apparitions and separate light curves, 
 numbers of thermal measurements from separate missions, and
 WISE diameters from \cite{Mainzer2011b, Masiero2011} 
 to be compared with diameters obtained in this work (see Table\ref{results}). 
 We also cite taxonomic type following
 \cite{Bus2002a,Bus2002b} and \cite{Tholen1989}, 
 for a consistency check with our values for albedo (consistent in all cases).

 Table \ref{results} summarises all the rotational and thermophysical properties of the targets studied here. 
 First the spin solution is presented, usually with its mirror counterpart. 
 The quality of the fit to light curves in the visible range is given in column 5. 
 The second part of the table presents the radiometric solution based on combined data from three infrared missions,
 the radiometric diameter, geometric albedo, thermal inertia, 
 and the reduced $\chi^2$ of modelled versus observed fluxes. Lastly, the table contains the average heliocentric distance at which thermal 
 measurements were taken, and thermal inertia reduced to one astronomical unit, using the formula \citep{Rozitis2018}: 
 
 \begin{equation}
 \Gamma_{1AU} = \Gamma(r)r^{\alpha}
 ,\end{equation}
 
 where the $\alpha$ exponent is equal to 0.75, which takes into account a radiative conduction term in thermal conductivity. 
 Different exponents are also possible \citep{Rozitis2018}, but here we opted for the most widely used value to facilitate comparison 
 with previous works \citep[see the discussion in][]{AliLagoa2020, Szakats2020}.

 In Appendix A we present the plots of $\chi^2_\text{red}$ versus thermal inertia for various combinations of surface roughness 
 and optimised size (Figs. \ref{108chi2} - \ref{995chi2}).  
 To transform various combinations of crater coverage and opening angle to rms of surface roughness, we used the formula no. 20 from \cite{Lagerros1998}. 
 In these figures, $f$ is the fraction of crater coverage, and the plots show the $\chi^2_\text{red}$ of the crater 
 opening angle that minimised the $\chi^2_\text{red}$ for that value of $f$. 
 The horizontal line is the acceptance threshold for $\chi^2_\text{red}$ values, depending classically on the number of IR measurements and best 
 $\chi^2_\text{red}$ value: we accept all the solutions with $\chi^2_\text{red}<(1+\sigma)$, where $\sigma=\sqrt{2\nu}/\nu$, with 
 $\nu$ being the number of degrees of freedom. 
 For a few targets with a value of best $\chi^2_\text{red}$ much below 1, probably due to unresolvable mutual parameter correlations, 
 we used an empirical approach by \cite{Hanus2015} to define that threshold: $\chi^2_\text{red}<(\chi^2_\text{min}+\sigma)$. 
 
 For each target we also present the fit to WISE W3 and W4 thermal light curves, whenever available 
 (Figures \ref{108thermal_lcW3} - \ref{995thermal_lcW4}). 
 Due to the scarce character of Akari and IRAS data (only 1 -- 3 points per band on average), the model fits to them are not shown. 
 The plots present the results for only one of two mirror pole solutions (the other pole gave very similar results, as indicated by 
 $\chi^2_\text{red}$ values from Table \ref{results}). 

\clearpage

\clearpage

\thispagestyle{empty}
\voffset=5cm

\begin{landscape}
\begin{table*}
\footnotesize\addtolength{\tabcolsep}{-1pt}
%  \flushleft
%\begin{small}
%\captionsetup{width=1.15\linewidth}
\caption{Spin parameters and thermophysical characteristics of asteroid models obtained in this work. 
 The columns contain asteroid name, J2000 ecliptic coordinates $\lambda_p$, $\beta_p$ of the spin solution, with mirror pole solution 
 in the second row, sidereal rotation period $P$, and the deviation 
 of model fit to those light curves (including fit to sparse data).
 The next part of the table details the radiometric solution for combined data: surface-equivalent size $D$, 
 geometric albedo $p_V$, thermal inertia $\Gamma$ in J\,m$^{-2}$\,s$^{-1/2}$\,K$^{-1}$ (SI) units, 
 and the reduced chi-square of the best-fit ($\chi^{2}_\text{red}$). 
 The last two columns give average heliocentric distance of thermal infrared observations $r_\text{hel}$ with the standard deviation,  
 and thermal inertia normalised to 1 AU $\Gamma_{\rm{AU}}$ calculated according to equation 1.
 Numbers in italics mark the pole solution of (667) Denise clearly rejected by occultation fitting.
 }
\label{results}
\begin{tabular}{lllllllllll}
\hline
    Asteroid   & \multicolumn{2}{c}{Pole}              &   P          &vis. dev&  D            &   p$_V$                   & $\Gamma$          & $\chi^{2}_\text{red}$ & $r_\text{hel}$ & $\Gamma_{1AU}$ \\
               & $\lambda_p [\deg]$ & $\beta_p [\deg]$ & [hours]      & [mag]  & [km]          &                           &  [SI units]       &     IR           &  [AU]     &  [SI units]    \\
\hline                                                                                                                                                          
(108) Hecuba   & $181 \pm 2 $ & $+42 \pm 5 $ & $14.25662 \pm 0.00003$ &  0.013 &$69^{+3}_{-1}$ & 0.24$^{+0.04}_{-0.01}$    &  35$^{+25}_{-30}$ & 1.08 & 3.18  $ \pm $ 0.10 & 85 \\    
               & $352 \pm 1 $ & $+39 \pm 6 $ & $14.25662 \pm 0.00003$ &  0.012 &$70 \pm 2$     & 0.24$^{+0.04}_{-0.01}$    &  40$ \pm 30$      & 1.10 & 3.18  $ \pm $ 0.10 & 95 \\  
\hline                                                                                                                                                  
(202) Chryseis &  $94 \pm 1 $ & $-49 \pm 4 $ & $23.67025 \pm 0.00006$ &  0.012 &$90^{+4}_{-3}$ & 0.22$^{+0.03}_{-0.01}$    &     < 180         & 0.35 & 2.96  $ \pm $ 0.15 & < 405 \\    
               & $261 \pm 1 $ & $-34 \pm 4 $ & $23.67028 \pm 0.00004$ &  0.012 &$90^{+3}_{-3}$ & 0.22$^{+0.01}_{-0.01}$    &     < 180         & 0.36 & 2.96  $ \pm $ 0.15 & < 405 \\  
\hline                                                                                                                                           
(219) Thusnelda& $300 \pm 10$ & $-66 \pm 10$ & $59.7105 \pm 0.0001$   &  0.014 &$44^{+2}_{-4}$ & 0.19$^{+0.04}_{-0.01}$    &     < 120         & 0.80 &       2.24 $ \pm $ 0.42 & < 220 \\    
\hline                                                                                                                                                  
(223) Rosa     &  $22 \pm 3 $ & $+18 \pm 18$ & $20.2772 \pm 0.0003$   &  0.012 &$69^{+9}_{-3}$ & 0.033$^{+0.006}_{-0.004}$ &     < 300         & 0.72 & 2.99  $ \pm $ 0.12 & < 680\\    
               & $203 \pm 2 $ & $+26 \pm 15$ & $20.2769 \pm 0.0003$   &  0.012 &$70^{+6}_{-2}$ & 0.032$^{+0.007}_{-0.003}$ &     < 300         & 0.78 & 2.99  $ \pm $ 0.12 & < 680\\      
\hline                                                                                                                                                  
(362) Havnia   &  $14 \pm 2 $ & $+33 \pm 2 $ & $16.92665 \pm 0.00003$ &  0.017 &$92^{+6}_{-5}$ & 0.044$^{+0.006}_{-0.004}$ &     < 180         & 0.80 & 2.64  $ \pm $ 0.04 & < 370 \\    
               & $208 \pm 8 $ & $+35 \pm 4 $ & $16.92668 \pm 0.00003$ &  0.017 &$91^{+8}_{-3}$ & 0.046$^{+0.004}_{-0.008}$ &     < 200         & 0.67 & 2.64  $ \pm $ 0.04 & < 410 \\  
\hline                                                                                                                                                  
(478) Tergeste & $  2 \pm 5 $ & $-38 \pm 8 $ & $16.10308 \pm 0.00004$ &  0.019 &$83 \pm 4$     & 0.16$^{+0.05}_{-0.01}$    &   2$^{+45}_{-1}$  & 0.94 &       3.05 $ \pm $ 0.10 &   5 \\    
               & $216 \pm 7 $ & $-62 \pm 4 $ & $16.10312 \pm 0.00004$ &  0.016 &$81^{+5}_{-4}$ & 0.18$^{+0.03}_{-0.02}$    &  26$ \pm 25$      & 0.88 &       3.05 $ \pm $ 0.10 &  60 \\  
\hline                                                                                                                                                  
(483) Seppina  & $127 \pm 3 $ & $+47 \pm 3 $ & $12.720968 \pm 0.000004$ & 0.019 &$73^{+5}_{-2}$ & 0.16$^{+0.04}_{-0.01}$   &  17$^{+23}_{-12}$ & 0.80 & 3.45  $ \pm $ 0.14 & 45 \\    
               & $356 \pm 4 $ & $+60 \pm 3 $ & $12.720977 \pm 0.000002$ & 0.019 &$74^{+4}_{-2}$ & 0.16$^{+0.04}_{-0.01}$   &  23$^{+17}_{-18}$ & 0.83 & 3.45  $ \pm $ 0.14 & 60 \\  
\hline                                                                                                                                                  
(501) Urhixidur&  $49 \pm 40 $& $+84 \pm 12 $& $13.17203 \pm 0.00002$ &  0.019 &$77^{+5}_{-2}$ & 0.051$^{+0.003}_{-0.008}$ &   4$^{+35}_{-2}$  & 0.53 &       3.20 $ \pm $ 0.32 & 10 \\    
               & $262 \pm 24 $& $+66 \pm 11 $& $13.17203 \pm 0.00001$ &  0.018 &$82^{+2}_{-4}$ & 0.050$^{+0.002}_{-0.007}$ &  13$^{+30}_{-11}$ & 0.53 &       3.20 $ \pm $ 0.32 & 31 \\  
\hline                                                                                                                                                  
(537) Pauly    &  $32 \pm 3 $ & $+43 \pm 6 $ & $16.29601 \pm 0.00002$ &  0.018 &$47^{+1}_{-2}$ & 0.26$^{+0.03}_{-0.02}$    &  11$^{+30}_{-10}$ & 0.70 &       2.96 $ \pm $ 0.45 & 25 \\    
               & $214 \pm 4 $ & $+60 \pm 9 $ & $16.29597 \pm 0.00001$ &  0.018 &$46 \pm 2$     & 0.25$^{+0.05}_{-0.02}$    &  13$^{+50}_{-12}$ & 0.74 &       2.96 $ \pm $ 0.45 & 29 \\  
\hline                                                                                                                                                  
(552) Sigelinde&   $8 \pm 24 $& $+73 \pm 9 $ & $17.14963 \pm 0.00001$ &  0.017 &$88^{+10}_{-5}$& 0.030$^{+0.011}_{-0.007}$ &   3$^{+55}_{-2}$  & 0.97 & 3.26  $ \pm $ 0.09 &  7 \\    
               & $189 \pm 18 $& $+60 \pm 17 $& $17.14962 \pm 0.00003$ &  0.017 &$91^{+7}_{-13}$& 0.029$^{+0.005}_{-0.007}$ &   2$^{+55}_{-1}$  & 1.13 & 3.26  $ \pm $ 0.09 &  5 \\  
\hline                                                                                                                                                                  
(618) Elfriede & $102 \pm 20 $& $+64 \pm 7 $ & $14.79565 \pm 0.00002$ &  0.015 &$145^{+15}_{-13}$& 0.047$^{+0.010}_{-0.003}$ &     < 350       & 0.28 & 3.32  $ \pm $ 0.10 & < 860 \\    
               & $341 \pm 13 $& $+49 \pm 6 $ & $14.79564 \pm 0.00002$ &  0.015 &$146^{+15}_{-16}$& 0.053$^{+0.002}_{-0.009}$ &     < 400       & 0.32 & 3.32  $ \pm $ 0.10 & < 980 \\  
\hline
(666) Desdemona& $ 10 \pm 4 $ & $+39 \pm 5$  & $14.60795 \pm 0.00008$ &  0.022 &$28.4^{+0.9}_{-0.8}$& 0.111$^{+0.007}_{-0.009}$ &      <  70   & 0.83 & 2.79  $ \pm $ 0.34 & < 150 \\    
               & $174 \pm 3 $ & $+36 \pm 11$ & $14.60796 \pm 0.00003$ &  0.022 &$28.3^{+0.9}_{-1.0}$& 0.116$^{+0.002}_{-0.014}$ &      < 100   & 0.77 & 2.79  $ \pm $ 0.34 & < 215 \\  
\hline
(667) Denise   & $15  \pm 25$ & $-83 \pm 6 $ & $12.68499 \pm 0.00003$ & 0.024 & $83^{+4}_{-2}$    &   0.051$ \pm 3$             &  13$^{+17}_{-8}$  & 1.19 & 3.36 $ \pm $ 0.38 & 32 \\    
               & {\it 237 $ \pm $ 3 }&{\it $-$23 $ \pm $ 6 }&{\it 12.68497 $ \pm $ 0.00003}&  0.025 & $82^{+5}_{-2}$ & 0.051$^{+0.002}_{-0.004}$ &   6$^{+24}_{-1}$  & 1.16 & 3.36 $ \pm $ 0.38 & 15 \\  
\hline
(780) Armenia  & $144 \pm 7 $ & $-44 \pm 11 $& $19.88453 \pm 0.00007$ &  0.014 & $98^{+2}_{-3}$ & 0.042$^{+0.005}_{-0.003}$ &     < 300         & 0.47 &  3.00 $ \pm $ 0.10 & < 680 \\    
               & $293 \pm 3 $ & $-23 \pm 6 $ & $19.88462 \pm 0.00009$ &  0.015 & $102^{+3}_{-2}$& 0.038$ \pm 0.003$         &     < 250         & 0.63 &  3.00 $ \pm $ 0.10 & < 570 \\    
\hline
(923) Herluga  & $218 \pm 9 $ & $-68 \pm 5 $ & $29.72820 \pm 0.00002$ &  0.022 & $36.2^{+0.4}_{-1.5}$ & 0.047$^{+0.004}_{-0.003}$ &  37$^{+15}_{-36}$ & 0.92 & 2.73  $ \pm $ 0.40 & 80 \\    
               & $334 \pm 7 $ & $-52 \pm 3 $ & $29.72826 \pm 0.00001$ &  0.023 & $34.1^{+0.8}_{-1.0}$ & 0.050$^{+0.002}_{-0.003}$ &  14$^{+35}_{-13}$ & 0.95 & 2.73  $ \pm $ 0.40 & 30 \\    
\hline
(995) Sternberga& $27 \pm 3 $ & $-20 \pm 6 $ & $11.19511 \pm 0.00012$ &  0.019 & $25.5^{+1.1}_{-1.4}$ & 0.22$^{+0.03}_{-0.04}$    &      < 100        & 0.85 & 2.73 $ \pm $ 0.30 & < 210 \\    
               & $222 \pm 4 $ & $-26 \pm 5 $ & $11.19512 \pm 0.00008$ &  0.019 & $25.2^{+1.1}_{-0.9}$ & 0.226$^{+0.005}_{-0.032}$ &      < 120        & 0.84 & 2.73 $ \pm $ 0.30 & < 250 \\    
\hline
\end{tabular}
%\end{small}
\end{table*}
\end{landscape}

\voffset=0.0cm
%\clearpage

 As a consistency check, we re-ran one of our previous targets, (478) Tergeste, now using the CITPM. In one of our earlier 
 works \citep{M2018}, this target was spin- and shape-modelled, and then the resulting models that best fitted the light curves 
 in the visible were applied in TPM procedures. In that work we obtained 
 thermal inertia in the range of 30 - 120 J\,m$^{-2}$\,s$^{-1/2}$\,K$^{-1}$ (SI units), 
 and reduced $\chi^2$ of models fit to infrared data of 2.18 and 1.53 for poles 1 and 2, respectively, 
 revealing a strong preference for one of the spin and shape solutions, but also problems with fitting all the thermal data. 
 New simultaneous optimisation on the same visible and infrared datasets performed here led to a somewhat different model. Most notably, 
 the reduced $\chi^2$ decreased substantially to 0.94 for pole 1, and 0.88 for pole 2, and so some preference for one spin solution 
 remained, and thermal inertia shifted to smaller values: 1 - 50 SI units. 
 To further check, we modelled the IR data using the new shape models with the classical TPM approach 
 \citep{Lagerros1996, Lagerros1997, Lagerros1998} and found a consistent solution.

 The fit to visible light curves remained similarly good with both approaches, and the spin 
 axis coordinates, size, and albedo agreed with the
original ones within the error bars. 
 In summary, the CITPM method enabled us to find a much better combination of spin, shape, and thermal parameters 
 than the two-step approach used originally. 

 The CITPM method provides models for several targets for which previous analyses with the classical TPM method failed; for example a unique and stable solution was found for (487) Seppina. For (666) Desdemona, we constrained 
 the size and albedo to a narrow range, while thermal inertia still remains uncertain. Furthermore, for two targets (667, 995), 
 additional calibrated data used in the CITPM improved the solution of inertia tensors, which were previously erroneous 
 (i.e. excessively stretched along the spin axis). 
 Also, we were able to find more precisely constrained
dimensions   along the spin
axis for the shape models for all the other targets, which is an area of frequent weakness in shape models based exclusively on relative photometry. 

 Independent confirmation of the robustness of our models also  comes from fitting the models to stellar occultation chords.
 The results of occultation fitting are presented in Table \ref{occult}, and in Figs. \ref{Havnia_occult} - \ref{Denise_occult} in Appendix C,  
 which show the instantaneous silhouette of the shape model on the $\eta, \xi$ sky plane scaled in kilometres. 
 Table \ref{occult_obs} in the Appendix D lists the occultation observers and sites.

\voffset=0.0cm
%\clearpage

\begin{table*}
%  \flushleft
\begin{small}
\begin{center}
\caption{Previously published spin parameters for targets studied here. 
 The columns contain asteroid name, J2000 ecliptic coordinates $\lambda_p$, $\beta_p$ of the spin solution, sidereal rotation period $P$, 
 and the reference. Values in italics denote solutions substantially differing from the ones obtained in the current paper.}
\label{previous}
\begin{tabular}{lllll}
\hline
    Asteroid   & \multicolumn{2}{c}{Pole}              &   P          & Reference\\
               & $\lambda_p [\deg]$ & $\beta_p [\deg]$ & [hours]      &         \\ 
\hline                                                                          
(108) Hecuba    & {\it 79} $\pm 1$  & {\it $+$13} $\pm 11$   &     --       &  \cite{Blanco1998} \\  
                & {\it 259} $\pm 7$ & {\it $-$6} $\pm 7 $   &    --        &  \cite{Blanco1998} \\
\hline
(219) Thusnelda & {\it 253} $\pm 4$ & $-69 \pm 4 $ & $59.712 \pm 0.002$     & \cite{ATLAS} \\
\hline
(362) Havnia    & {\it 96}$^{+3}_{-2}$ & $+39^{+7}_{-2} $  & {\it 16.918773}$^{+0.00003}_{0.000006}$ & \cite{Wang2015} \\
                & {\it 273}$^{+1}_{-6}$& $+40^{+1}_{-10}$  & {\it 16.918935}$^{+0.00001}_{0.00004}$ & \cite{Wang2015} \\
\hline
(478) Tergeste  & $  2 \pm 2 $ & $-42 \pm 3$  & $16.10308 \pm 0.00003$ & \cite{M2018}\\
                & $216 \pm 6 $ & $-56 \pm 4$  & $16.10308 \pm 0.00003$ & \cite{M2018}\\
\hline
(483) Seppina   &   --         & $+42 \pm 20$ & $12.72081 \pm 0.00006$ & \cite{ATLAS} \\
\hline
(537) Pauly     &{\it 290} $\pm 31$ & $+40 \pm 31$ &         -              & \cite{Blanco2000}\\
                & $ 31 \pm 12$ & $+32 \pm 10$  & $16.2961 \pm 0.0005$   & \cite{Hanus2016}\\ 
                & $211 \pm 16$ & $+51 \pm 10$  & $16.2961 \pm 0.0005$   & \cite{Hanus2016}\\ 
\hline
(552) Sigelinde &   --         & $+48 \pm 19$ & $17.1494 \pm 0.0002$   & \cite{ATLAS} \\
\hline
(618) Elfriede  & $113 \pm 3 $ & $+54 \pm 3$  & $14.7952 \pm 0.0001$   & \cite{ATLAS} \\
                & $323 \pm 1 $ &{\it $+$25} $\pm 3$& $14.7952 \pm 0.0001$   & \cite{ATLAS} \\
\hline
(666) Desdemona &    --        & $+12 \pm 22$ & $14.6080 \pm 0.0002$   & \cite{ATLAS} \\
\hline
(667) Denise    & $40 \pm 6  $ & $-86 \pm 3$  & $12.6848 \pm 0.0002$   & \cite{ATLAS} \\
\hline
(923) Herluga   & $188 \pm 5 $ & $-60 \pm 5$  & $29.7282 \pm 0.0007$   & \cite{ATLAS} \\
\hline
\end{tabular}
\end{center}
\end{small}
\end{table*}

  Spin and shape solutions had already been determined and
published in the literature for some of our targets, 
  while in some cases only some of the parameters were available.  
  In Table \ref{previous} we cite their spin axis coordinates and sidereal periods, if available, together with their reference. 
  Comparison with our results in Table~\ref{results} shows a general agreement, with the exception of (108) Hecuba modelled 
  by \cite{Blanco1998}, (362) Havnia modelled by \cite{Wang2015}, and (537) Pauly modelled by \cite{Blanco2000} based on different shape approximations.  
  Parameters strongly differing from the solutions obtained in this work are marked in italics in Table \ref{previous}. 
  Within consistent solutions, the differences in sidereal periods 
are sometimes of the order of a few 10$^{-4}$ h, 
  which may appear small, but might be noticeable after a few apparitions. 
   In the sections below, we focus on a few specific targets in more detail.

 \subsection{(362) Havnia}
 
 There were problems with some photometric data for this target. Firstly, data obtained by \cite{Harris1980} were 
 published in the APC archive as a composite light curve, with an incorrect period of 18 hours. As a consequence, only one 
 out of three light curves could be used, the one with original timings. This is a general problem with some early 
 asteroid light curves in the archives, and special attention must be paid when using them.  
 Other problems were caused by Super WASP data. Although in many cases  these serendipitously gathered data provided good 
 light curves from desired geometries, in this case their intrinsic noise made it impossible to find a unique
 spin and shape solution. After removing most of the Super WASP light curves for Havnia and keeping only the five best ones 
 (Fig. \ref{362composit2006}), the uniqueness of the solution greatly improved. 
 This demonstrates that the light curve inversion method is quite sensitive to noise in the data. 
 
 A spin and shape model of Havnia previously published by \cite{Wang2015}  was based on a light-curve inversion 
 using the Monte Carlo method on data from four apparitions (see Table \ref{previous}), while our model was based on (visible) light curves from seven apparitions. 
 The model by \cite{Wang2015} agrees with the model obtained here only in spin axis latitude 
 (see Table \ref{results}), whereas the longitudes are substantially different. Sidereal periods might appear similar at first sight, 
 but they would lead to a large divergence of extrema timings over just two apparitions. 

 Our model is characterised by a rather wide range of thermal inertia values due to a poor infrared dataset 
 (only data from Akari and WISE W4 were available; see N values in Table \ref{ancillary_info}), but Figure \ref{362chi2} 
 shows a clear minimum around 
 $\Gamma$=100 SI units. Unfortunately, all WISE W3 data had to be removed because of partial saturation. 
 Even keeping only their best subset led to divergence. 

 There is a four-chord stellar occultation from the year 2017 available in the PDS archive. 
 Both of our spin and shape solutions fit  this event very well, with all chords crossing close to the centre of the body 
 (see Fig. \ref{Havnia_occult}), resulting in volume-equivalent sizes a few percent smaller than the sizes provided by the CITPM method 
 (compare D values in Table \ref{occult}, and \ref{results}). The small $\pm 1$ km error in the occultation diameter is only a formal 
 uncertainty determined via bootstrapping separate chords and repeating the fitting procedure multiple times. However, the real 
 uncertainty must be larger because of the uncertainty on the shape model itself.

\begin{table}[h]
\begin{center}
\begin{tabular}{ccc}
\hline
&&\\
  Target      &  Pole 1          &     Pole 2 \\
&&\\
\hline
&&\\
  362 Havnia  &  84 $\pm$ 1 km   &   88 $\pm$ 1 km \\      
&&\\
\hline
&&\\
  618 Elfriede& 145 $\pm$ 7 km   &  155 $\pm$ 2 km \\
&&\\
\hline
&&\\
  667 Denise  &  83 $\pm$ 2 km   &    rejected \\ 
&&\\
\hline
\end{tabular}
\caption{Diameters of equivalent volume spheres for CITPM shape models fitted to stellar occultations.}
\label{occult}
\end{center}
\end{table}

 \subsection{(537) Pauly}

  Spin and shape solutions for (537) Pauly have already been published by \cite{Blanco2000} and \cite{Hanus2016}. 
  The results from the latter work are consistent with ours (see Tables~\ref{results} and \ref{previous}), although our model of Pauly is made using many more dense light curves 
  and also a richer set of thermal data (+ 9 Akari points), and via simultaneous optimisation of both data types. 
  Later, (537) Pauly was also analysed with the TPM via the data bootstrapping method \citep{Hanus2018}. 
  Our size determinations (46 $\pm$ 2 km, and 47 $\pm$ 4 km) are somewhat larger than 
  40.7 $\pm$ 0.8 km by \cite{Hanus2018}, but the thermal inertia and albedo values agree.  
  Our $\chi^2_\text{red}$ 
  IR residuals are smaller than in the previous model (0.7 vs. 1.1). 
  The difference in size might stem from the elongated shape of this target, and 
  the smaller set of infrared measurements in \cite{Hanus2018}, capturing the target within a limited range of rotation phases, 
  which might have led to underestimation of the size in previous study.

 \subsection{(618) Elfriede}

  There were as many as four different stellar occultations by this target, each containing from two to four chords (Fig. \ref{Elfriede_occult} 
  in the Appendix C). 
  However, these data did not help us reject any of our two models and we take pole 2 ($\lambda_p=341\deg, \beta_p=+49\deg$)
  as the preferred solution based simply on its slightly lower $\chi^2$.

 In this case, the occultation size agrees exactly for pole 1 with the radiometric size, while for pole 2 it is 
 a few percent larger (see Tables \ref{occult} and \ref{results}), but still within the radiometric error bars. 
 Our results, though self-consistent, are in disagreement with most previous size determinations for 618 Elfriede. 
 The occultation-determined size for pole 2 (155 $\pm$ 2 km) is almost 30\% larger than Akari (121.54 km) and IRAS (120.29 km) determinations 
 \citep{Usui2011, Tedesco2004}, 
 and 18\% larger than  the diameter given by WISE \citep[131.165 km][]{Mainzer2011b}. 
 For pole 1, the size disagreement is less pronounced (20\% and 11\% respectively) and is even compatible with the WISE diameter 
 within the error bars.

 In summary, as the present study is the first to take a comprehensive and multi-technique approach to analysing this target (rich photometric set 
 simultaneously combined with infrared data from three missions, plus independent occultation fitting), the size determined   here 
 (145 - 155 km) can probably be considered the most reliable.

 \subsection{(667) Denise}
 
 For asteroid (667) Denise there were three good stellar occultations ---with one containing as many as ten positive chords--- 
 thanks to a very successful European campaign (observers are acknowledged in Table \ref{occult_obs}). 
 Although both pole solutions are formally acceptable from the thermophysical point of view (both present in Table \ref{results}), 
 the occultation fitting clearly enabled us to reject the solution for pole 2 (see Fig. \ref{Denise_occult} in the Appendix C), which is marked 
 with italics in Table \ref{results}. 
 The size determined from occultations ($83 \pm 2$ km) is the same as the radiometric size ($83^{+4}_{-2}$ km). 
 The CITPM method proved to be robust and accurate, and provided the most accurate 
 parameters in the case of dense
stellar occultation chords.

\section{Conclusions and future work}

 We fully characterised spin, shape, and thermal properties of 16 main-belt asteroids from the group that until recently
 has been neglected because of observing selection effects.
 The multi-apparition targeted observing campaign together with good-quality infrared data, especially from the WISE spacecraft,  
 led to consistent spin and shape models accompanied by precise size and albedo determinations, and thermal inertia
 being determined for most of the targets for the first time. Thanks to simultaneous use of both visible and infrared 
 data, our shape models are optimal in terms of reproducing both types of data well. Also, the CITPM 
 gained additional evidence for its robustness, providing an optimal solution in one of the cases, 
 as confirmed by an independent method.
 The set contains two updated models (478 Tergeste, and 537 Pauly), and a few targets with partial solutions  
 due to the scarcity of infrared data. 

 With this work we increase the number
 of slow rotators with thermal inertia determined from detailed thermophysical modelling  by 40\%. It is necessary to 
 enlarge the pool of such well-studied targets so that 
 we can gain more insight into different asteroid groups and families separately and explore links between thermal 
 properties, surface material properties, and family formation ages \citep{HarrisDrube2020}. 
 Most targets presented here do not belong to any collisional family (with the exception of 923 Herluga and 995 Sternberga, 
 both from the Eunomia family, and also 618 Elfriede and 780 Armenia, each having their own small, compact family),  
 and so their low thermal inertia was expected. 

 Our target sizes span the range from a few tens of kilometres to over 100 km, with most of the determinations being within 10\% of 
 previous determinations based on WISE data only, and the NEATM thermal model \citep{Harris1998}. 
 Sizes determined for a few targets (223, 552, 618) differ by more, although our approach (including infrared data combined 
 with spin and shape models) has been shown to be robust. We therefore consider our results to be most reliable. 
 Furthermore, obtained albedo values agree with previously published taxonomic classifications. 
 
 The thermal inertia values determined here are < 100 SI units for most targets, indicating the presence of a thick layer of insulating regolith on most of these bodies. 
 These values of thermal inertia reduced to 1 AU display no trend with size, because our current targets 
 are well within the size range where largely different thermal inertias have been found 
 in previous works \citep[see fig. 7 in][]{Hanus2018}. The correlation between thermal inertia and size found by \cite{Delbo2007} could only 
 be evident if our sample also contained asteroids smaller than 10 km, these being too faint for our photometric campaign on small telescopes.
 
 We also found no evidence to support the hypothesis that thermal inertia increases with rotation period 
 \citep[e.g.][]{HarrisDrube2016}. Our results are in agreement with those of \citep{M2019} and \cite{Hanus2018}. 
 \cite{Biele2019} showed that a fine-grained, highly porous surface layer of just a few millimetres thick can hide thermal signatures 
 of denser, more thermally conductive layers due to its relatively small thermal skin depth ($d_s$) of a few millimetres, while 
 to see signatures of the denser layers would require probing a centimetre range. 
 However, despite their longer rotation periods (11 to 59 hours) compared to the typical light-curve inversion and TPM targets 
 found in the literature, the thermal skin depths of our targets calculated according to the formula given by \cite{Spencer1989} 
 still lie in the  range of a few millimetres. The cases with large thermal inertia error bars could still be compatible with 
 $d_s$ up to 3.5 cm, however all the values below it are equally 
 possible, and so this cannot be used for drawing firm conclusions.   
 
 Furthermore, we did not find any correlation between thermal inertia and spin axis inclination, or any specific problems with fitting 
 more inclined targets, which must experience seasonal cycles of heating and cooling. 
 However, our thermal inertia determinations, as is often the case, are burdened with large uncertainties. 
 It is possible that the trend linking thermal inertia and rotation period simply eludes us in our investigations, 
 as precise thermal inertia determinations might be hampered by slow rotation, decreasing the thermal lag.  
 For future studies, it will be beneficial to focus on targets with thermal measurements from WISE spacecraft 
 obtained at epochs separated  in time by as much as possible (longer than $\sim$100 days). 
 This should help to constrain thermal inertia better thanks to more
 varied viewing geometries, enabling comparison of thermal flux from for example pre- and post-opposition geometries. 
 
 Our scaled spin and shape models and their thermal parameters are available in the new version of DAMIT 
 (Database for Asteroid Models from Inversion Techniques) \citep{DAMIT}\footnote{https://astro.troja.mff.cuni.cz/projects/damit/}, 
 and data tables with photometry in the visible are available via the CDS (Centre de Donn{\'e}es astronomiques de Strasbourg) archive.

\flushleft

\begin{acknowledgements}
This work was was initiated with the support from 
the National Science Centre, Poland, through grant no. 2014/13/D/ST9/01818; 
and from the European Union's Horizon 2020 Research and Innovation Programme, under Grant Agreement no 687378 (SBNAF).
The work of J.D. was supported by the grant 20-08218S of the Czech Science Foundation.
A.P. and R.S. have been supported by the K-125015 grant of the National Research, Development and Innovation Office (NKFIH), Hungary.
This project has been supported by the Lend{\"u}let grant LP2012-31 of the Hungarian Academy of Sciences.
This project has been supported by the GINOP-2.3.2-15-2016-00003 grant of the Hungarian National Research, Development and Innovation 
Office (NKFIH).
L. M. was supported by the Premium Postdoctoral Research Program of the Hungarian Academy of Sciences. 
The research leading to these results has received funding from the LP2018-7/2020 Lend{\"u}let grant of the Hungarian Academy of Sciences.
The work of T.S.-R. was carried out through grant APOSTD/2019/046 by Generalitat
Valenciana (Spain). This work was supported by the MINECO (Spanish Ministry of
Economy) through grant RTI2018-095076-B-C21 (MINECO/FEDER, UE).

This article is based on observations obtained at the Observat{\'o}rio Astron{\^o}mico do Sert{\~a}o de Itaparica (OASI, Itacuruba) 
of the Observat{\'o}rio Nacional, Brazil. F.M. would like to thank the financial support given by FAPERJ (Process E-26/201.877/2020). 
E.R., P.A., H.M., M.E. and J.M. would like to thank CNPq and CAPES (Brazilian agencies) for their support through diverse fellowships. 
Support by CNPq (Process 305409/2016-6) and FAPERJ (Process E-26/202.841/2017) is acknowledged by D.L.

       The Joan Or{\'o} Telescope (TJO) of the Montsec Astronomical Observatory (OAdM) 
       is owned by the Catalan Government and operated by the Institute for Space Studies of Catalonia (IEEC).
       This article is based on observations made in the Observatorios de Canarias del IAC with the 
       0.82m IAC80 telescope operated on the island of Tenerife by the Instituto de Astrof{\'i}sica de Canarias (IAC) 
       in the Observatorio del Teide. 
This article is based on observations made with the SARA telescopes (Southeastern Association for Research in Astronomy),
whose nodes are located at the Observatorios de Canarias del IAC on the island of La Palma in the Observatorio 
del Roque de los Muchachos; Kitt Peak, AZ under the auspices of the National Optical Astronomy Observatory (NOAO);  
and Cerro Tololo Inter-American Observatory (CTIO) in La Serena, Chile.
This project uses data from the SuperWASP archive. The WASP project is currently funded and operated by Warwick University 
and Keele University, and was originally set up by Queen's University Belfast, the Universities of Keele, St. Andrews, 
and Leicester, the Open University, the Isaac Newton Group, the Instituto de Astrofisica de Canarias, 
the South African Astronomical Observatory, and by STFC.
TRAPPIST-South is a project funded by the Belgian Fonds
(National) de la Recherche Scientifique (F.R.S.-FNRS) under
grant PDR T.0120.21. TRAPPIST-North is a project funded by
the University of Li{\`e}ge, in collaboration with the Cadi Ayyad
University of Marrakech (Morocco). E. Jehin is FNRS Senior
Research Associate.

Funding for the Kepler and K2 missions are
provided by the NASA Science Mission Directorate.
The data presented in this paper were
obtained from the Mikulski Archive for Space
Telescopes (MAST). STScI is operated by the
Association of Universities for Research in Astronomy,
Inc., under NASA contract NAS5-
26555. Support for MAST for non-HST data is
provided by the NASA Office of Space Science
via grant NNX09AF08G and by other grants
and contracts.

Data from Pic du Midi Observatory have been obtained with the 0.6-m telescope, a facility operated by 
Observato{\'i}re Midi Pyr{\'e}n{\'e}es and Association T60, an amateur association. 

We acknowledge the contributions of the occultation observers
who have provided the observations in the dataset. Most of
those observers are affiliated with one or more of:
European Asteroidal Occultation Network (EAON), International Occultation Timing Association (IOTA),
International Occultation Timing Association  European Section (IOTA/ES), Japanese Occultation Information Network (JOIN), and 
Trans Tasman Occultation Alliance (TTOA).

\end{acknowledgements}

 % Acknowledgements:
 % SARA
 % OAdM 
 % IAC80
 % Mercator - nie
 % spr. jakie jeszcze?

 \bibliographystyle{aa}
\bibliography{bibliography}

\newpage 

\onecolumn
\begin{appendix}
\section{Chi-squared plots vs. thermal inertia}
 This section contains 
 plots of $\chi^2_\text{red}$ versus thermal inertia for various combinations of surface roughness and optimised size (Figures \ref{108chi2} - \ref{995chi2}).
    \begin{table*}[h!]
    \centering
%\vspace{0.5cm}
\begin{tabularx}{\linewidth}{XX}
 \includegraphics[width=0.35\textwidth]{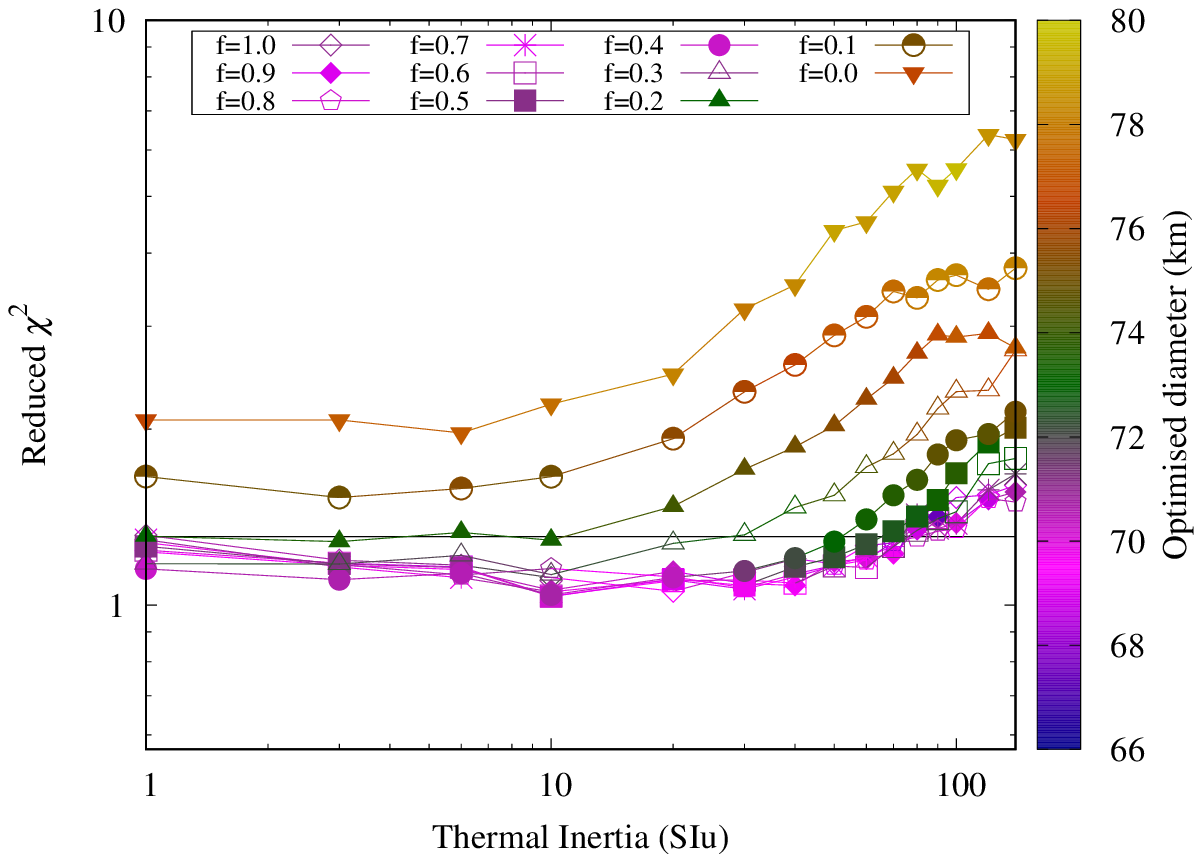}
\captionof{figure}{Reduced $\chi^2$ values vs. thermal inertia for various combinations of surface roughness (symbol coded) 
and optimised diameters (colour coded) for asteroid (108) Hecuba.}
\label{108chi2}
&
 \includegraphics[width=0.35\textwidth]{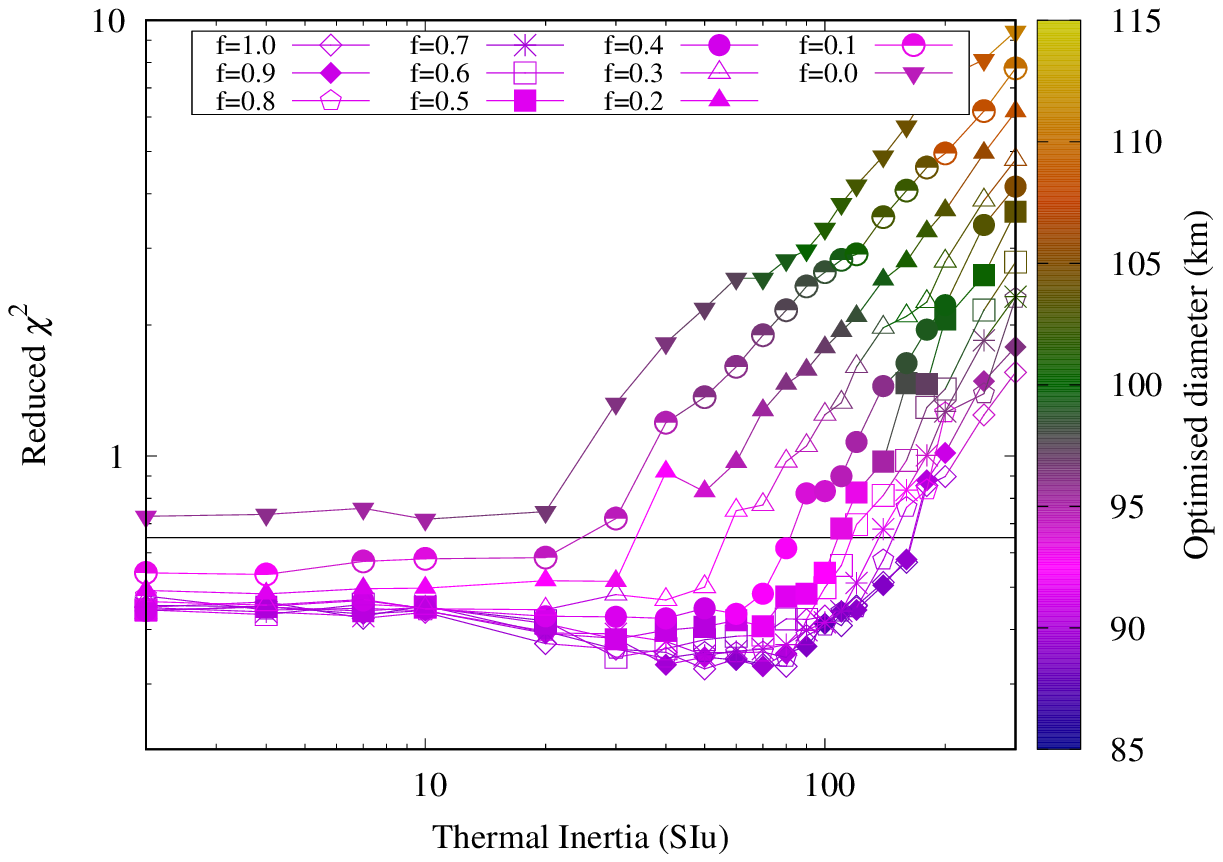}
\captionof{figure}{Reduced $\chi^2$ values vs. thermal inertia for (202) Chryseis}
\label{202chi2}
\\
 \includegraphics[width=0.35\textwidth]{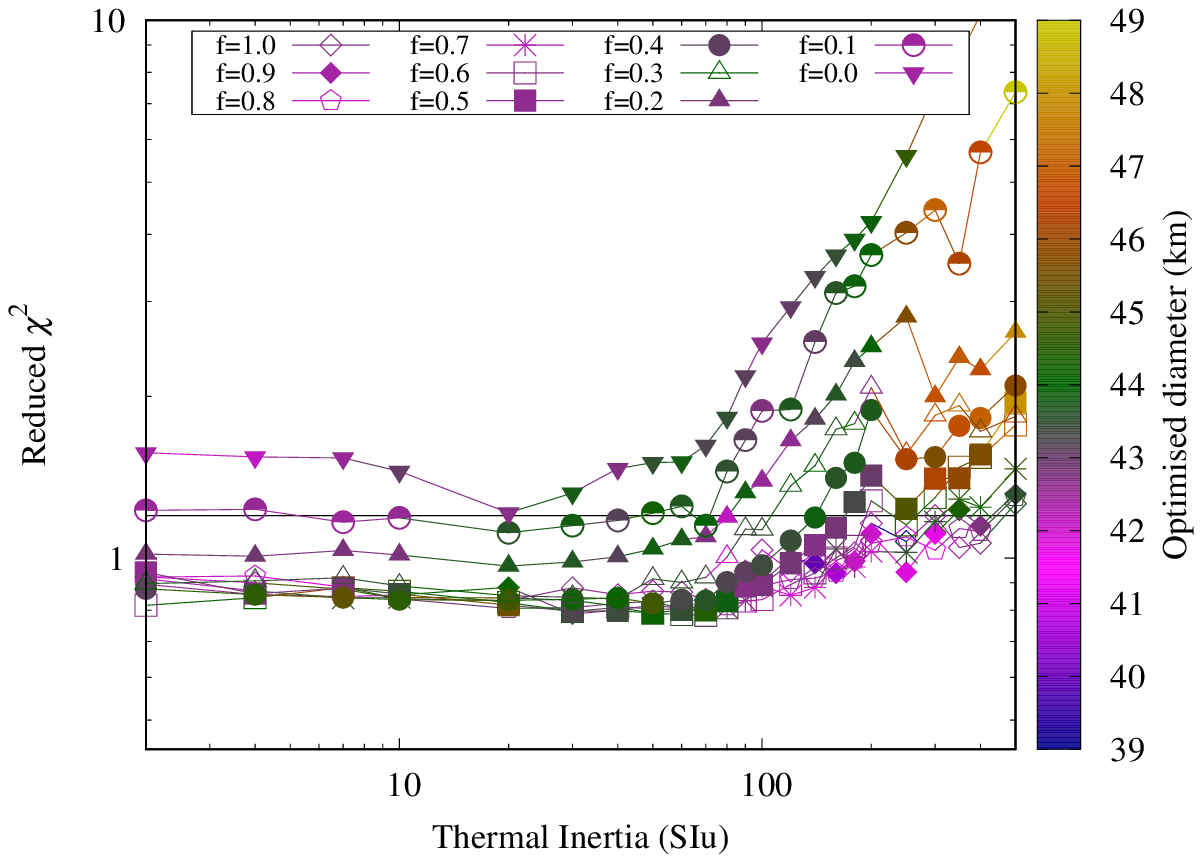}
\captionof{figure}{Reduced $\chi^2$ values vs. thermal inertia for (219) Thusnelda}
\label{219chi2}
&
 \includegraphics[width=0.35\textwidth]{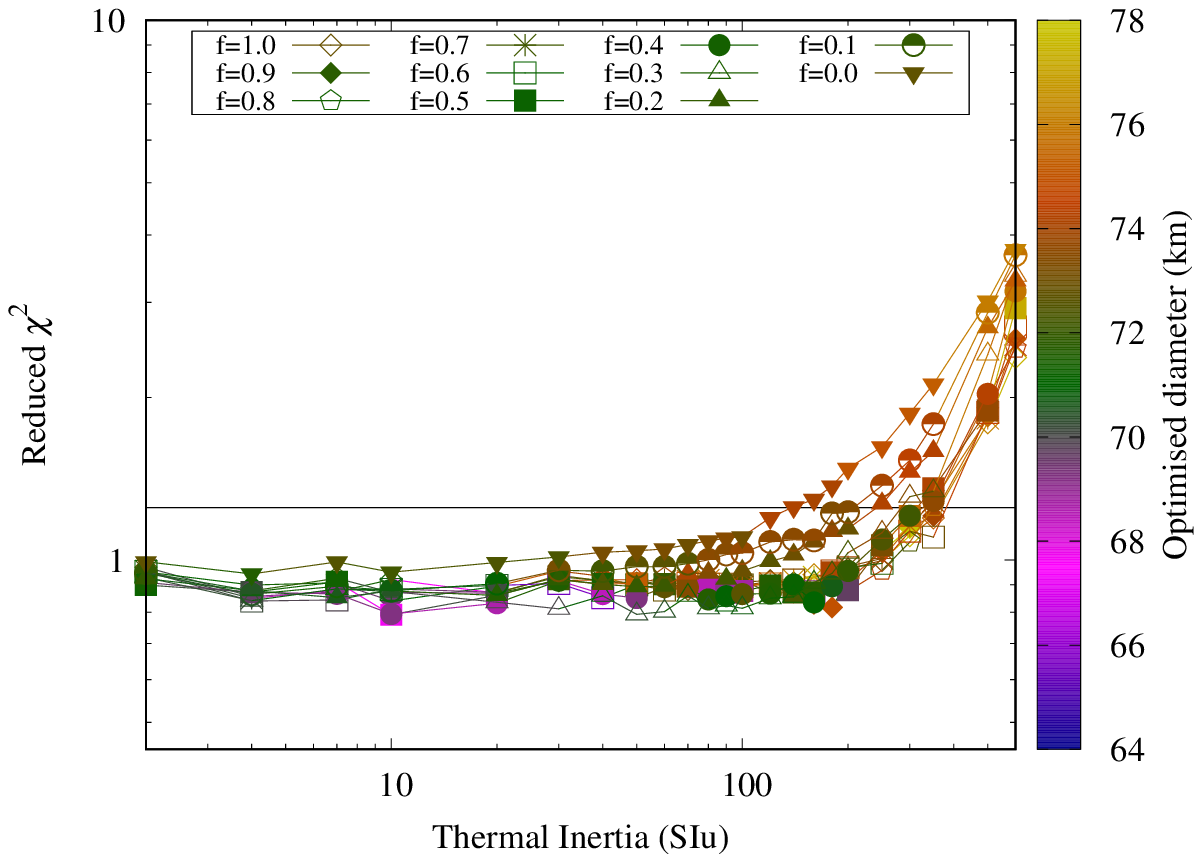}
\captionof{figure}{Reduced $\chi^2$ values vs. thermal inertia for (223) Rosa}
\label{223chi2}
\\
 \includegraphics[width=0.35\textwidth]{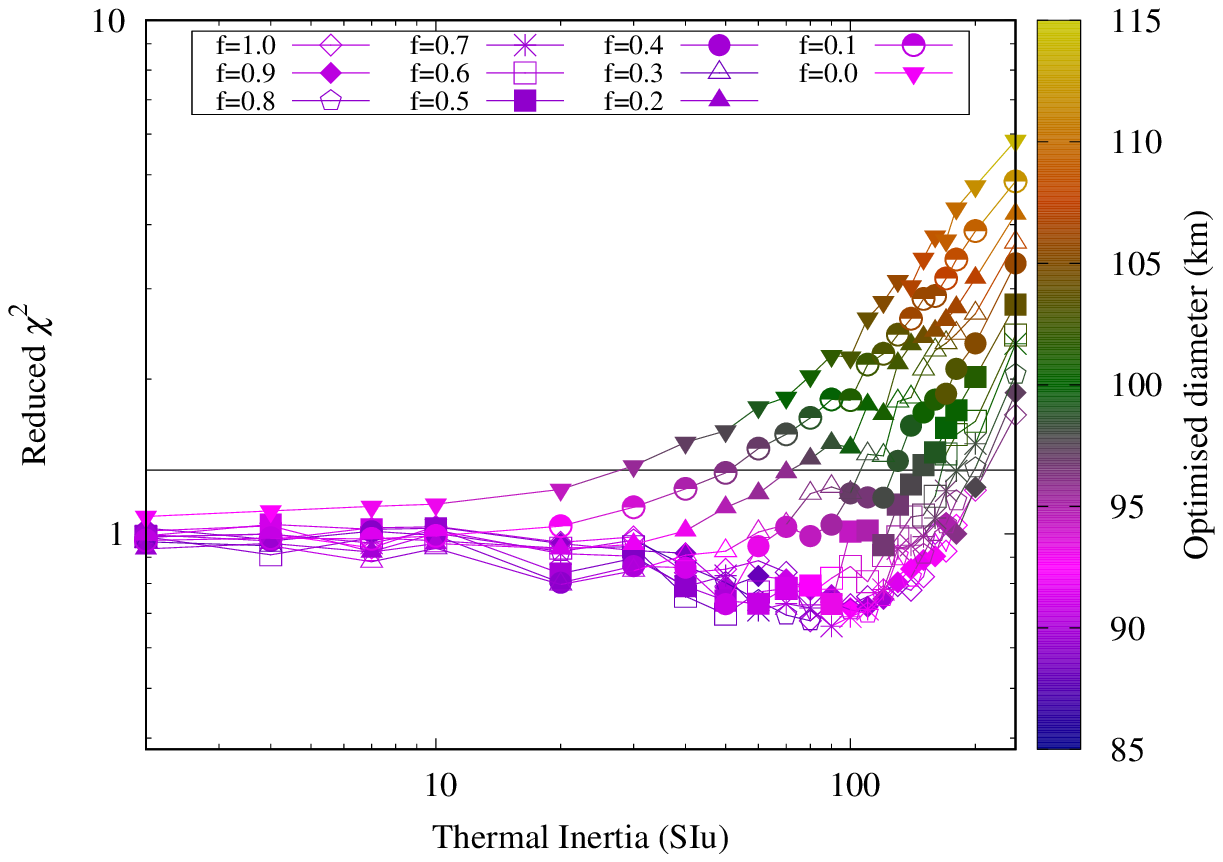}
\captionof{figure}{Reduced $\chi^2$ values vs. thermal inertia for (362) Havnia}
\label{362chi2}
&
 \includegraphics[width=0.35\textwidth]{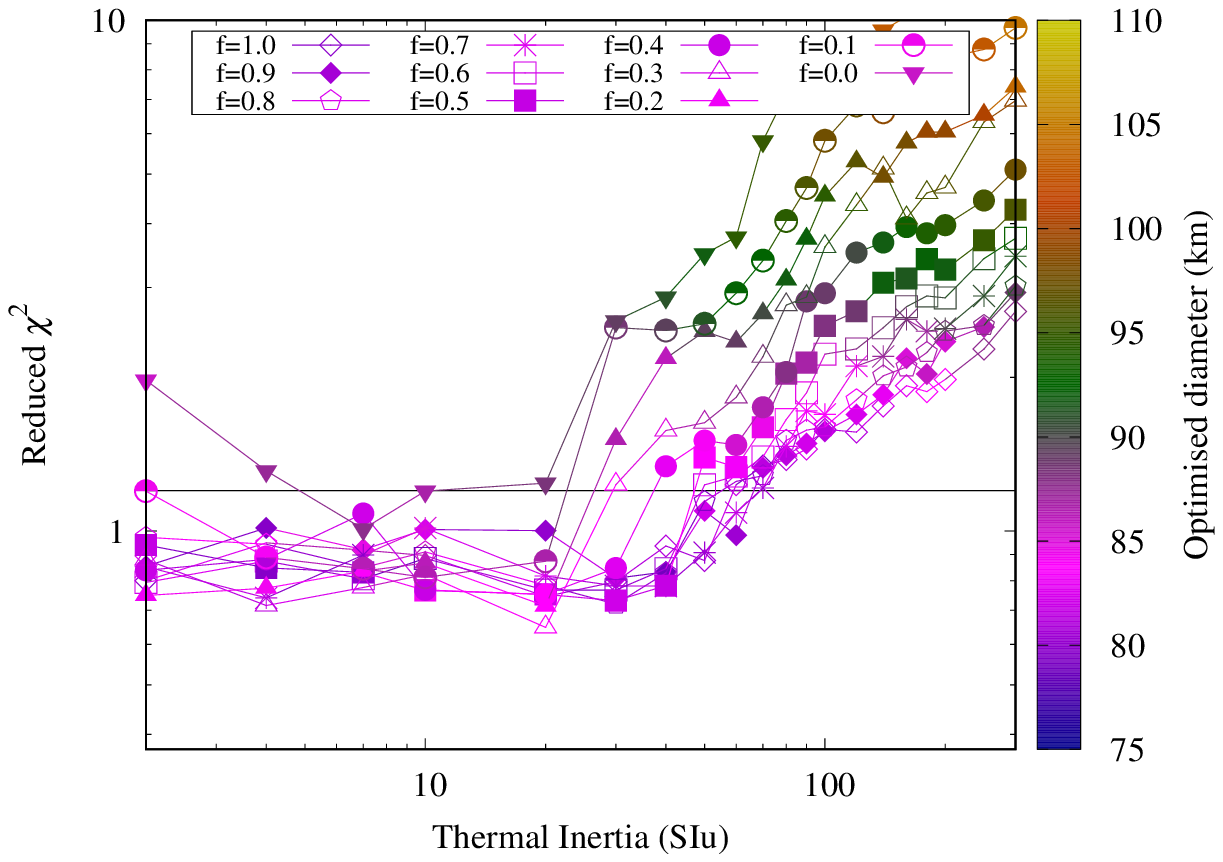}
\captionof{figure}{Reduced $\chi^2$ values vs. thermal inertia for (478) Tergeste}
\label{478chi2}
\\
 \includegraphics[width=0.35\textwidth]{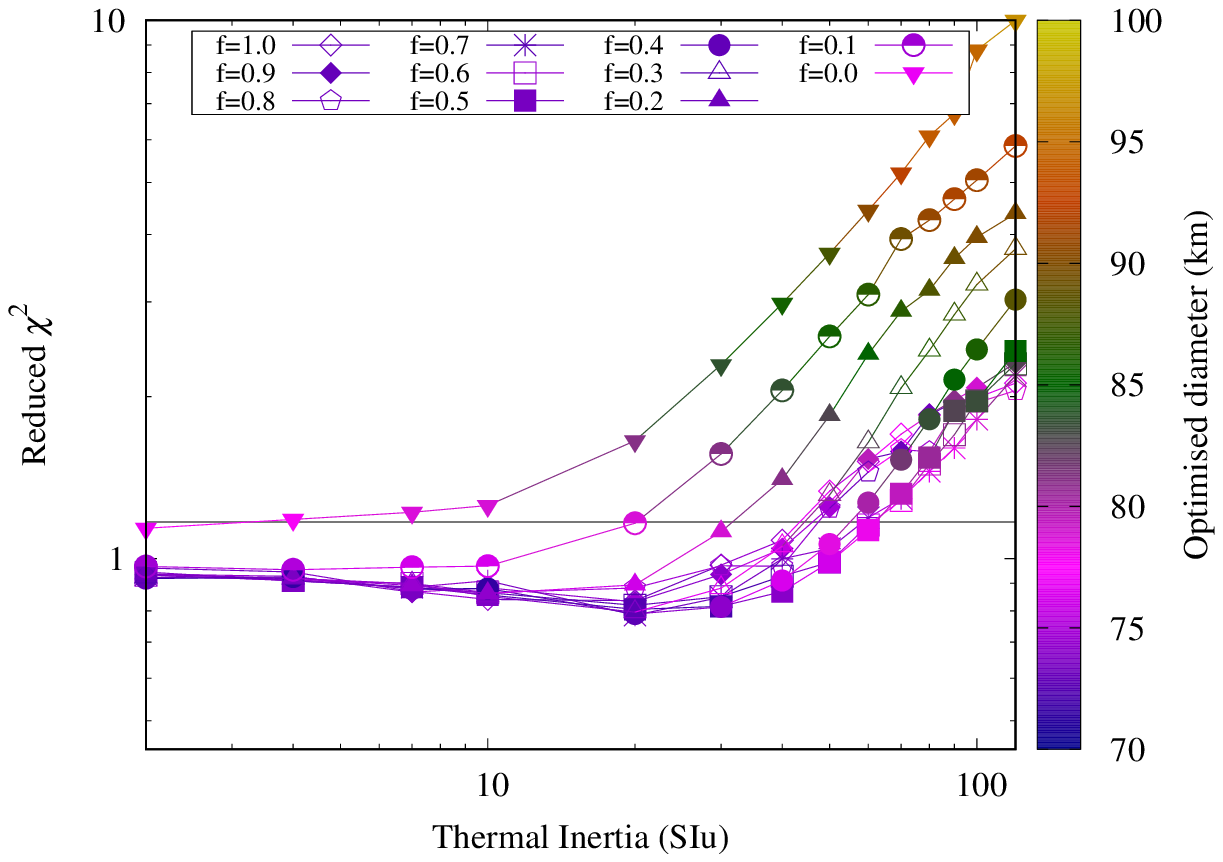}
\captionof{figure}{Reduced $\chi^2$ values vs. thermal inertia for (483) Seppina}
\label{483chi2}
&
 \includegraphics[width=0.35\textwidth]{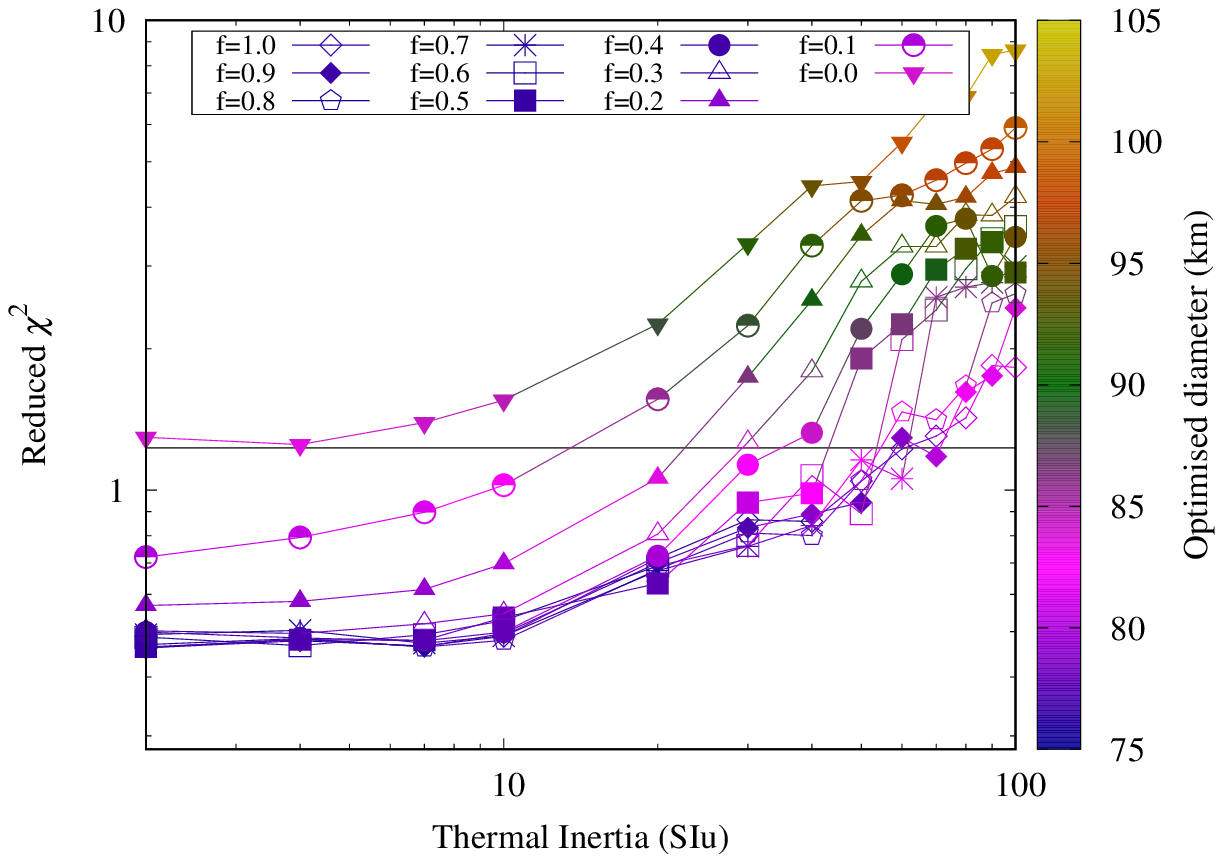}
\captionof{figure}{Reduced $\chi^2$ values vs. thermal inertia for (501) Urhixidur}
\label{501chi2}
\\
\end{tabularx}
    \end{table*}

    \begin{table*}[ht]
    \centering
%\vspace{0.5cm}
\begin{tabularx}{\linewidth}{XX}
 \includegraphics[width=0.40\textwidth]{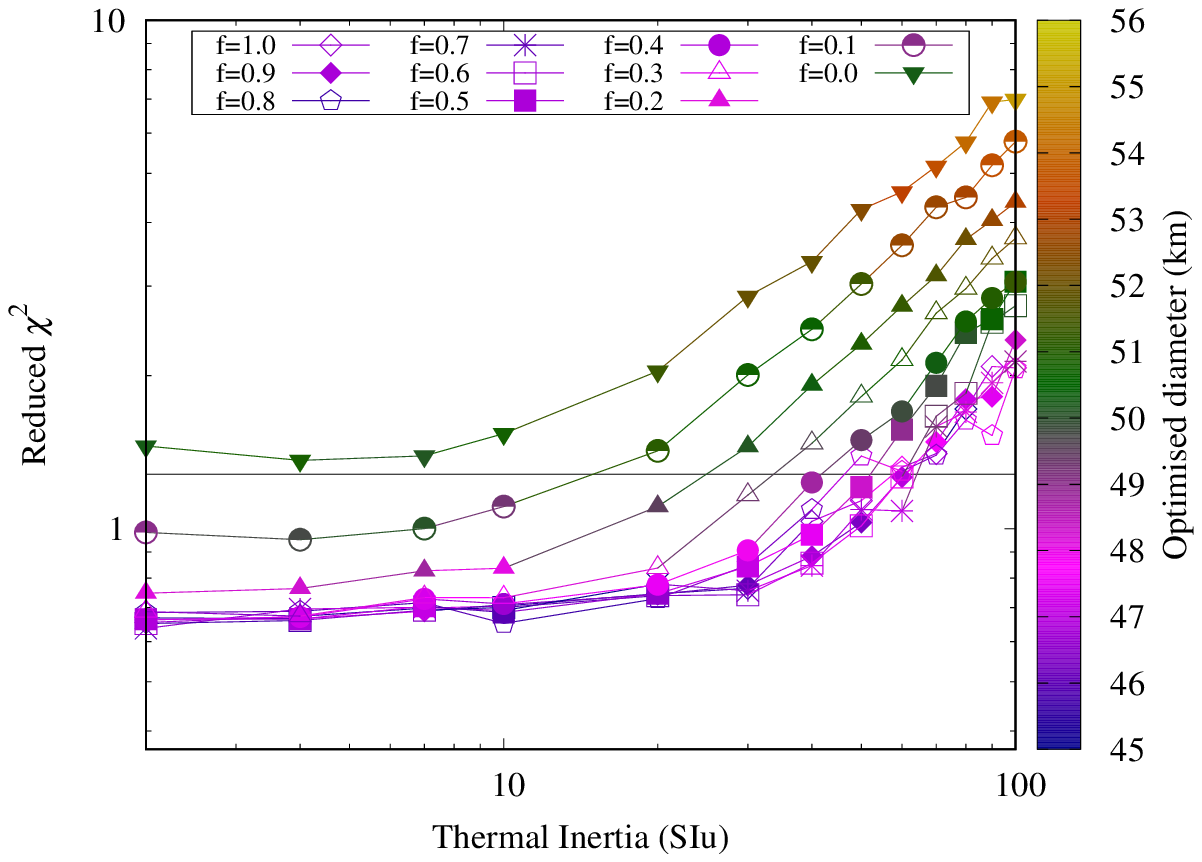}
\captionof{figure}{Reduced $\chi^2$ values vs. thermal inertia for (537) Pauly}
\label{537chi2}
&
 \includegraphics[width=0.40\textwidth]{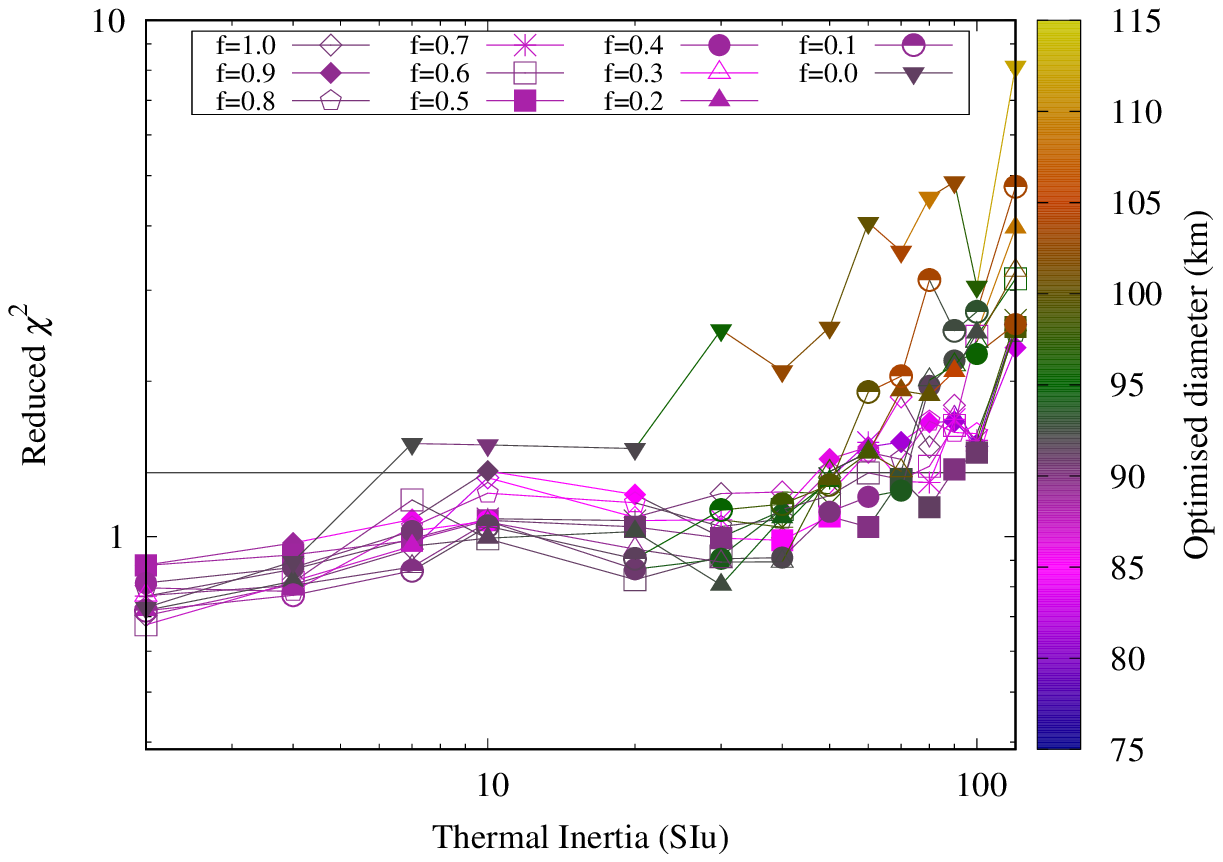}
\captionof{figure}{Reduced $\chi^2$ values vs. thermal inertia for (552) Sigelinde}
\label{552chi2}
\\
 \includegraphics[width=0.40\textwidth]{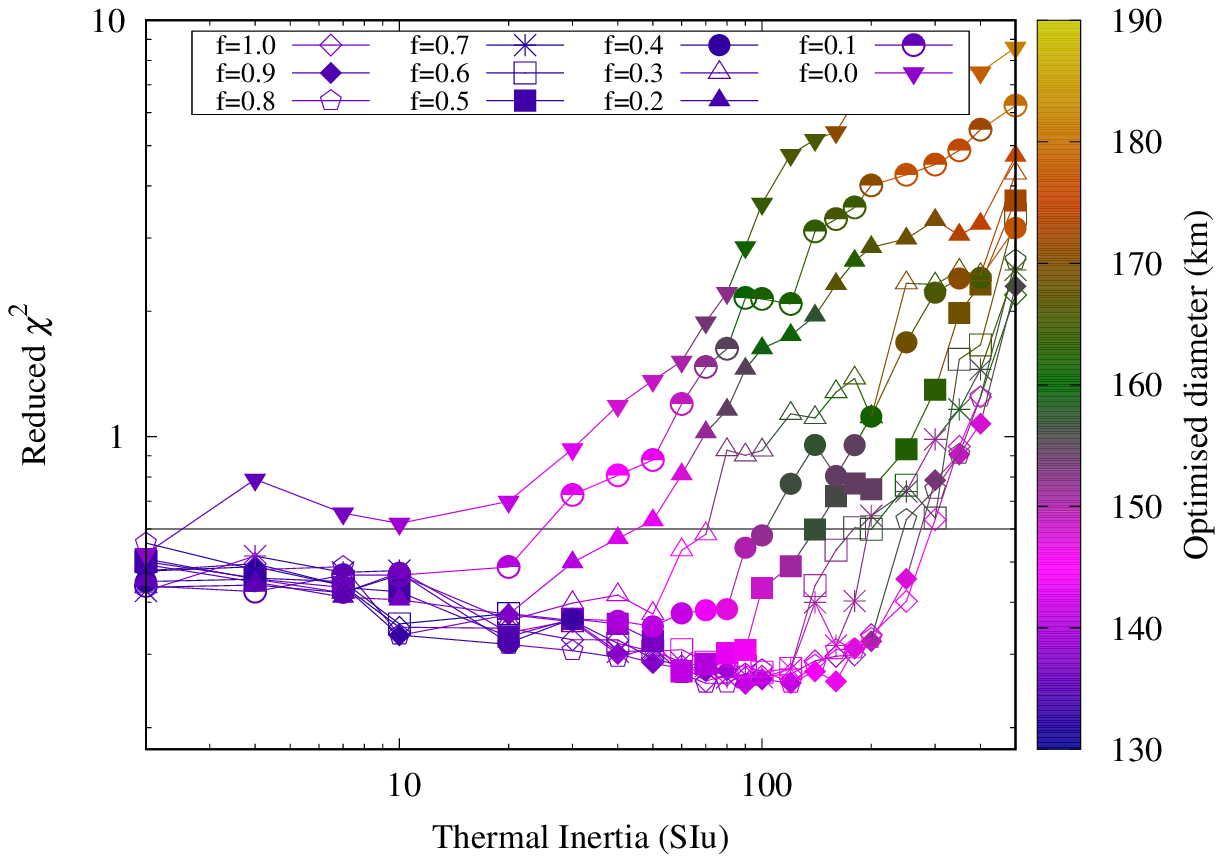}
\captionof{figure}{Reduced $\chi^2$ values vs. thermal inertia for (618) Elfriede}
\label{618chi2}
&
 \includegraphics[width=0.40\textwidth]{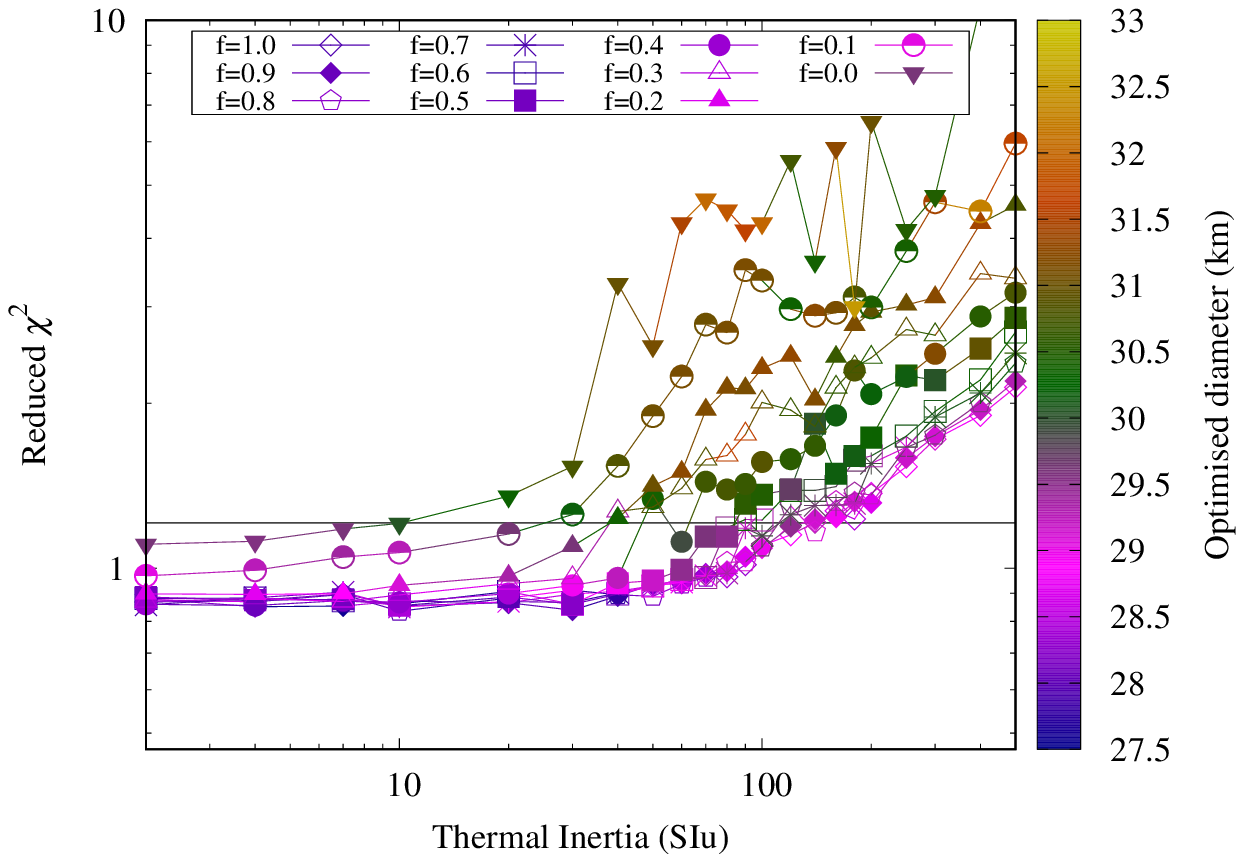}
\captionof{figure}{Reduced $\chi^2$ values vs. thermal inertia for (666) Desdemona}
\label{666chi2}
\\
 \includegraphics[width=0.40\textwidth]{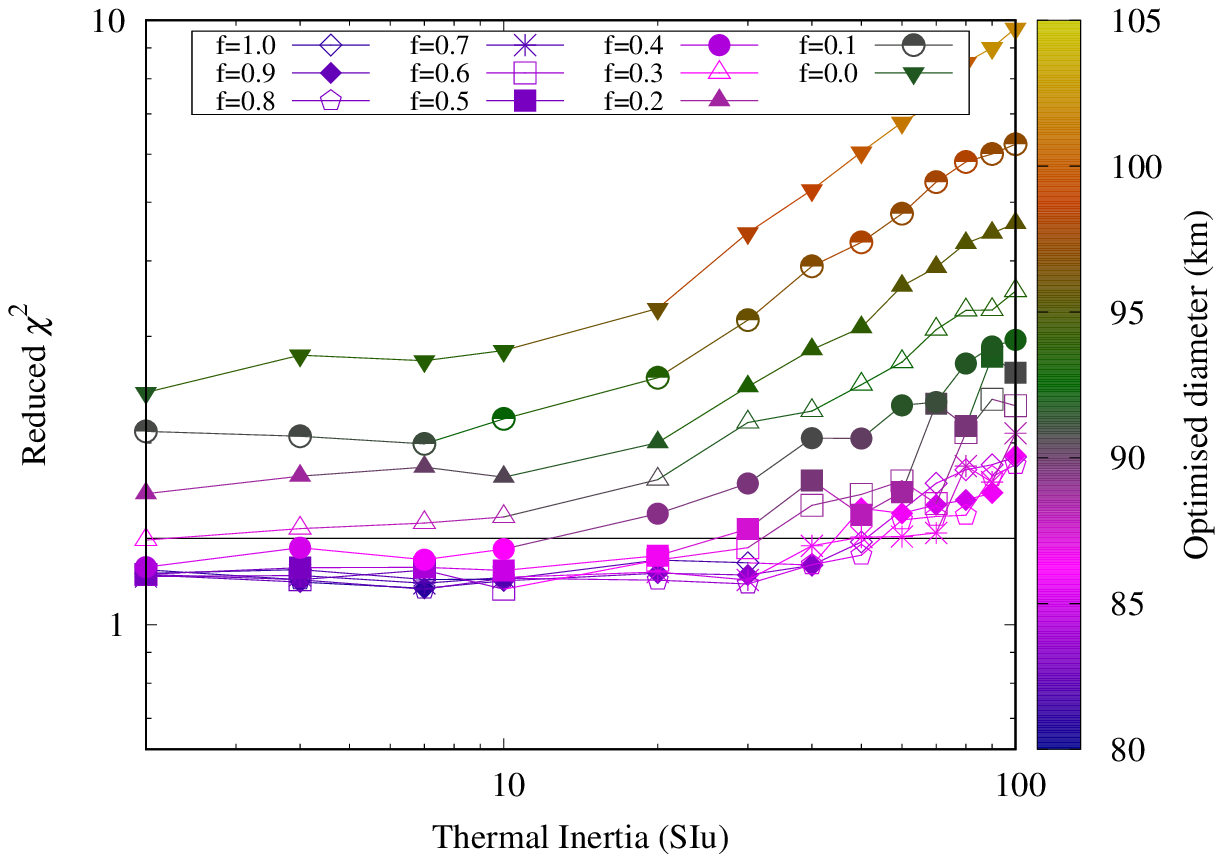}
\captionof{figure}{Reduced $\chi^2$ values vs. thermal inertia for (667) Denise}
\label{667chi2}
&
 \includegraphics[width=0.40\textwidth]{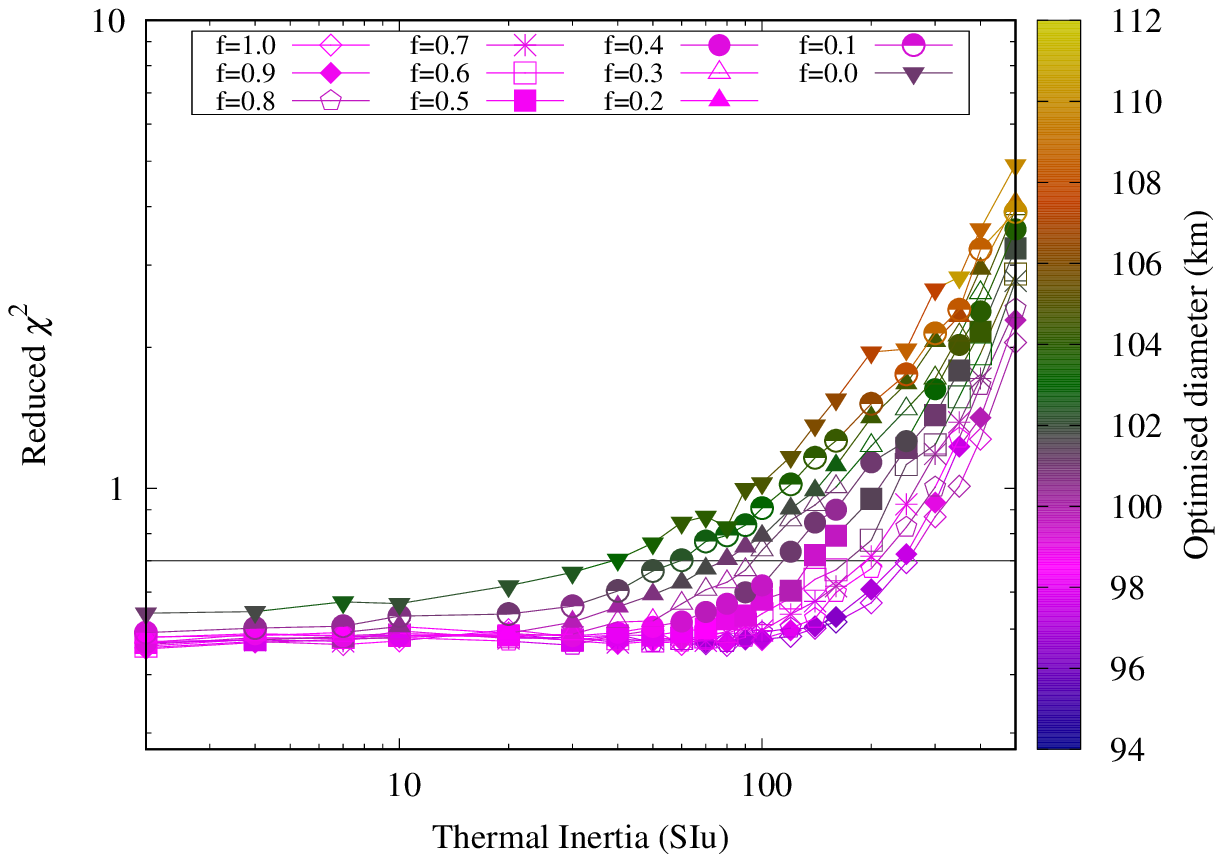}
\captionof{figure}{Reduced $\chi^2$ values vs. thermal inertia for (780) Armenia}
\label{780chi2}
\\
 \includegraphics[width=0.40\textwidth]{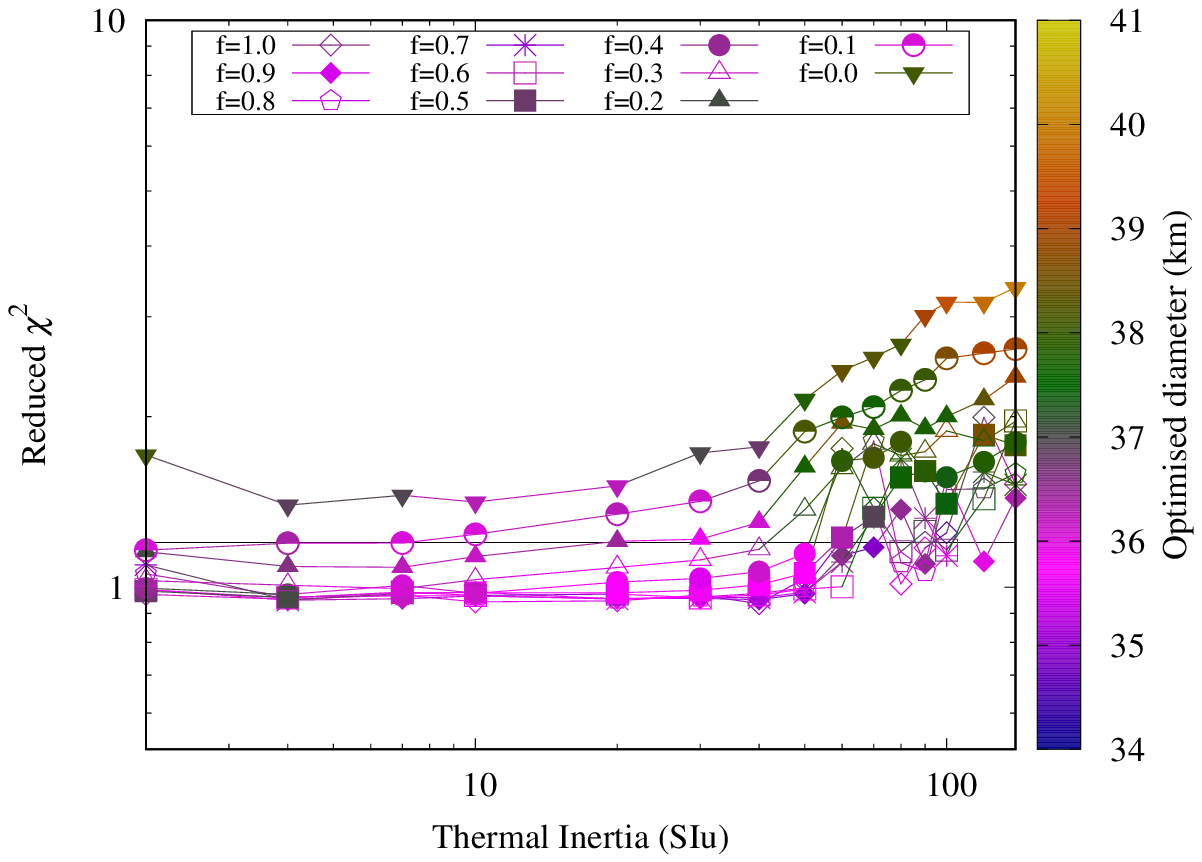}
\captionof{figure}{Reduced $\chi^2$ values vs. thermal inertia for (923) Herluga}
\label{923chi2}
&
 \includegraphics[width=0.40\textwidth]{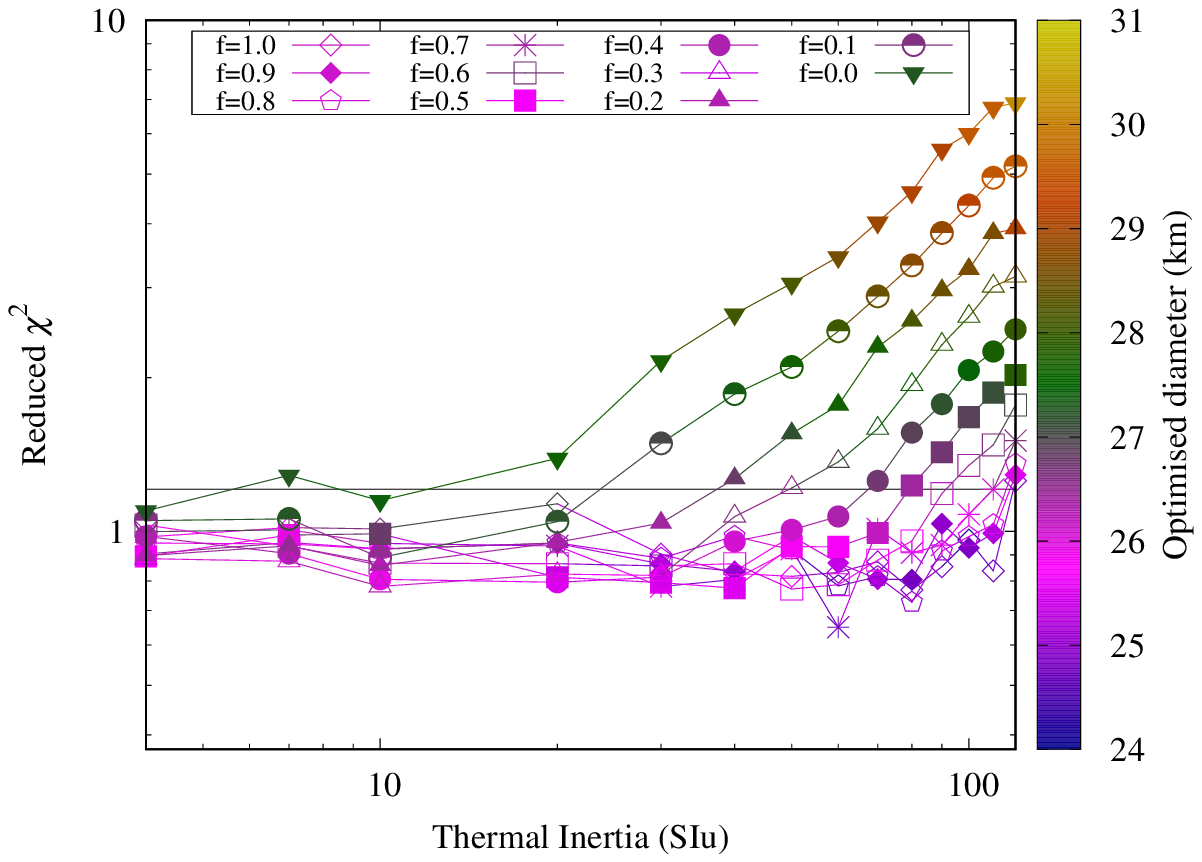}
\captionof{figure}{Reduced $\chi^2$ values vs. thermal inertia for (995) Sternberga}
\label{995chi2}
\\
\end{tabularx}
    \end{table*}

\clearpage
\section{Thermal light curves}
 Model fits to WISE thermal light curves (Figures \ref{108thermal_lcW3} - \ref{995thermal_lcW4}).
    \begin{table*}[h!]
    \centering
\vspace{0.5cm}
\begin{tabularx}{\linewidth}{XXX}
 \includegraphics[width=0.3\textwidth]{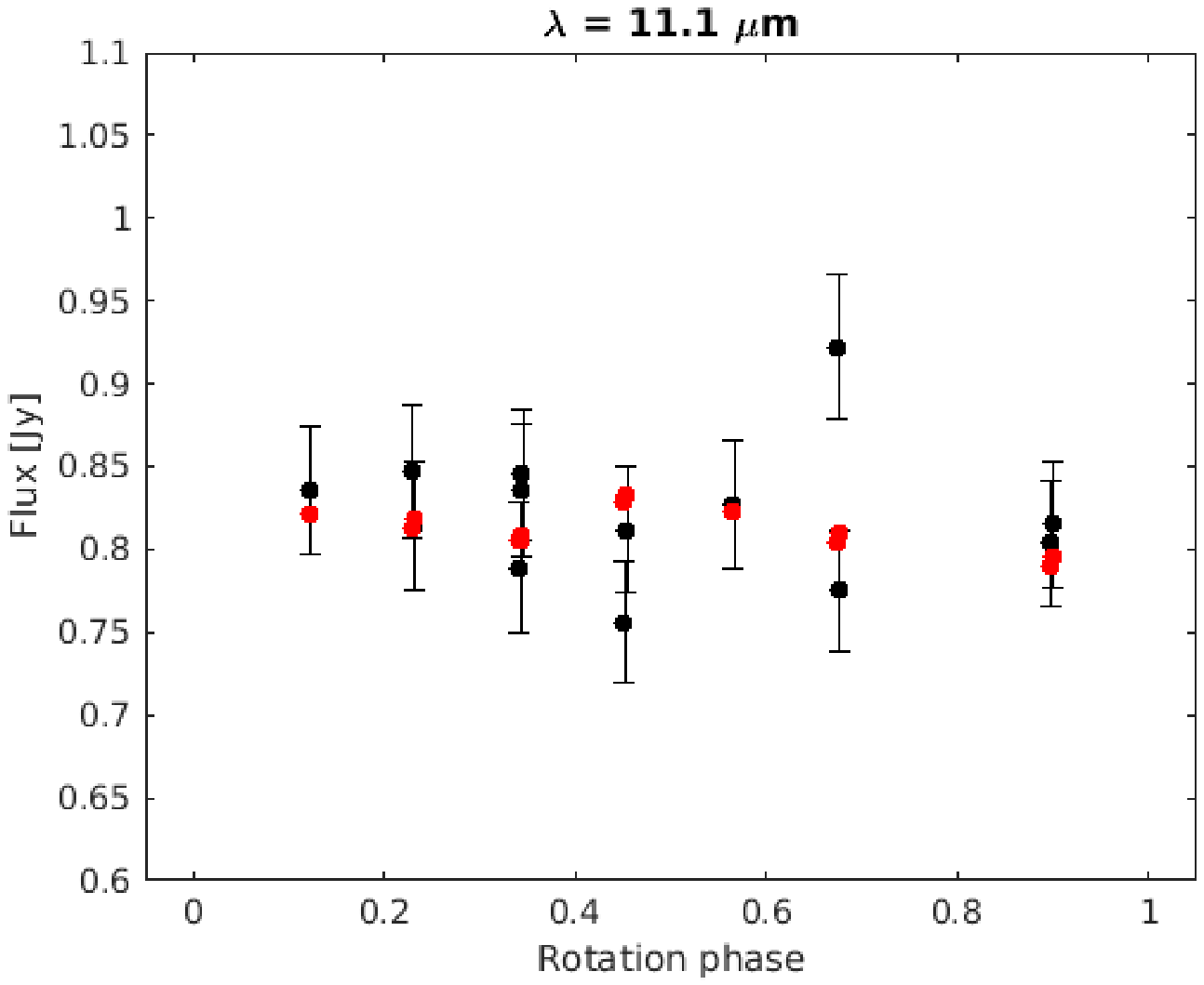}
\captionof{figure}{Infrared model fluxes (red circles) compared to measured fluxes in W3 band of WISE spacecraft (black circles) for
asteroid (108) Hecuba.}
\label{108thermal_lcW3}
&
 \includegraphics[width=0.3\textwidth]{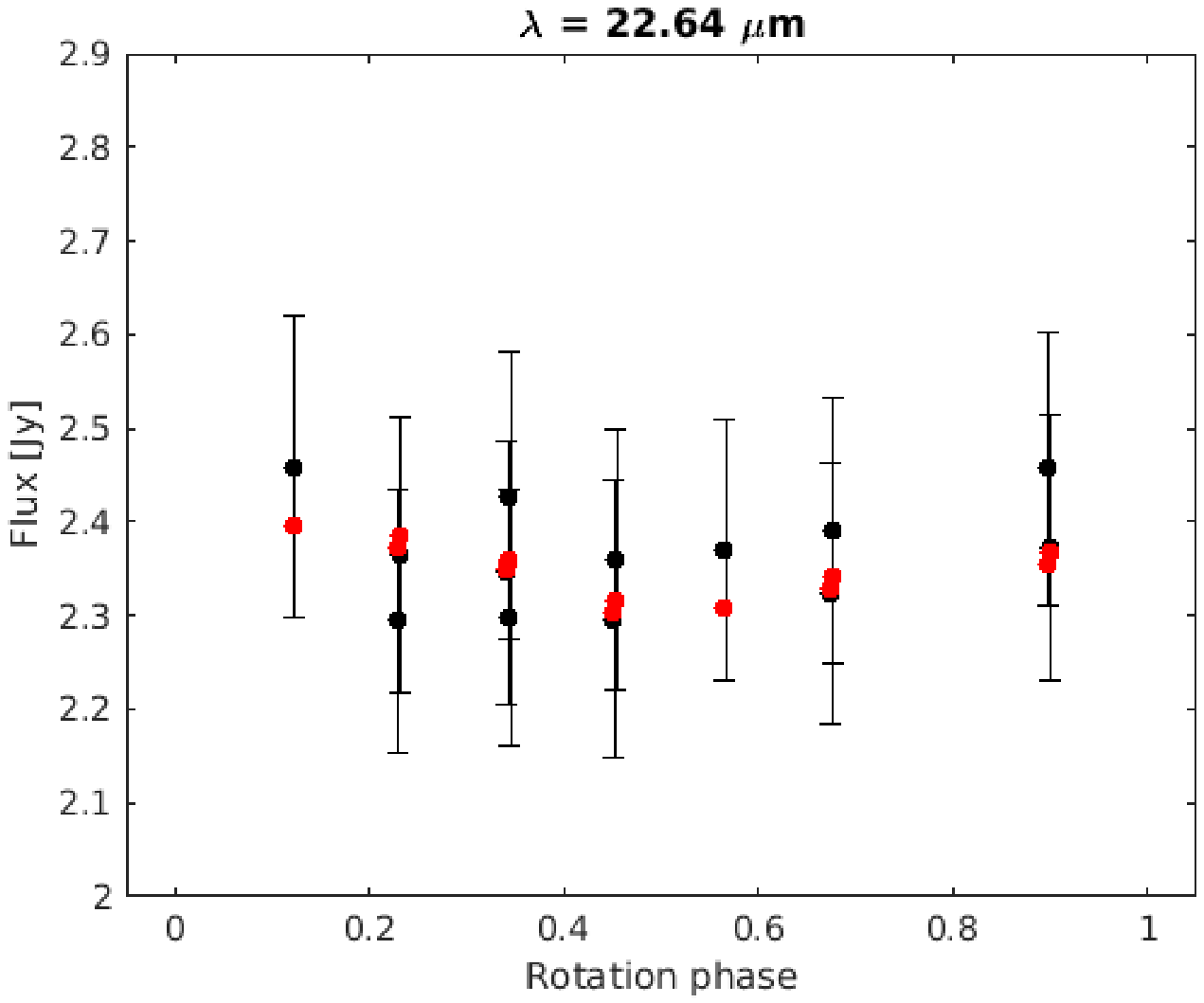}
\captionof{figure}{(108) Hecuba, thermal light curve in W4 band.}
\label{108thermal_lcW4}
&
 \includegraphics[width=0.3\textwidth]{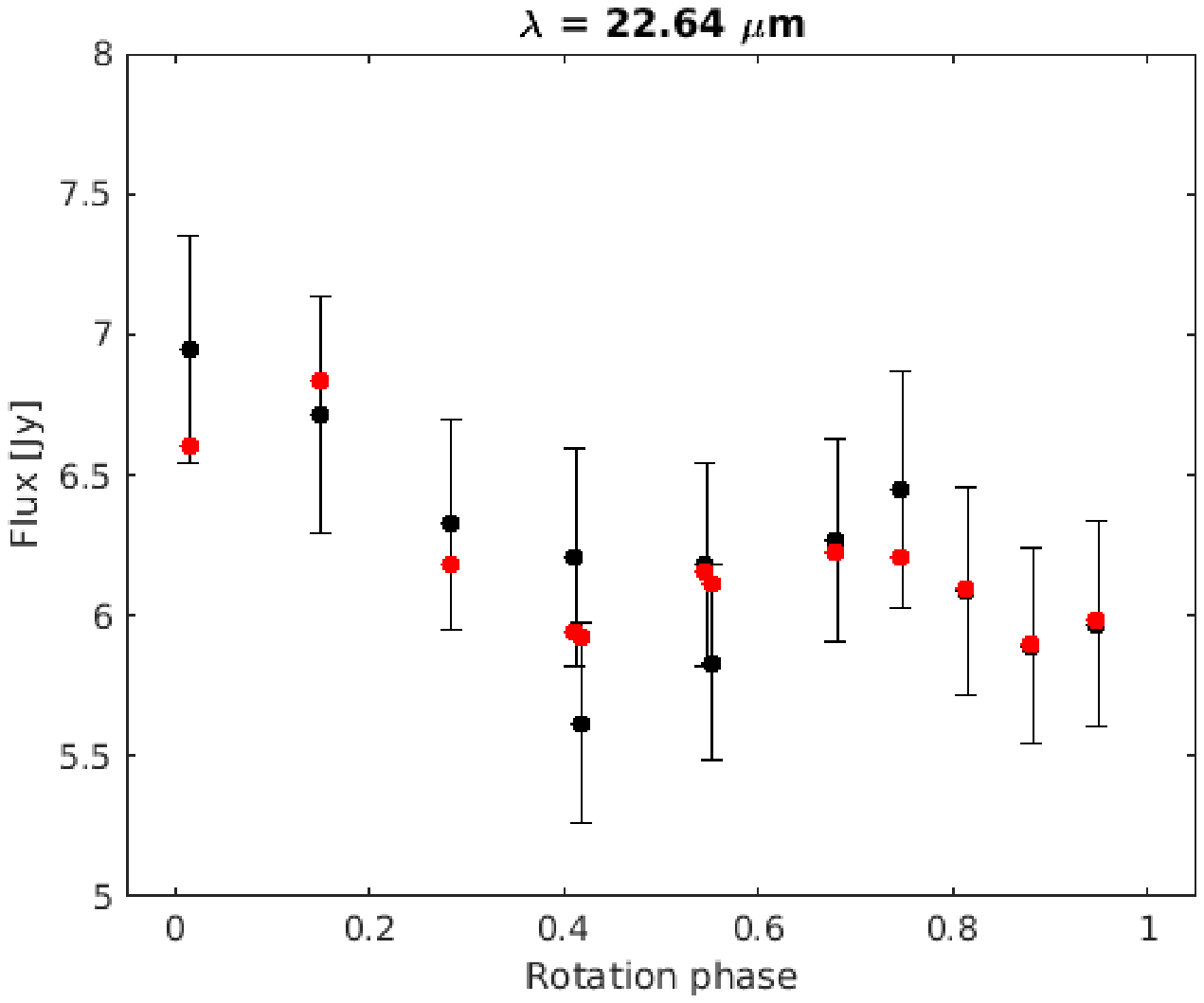}
\captionof{figure}{(202) Chryseis}
\label{202thermal_lc}
\\
 \includegraphics[width=0.3\textwidth]{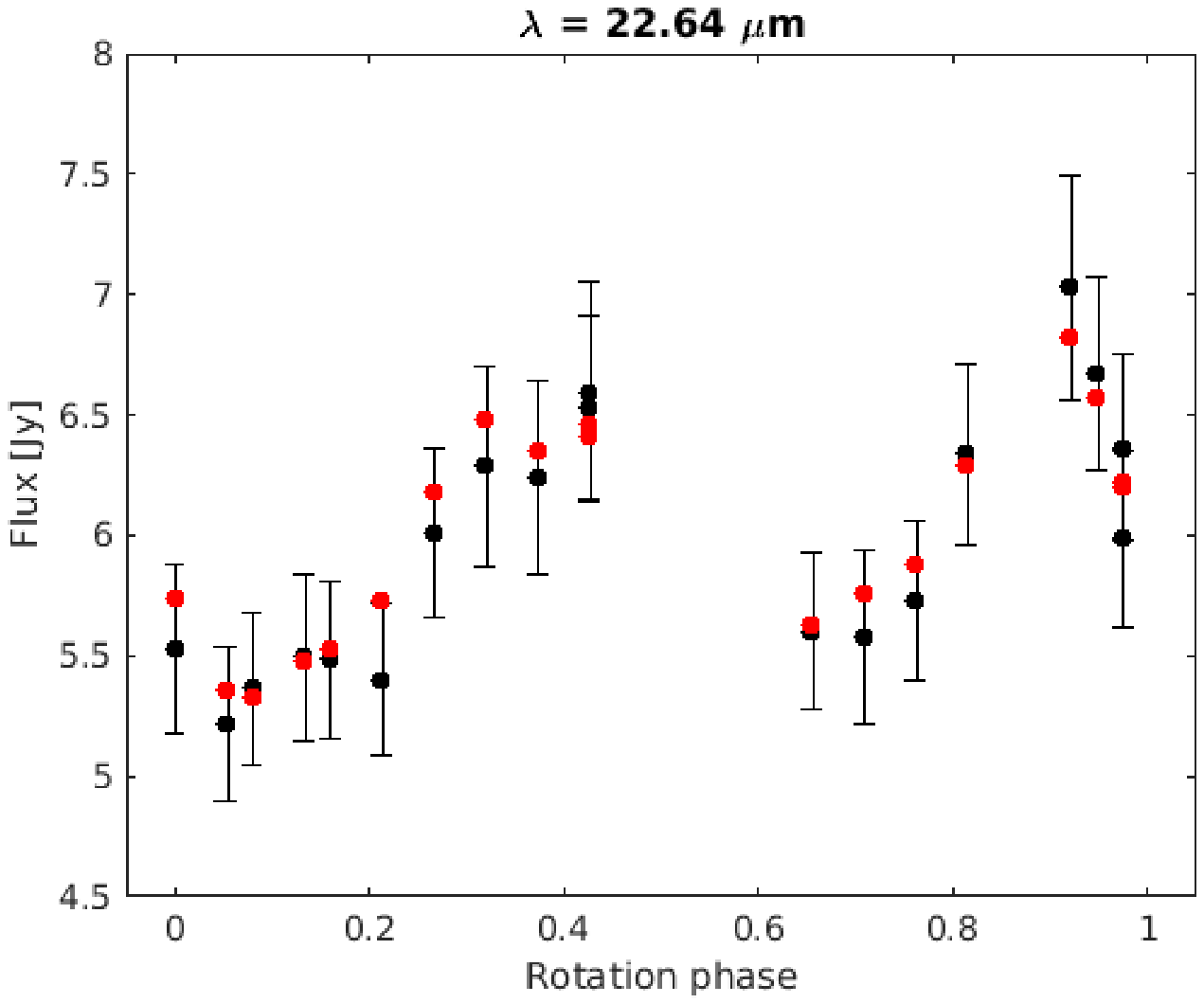}
\captionof{figure}{(219) Thusnelda}
\label{219thermal_lc}
&
 \includegraphics[width=0.3\textwidth]{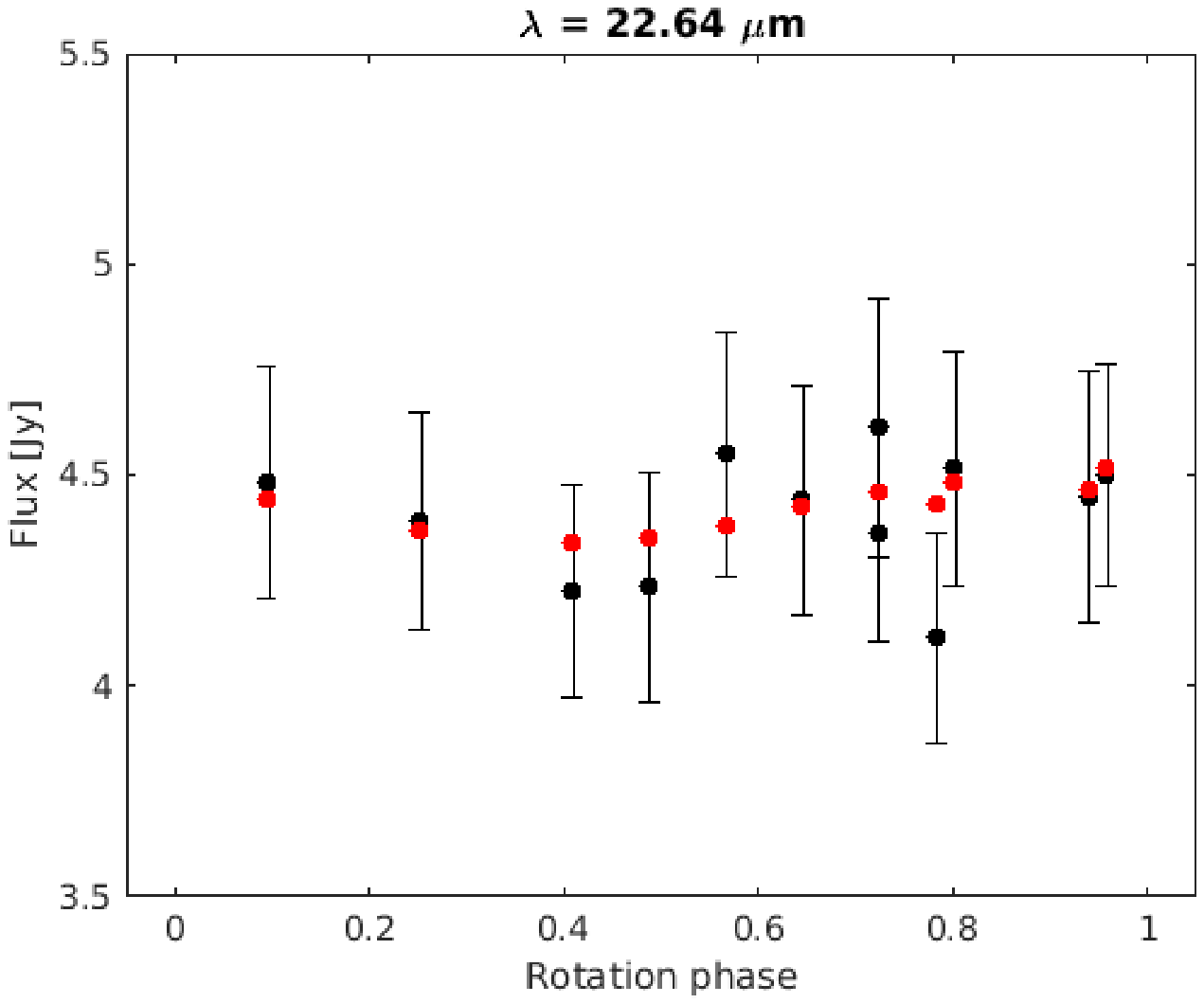}
\captionof{figure}{(223) Rosa}
\label{223thermal_lc}
&
 \includegraphics[width=0.3\textwidth]{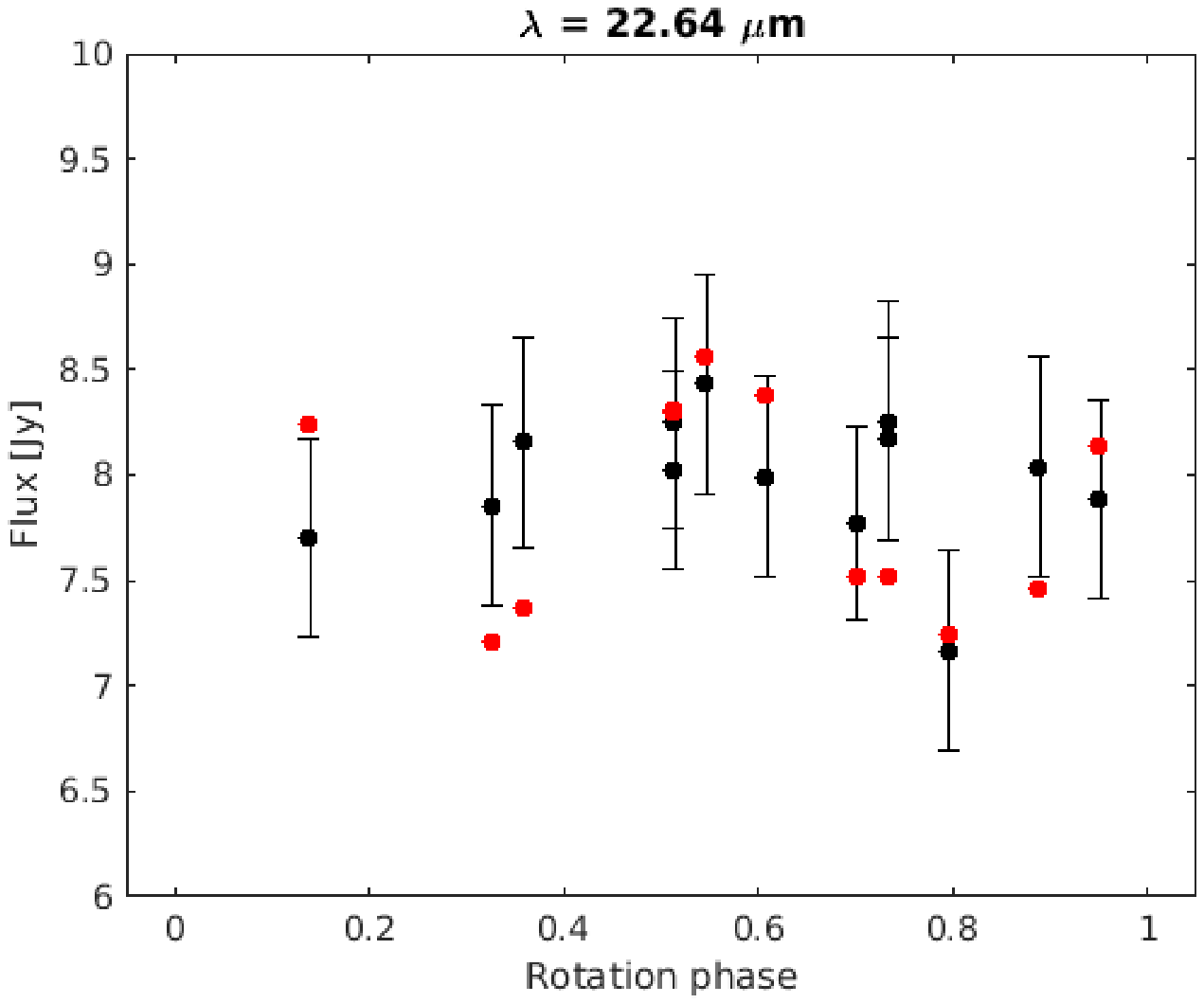}
\captionof{figure}{(362) Havnia}
\label{362thermal_lc}
\\
 \includegraphics[width=0.3\textwidth]{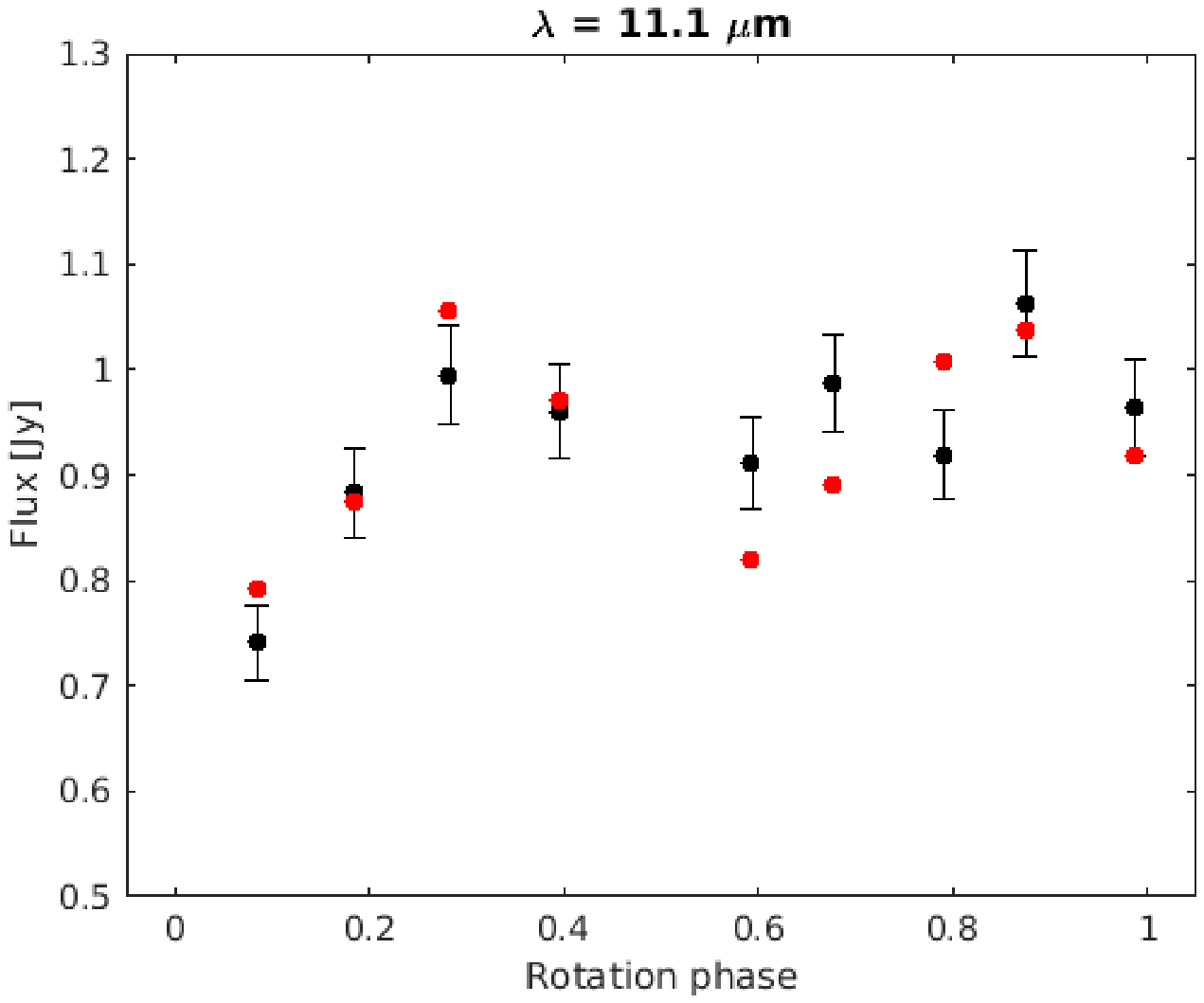}
\captionof{figure}{(478) Tergeste}
\label{478thermal_lcW3}
&
 \includegraphics[width=0.3\textwidth]{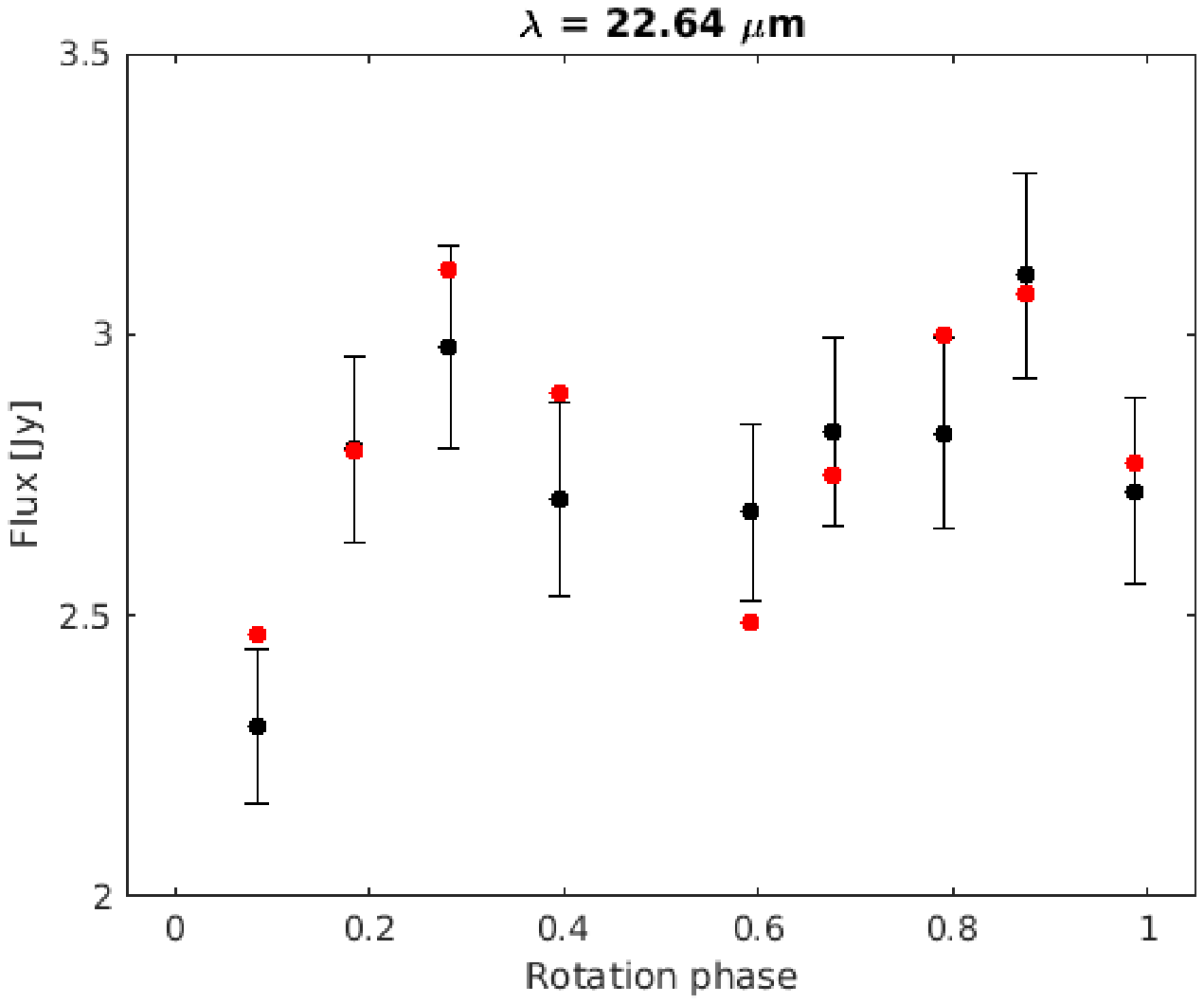}
\captionof{figure}{(478) Tergeste}
\label{478thermal_lcW4}
&
 \includegraphics[width=0.3\textwidth]{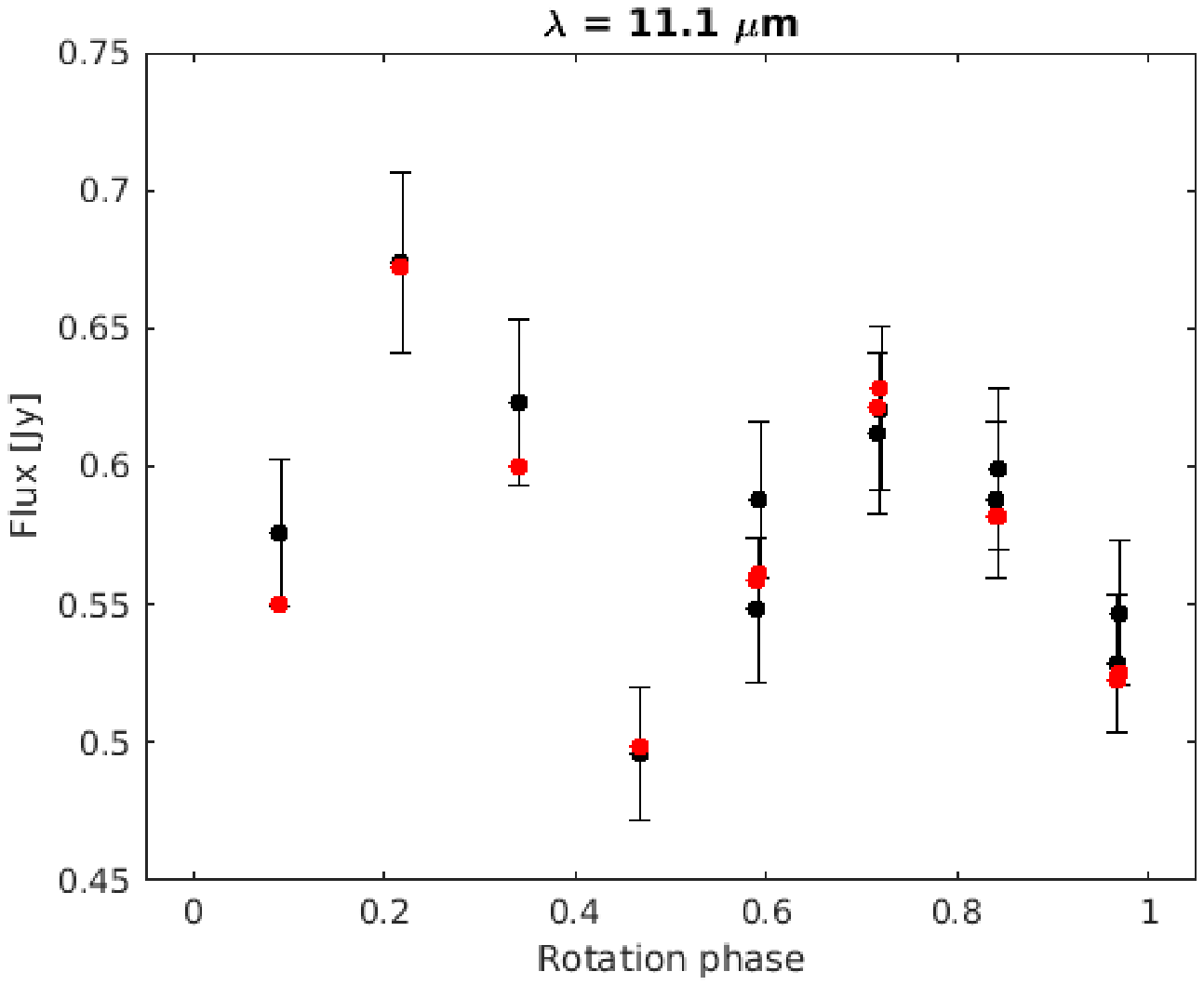}
\captionof{figure}{(483) Seppina}
\label{483thermal_lcW3}
\\
 \includegraphics[width=0.3\textwidth]{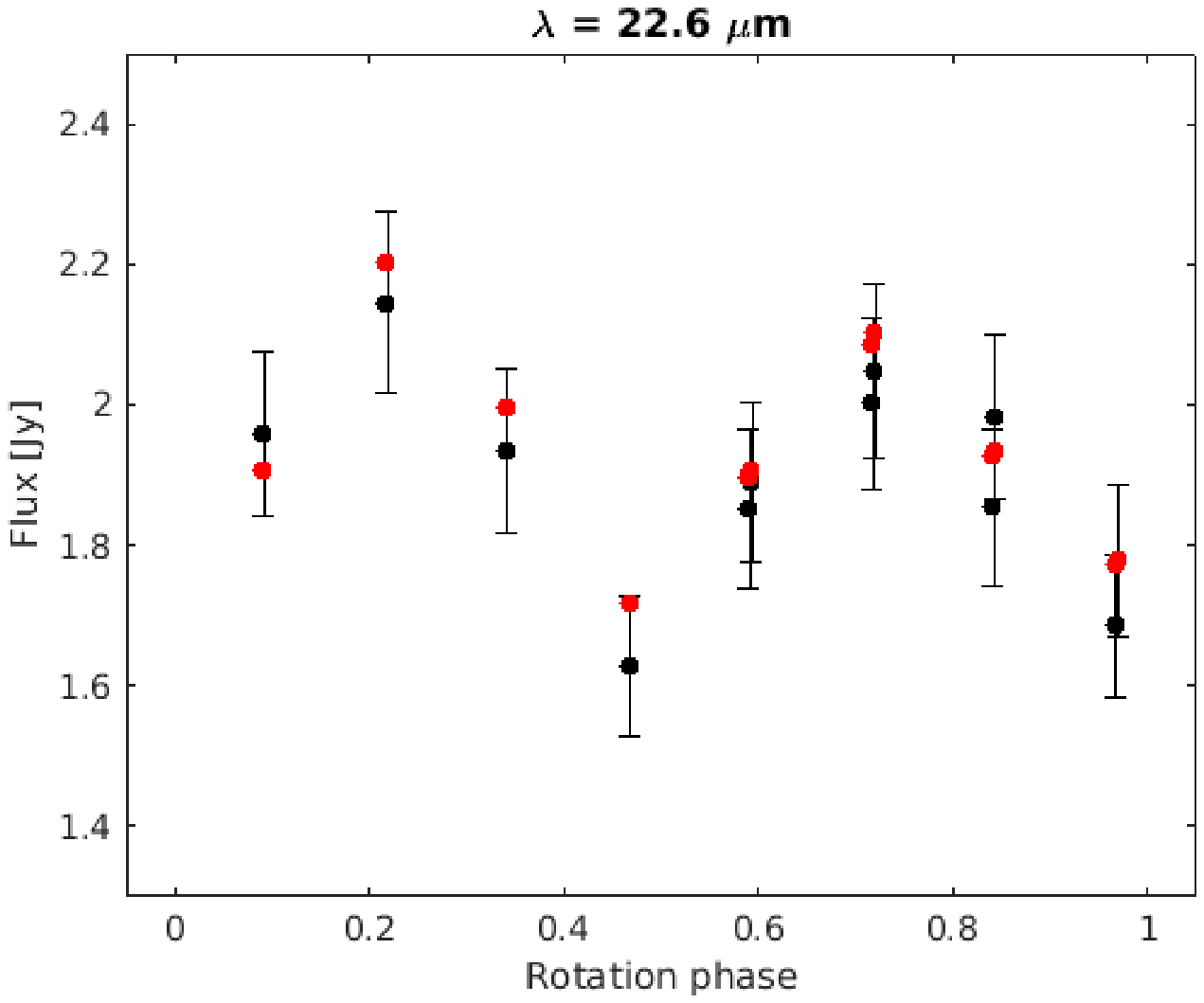}
\captionof{figure}{(483) Seppina}
\label{483thermal_lcW4}
&
 \includegraphics[width=0.3\textwidth]{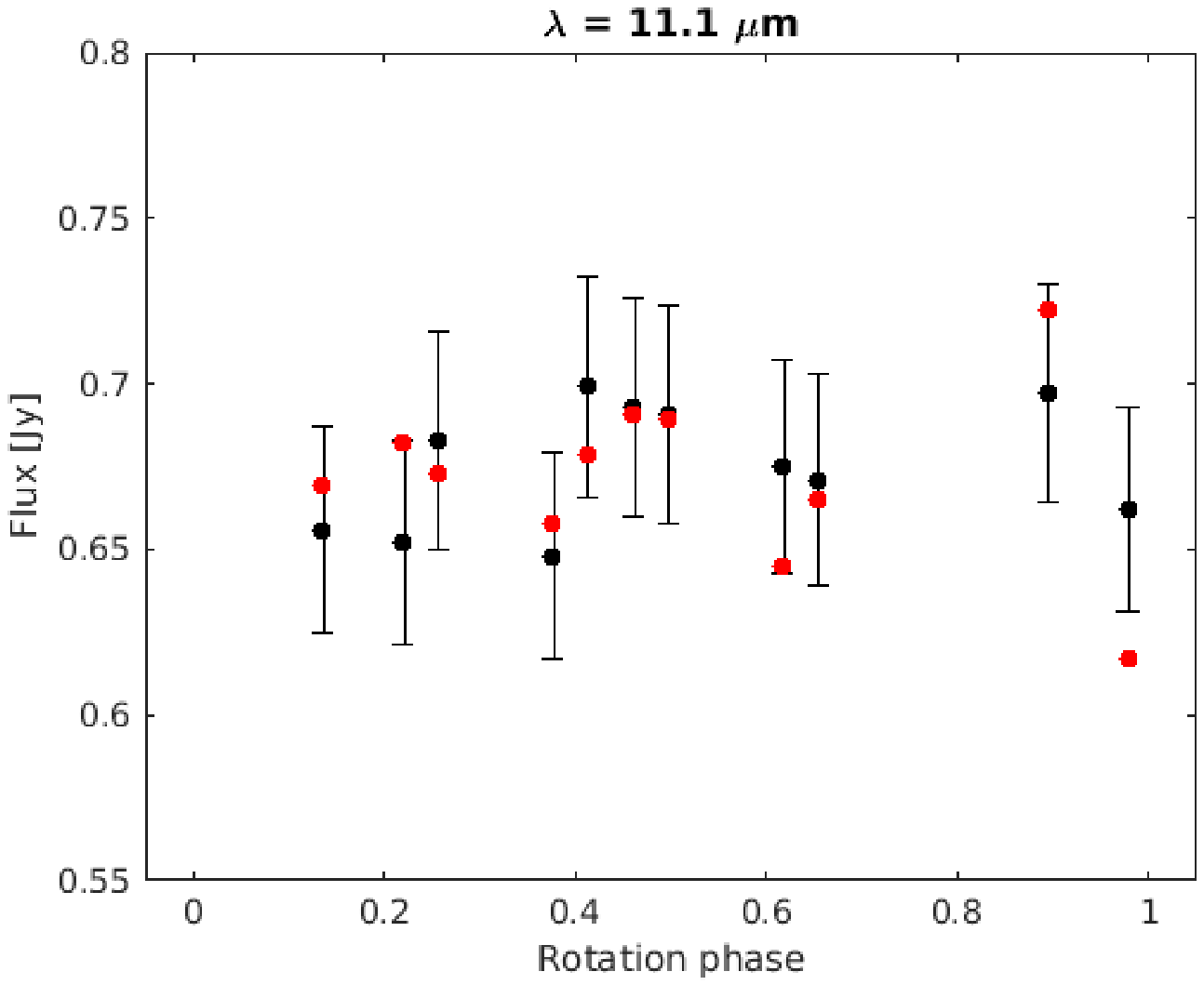}
\captionof{figure}{(501) Urhixidur}
\label{501thermal_lcW3}
&
 \includegraphics[width=0.3\textwidth]{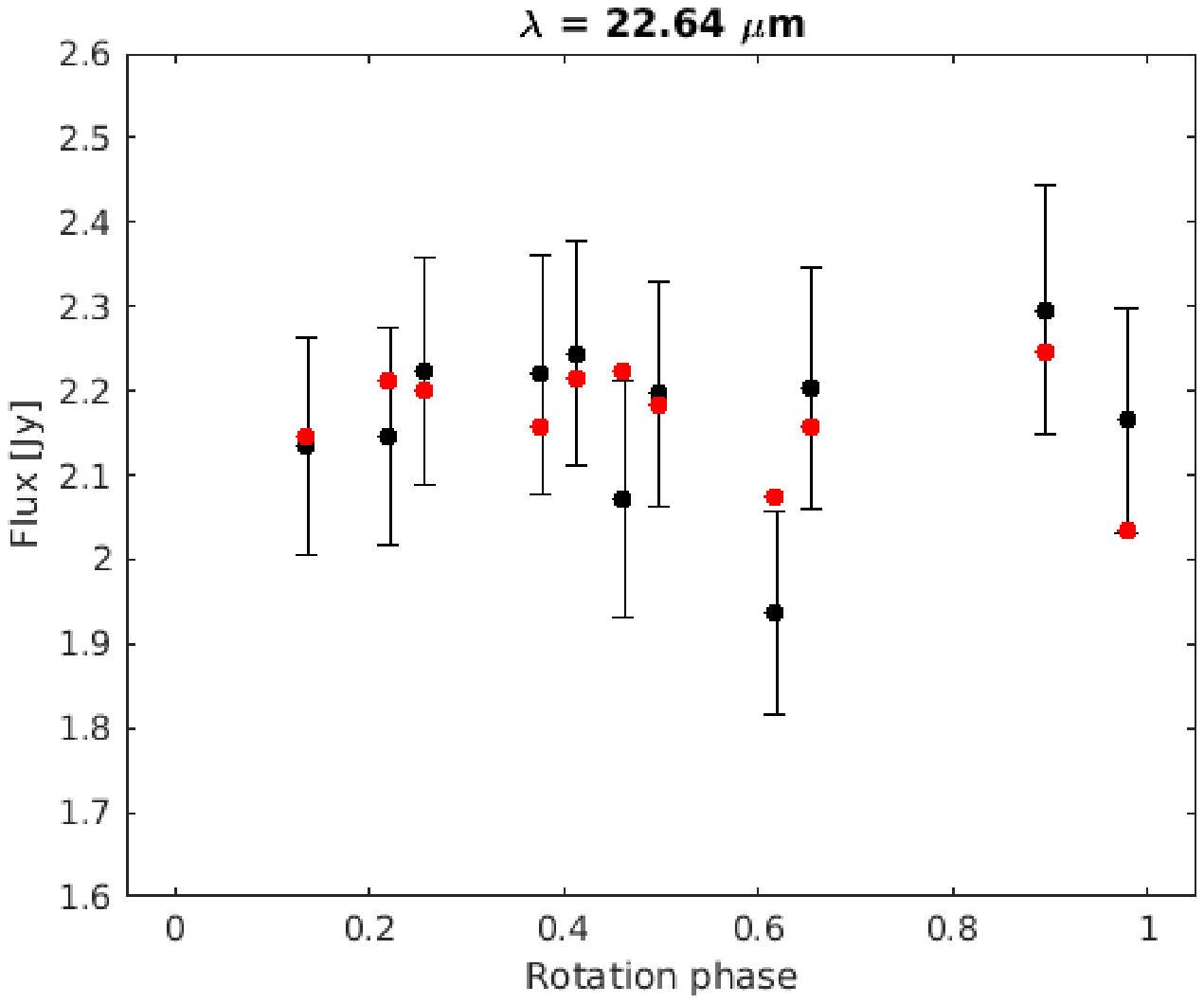}
\captionof{figure}{(501) Urhixidur}
\label{501thermal_lcW4}
\\
\end{tabularx}
    \end{table*}

    \begin{table*}[h]
    \centering
\vspace{0.5cm}
\begin{tabularx}{\linewidth}{XXX}
 \includegraphics[width=0.33\textwidth]{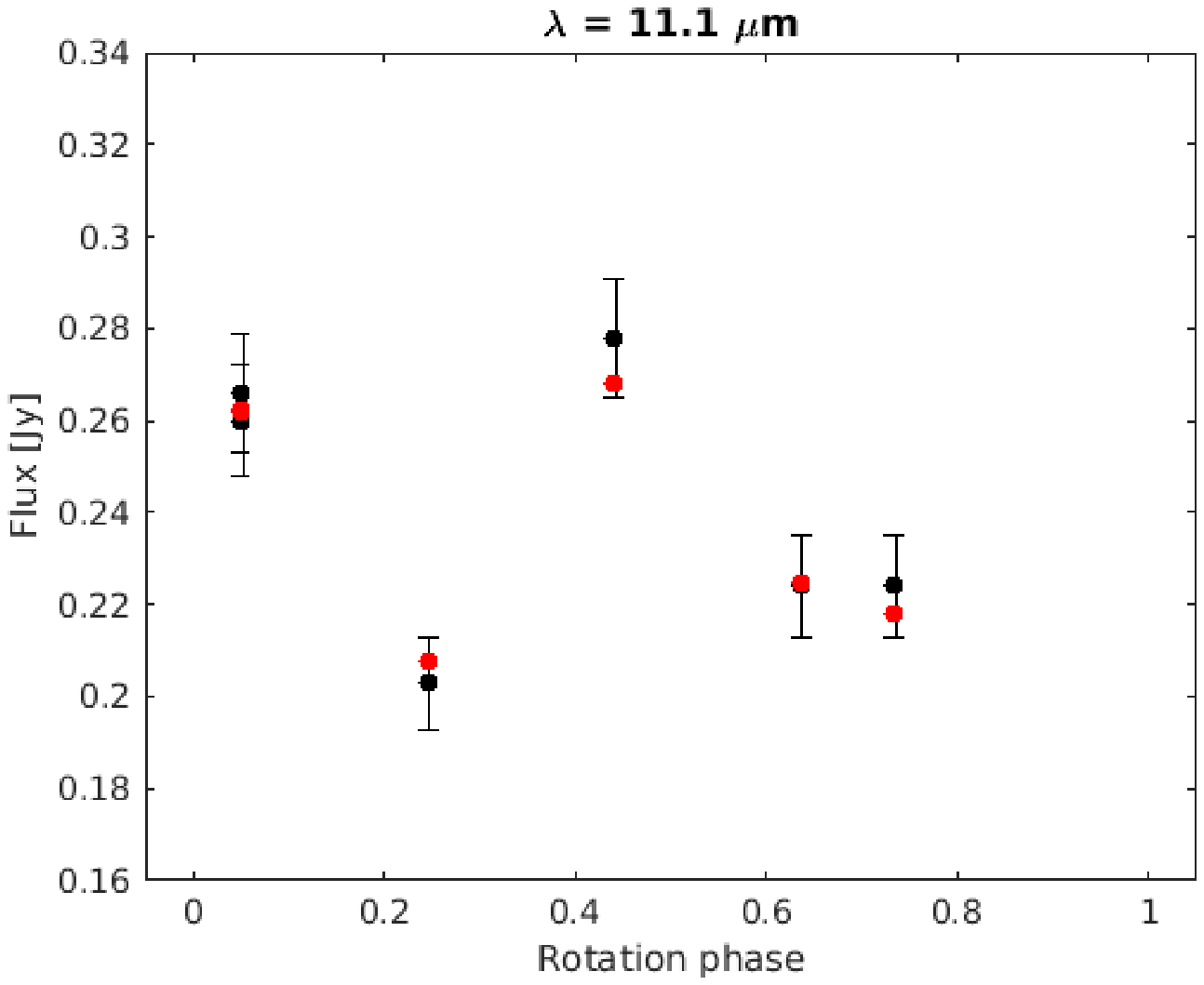}
\captionof{figure}{(537) Pauly}
\label{537thermal_lcW3}
&
 \includegraphics[width=0.33\textwidth]{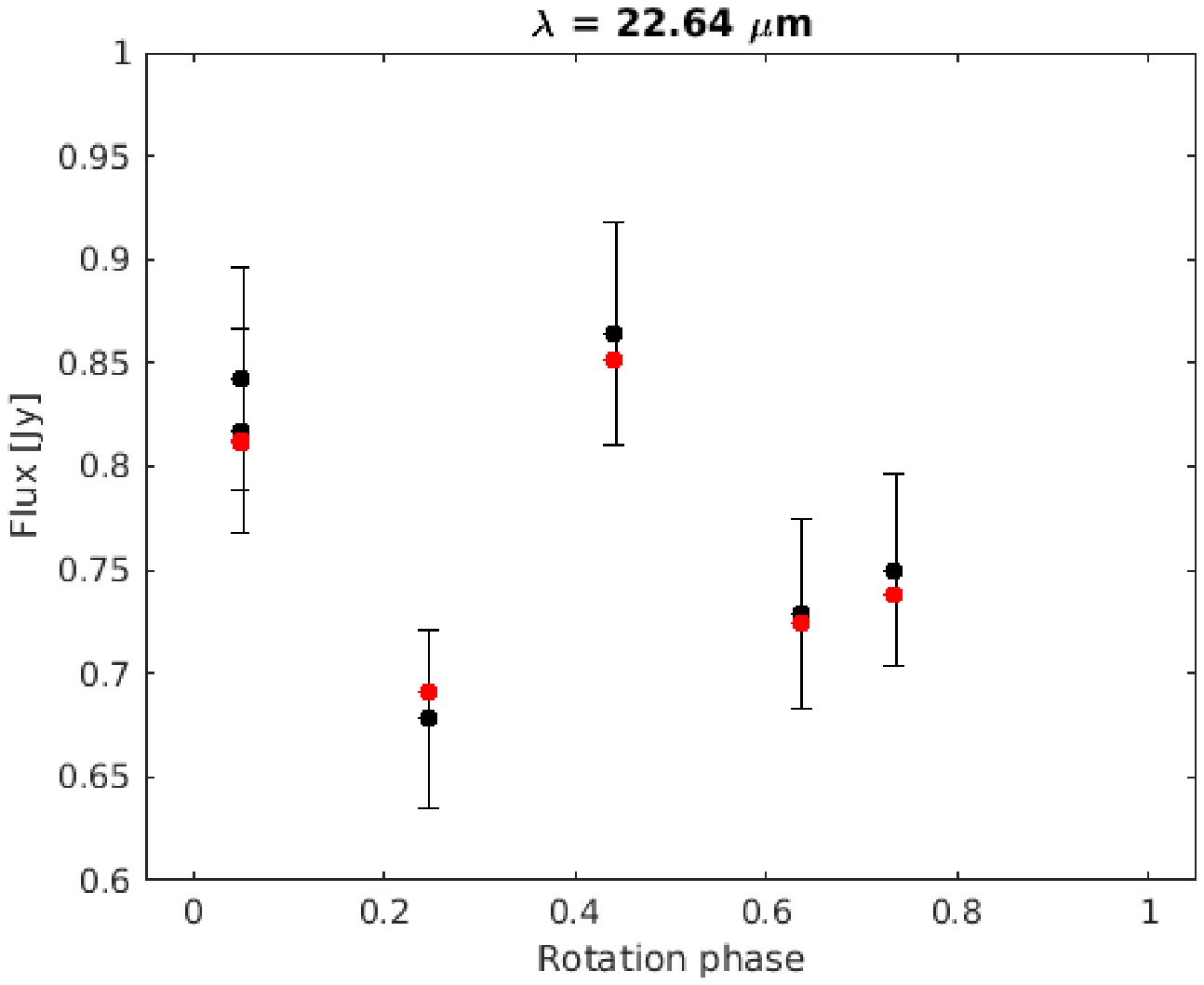}
\captionof{figure}{(537) Pauly}
\label{537thermal_lcW4}
&
 \includegraphics[width=0.33\textwidth]{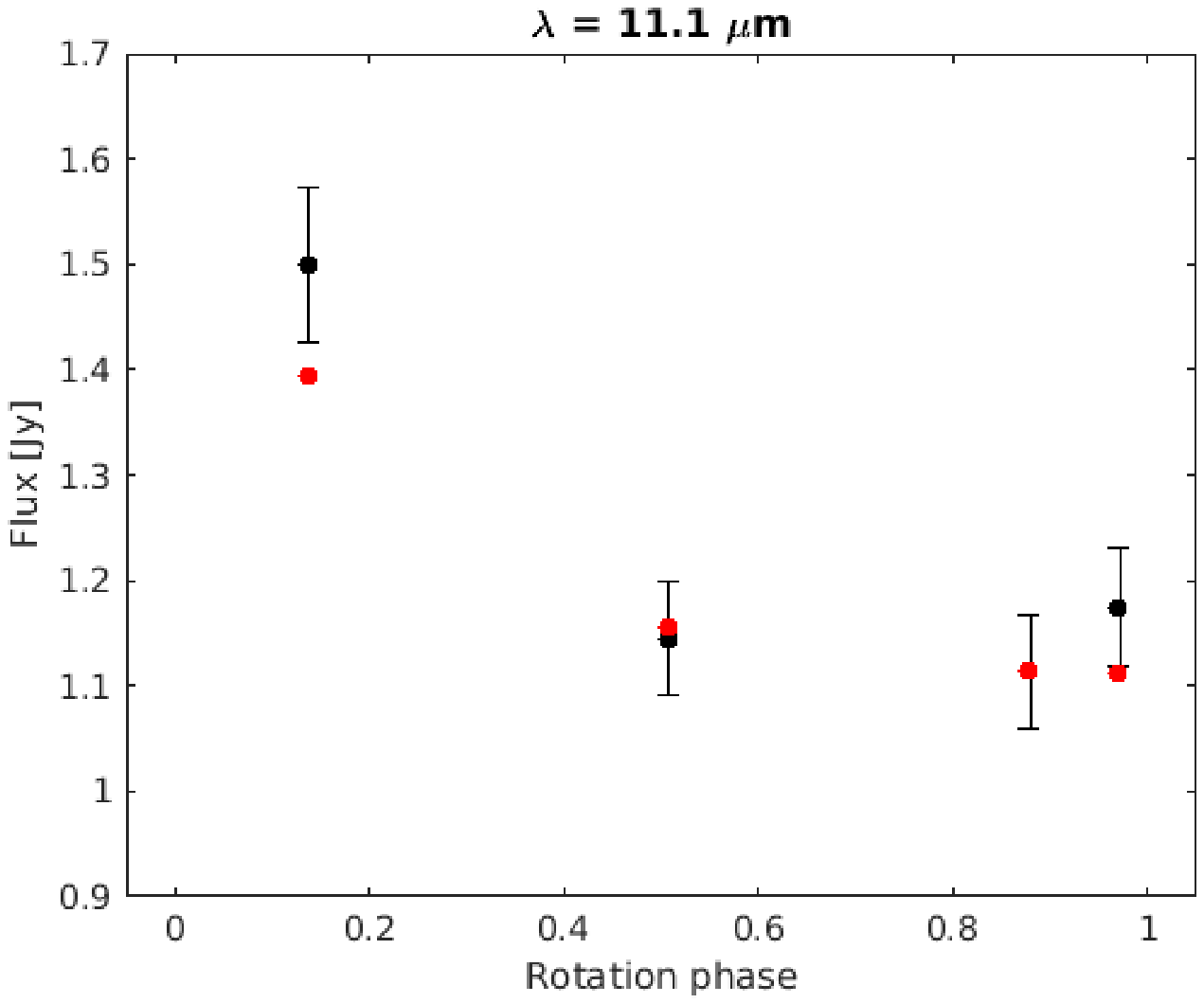}
\captionof{figure}{(552) Sigelinde}
\label{552thermal_lc}
\\
 \includegraphics[width=0.33\textwidth]{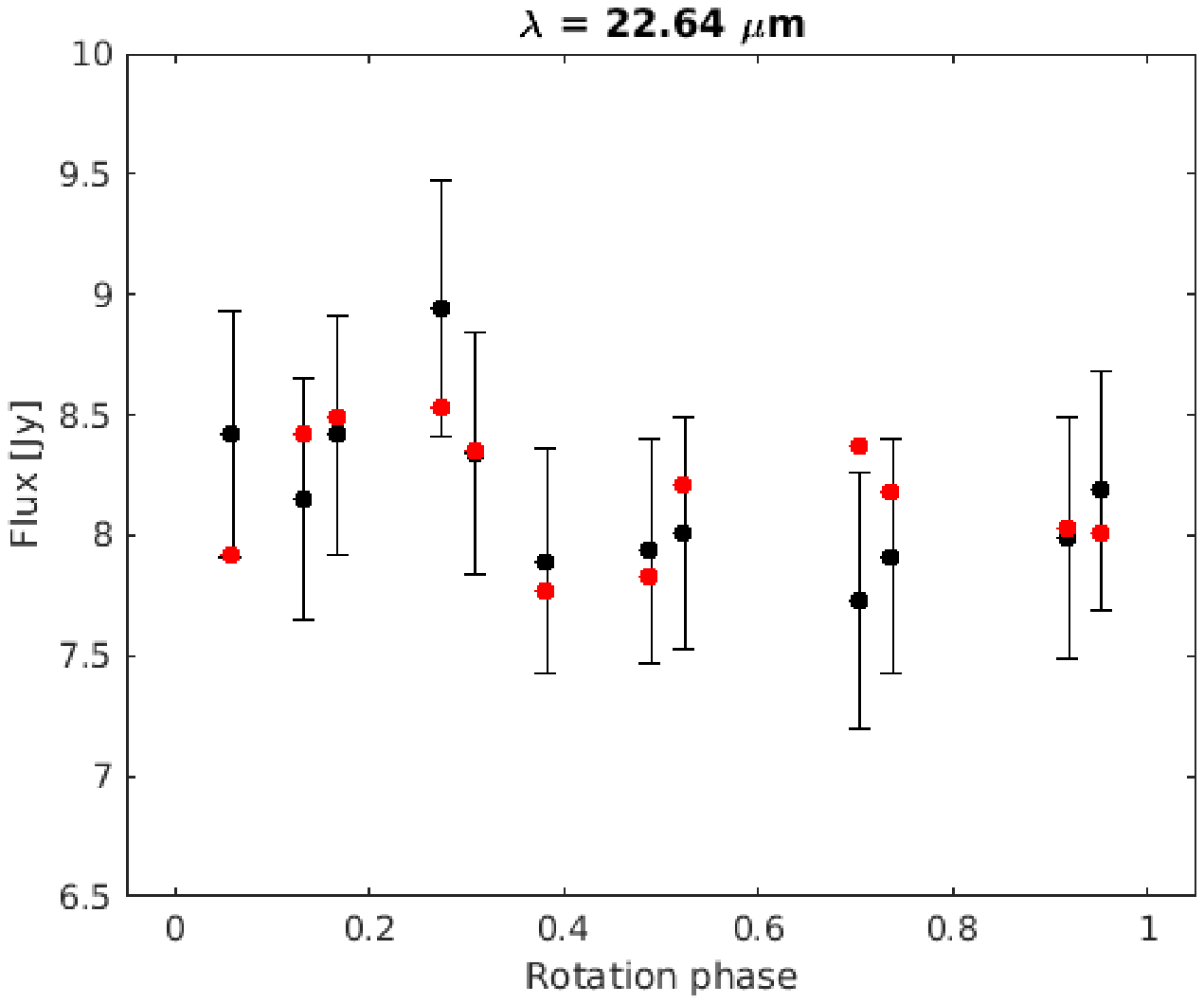}
\captionof{figure}{(618) Elfriede}
\label{618thermal_lc}
&
 \includegraphics[width=0.33\textwidth]{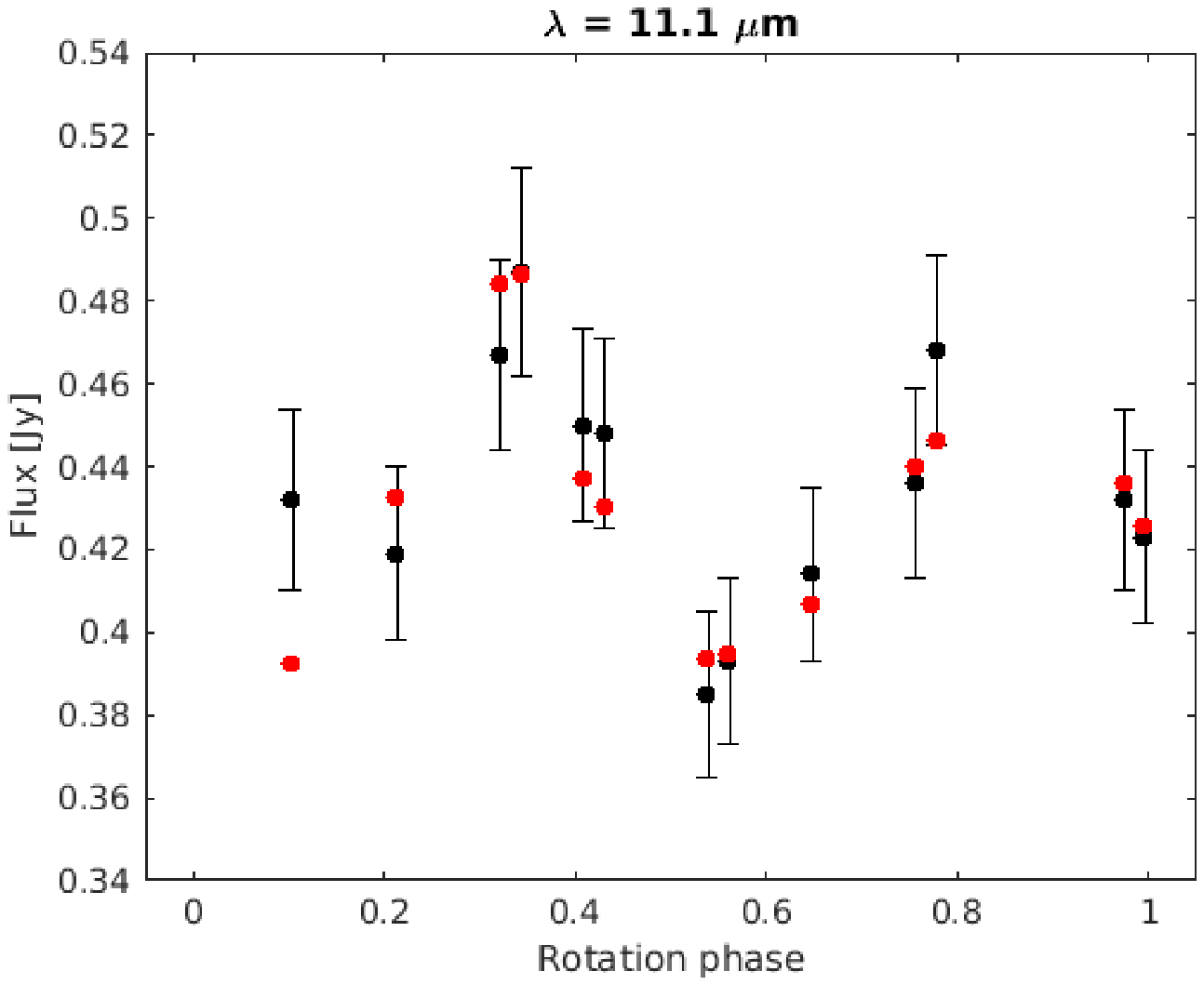}
\captionof{figure}{(666) Desdemona}
\label{666thermal_lcW3}
&
 \includegraphics[width=0.33\textwidth]{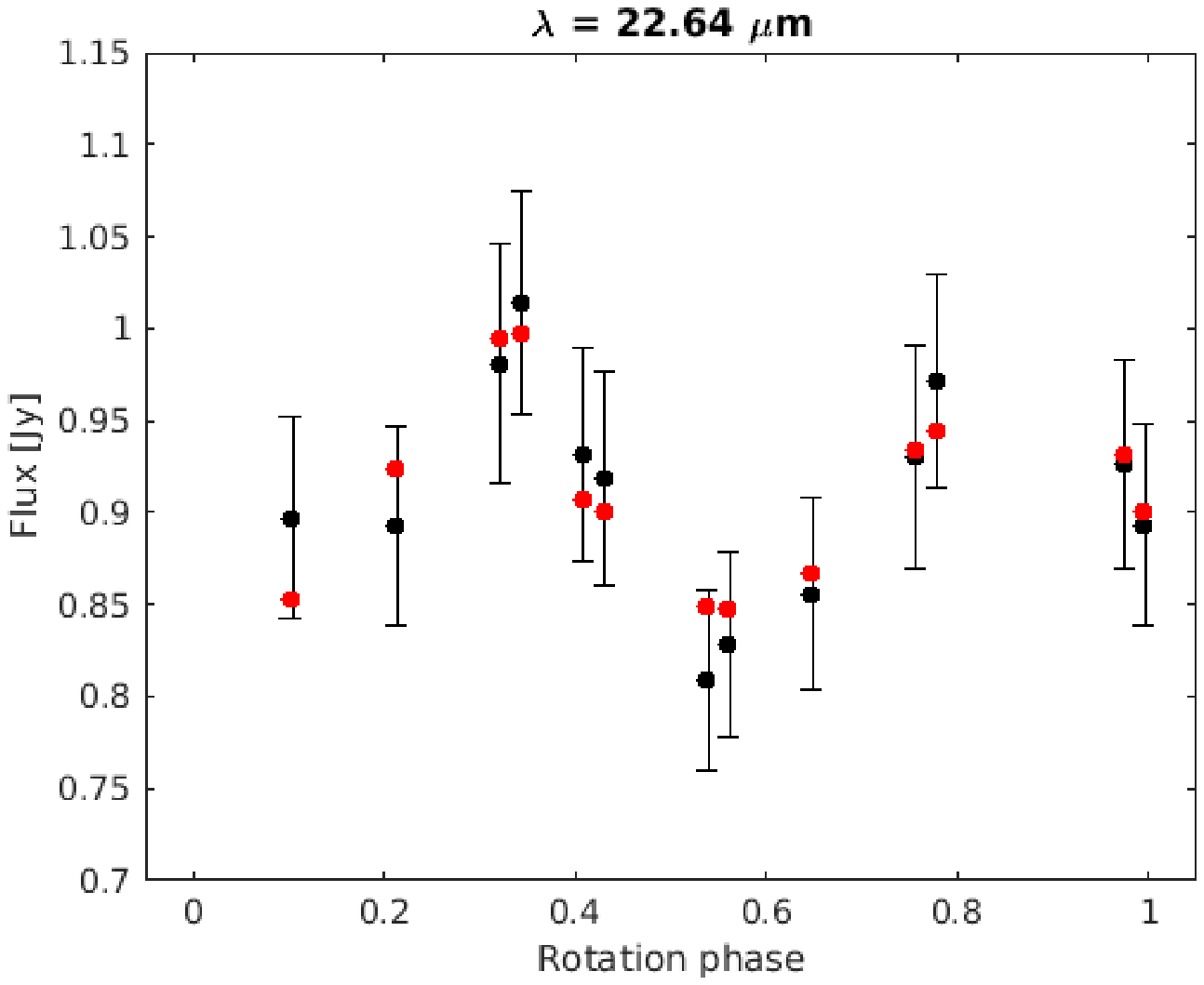}
\captionof{figure}{(666) Desdemona}
\label{666thermal_lcW4}
\\
 \includegraphics[width=0.33\textwidth]{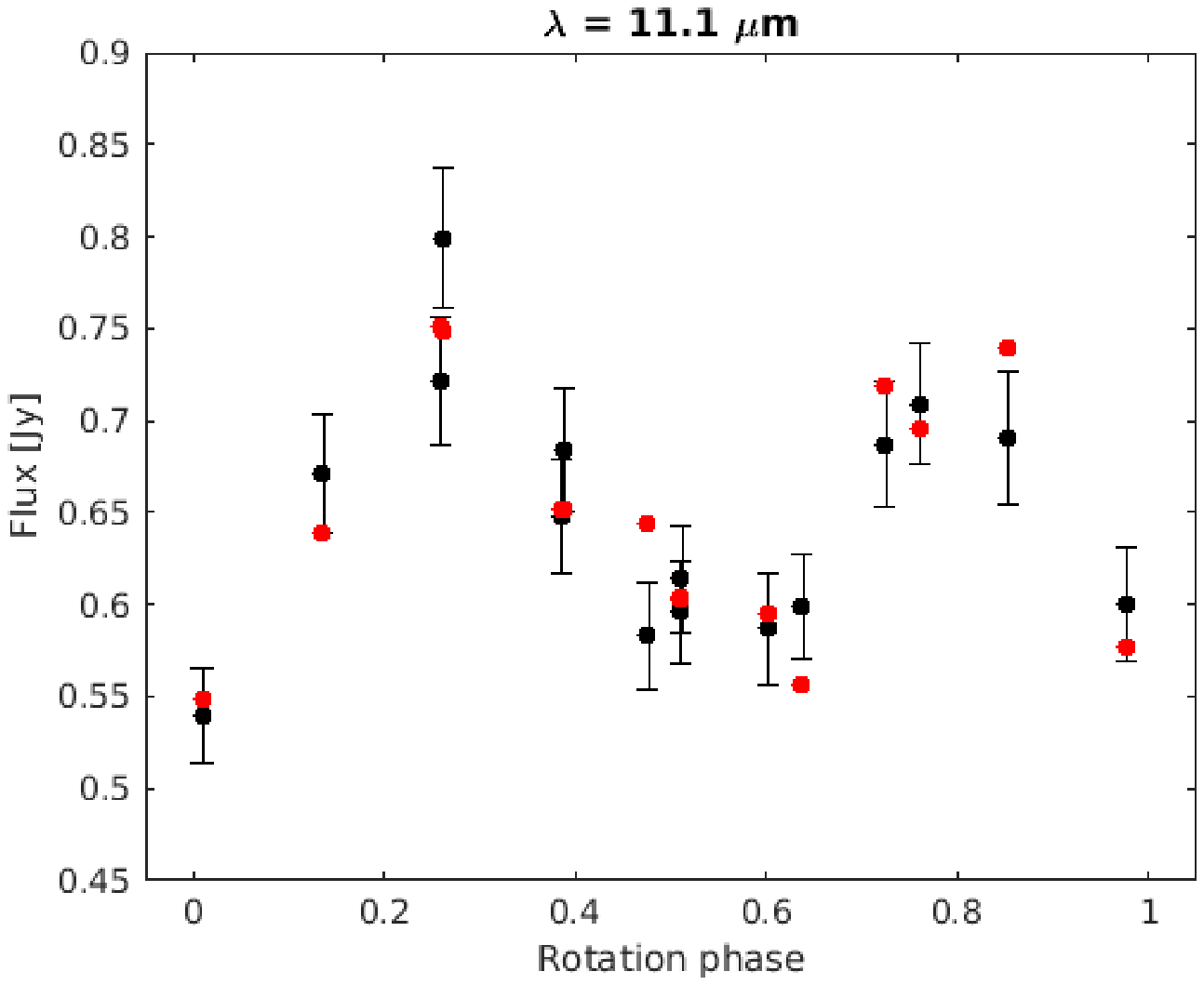}
\captionof{figure}{(667) Denise}
\label{667thermal_lcW3}
&
 \includegraphics[width=0.33\textwidth]{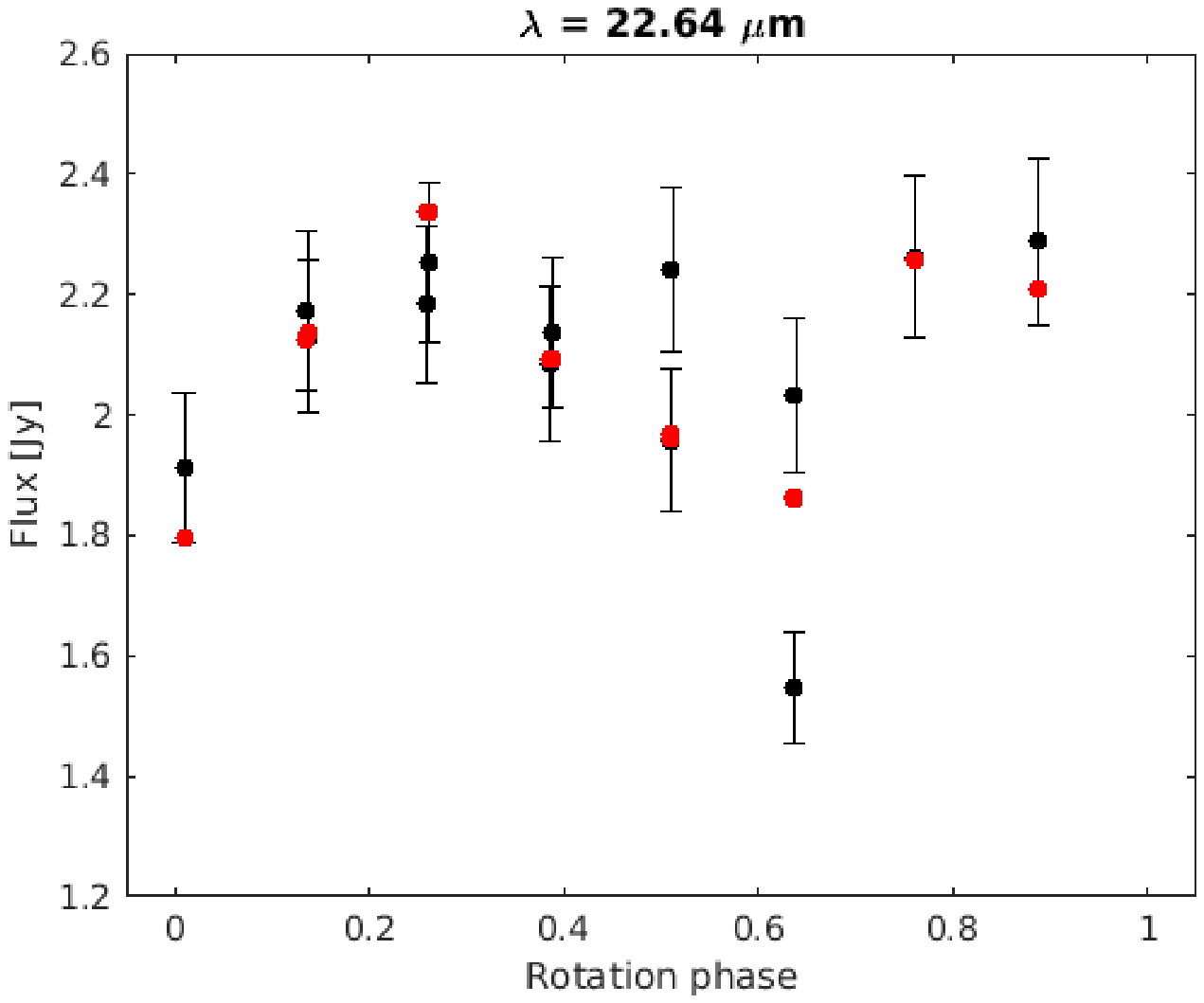}
\captionof{figure}{(667) Denise}
\label{667thermal_lcW4}
&
 \includegraphics[width=0.33\textwidth]{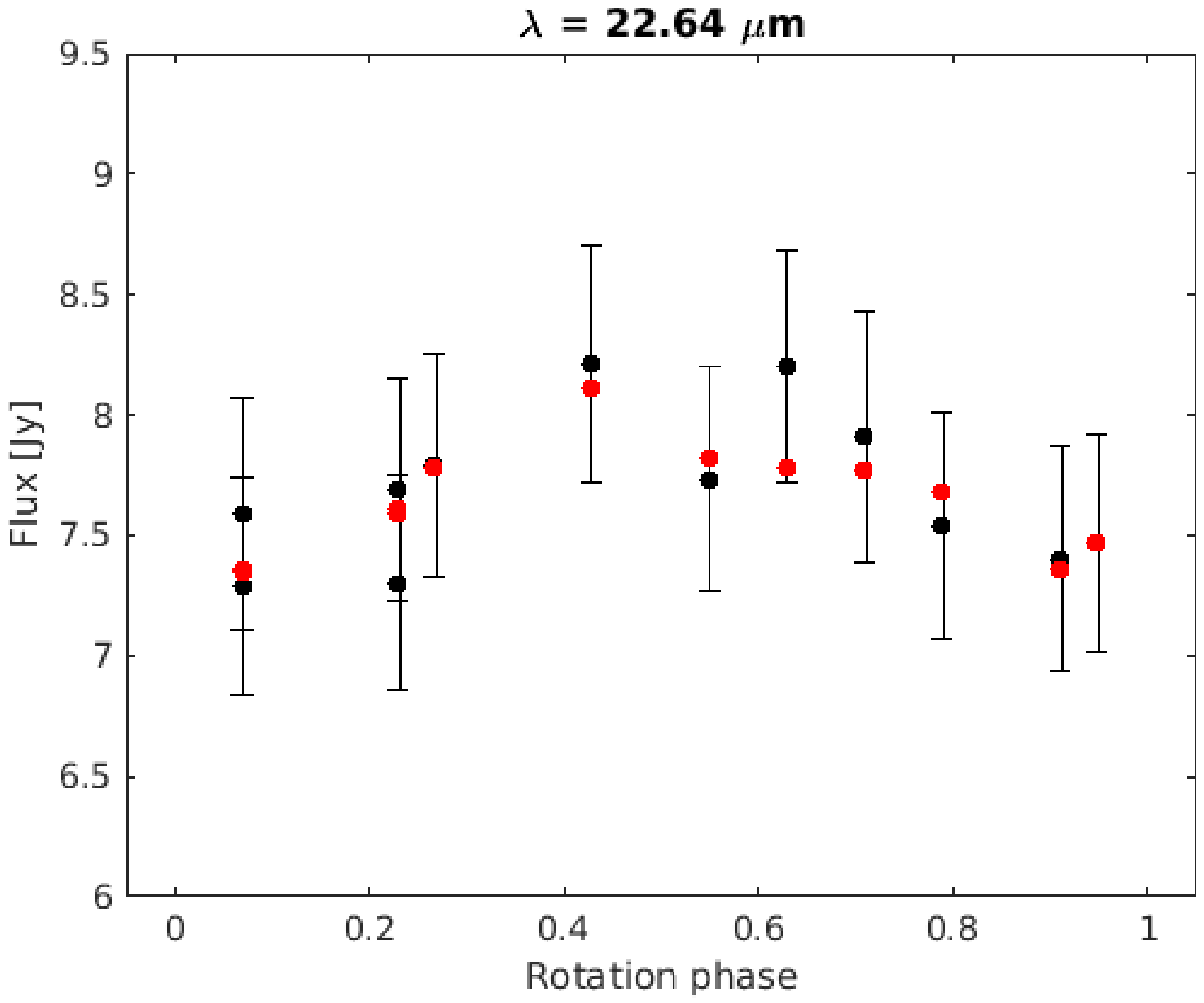}
\captionof{figure}{(780) Armenia}
\label{780thermal_lc}
\\
 \includegraphics[width=0.33\textwidth]{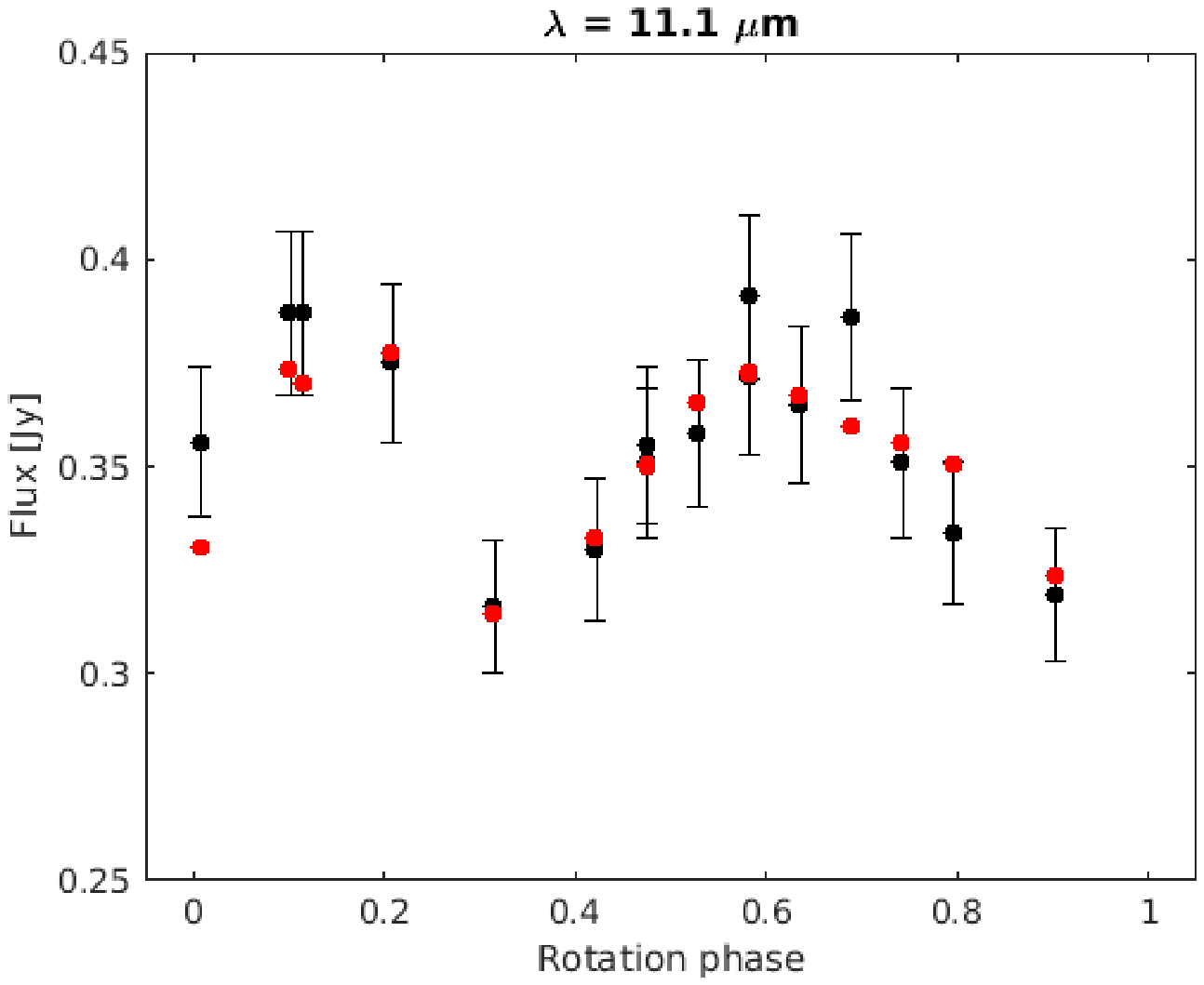}
\captionof{figure}{(923) Herluga}
\label{923thermal_lcW3}
&
 \includegraphics[width=0.33\textwidth]{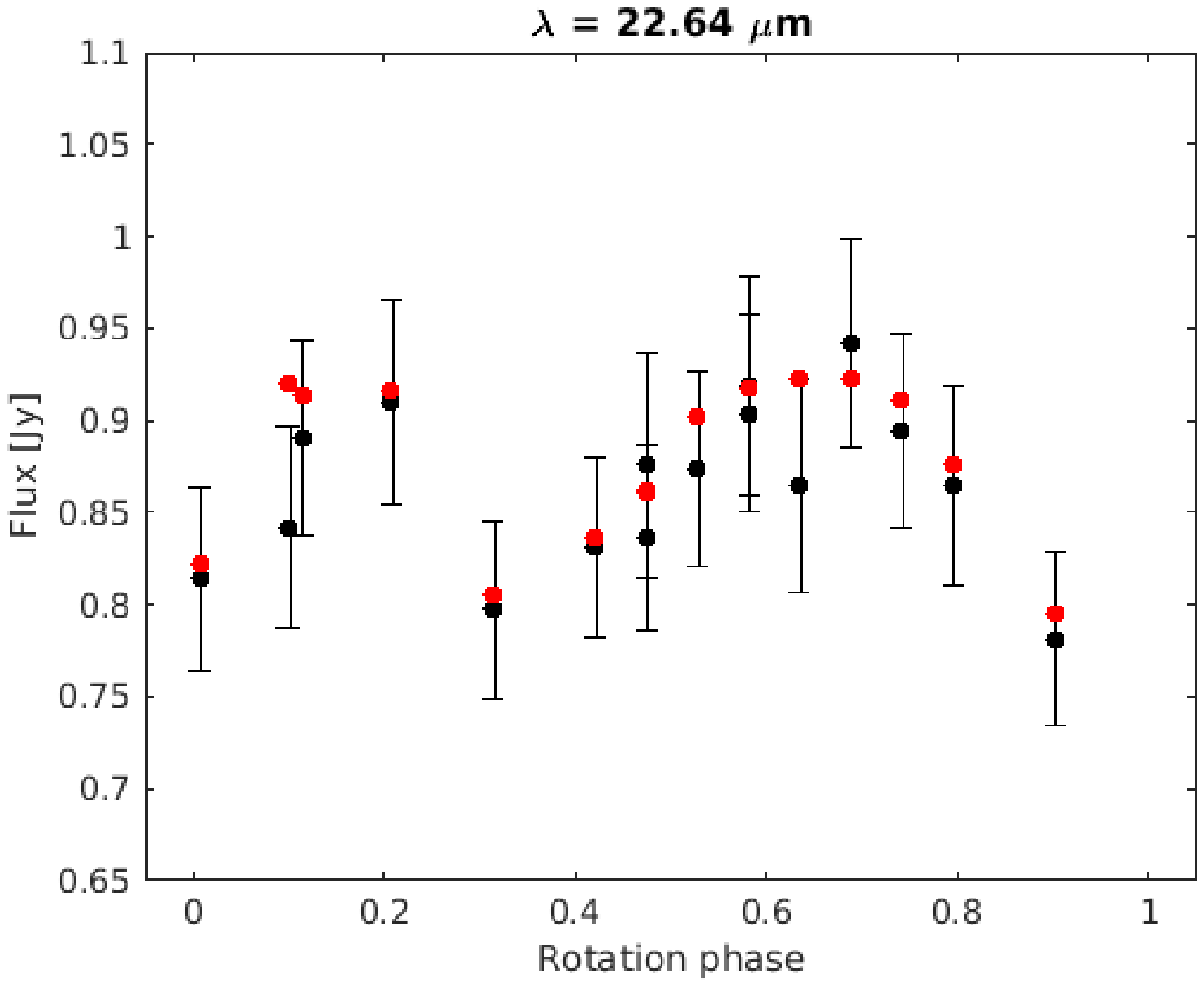}
\captionof{figure}{(923) Herluga}
\label{923thermal_lcW4}
&
 \includegraphics[width=0.33\textwidth]{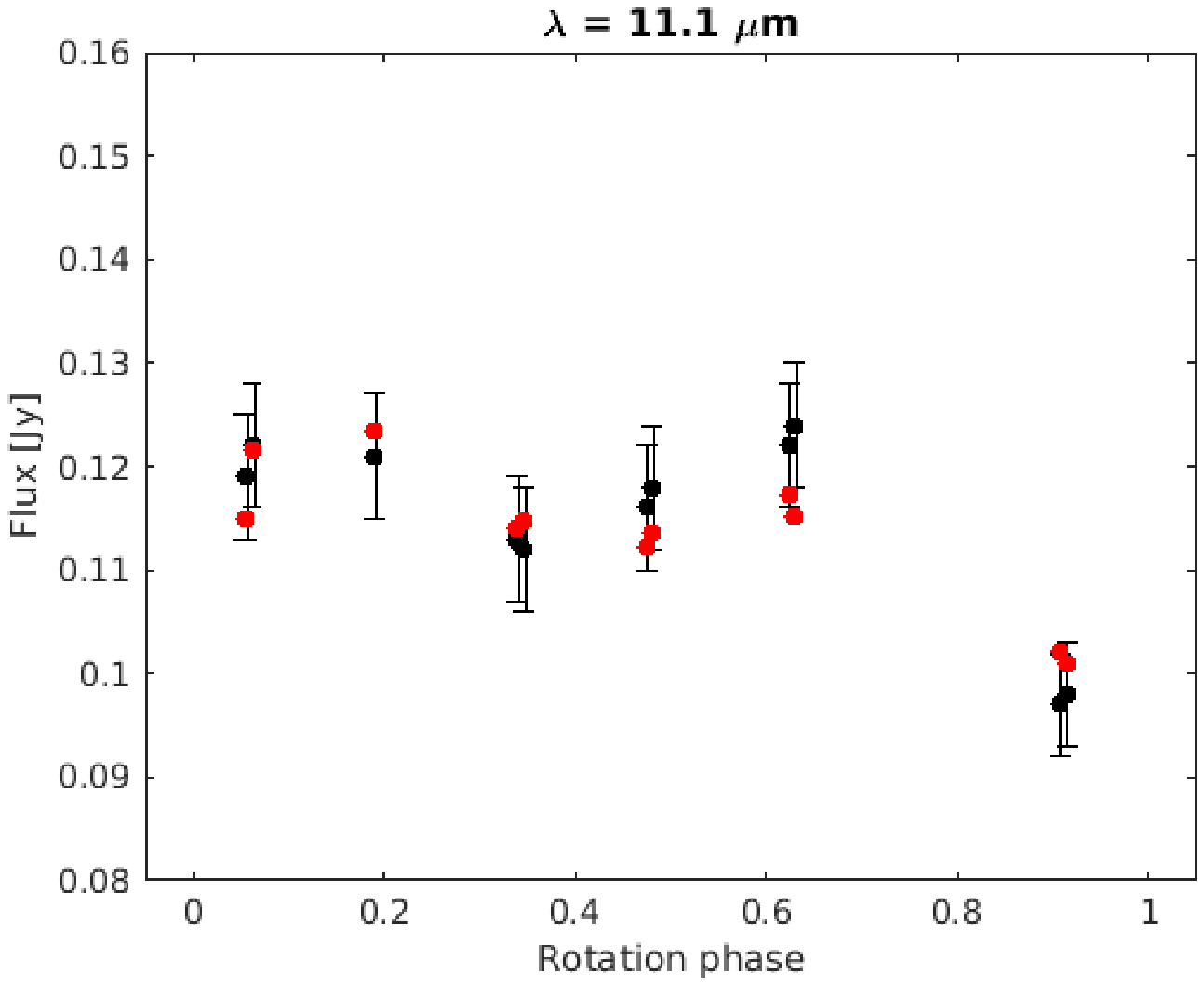}
\captionof{figure}{(995) Sternberga}
\label{995thermal_lcW3}
\\
\end{tabularx}
    \end{table*}
    \begin{table*}[t!]
    \centering
\vspace{0.5cm}
\begin{tabularx}{\linewidth}{XXX}
 \includegraphics[width=0.33\textwidth]{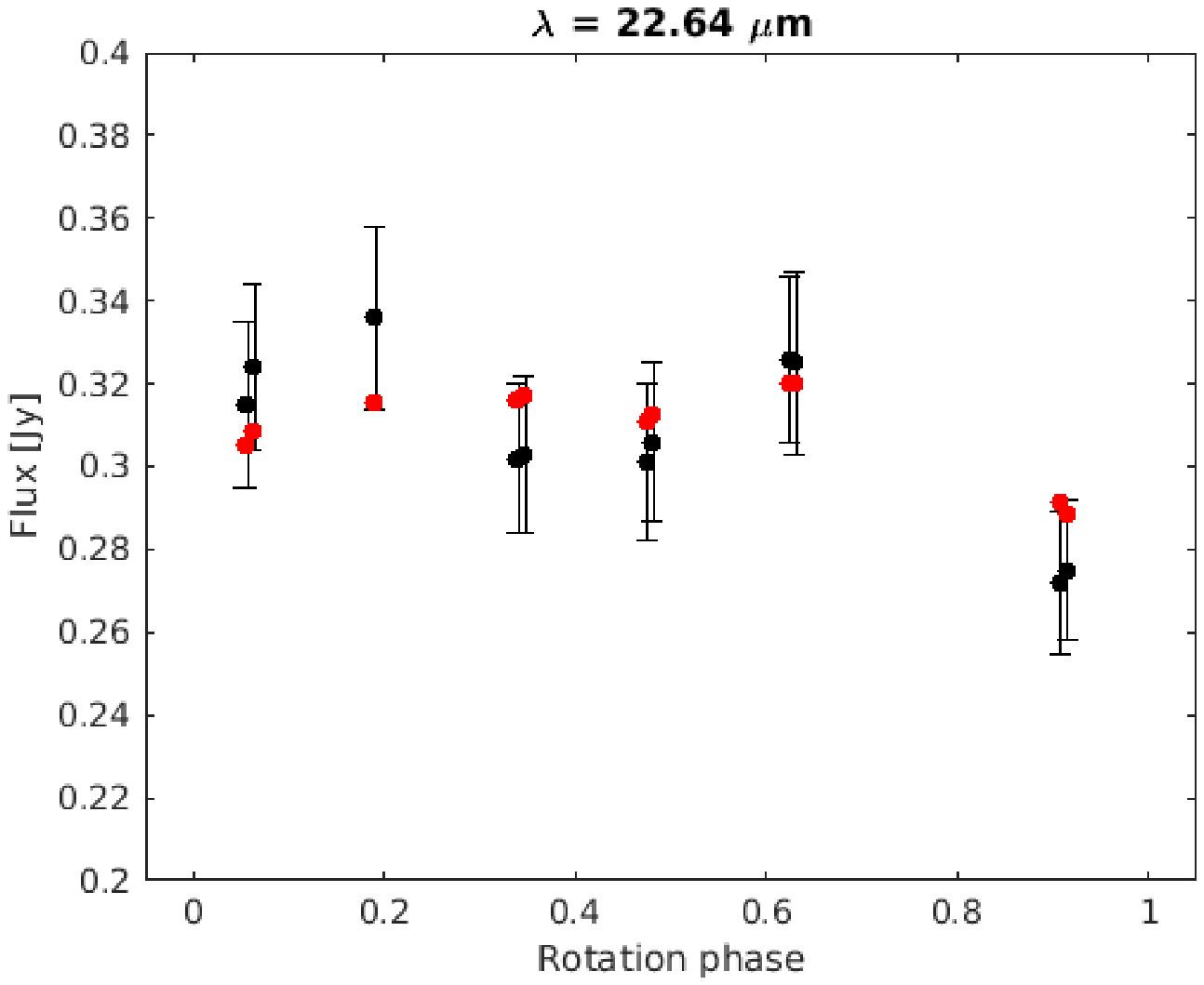}
\captionof{figure}{(995) Sternberga}
\label{995thermal_lcW4}
&
&
\\
\end{tabularx}
    \end{table*}

\clearpage
%\newpage

\vspace{2cm}
\section{Occultation fits}
 Instantaneous silhouettes of shape models from this work fitted to occultation timing chords.
\begin{figure}[h!]
\begin{center}
%\begin{tabular}{cc}
 \includegraphics[width=0.44\textwidth]{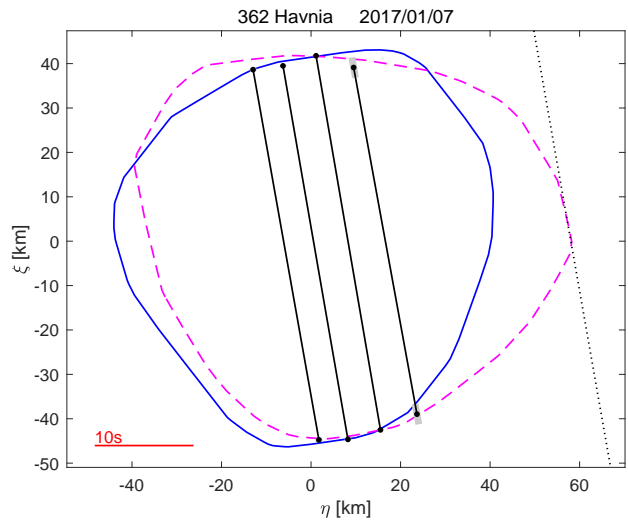}
%\end{tabular}
\caption{CITPM shape models of asteroid (362) Havnia fitted to a stellar occultation 
    from 7. January 2017. In all the figures, north is up and west is right. 
 The blue solid contour and the magenta dashed contour represent the model for pole 1 and pole 2, respectively.
 Black lines in those figures mark occultation shadow chords calculated from occultation timings, with timing uncertainties shown 
 at the extremities of each chord. The scale in seconds is given for reference as a red line. 
 Negative (no occultation) chords are marked with dotted lines, while visual observations (as opposed to video or photoelectric) 
 are marked with dashed lines. See Table \ref{occult} for diameters of equivalent volume spheres.}
\label{Havnia_occult}
\end{center}
\end{figure}

\begin{figure*}[h]
\begin{tabular}{cc}
 \includegraphics[width=0.45\textwidth]{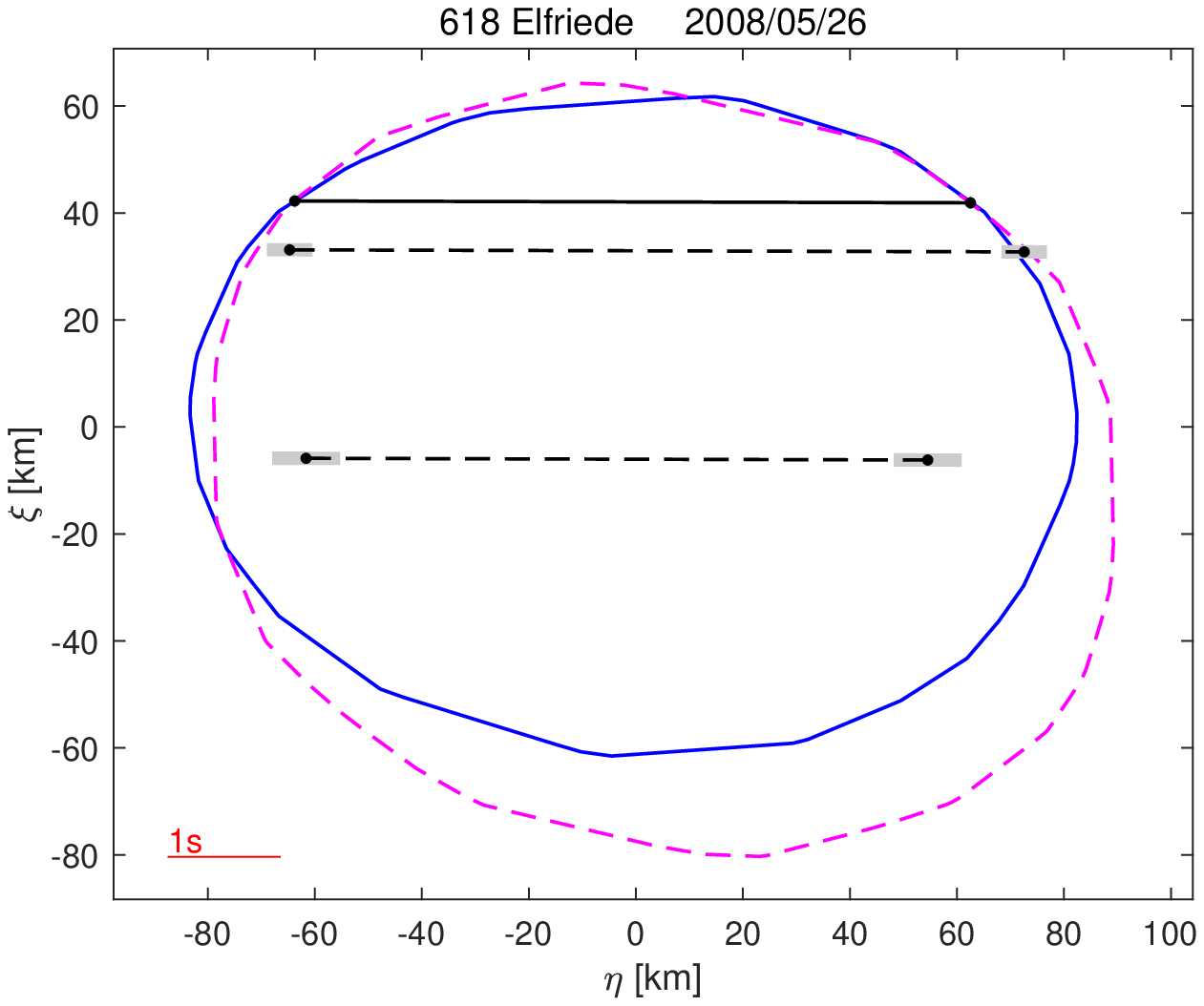}&
 \vspace{0.1cm} \includegraphics[width=0.45\textwidth]{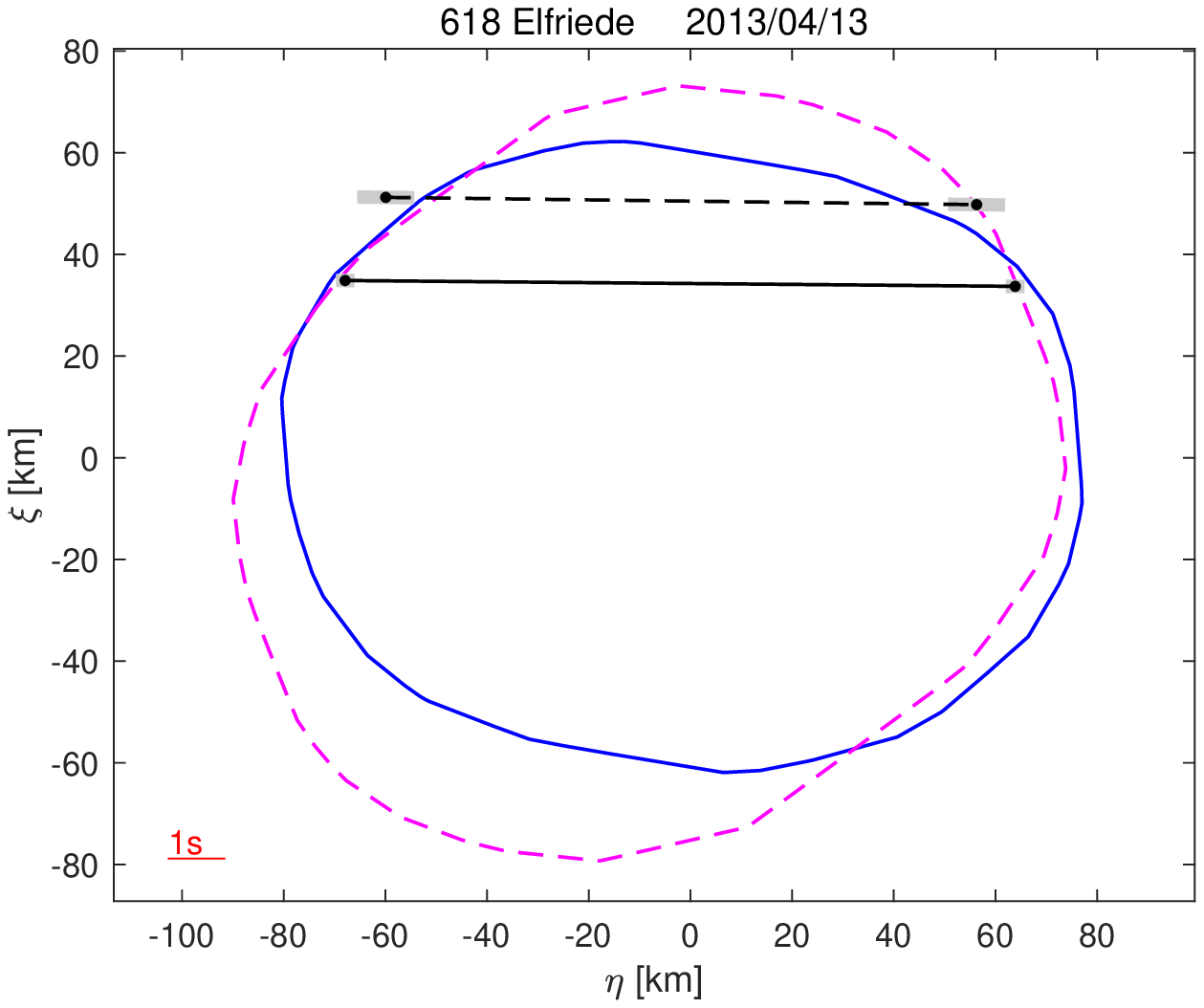}\\
 \includegraphics[width=0.45\textwidth]{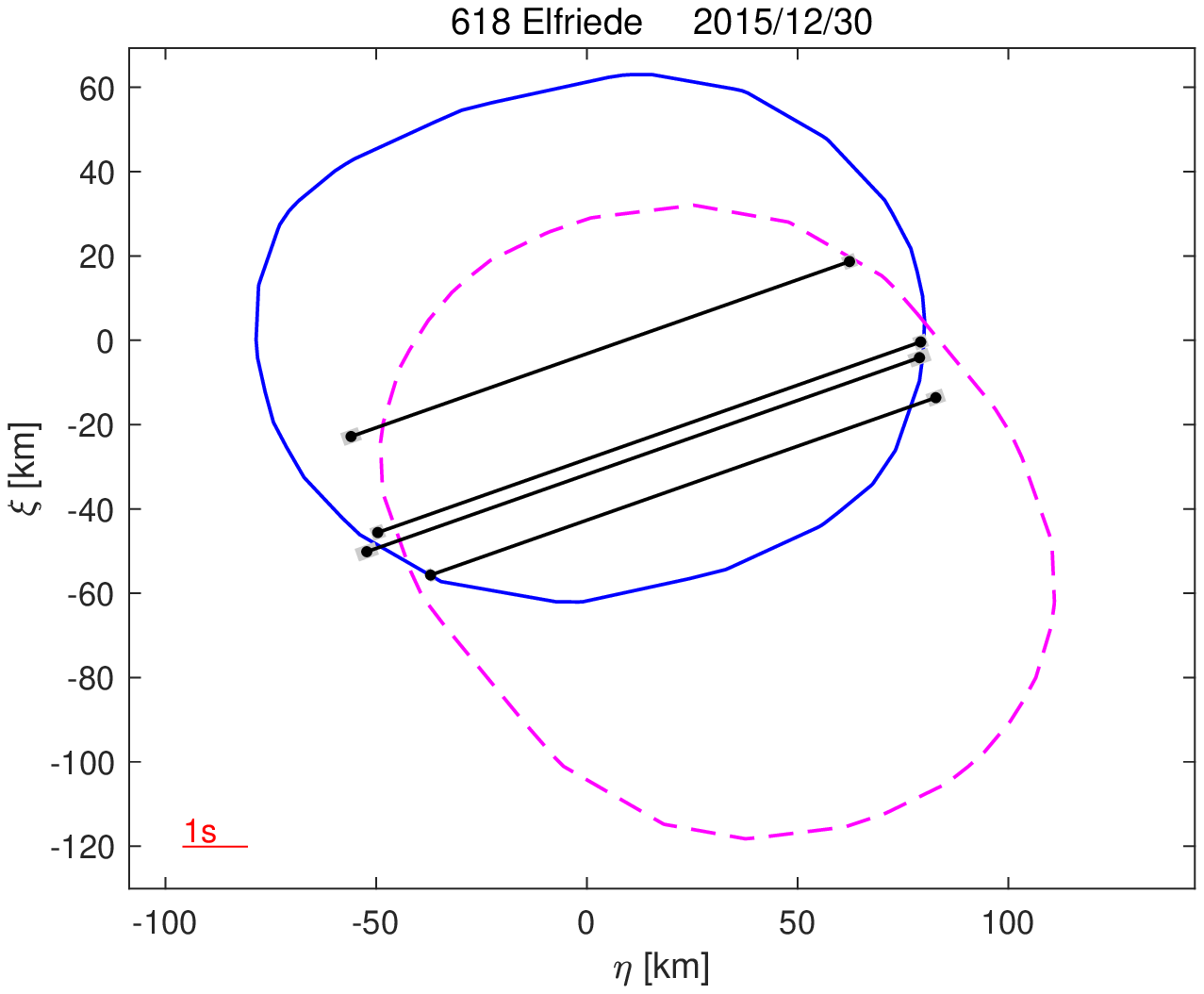}& 
 \includegraphics[width=0.45\textwidth]{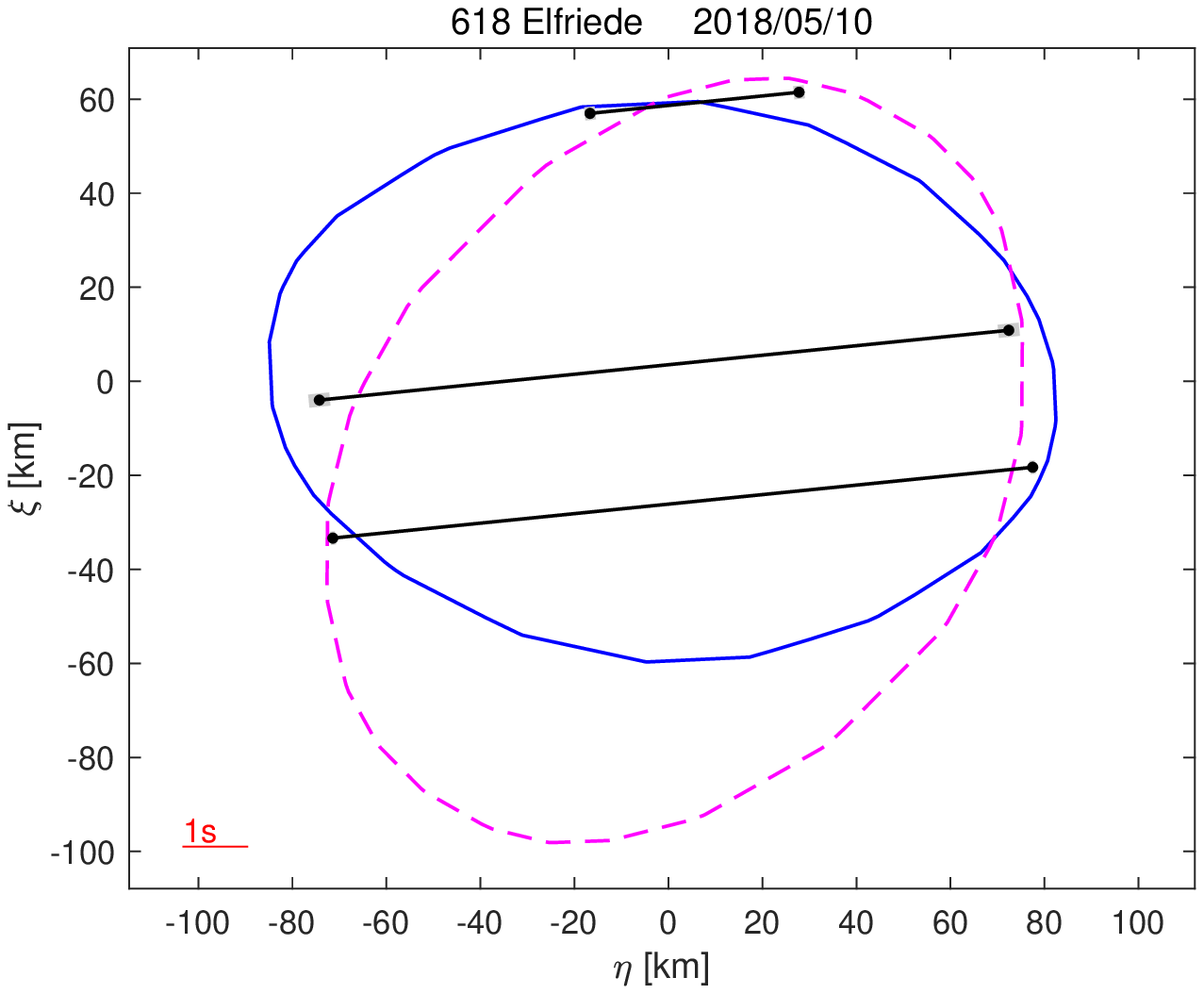}\\ 
\end{tabular}
\caption{CITPM shape models of (618) Elfriede fitted to stellar occultations from 26 May 2008, 13 April 2013, 30 December 2015, 
 and 10 May 2018. The visual, southernmost chord  in the first event probably has an underestimated duration.
 See Table \ref{occult} for diameters of equivalent volume spheres. See caption of Fig. \ref{Havnia_occult} for description of the figure.}
\label{Elfriede_occult}
\end{figure*}

\begin{figure*}[h]
\begin{tabular}{ccc}
 \includegraphics[width=0.33\textwidth]{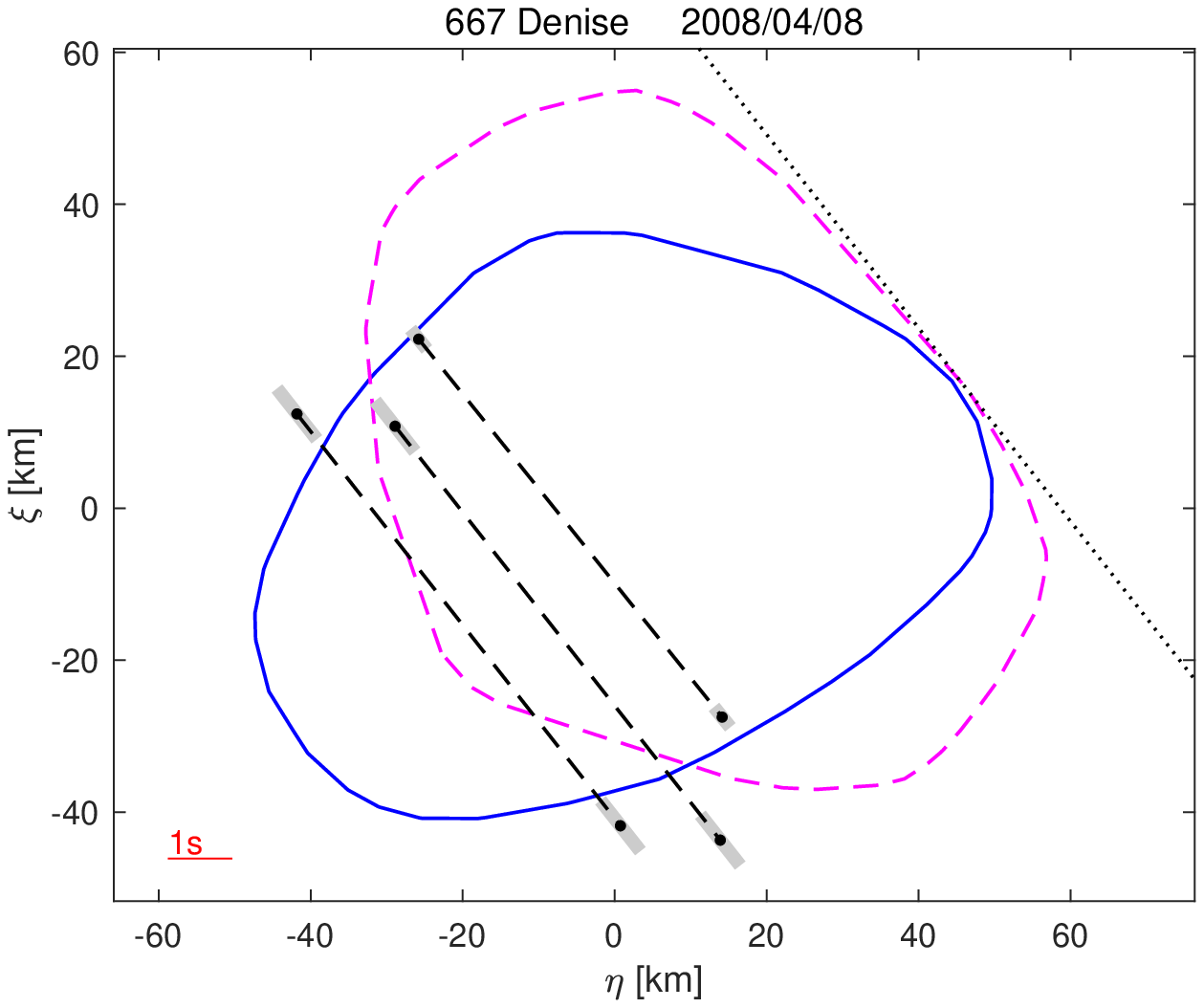}&
 \includegraphics[width=0.33\textwidth]{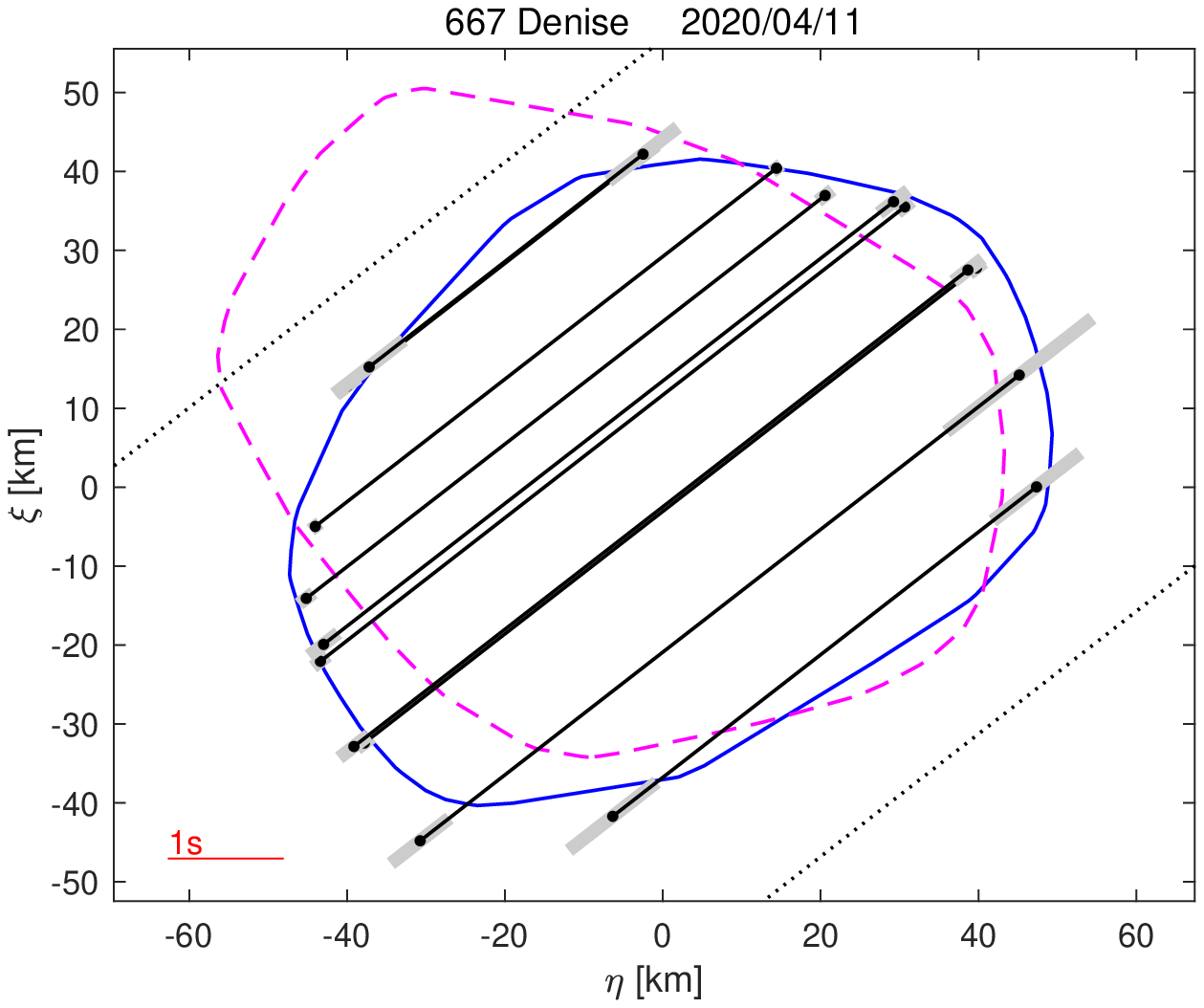}&
 \includegraphics[width=0.33\textwidth]{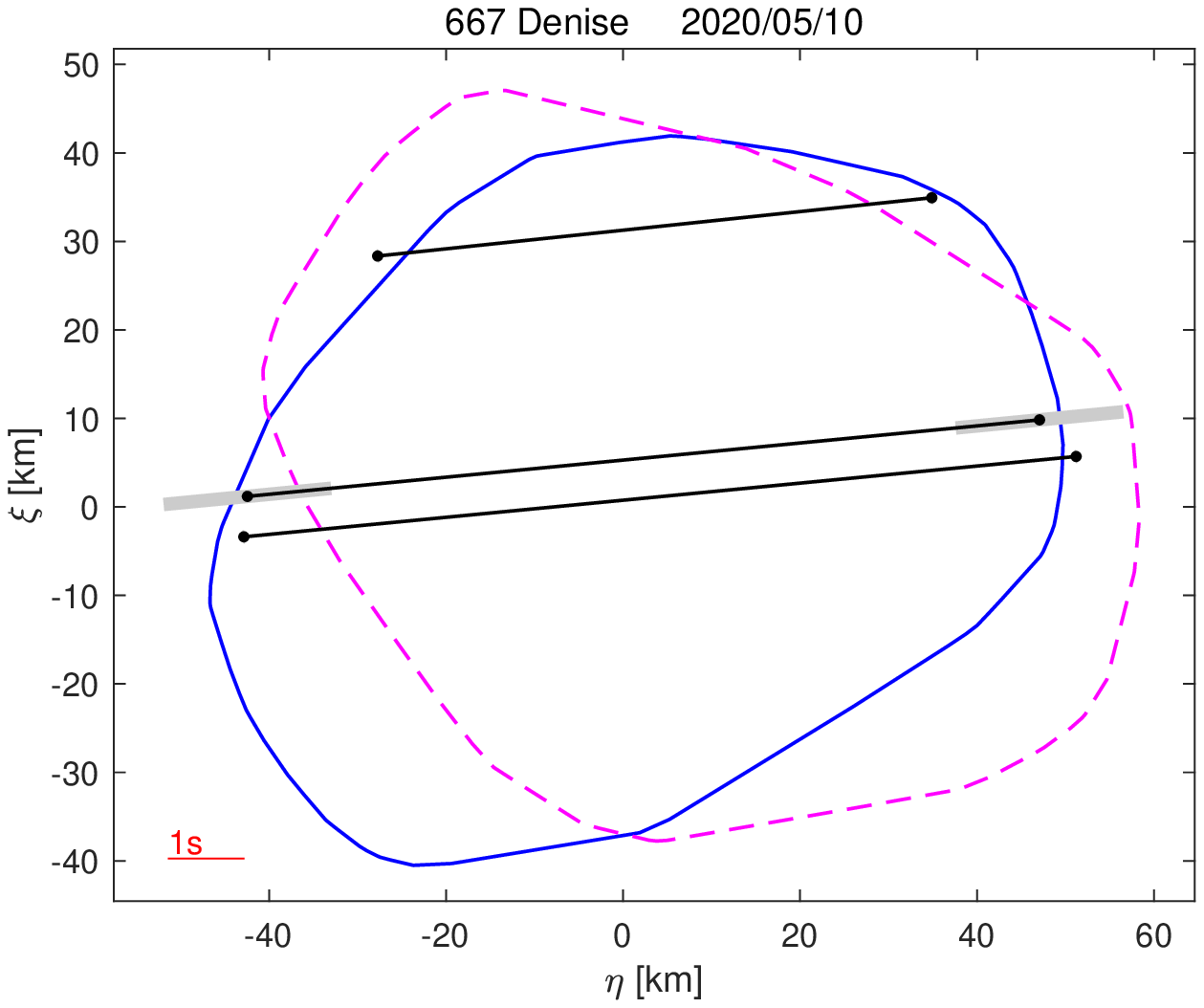}
\end{tabular}
\caption{CITPM shape models of (667) Denise fitted to stellar occultations from 8 April 2008, 11 April 2020, and 10 May 2020. 
 The pole 1 solution (blue contour) is clearly preferred over pole 2 (dashed magenta contour). 
 See Table \ref{occult} for equivalent volume sphere diameter for the preferred pole solution. See caption of Fig. \ref{Havnia_occult} 
 for description of the figure.}
\label{Denise_occult}
\end{figure*}

\clearpage

\section{Observational details}
 Details of all light curve observations used for the modelling (Table \ref{obs}), and the 
 list of stellar occultation observers and sites (Table\ref{occult_obs}).
\begin{table*}[h!]
\begin{scriptsize}
\noindent 
\caption{{\small Details of all visible photometric observations: observing dates, number of light curves, ecliptic longitude of the target, 
 sun-target-observer phase angle, observer's name (or paper citation in case of published data), and the observing site. 
 Some data come from robotic telescopes, and so they have no observer specified. 
 For data from the TESS spacecraft, the number of light curves denotes the number of days of continuous observations.
 CSSS stands for Center for Solar System Studies, PTF - Palomar Transient Factory, GMARS - Goat Mountain Astronomical Research Station, 
 ESO - European Southern Observatory, SOAO - Sobaeksan Optical Astronomy Observatory, 
 BOAO - Bohyunsan Optical Astronomy Observatory, LOAO - Lemonsan Optical Astronomy Observatory, 
 OASI - Observat{\'o}rio Astron{\^o}mico do Sert{\~a}o de Itaparica, CTIO - Cerro Tololo Interamerican Observatory, 
 ORM - Roque de los Muchachos Observatory. 
 }}
\begin{tabularx}{\textwidth}{ccccll}
\hline
 Date                    & N$_{lc}$ & $\lambda$ & Phase angle & Observer & Site \\
                         &      & [deg]     &  [deg]      &          &  \\
\hline
&&&&&\\
\multicolumn{6}{l}{(108) Hecuba}\\
&&&&&\\
\hline
 2007 03 04 - 2007 03 09 &   5  &    126    &  11 - 12    & \cite{Warner2007a} & CSSS - Palmer Divide Station, USA \\
 2011 11 30              &   1  &     68    &     2       & T. Kundera & Suhora, Poland\\
 2012 01 23 - 2012 03 05 &   7  &  61 -  66 &  15 - 17    & \cite{Waszczak2015} & PTF, USA\\ 
 2012 12 27 - 2013 02 26 &   8  & 140 - 149 &   2 - 14    & \cite{Pilcher2013} & Organ Mesa, USA \\
 2014 04 20 - 2014 04 17 &   9  & 225 - 230 &   2 -  6    & \cite{Pilcher2014} & Organ Mesa, USA\\
 2015 07 02              &   1  &    310    &     9       & A. Marciniak & Teide, Spain \\
 2015 08 05 - 2015 08 07 &   3  &    304    &   3 -  5    & M. {\.Z}ejmo & Adiyaman, Turkey\\
 2015 08 10 - 2015 09 13 &   6  & 299 - 303 &   5 - 14    & -  & Montsec, Spain\\
 2016 08 28 - 2017 01 08 &   7  &   7 -  19 &   7 - 16    & A. Marciniak, R. Hirsch, K. {\.Z}ukowski, M. Butkiewicz - B\k{a}k  & Borowiec, Poland\\
 2017 11 03 - 2017 11 15 &   5  &  85 -  86 &   9 - 12    & - & Montsec, Spain \\   
 2018 02 28 - 2018 03 02 &   2  &  75 -  76 &    18       & J. Horbowicz, K. {\.Z}ukowski & Borowiec, Poland \\
 2019 01 19 - 2019 03 30 &   3  & 151 - 162 &   5 - 13    & V. Kudak, V. Perig & Derenivka, Ukraine\\
 2019 03 23              &   1  &    152    &     10      & M. {\.Z}ejmo & Suhora, Poland \\
 2019 04 01              &   1  &    151    &     12      & E. Pak\v{s}tien{\.e} & Moletai, Lithuania\\
\hline
&&&&&\\
\multicolumn{6}{l}{(202) Chryseis}\\
&&&&&\\
\hline
 2011 01 19 - 2011 04 01 &  15  & 140 - 151 &   1 - 16    & \cite{Stephens2011} & GMARS, USA; Organ Mesa, USA; Hamanowa, \\
                         &      &           &             &                     & Japan; Bigmuskie, Italy\\ 
 2013 07 31              &   1  &    311    &     1       & - & Montsec, Spain\\
 2014 09 05              &   1  &     24    &    12       & P. Kankiewicz & Kielce, Poland\\
 2014 09 16 - 2014 10 05 &   2  &  19 - 23  &   4 - 9     & A. Marciniak & Borowiec, Poland\\
 2014 10 10 - 2014 10 26 &   2  &  15 - 18  &   3 - 6     & G. Stachowski, W. Og{\l}oza & Suhora, Poland \\
 2014 10 31 - 2014 11 20 &   3  &  12 - 14  &   8 - 13    & - & Montsec, Spain\\
 2014 11 03              &   1  &    14     &     9       & S. Urakawa & Bisei, Japan\\
 2015 10 27 - 2016 01 28 &   6  & 102 - 112 &   9 - 20    & R. Hirsch, I. Konstanciak, A. Marciniak, P. Kulczak & Borowiec, Poland\\
 2015 11 08              &   1  &   112     &    19       & P. Kankiewicz & Kielce, Poland\\
 2016 01 10 - 2016 02 23 &   3  & 100 - 106 &   3 - 17    & F. Pilcher & Organ Mesa, USA\\
 2017 03 20 - 2017 05 28 &   6  & 202 - 213 &   4 - 14    & F. Pilcher & Organ Mesa, USA\\
 2017 03 24 - 2017 04 18 &   3  & 208 - 212 &   4 - 10    & - & Montsec, Spain\\
 2017 03 24              &   1  &   212     &    10       & V. Kudak, V. Perig & Derenivka, Ukraine\\
 2017 03 27              &   1  &   212     &     9       & A. Marciniak & Borowiec, Poland\\
 2018 07 19 - 2018 09 13 &  10  & 280 - 284 &   4 - 16    & - & Montsec, Spain\\
 2019 08 23 - 2019 09 18 &   5  & 350 - 355 &   3 - 7     & S. Fauvaud & Le Bois de Bardon, France\\
 2019 08 25 - 2019 10 14 &   3  & 346 - 355 &   4 - 10    & W. Og{\l}oza & Adiyaman, Turkey\\
 2019 09 18 - 2019 09 20 &   3  &   350     &     3       & - & Montsec, Spain\\
 2019 10 05              &   1  &   347     &     7       & K. Kami{\'n}ski & Winer, USA\\
 2019 10 07              &   1  &   346     &     8       & V. Kudak, V. Perig & Derenivka, Ukraine\\
 2019 10 14              &   1  &   346     &    10       & R. Szak{\'a}ts & Piszk{\'e}stet\H{o}, Hungary\\
\hline
&&&&&\\
\multicolumn{6}{l}{(219) Thusnelda}\\
&&&&&\\
\hline
 1981 08 21 - 1981 09 27 &   7  & 340 - 347 &  8 - 14     & \cite{Harris1992} & Table Mountain, USA\\ 
 1981 09 02 - 1981 09 06 &   5  & 344 - 345 &  8 - 9      & \cite{Lagerkvist1982} & ESO, Chile\\  
 2013 05 25 - 2013 06 30 &  14  & 236 - 238 &  6 - 21     & - & Montsec, Spain\\
 2013 05 27 - 2013 06 26 &   4  & 236 - 238 &  7 - 19     & F. Pilcher & Organ Mesa, USA\\
 2014 10 10 - 2014 12 24 &  26  &  81 - 95  &  7 - 26     & \cite{M2015} & Suhora, Poland; Borowiec, Poland, Organ Mesa, USA;\\
                         &      &           &             &              &  Winer, USA; Montsec, Spain; Bisei, Japan\\
 2015 01 07              &   1  &    81     &    12       & F. Pilcher & Organ Mesa, USA\\
 2016 02 04 - 2016 03 06 &   5  & 181 - 186 &  5 - 15     & - & Montsec, Spain\\
 2016 02 08 - 2016 02 27 &   4  & 183 - 186 &  9 - 14     & K. Kami{\'n}ski & Winer, USA\\
 2016 02 27 - 2016 03 31 &   4  & 175 - 183 &  2 -  9     & F. Pilcher & Organ Mesa, USA\\ 
 2016 04 02 - 2016 04 21 &   3  & 174 - 171 &  7 - 14     & P. Kulczak, K. {\.Z}ukowski, R. Hirsch & Borowiec, Poland\\
 2017 05 26 - 2016 05 27 &   2  &   311     &   27        & M.-J. Kim & SOAO, South Korea\\
 2017 06 23              &   1  &   314     &   22        & R. Szak{\'a}ts & Piszk{\'e}stet\H{o}, Hungary\\
 2017 07 03 - 2017 07 04 &   2  &   314     &  18 - 19    & R. Duffard & La Sagra, Spain\\
 2017 07 13 - 2017 08 16 &   4  & 306 - 313 &  12 - 15    & S. Brincat & Flarestar, Malta\\
 2017 07 13 - 2017 09 08 &  10  & 304 - 313 &  15 - 23    & - & Montsec, Spain\\
 2018 11 15 - 2019 02 08 &   2  & 125 - 135 &   8 - 22    & V. Kudak, V. Perig & Derenivka, Ukraine\\
 2018 12 10 - 2019 02 14 &  14  & 124 - 137 &   7 - 18    & - & Montsec, Spain\\
 2019 01 04 - 2019 01 31 &   6  & 127 - 134 &   6 - 12    & F. Pilcher & Organ Mesa, USA\\
 2019 01 30 - 2019 02 06 &   2  & 126 - 127 &   6 - 7     & A. Marciniak & Borowiec, Poland\\
\hline
\end{tabularx}
\label{obs}
\end{scriptsize}
\end{table*}

\begin{table*}[h]
\begin{scriptsize}
\noindent 
\begin{tabularx}{\textwidth}{ccccll}
\hline
 Date                    & N$_{lc}$ & $\lambda$ & Phase angle & Observer & Site \\
                         &      & [deg]     &  [deg]      &          &  \\
\hline
&&&&&\\
\multicolumn{6}{l}{(223) Rosa}\\
&&&&&\\
\hline
 2007 03 25 - 2007 04 14 &   4  & 166 - 169 &   5 - 12    & \cite{Warner2007b} & CSSS, USA\\ 
 2011 12 30 - 2012 02 10 &   8  & 121 - 129 &   2 - 10    & \cite{Pilcher2012} & Organ Mesa, USA\\
 2015 09 07 - 2015 10 09 &   3  & 344 - 350 &   1 - 10    & S. Fauvaud & Le Bois de Bardon, France\\
 2015 09 01 - 2015 09 03 &   2  &    347    &     4       & S. Fauvaud, M. Fauvaud, F. Richard & Pic du Midi, France\\
 2015 11 22              &   1  &    344    &    18       & D. Oszkiewicz & Lowell, USA\\
 2016 11 01 - 2016 12 18 &   6  &  72 - 82  &   1 - 14    & - & Montsec, Spain\\
 2016 12 28              &   1  &     72    &     9       & J. Horbowicz & Borowiec, Poland\\ 
 2018 02 16 - 2018 02 25 &   2  & 184 - 185 &   9 - 12    & R. Hirsch &  Borowiec, Poland\\ 
 2018 03 23              &   1  &    179    &     2       & V. Kudak & Derenivka, Ukraine\\
 2018 04 12 - 2018 04 19 &   8  & 174 - 175 &   9 - 11    & - & Kepler Space Observatory\\
 2019 05 15 - 2019 06 01 &   5  & 257 - 250 &   2 -  7    & - & Montsec, Spain\\
 2019 05 23              &   1  &   259     &     5       & K. Kami{\'n}ski & Winer, USA\\
 2019 07 11              &   1  &   250     &    13       & M. Ferrais, E. Jehin & TRAPPIST-South, Chile\\
 2020 07 19 - 2020 10 19 &   9  & 312 - 322 &   1 - 16    & M. Ferrais, E. Jehin & TRAPPIST-North, Morocco\\
 2020 08 23 - 2020 08 25 &   4  &   316     &   4 -  5    & F. Monteiro, M. Evangelista-Santana, E. Rond{\'o}n, P. Arcoverde, & OASI, Itacuruba, Brasil\\
                         &      &           &             & J. Michimani-Garcia, D. Lazzaro, T. Rodrigues & \\
 2020 10 18 - 2020 10 19 &   2  &   313     &    16       & M. Ferrais, E. Jehin & TRAPPIST-South, Chile\\
\hline
&&&&&\\
\multicolumn{6}{l}{(362) Havnia}\\
&&&&&\\
\hline
 1978 12 04              &   1  &     63    &      5      & \cite{Harris1980} & Table Mountain, USA\\
 2006 06 05 - 2006 08 26 &   5  & 299 - 312 &    9 - 19   & - & SuperWASP \\ 
 2009 04 06 - 2009 04 20 &   6  & 180 - 186 &    5 - 11   & \cite{Stephens2009} & Rancho Cucamonga, USA\\ 
 2015 10 28 - 2015 12 10 &   3  &  24 -  30 &    2 - 19   & M. Butkiewicz - B\k{a}k, A. Marciniak, P. Kulczak & Borowiec, Poland\\
 2015 11 29 - 2015 12 01 &   2  &     24    &     16      & - & Montsec, Spain\\
 2016 12 22 - 2017 03 03 &   4  & 152 - 160 &    6 - 21   & J. Horbowicz, A. Marciniak, K. {\.Z}ukowski, M. Butkiewicz - B\k{a}k & Borowiec, Poland\\
      2017 01 19         &   1  &    161    &     15      & V. Kudak, V. Perig & Derenivka, Ukraine\\
 2017 01 25 - 2017 01 31 &   7  & 159 - 160 &   11 - 13   & T. Polakis, B. Skiff & Command Module, USA\\
      2018 07 26         &   1  &    252    &     17      & A. Marciniak & CTIO, Chile\\     
 2019 08 29 - 2019 09 19 &   2  &  17 - 20  &    8 - 16   & S. Fauvaud & Le Bois de Bardon, France\\
      2019 09 04         &   1  &     19    &     14      & V. Kudak, V. Perig & Derenivka, Ukraine\\
      2019 09 21         &   1  &     16    &      7      & J. Skrzypek & Borowiec, Poland\\ 
 2019 09 28 - 2019 10 15 &   2  &  11 - 15  &    4 - 5    & W. Og{\l}oza & Adiyaman, Turkey\\
 2019 10 17 - 2020 01 10 &   2  &  10 - 14  &    6 - 13   & R. Szak{\'a}ts & Piszk{\'e}stet\H{o}, Hungary\\
\hline
&&&&&\\
\multicolumn{6}{l}{(483) Seppina}\\
&&&&&\\
\hline
 1986 07 11 - 1986 07 27 &   6  & 268 - 270 &    9 - 12   & \cite{Zappala1989} & ESO, La Silla, Chile\\
 2005 06 25 - 2005 07 11 &   2  & 264 - 266 &    8 - 10   & F. Manzini & Sozzago, Italy\\ 
 2005 07 04              &   1  &    265    &      9      & G. Farroni, P. Pinel & Saint-Avertin, France\\
 2005 07 10 - 2005 07 30 &   4  & 262 - 264 &   10 - 13   & R. Roy & Blauvac, France\\
 2005 07 12              &   1  &    264    &     11      & L. Bernasconi & Engarouines, France\\
 2006 08 21              &   1  &    348    &      6      & L. Brunetto & Le Florian, France\\
 2013 10 08 - 2013 12 23 &   4  &  25 -  34 &   5 - 16    & K. Sobkowiak, D. Oszkiewicz, A. Marciniak & Borowiec, Poland\\
 2013 12 16 - 2013 12 17 &   4  &  25 -  33 &   5 - 15    & F. Pilcher & Organ Mesa, USA\\
 2015 01 20 - 2015 03 22 &   3  &  96 -  97 &   9 - 16    & K. Kami{\'n}ski & Winer, USA\\
 2015 02 12 - 2015 03 18 &   2  &  94 -  95 &  13 - 16    & A. Marciniak, J. Horbowicz, M. Figas & Borowiec, Poland\\
 2016 01 03 - 2016 04 01 &   5  & 155 - 167 &   5 - 14    & P. Kulczak, A. Marciniak, R. Hirsch, M. Butkiewicz - B\k{a}k & Borowiec, Poland\\
 2017 04 01 - 2017 05 29 &   9  & 214 - 224 &   5 - 10    & R. Hirsch, K. {\.Z}ukowski, J. Horbowicz, A. Marciniak, J. Skrzypek & Borowiec, Poland\\
 2018 07 19 - 2018 08 15 &  14  & 287 - 291 &   7 - 12    & - & Montsec, Spain\\
\hline
&&&&&\\
\multicolumn{6}{l}{(501) Urhixidur}\\
&&&&&\\
\hline
 1990 08 22 - 1990 08 29 &   6  & 327 - 329 &   6 -  7   & \cite{Lagerkvist1992} & ESO, La Silla, Chile\\
 2013 09 06 - 2013 12 30 &   5  &  18 -  28 &   6 - 20   & R. Hirsch, T. Santana-Ros, D. Oszkiewicz, A. Marciniak & Borowiec, Poland\\
 2014 10 29 - 2015 03 17 &   6  & 104 - 117 &   8 - 18   & R. Hirsch, A. Marciniak, I. Konstanciak, J. Horbowicz & Borowiec, Poland\\
 2015 01 21 - 2015 04 29 &   2  & 109 - 111 &   8 - 16   & K. Kami{\'n}ski & Winer, USA\\
 2015 03 22 - 2015 03 23 &   2  &   105     &     17     & W. Og{\l}oza, A. Marciniak, V. Kudak & Suhora, Poland\\
 2016 02 06 - 2016 04 29 &   4  & 163 - 176 &   2 - 14   & A. Marciniak, R. Hirsch & Borowiec, Poland\\
 2016 02 13 - 2016 03 01 &   4  & 172 - 175 &   3 -  8   & K. Kami{\'n}ski & Winer, USA\\
 2017 02 08 - 2017 03 09 &   2  & 231 - 233 &  15 - 16   & A. Marciniak & CTIO, Chile\\
 2017 05 04              &   1  &   227     &     7      & F. Monteiro & OASI, Brasil\\ 
 2018 08 04 - 2018 08 22 &  18  & 312 - 315 &   8 - 9    & \cite{Pal2020} & TESS Spacecraft\\ 
 2018 08 14              &   1  &   313     &     8      & A. Marciniak & CTIO, Chile\\
 2018 09 14 - 2018 09 15 &   2  &   309     &    15      & F. Monteiro, E. Rond\'on, M. Evangelista-Santana, P. Arcoverde, & OASI, Brasil\\ 
                         &      &           &            & D. Lazzaro, T. Rodrigues & \\
 2019 08 12 - 2019 08 19 &   4  & 58 - 59   &    21      & W. Og{\l}oza & Adiyaman, Turkey\\
 2019 10 11 - 2019 12 15 &   2  & 52 - 64   &  12 -  15  & R. Szak{\'a}ts, V. Kecskem{\'e}thy & Piszk{\'e}stet\H{o}, Hungary\\
 2019 10 12 - 2019 10 15 &   2  &    64     &  14 -  15  & J. Skrzypek, M. Paw{\l}owski & Borowiec, Poland\\
\hline
&&&&&\\
\multicolumn{6}{l}{(537) Pauly}\\
&&&&&\\
\hline
 1984 05 10              &   1  &    211    &      8      & \cite{Weidenschilling1990} & Kitt Peak, USA\\
 1985 09 08 - 1985 09 12 &   4  & 354 - 355 &      6      & \cite{Barucci1992} & ESO, Chile\\
 1989 04 16              &   1  &    182    &      7      & \cite{Weidenschilling1990} & Kitt Peak, USA\\
 2016 02 24 - 2016 03 10 &   4  & 202 - 204 &  10 -  13   & K. Kami{\'n}ski & Winer, USA\\
 2016 03 18 - 2016 05 09 &   4  & 191 - 201 &   8 -  12   & M. Butkiewicz - B\k{a}k, A. Marciniak, P. Kulczak & Borowiec, Poland\\
 2017 08 03 - 2017 09 24 &   9  & 311 - 316 &   2 -  19   & - &  Montsec, Spain\\
 2018 10 09 - 2018 11 29 &   5  &  65 -  73 &   4 -  15   & K. {\.Z}ukowski, A. Marciniak, M. K. Kami{\'n}ska, J. Krajewski, M. Paw{\l}owski & Borowiec, Poland\\
 2018 11 26 - 2018 12 10 &  16  &  62 -  66 &   4 -   6   & \cite{Pal2020} & TESS Spacecraft\\ 
 2019 11 24 - 2019 12 17 &   3  & 126 - 128 &  10 -  14   & W. Og{\l}oza & Adiyaman, Turkey\\
 2019 11 27              &   1  &    128    &     14      & M.-J. Kim, D.-H. Kim & SOAO, South Korea\\
 2019 12 05              &   1  &    127    &     12      & A. Marciniak & Borowiec, Poland\\
 2020 01 15              &   1  &    122    &      2      & V. Kudak, V. Perig & Derenivka, Ukraine\\
\hline
\end{tabularx}
%\caption{Table \ref{obs} continued.}
\end{scriptsize}
\end{table*}

\clearpage

\begin{table*}[h]
\begin{scriptsize}
\noindent 
\begin{tabularx}{\textwidth}{ccccll}
\hline
 Date                    & N$_{lc}$ & $\lambda$ & Phase angle & Observer & Site \\
                         &      & [deg]     &  [deg]      &          &  \\
\hline 
&&&&&\\
\multicolumn{6}{l}{(552) Sigelinde}\\
&&&&&\\
\hline
 2008 04 11 - 2008 05 03 &   8  & 245 - 247 &   7 - 14    & \cite{Oey2009} & Leura, Australia\\
 2010 08 13 - 2011 01 26 &  11  &  42 -  52 &   2 - 18    & \cite{Waszczak2015} & Palomar Transient Factory, USA\\
 2015 08 25 - 2015 10 02 &   5  &   6 -  12 &   3 - 12    & A. Marciniak, R. Hirsch, P. Kulczak & Borowiec, Poland\\ 
 2016 11 20 - 2016 12 05 &   2  &  74 -  76 &   1 -  5    & K. {\.Z}ukowski, R. Hirsch & Borowiec, Poland\\ 
 2017 01 15              &   1  &     67    &    12       & S. Geier & ORM, Spain\\
 2017 01 25              &   1  &     67    &    14       & M.-J. Kim, D.-H. Kim & SOAO, South Korea\\
 2017 02 08              &   1  &     67    &    16       & A. Marciniak & CTIO, Chile\\
 2017 02 10 - 2017 02 11 &   2  &     67    &    16       & M.-J. Kim, D.-H. Kim & BOAO, South Korea\\
 2018 02 23 - 2018 04 09 &   3  & 132 - 136 &   6 - 17    & A. Marciniak, R. Hirsch, K. {\.Z}ukowski & Borowiec, Poland\\ 
 2018 03 14 - 2018 03 16 &   2  &    133    &    12       & M.-J. Kim, D.-H. Kim & BOAO, South Korea\\
 2018 03 19              &   1  &    133    &    13       & K. Kami{\'n}ski & Winer, USA\\
 2019 04 26 - 2019 04 29 &   4  & 224 - 225 &   3 - 4     & K. Kami{\'n}ski & Winer, USA\\  
 2019 05 11              &   1  &    222    &     4       & - & Montsec, Spain\\
 2019 04 26 - 2019 05 19 &  23  & 220 - 225 &   3 - 7     & \cite{Pal2020} & TESS Spacecraft\\ 
\hline
&&&&&\\
\multicolumn{6}{l}{(618) Elfriede}\\
&&&&&\\
\hline
 1984 05 12              &   1  &    236    &     6       & \cite{Weidenschilling1990} & Kitt Peak, USA\\
 1989 04 16 - 1989 04 17 &   2  &    183    &     9       & \cite{Weidenschilling1990} & Kitt Peak, USA\\
 2004 12 03 - 2004 12 11 &   6  &    125    &  12 - 14    & L. Bernasconi & Engarouines, France\\
 2006 05 12 - 2006 06 02 &   7  & 175 - 176 &  15 - 17    & \cite{Warner2006} & Palmer Divide, USA\\
 2014 10 01 - 2014 12 09 &  12  &   4 - 10  &   8 - 18    & - & Montsec, Spain\\
 2015 10 10 - 2016 01 27 &   5  &  86 - 98  &   4 - 18    & - & Montsec, Spain\\
 2016 01 22              &   1  &    87     &    10       & A. Marciniak & Borowiec, Poland\\
 2016 12 29 - 2017 04 09 &   4  & 150 - 163 &   8 - 15    & J. Horbowicz, K. {\.Z}ukowski & Borowiec, Poland\\
 2017 01 25              &   1  &   162     &    10       & M.-J. Kim, D.-H. Kim & SOAO, South Korea\\
 2017 02 04 - 2017 03 14 &   2  & 153 - 160 &     8       & W. Og{\l}oza, M. {\.Z}ejmo & Suhora, Poland\\
 2017 02 22 - 2017 02 25 &   4  &    157    &     5       & T. Polakis, B. Skiff & Command Module, USA\\
 2017 03 02 - 2017 03 17 &   9  & 153 - 156 &   5 -  8    & \cite{Klinglesmith2017} & Socorro, USA\\
 2018 02 27 - 2018 05 09 &   5  & 226 - 232 &   7 - 17    & K. {\.Z}ukowski, J. Horbowicz, A. Marciniak, J. Skrzypek & Borowiec, Poland\\
 2019 07 22 - 2019 07 26 &   5  &    306    &   2 -  3    & W. Og{\l}oza & Adiyaman, Turkey\\
\hline
&&&&&\\
\multicolumn{6}{l}{(666) Desdemona}\\
&&&&&\\
\hline
 2013 10 02 - 2014 02 14 &   8  &  82 -  95 &   9 - 29    & \cite{M2015}  & Borowiec, Poland; Winer, USA\\
 2014 12 31 - 2015 03 14 &  15  & 192 - 196 &   5 - 19    & - & Montsec, Spain\\
 2015 02 11 - 2015 03 31 &   2  & 186 - 197 &   6 - 15    & K. Kami{\'n}ski & Winer, USA\\
 2015 03 18              &   1  &    191    &     5       & M. Figas & Borowiec, Poland\\
 2016 04 16              &   2  &    264    &    16       & S. Geier & Kitt Peak, USA\\
 2016 04 29              &   1  &    264    &    13       & S. Geier & ORM, Spain\\
 2016 05 01              &   1  &    263    &    13       & - & Montsec, Spain\\
 2016 06 30 - 2016 07 05 &   4  & 251 - 252 &  10 - 11    & R. Duffard, N. Morales & La Sagra, Spain\\
 2016 07 23              &   1  &    249    &    17       & A. Marciniak & Teide, Spain\\ 
 2017 09 16 - 2017 09 22 &   7  &  57 - 58  &  27 - 25    & T. Polakis, B. Skiff & Tempe, USA\\
 2017 09 18 - 2018 01 08 &   5  &  48 - 58  &   5 - 25    & J. Horbowicz, K. {\.Z}ukowski, R. Hirsch & Borowiec, Poland\\
 2019 01 07              &   1  &    184    &    19       & R. Duffard, N. Morales & La Sagra, Spain\\
 2019 02 01 - 2019 04 07 &   6  & 172 - 184 &   3 - 15    & - & Montsec, Spain\\
 2019 02 07              &   1  &    184    &    14       & Cs. Kalup & Piszk{\'e}stet\H{o}, Hungary\\
 2019 04 01 - 2019 04 03 &   2  &    173    &   6 -  7    & M. Paw{\l}owski, J. Krajewski & Borowiec, Poland\\
\hline
&&&&&\\
\multicolumn{6}{l}{(667) Denise}\\
&&&&&\\
\hline
 2014 03 28 - 2014 05 19 &   5  & 166 - 169 &   8 - 19    & R. Hirsch, K. Sobkowiak, I. Konstanciak, P. Trela & Borowiec, Poland\\
 2015 03 23 - 2015 03 24 &   2  &    252    &     15      & W. Og{\l}oza, A. Marciniak. V. Kudak & Suhora, Poland\\
 2015 04 21 - 2015 04 23 &   2  & 250 - 251 &     12      & J. Horbowicz, A. Marciniak & Borowiec, Poland\\
 2015 05 31              &   1  &    243    &      9      & K. Kami{\'n}ski & Winer, USA\\
 2015 06 27              &   1  &    238    &     12      & A. Marciniak & Teide, Spain\\
 2016 07 23 - 2016 07 26 &   3  &    298    &   5 -  6    & - & Montsec, Spain\\
 2016 07 30 - 2016 08 20 &   4  & 293 - 297 &   6 - 10    & R. Szak{\'a}ts, E. Vereb{\'e}lyi & Piszk{\'e}stet\H{o}, Hungary\\
 2016 08 26              &   1  &    293    &     11      & S. Geier & ORM, Spain\\
 2016 08 31              &   1  &    292    &     12      & K. {\.Z}ukowski & Borowiec, Poland\\ 
 2017 08 07              &   1  &    359    &     12      & W. Og{\l}oza & Suhora, Poland\\
 2017 08 27              &   1  &    357    &      8      & A. Marciniak & Teide, Spain\\
 2017 08 31 - 2017 09 18 &   8  & 352 - 356 &   4 -  6    & - & Montsec, Spain\\
 2018 11 23 - 2019 01 27 &   9  &  96 - 108 &  15 - 18    & - & Montsec, Spain\\
 2019 02 18              &   1  &    94     &     19      & M. K. Kami{\'n}ska & Borowiec, Poland\\ 
\hline
\end{tabularx}
%\caption{Table \ref{obs} continued.}
\end{scriptsize}
\end{table*}

\clearpage

\begin{table*}[h]
\begin{scriptsize}
\noindent 
\begin{tabularx}{\textwidth}{ccccll}
\hline
 Date                    & N$_{lc}$ & $\lambda$ & Phase angle & Observer & Site \\
                         &      & [deg]     &  [deg]      &          &  \\
\hline
&&&&&\\
\multicolumn{6}{l}{(780) Armenia}\\
&&&&&\\
\hline
 2004 06 07 - 2004 07 15 &   3  & 256 - 263 &   8 - 13    & J.-G. Bosch & Collonges, France\\
 2009 05 02 - 2009 06 02 &  15  & 217 - 223 &   8 - 12    & \cite{Benishek2009} & Organ Mesa, USA; Belgrade, Serbia\\
 2010 07 20 - 2010 08 31 &   2  & 298 - 306 &   5 - 13    & R. Roy & Blauvac, France\\
 2014 02 25 - 2014 05 31 &   8  & 182 - 194 &  11 - 15    & J. Horbowicz, A. Marciniak, I. Konstanciak, D. Oszkiewicz, & Borowiec, Poland\\ 
                         &      &           &             & T. Santana - Ros, K. Sobkowiak & \\
 2015 04 16 - 2015 05 30 &   4  & 265 - 268 &   9 - 16    & P. Kulczak, A. Marciniak & Borowiec, Poland\\ 
 2015 05 21 - 2015 06 22 &   7  & 260 - 266 &   8 - 11    & K. Kami{\'n}ski & Winer, USA\\
 2015 06 17 - 2015 07 06 &  11  & 257 - 261 &   8 - 11    & - & Montsec, Spain\\
 2016 08 02              &   1  &    355    &     8       & A. Marciniak & CTIO, Chile\\
 2016 10 08 - 2016 10 15 &   2  &    346    &  11 - 13    & - & Montsec, Spain\\
 2016 10 10 - 2016 11 08 &  12  & 345 - 346 &  11 - 18    & B. Skiff & Lowell, USA\\
 2016 12 04 - 2016 12 05 &   2  &   348     &    20       & T. Polakis, B. Skiff & Command Module, USA\\
 2017 12 15 - 2017 01 21 &   3  &  84 -  91 &   8 - 12    & F. Monteiro, H. Medeiros, E. Rond\'on, P. Arcoverde,  & OASI, Brasil\\
                         &      &           &             & D. Lazzaro, T. Rodrigues & \\
 2017 12 17 - 2018 02 13 &   3  &  83 -  90 &   7 - 16    & M.-J. Kim, D.-H. Kim & SOAO, South Korea\\
 2018 01 03 - 2018 01 05 &   2  &  86 -  87 &   8 -  9    & M.-J. Kim, D.-H. Kim & LOAO, USA\\
 2018 01 24 - 2018 03 27 &   2  &  84 -  88 &  13 - 18    & K. Kami{\'n}ski & Winer, USA\\
 2018 02 08 - 2018 02 16 &   2  &     83    &  16 - 17    & M. Butkiewicz - B\k{a}k, R. Hirsch & Borowiec, Poland\\ 
 2018 12 04 - 2019 01 06 &   3  & 163 - 165 &  14 - 17    & M.-J. Kim, D-H. Kim & SOAO, South Korea\\
 2019 01 19 - 2019 03 30 &   6  & 152 - 164 &   2 - 12    & R. Hirsch, J. Krajewski, A. Marciniak, K. {\.Z}ukowski, J. Skrzypek & Borowiec, Poland\\ 
 2019 02 22              &   1  &    159    &     2       & V. Kudak, V. Perig & Derenivka, Ukraine\\
\hline
&&&&&\\
\multicolumn{6}{l}{(923) Herluga}\\
&&&&&\\
\hline
 2008 10 07 - 2008 11 29 &   8  &  39 -  49 &   8 - 16    & \cite{Brinsfield2009} & Via Capote, USA\\
 2012 10 19 - 2012 10 31 &   2  & 354 - 355 &  14 - 19    & R. Hirsch, J. Nadolny & Borowiec, Poland\\ 
 2014 03 14 - 2014 04 18 &   7  & 149 - 152 &   9 - 18    & - & Montsec, Spain\\
 2015 03 17 - 2015 03 27 &   6  & 227 - 238 &   7 - 16    & - & Montsec, Spain\\
 2015 04 18 - 2015 06 22 &   3  & 223 - 235 &   4 - 14    & K. Kami{\'n}ski & Winer, USA\\
 2016 07 25 - 2016 09 10 &  12  & 330 - 320 &   8 - 14    & \cite{M2018} & Montsec, Spain; ORM, Spain; Borowiec, Poland\\ 
 2018 03 18 - 2018 03 30 &   3  &    126    &  18 - 20    & K. Kami{\'n}ski & Winer, USA\\
 2019 03 30 - 2019 04 01 &   3  &    220    &   9 - 10    & R. Szak{\'a}ts & Piszk{\'e}stet\H{o}, Hungary\\
 2019 04 26 - 2019 05 11 &   7  & 211 - 215 &   2 -  7    & - & Montsec, Spain\\
\hline
&&&&&\\
\multicolumn{6}{l}{(995) Sternberga}\\
&&&&&\\
\hline
 1989 01 07 - 1989 01 12 &   4  & 124 - 126 &  8 -  9     & \cite{Barucci1992} & ESO, La Silla, Chile\\
 2007 03 21 - 2007 03 24 &   2  & 211 - 212 &  9 - 10     & - & Super WASP \\
 2012 06 30 - 2012 07 15 &  10  & 292 - 292 &  9 - 11     & \cite{Stephens2013} & Racho Cucamonga, USA\\
 2013 11 13 - 2014 02 04 &   4  &  79 - 93  &  7 - 18     & A. Marciniak, I. Konstanciak, P. Trela, J. Horbowicz, R. Hirsch & Borowiec, Poland\\ 
 2013 12 06 - 2013 12 15 &   5  &  87 - 89  &  5 -  7     & F. Pilcher & Organ Mesa, USA\\
 2014 01 02 - 2014 02 21 &   4  &  80 - 82  &  9 - 20     & K. Kami{\'n}ski & Winer, USA\\
 2015 01 01 - 2015 03 16 &  15  & 169 - 178 &  5 - 18     & - & Montsec, Spain\\
 2015 02 11              &   1  &    177    &    11       & K. Kami{\'n}ski & Winer, USA\\
 2015 02 25              &   1  &    174    &     7       & F. Pilcher & Organ Mesa, USA\\
 2015 03 23              &   1  &    168    &     7       & R. Hirsch & Borowiec, Poland\\ 
 2016 05 04 - 2016 07 10 &  24  & 252 - 265 &  5 - 15     & \cite{M2018} & Lowell, USA; Teide, Spain; Derenivka, Ukraine;\\
                         &      &           &             &              &  Command Module, USA; La Sagra, Spain; \\
                         &      &           &             &              & Montsec, Spain; Bardon, France\\ 
 2017 10 24              &   1  &     66    &    14       & V. Kudak, V. Perig & Derenivka, Ukraine\\
 2017 12 07 - 2018 02 06 &   3  &  53 - 56  &  8 - 22     & M. Butkiewicz - B\k{a}k, R. Hirsch, J. Skrzypek & Borowiec, Poland\\ 
\hline
\end{tabularx}
%\caption{Table \ref{obs} continued.}
\end{scriptsize}
\end{table*}

% Dla Super WASP (Havnia) niech Johny Grice doda kraj (North - Hawaje lub South - RPA)

% Wykluczone z tej tabeli:
% - Thusnelda 2014 (M2015)
% - Seppina 2004 - usuniêty z modelowania
% - Desdemona 2013/2014 (M 2015)
% - Herluga 2016 (M 2018)
% - Sternberga 2016 (M 2018) i 2004 (usuniêty z modelowania)

\clearpage

\begin{table}[h!]
\begin{tabular}{ccc}
\hline
&\\
\multicolumn{2}{c}{(362) Havnia, 2017-01-07}\\
&\\
\hline
P. Maley & Gila Bend, AZ\\
C. Wiesenborn & Boulder City, NV\\
W. Thomas & Florence, AZ\\
T. George & Scottsdale, AZ\\
\hline
&\\
\multicolumn{2}{c}{(618) Elfriede, 2008-05-26}\\
&\\
\hline
D. Breadsell & Toowoomba, Qld, AU\\       
J. Bradshaw & Samford, Qld, AU\\          
P. Anderson & Range Observatory, Qld, AU\\
\hline
&\\
\multicolumn{2}{c}{(618) Elfriede, 2013-04-13}\\
&\\
\hline
D. Herald & Murrumbateman, NSW\\
J. Drummond & Patutahi, Gisborne, NZ\\
\hline
&\\
\multicolumn{2}{c}{(618) Elfriede, 2015-12-30}\\
&\\
\hline
J. Rovira & ES\\
R. Naves & ES\\
C. Perello, A. Selva & ES\\
C. Schnabel & ES\\
\hline
&\\
\multicolumn{2}{c}{(618) Elfriede, 2018-05-10}\\
&\\
\hline
J. Broughton & Woodburn, NSW, AU\\
J. Broughton & Grafton, NSW, AU\\
J. Broughton & Mullaway, NSW, AU\\
\hline
&\\
\multicolumn{2}{c}{(667) Denise, 2008-04-08}\\
&\\
\hline
R. Nugent & Pontotoc, TX\\
G. Nason & Tobermory, ONT, CA\\
M. McCants & Kingsland, TX\\
P. Maley, D. Weber & Horseshoe Bay, TX\\
\hline
&\\
\multicolumn{2}{c}{(667) Denise, 2020-04-11}\\
&\\
\hline
S. Meister & CH\\
A. Schweizer & CH\\
C. Ellington & DE\\
S. Sposetti & CH\\
A. Manna & CH\\
A. Ossola & CH\\
O. Schreurs & BE\\
M. Bigi & IT\\
P. Baruffetti & IT\\
F. Van Den Abbeel & BE\\
J. Bourgeois & BE\\
R. Boninsegna & BE\\
\hline
&\\
\multicolumn{2}{c}{(667) Denise, 2020-05-10}\\
&\\
\hline
K. Hanna & MT\\
K. Green & CT\\
R. Kamin & PA\\
S. Conard & MD\\
K. Getrost & OH\\
A. Scheck & MD\\
A. Caroglanian & MD\\
J. Massura & IN\\
J. Harris & VA\\
C. Anderson, K. Thomason & ID\\
M. Wasiuta, B. Billard & VA\\
B. Billard & VA\\
\hline
\end{tabular}
\caption{List of stellar occultation observers and locations of the observing sites.}
\label{occult_obs}
\end{table}

\clearpage

% Dodaj±c obiekt uzupe³niæ:
% - tabelê z wynikami
% - tabelê z obserwacjami
% - referencje
% - sk³adanki
% - thermal lightcurves
% - liczbê obiektów w intro i conclusions

 %Plots:
 % Chisq ver Gamma for different sizes (colour coded) and roughnesses (symbol coded) for one pole. 

\newpage
\section{Visible light curves}
 Composite light curves in the visible, with the new data of target asteroids 
 (Figures \ref{108composit2015} - \ref{995composit2017}). 
%\clearpage
%\vspace{1.5cm}
%Composits 1
    \begin{table*}[h]
    \centering
\vspace{0.5cm}
\begin{tabularx}{\linewidth}{XX}
\includegraphics[width=0.46\textwidth]{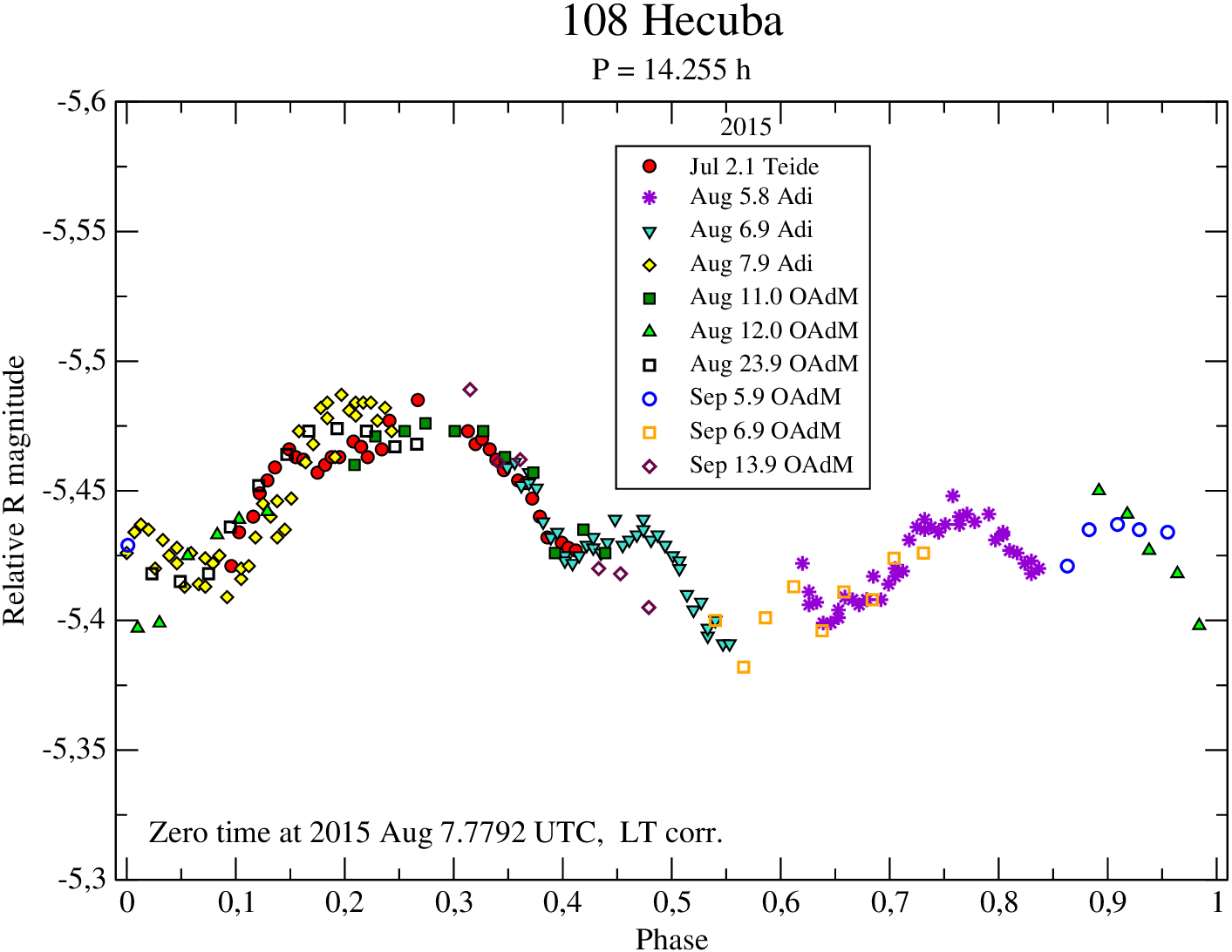} 
\captionof{figure}{Composite light curve of (108) Hecuba from the year 2015.}
\label{108composit2015}
&
\includegraphics[width=0.46\textwidth]{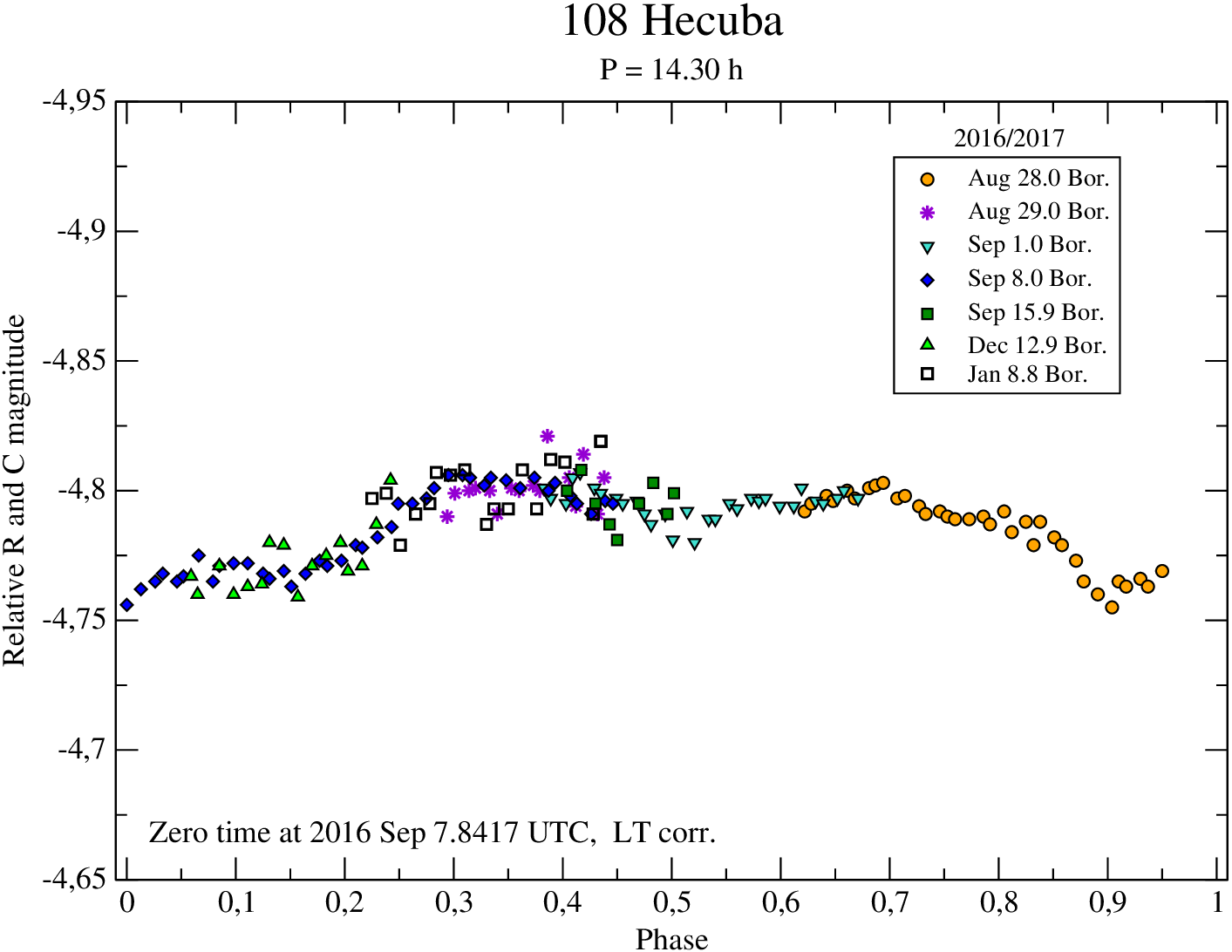} 
\captionof{figure}{Composite light curve of (108) Hecuba from the years 2016-2017.}
\label{108composit2016}
\\
\includegraphics[width=0.46\textwidth]{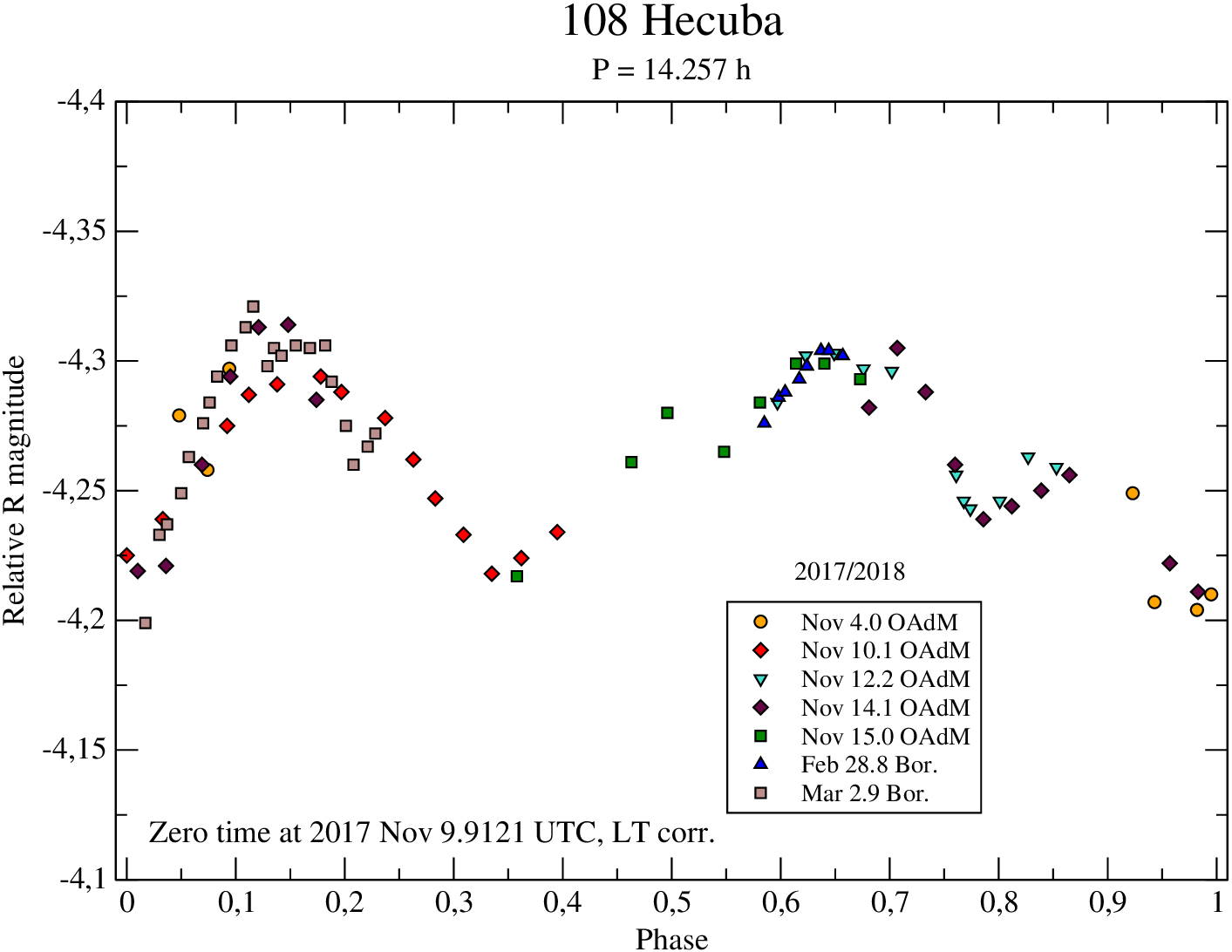} 
\captionof{figure}{Composite light curve of (108) Hecuba from the years 2017-2018.}
\label{108composit2017}
&
\includegraphics[width=0.46\textwidth]{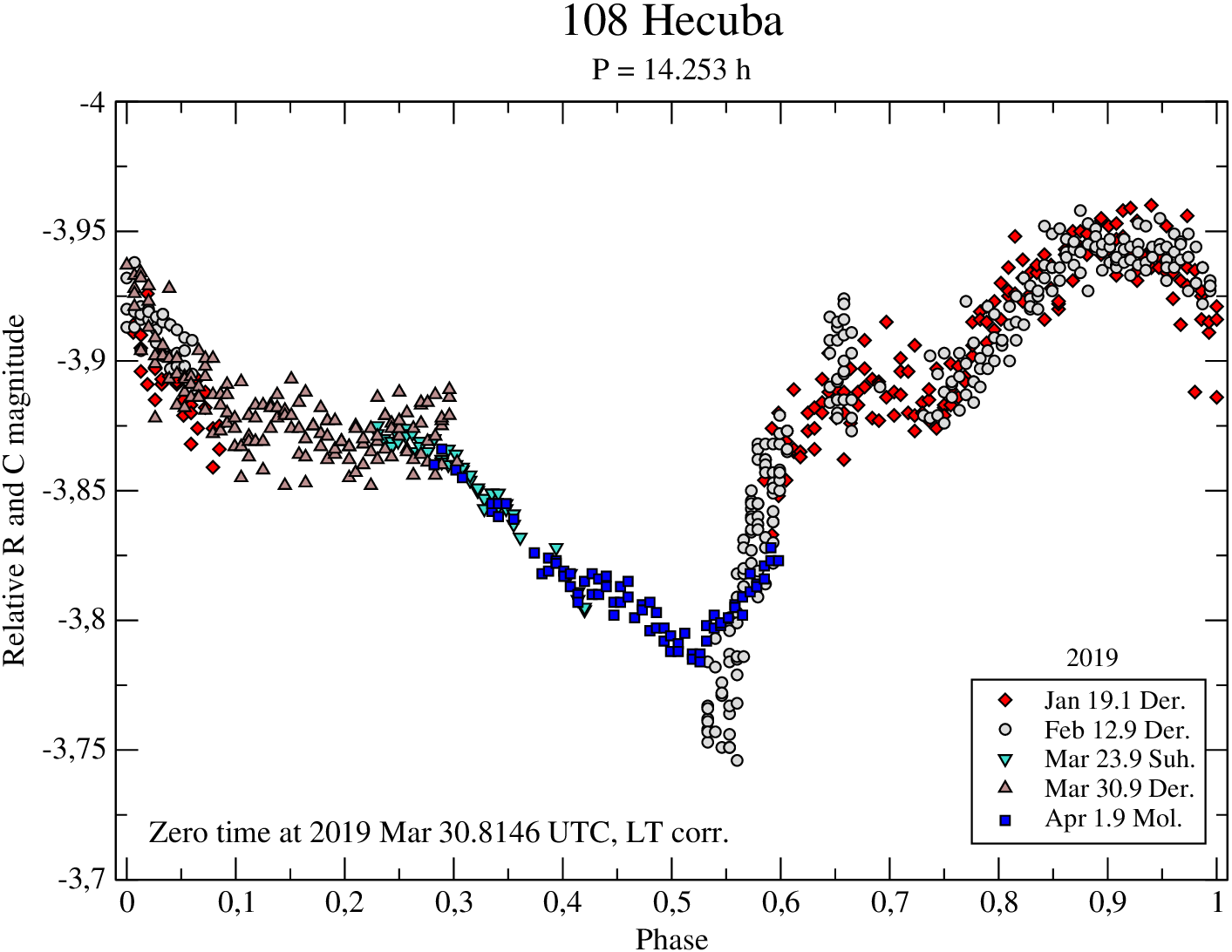} 
\captionof{figure}{Composite light curve of (108) Hecuba from the year 2019.}
\label{108composit2019}
\\
\includegraphics[width=0.46\textwidth]{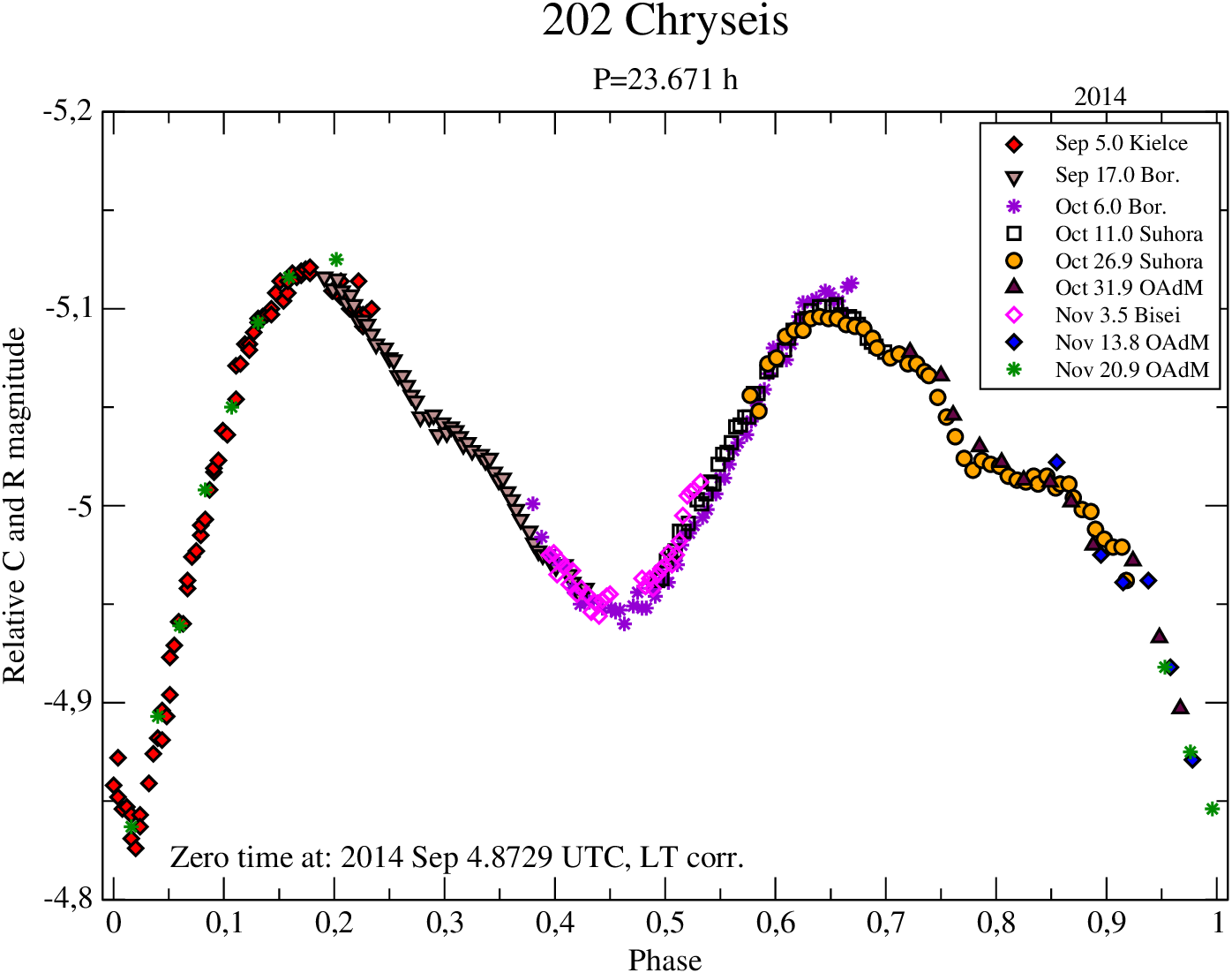} 
\captionof{figure}{Composite light curve of (202) Chryseis from the year 2014.}
\label{202composit2014}
&
\includegraphics[width=0.46\textwidth]{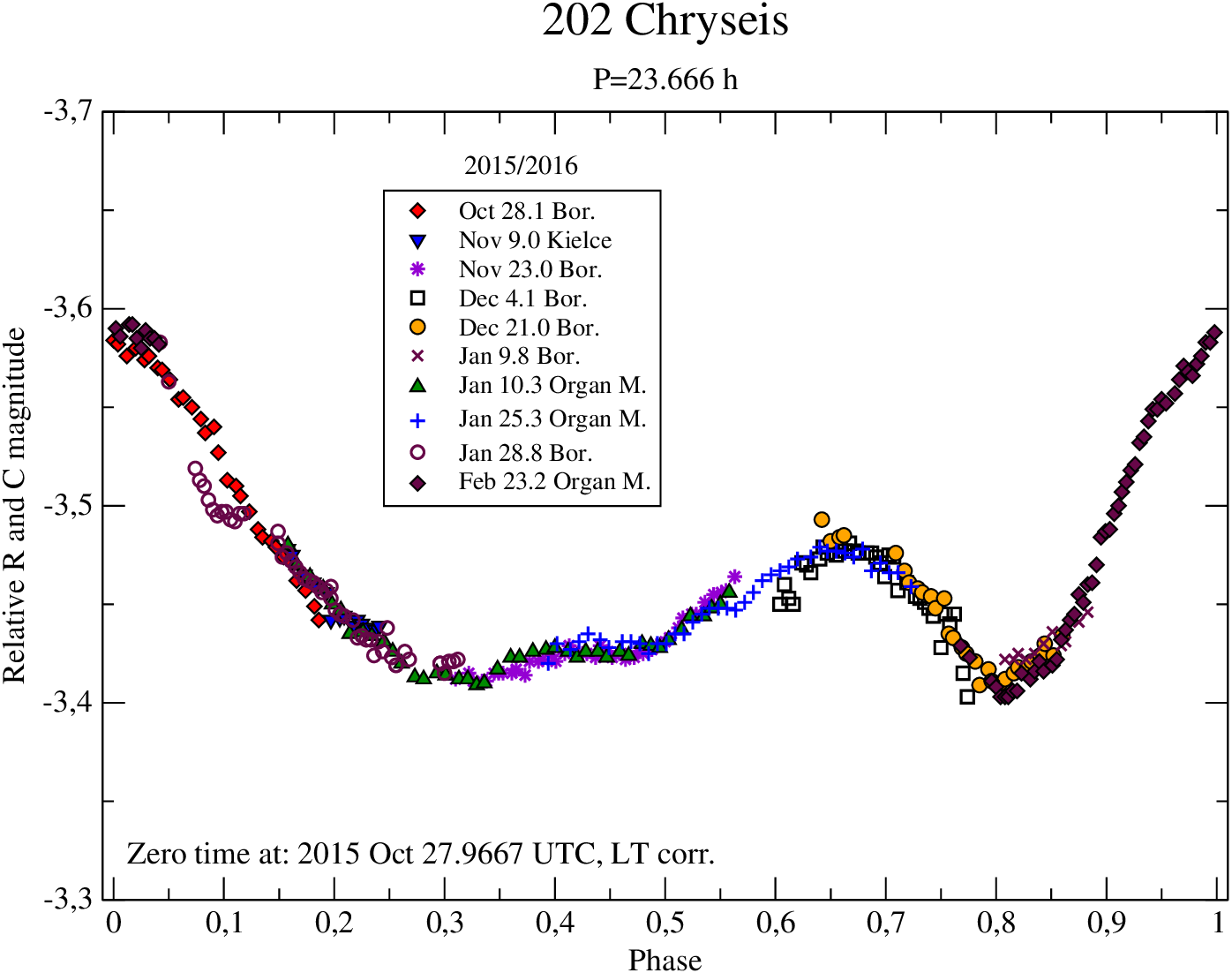} 
\captionof{figure}{Composite light curve of (202) Chryseis from the years 2015-2016.}
\label{202composit2015}
\\
\end{tabularx}
%\caption{A table with figures}
%\label{tab:mytable}
    \end{table*}%

%\clearpage
%\vspace{0.5cm}

%Composits 2
%\vspace{1.5cm}

    \begin{table*}[ht]
    \centering
\vspace{0.5cm}
\begin{tabularx}{\linewidth}{XX}
\includegraphics[width=0.46\textwidth]{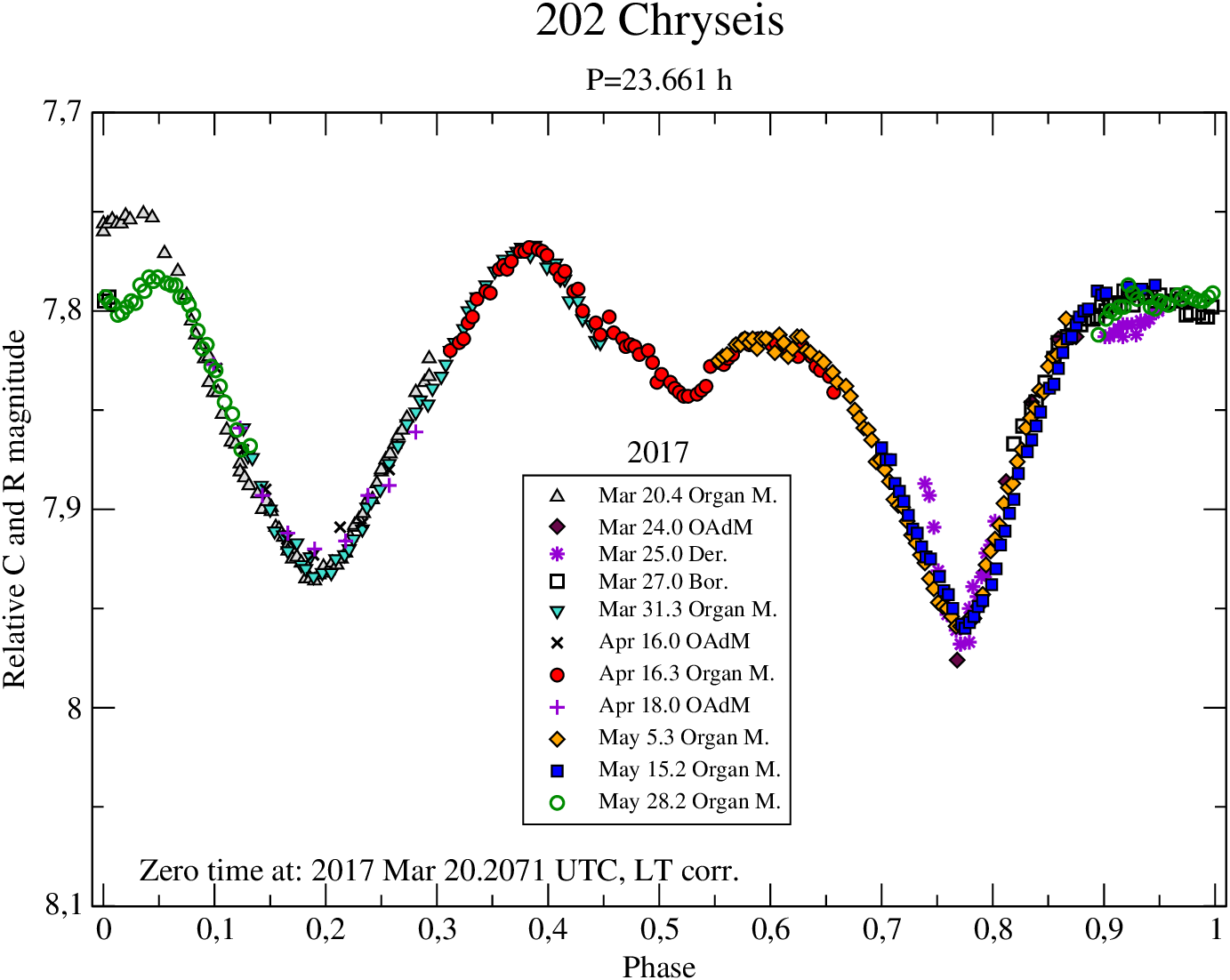} 
\captionof{figure}{Composite light curve of (202) Chryseis from the year 2017.}
\label{202composit2017}
&
\includegraphics[width=0.46\textwidth]{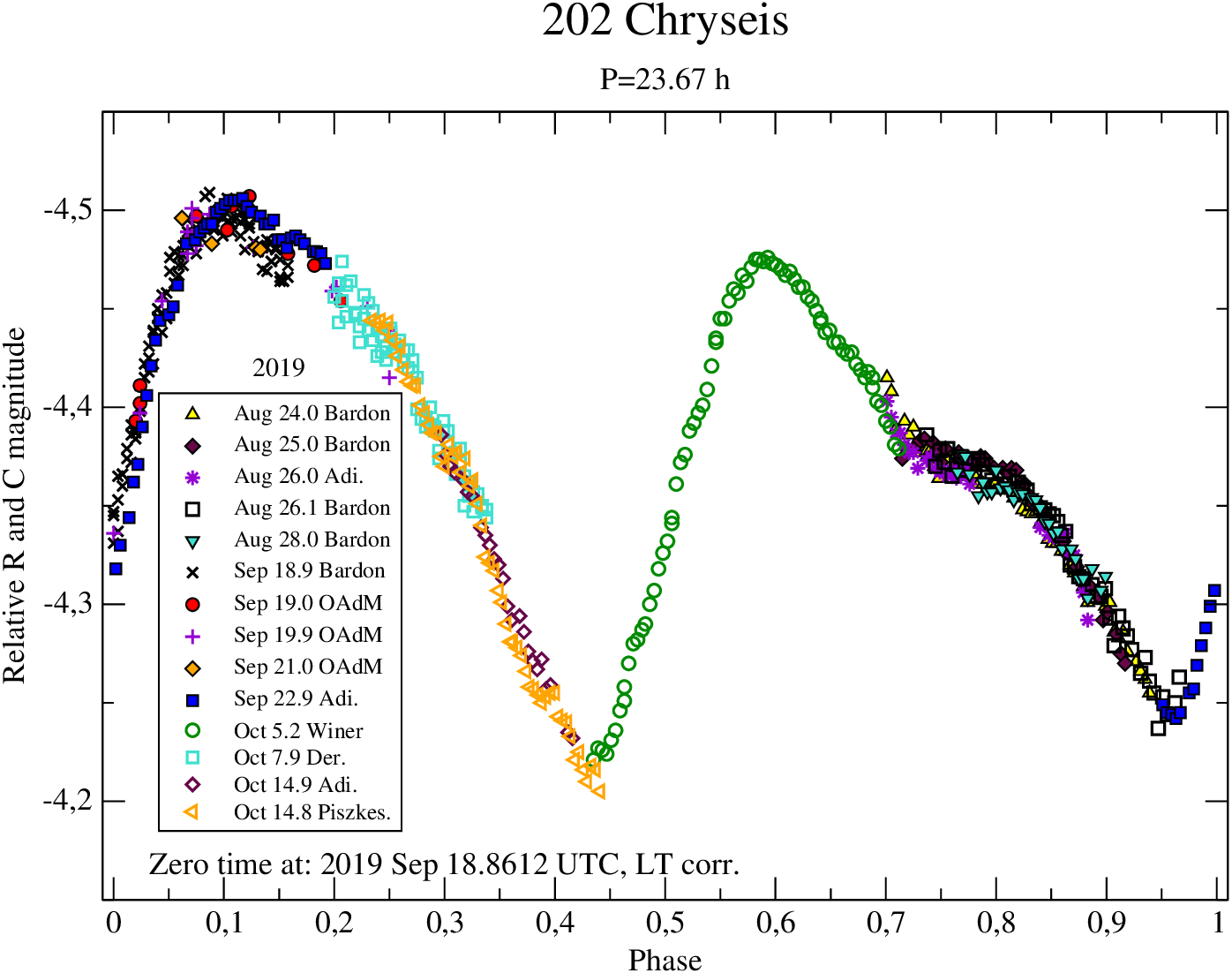} 
\captionof{figure}{Composite light curve of (202) Chryseis from the year 2019.}
\label{202composit2019}
\\
\includegraphics[width=0.46\textwidth]{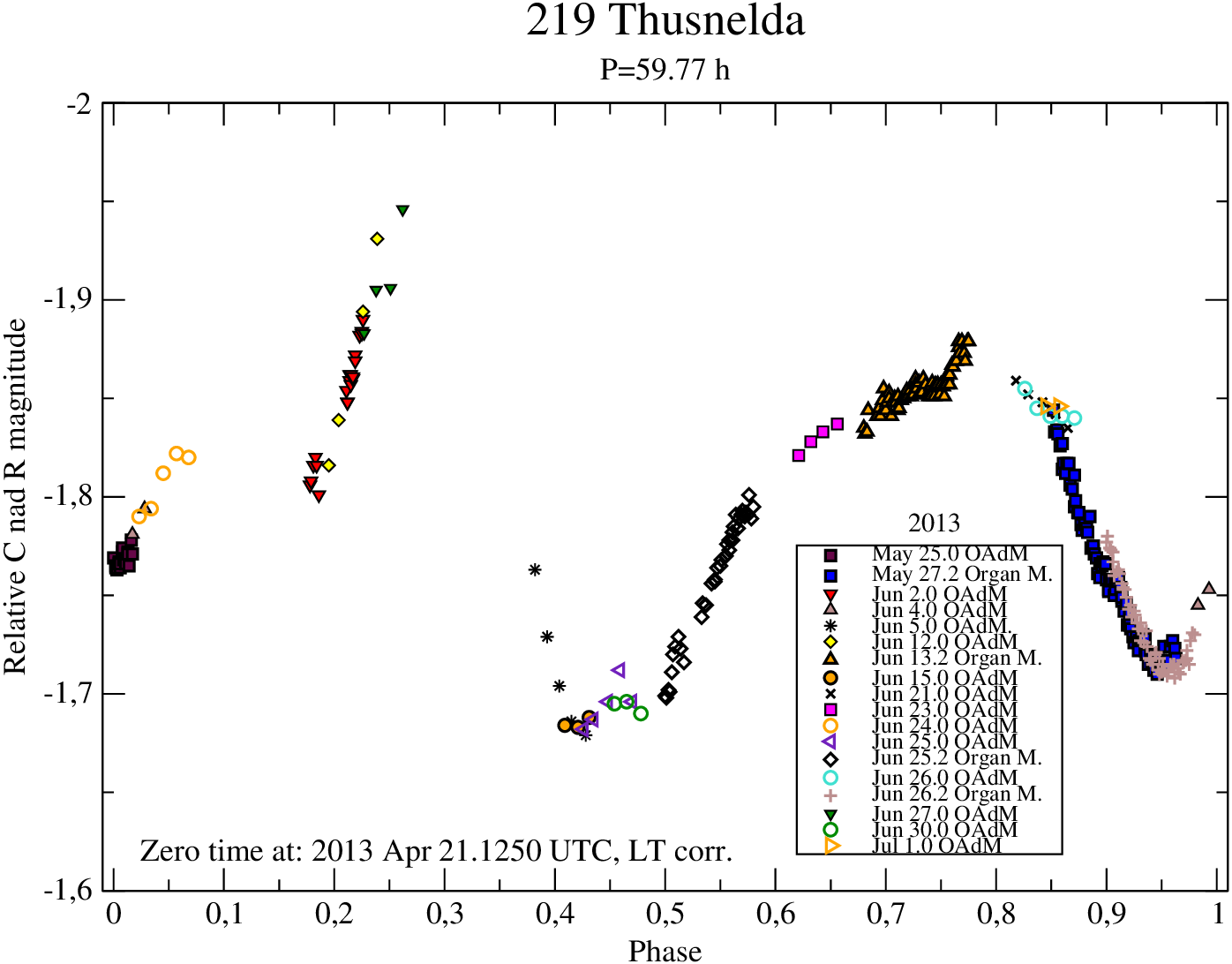} 
\captionof{figure}{Composite light curve of (219) Thusnelda from the year 2013.}
\label{219composit2013}
&
\includegraphics[width=0.46\textwidth]{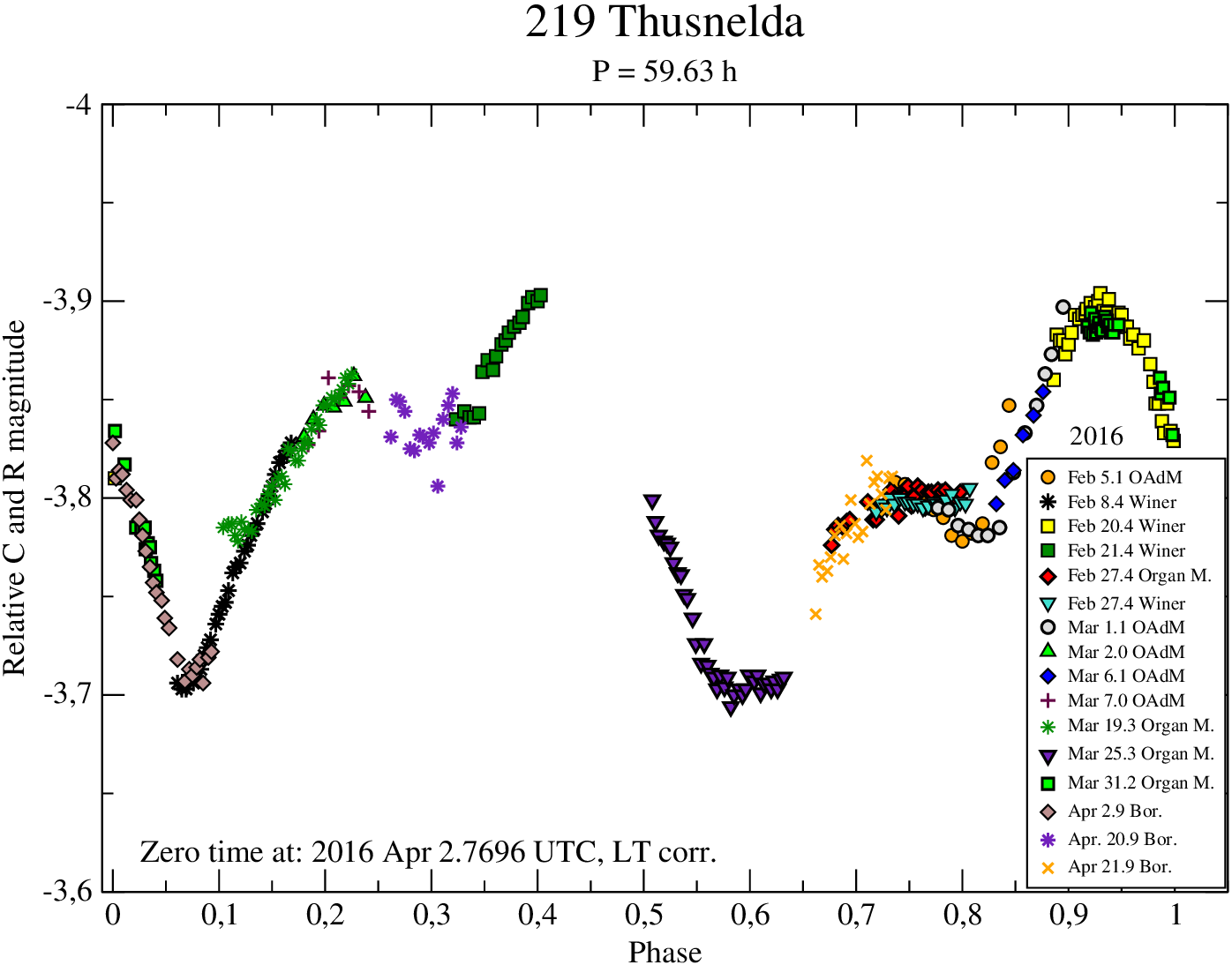} 
\captionof{figure}{Composite light curve of (219) Thusnelda from the year 2016.}
\label{219composit2016}
\\
\includegraphics[width=0.46\textwidth]{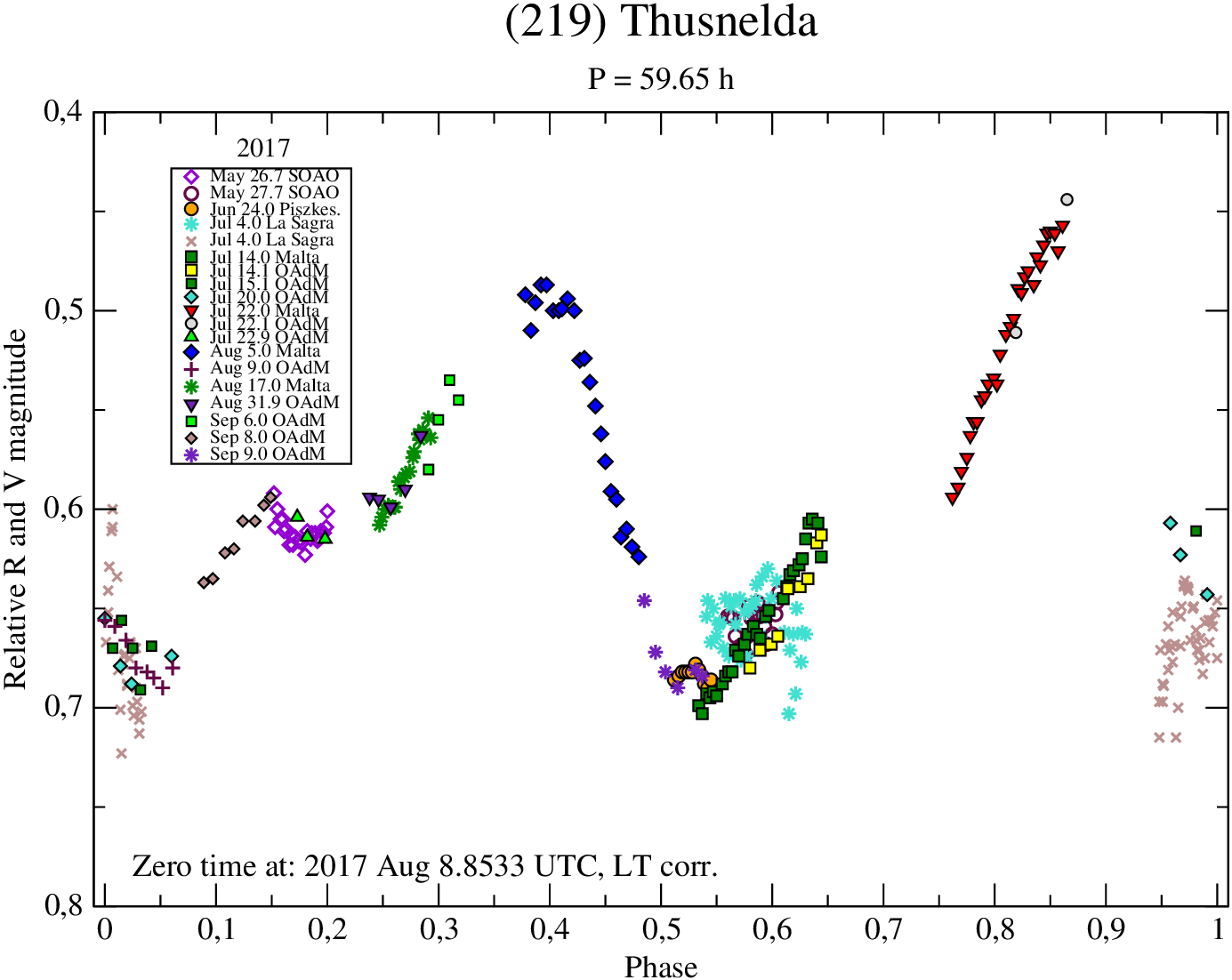} 
\captionof{figure}{Composite light curve of (219) Thusnelda from the year 2017.}
\label{219composit2017}
&
\includegraphics[width=0.46\textwidth]{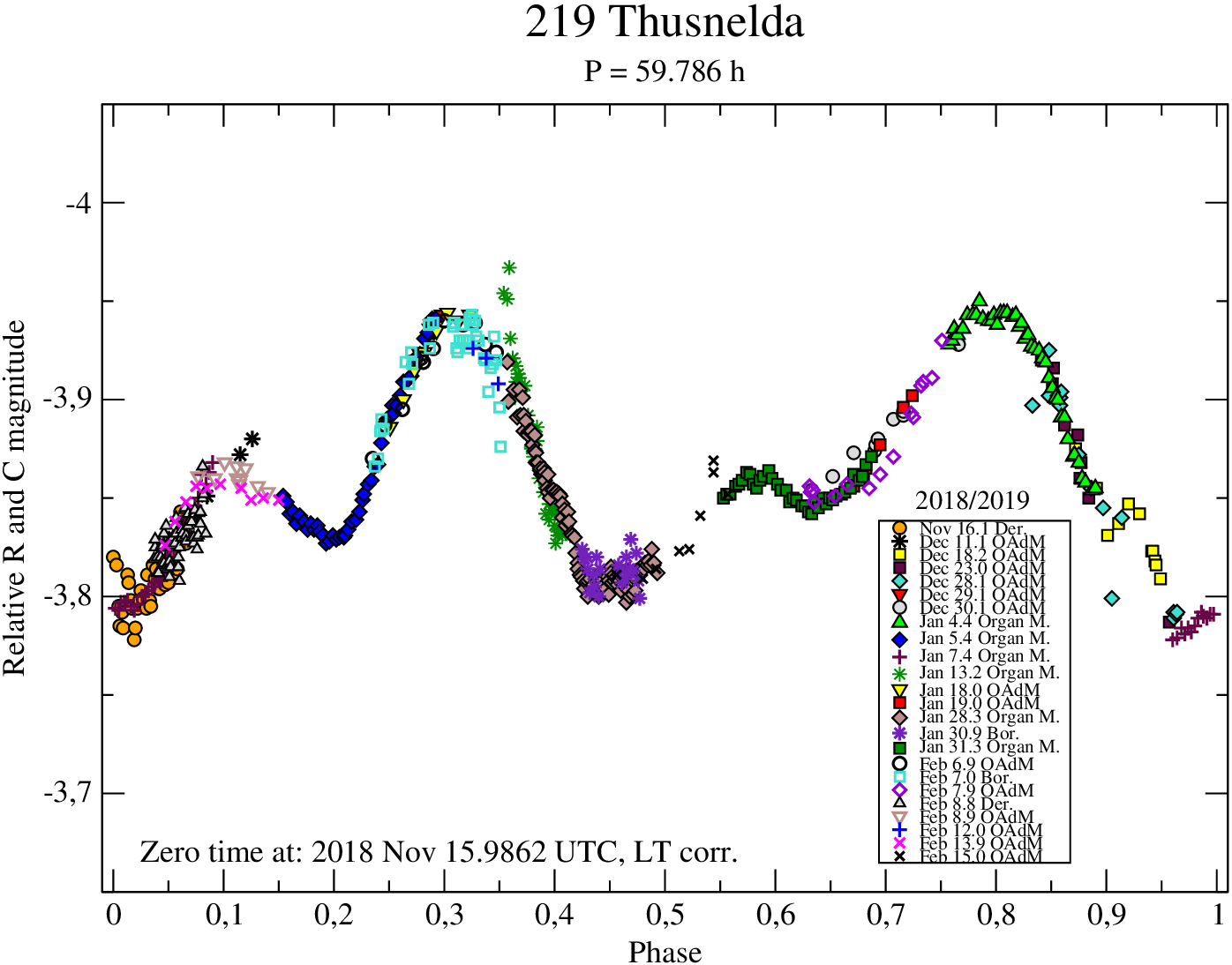} 
\captionof{figure}{Composite light curve of (219) Thusnelda from the years 2018-2019.}
\label{219composit2019}
\\
\end{tabularx}
    \end{table*}

%\clearpage
%\vspace{0.5cm}

%Composits 3
%\vspace{1.5cm}

    \begin{table*}[ht]
    \centering
\vspace{0.5cm}
\begin{tabularx}{\linewidth}{XX}
\includegraphics[width=0.44\textwidth]{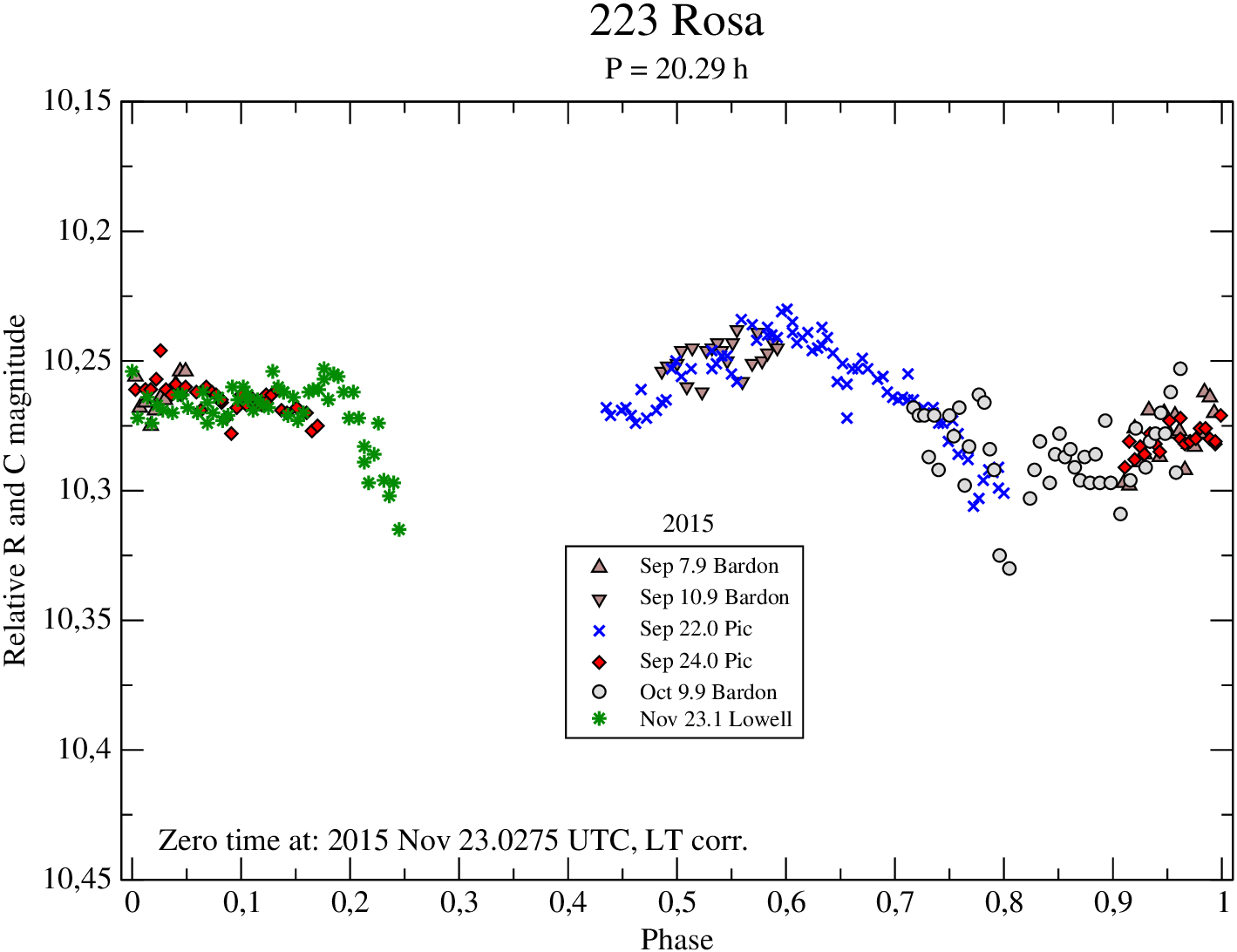} 
\captionof{figure}{Composite light curve of (223) Rosa from the year 2015.}
\label{223composit2015}
&
\includegraphics[width=0.44\textwidth]{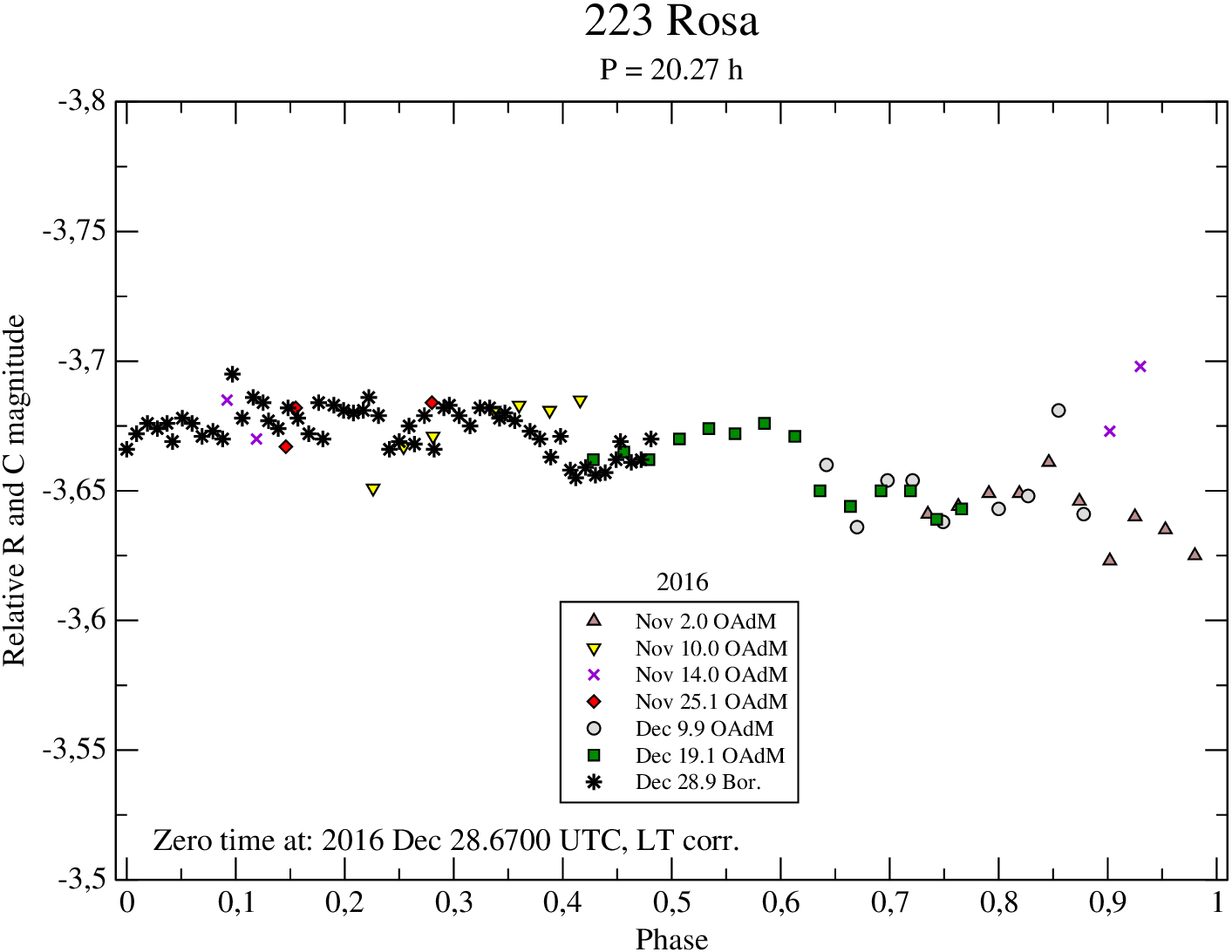} 
\captionof{figure}{Composite light curve of (223) Rosa from the year 2016.}
\label{223composit2016}
\\
\includegraphics[width=0.44\textwidth]{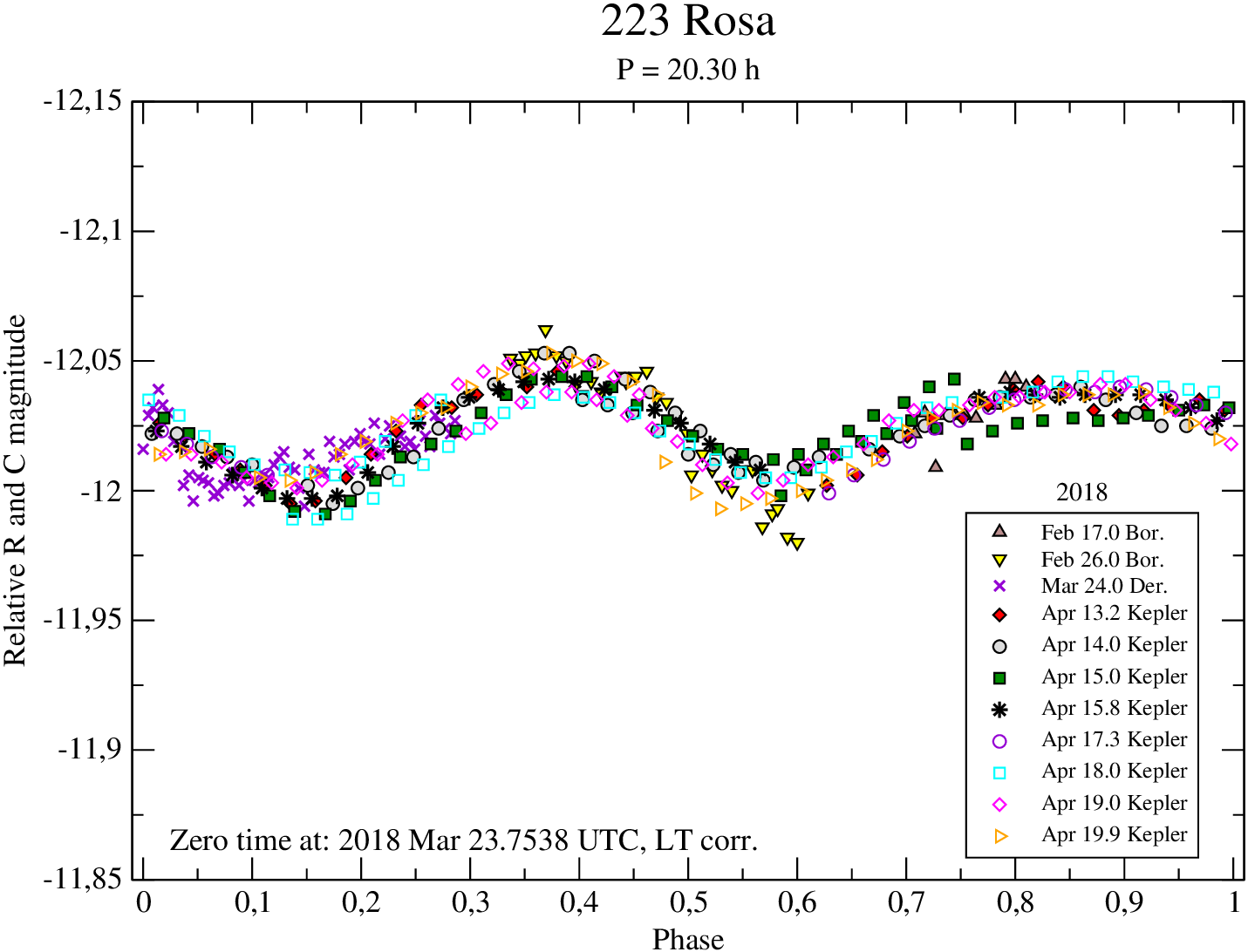} 
\captionof{figure}{Composite light curve of (223) Rosa from the year 2018.}
\label{223composit2018}
&
\includegraphics[width=0.44\textwidth]{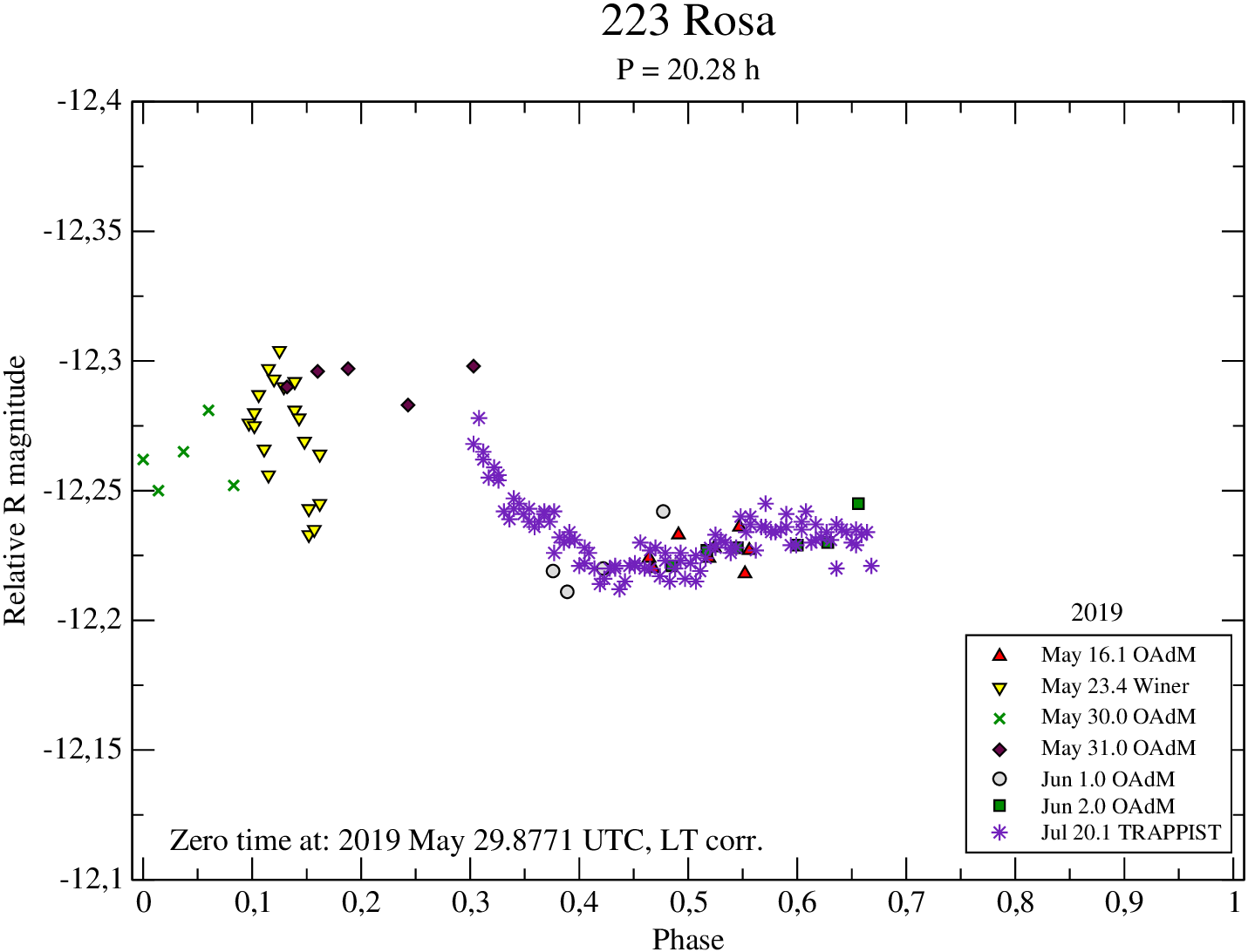} 
\captionof{figure}{Composite light curve of (223) Rosa from the year 2019.}
\label{223composit2019}
\\
\includegraphics[width=0.44\textwidth]{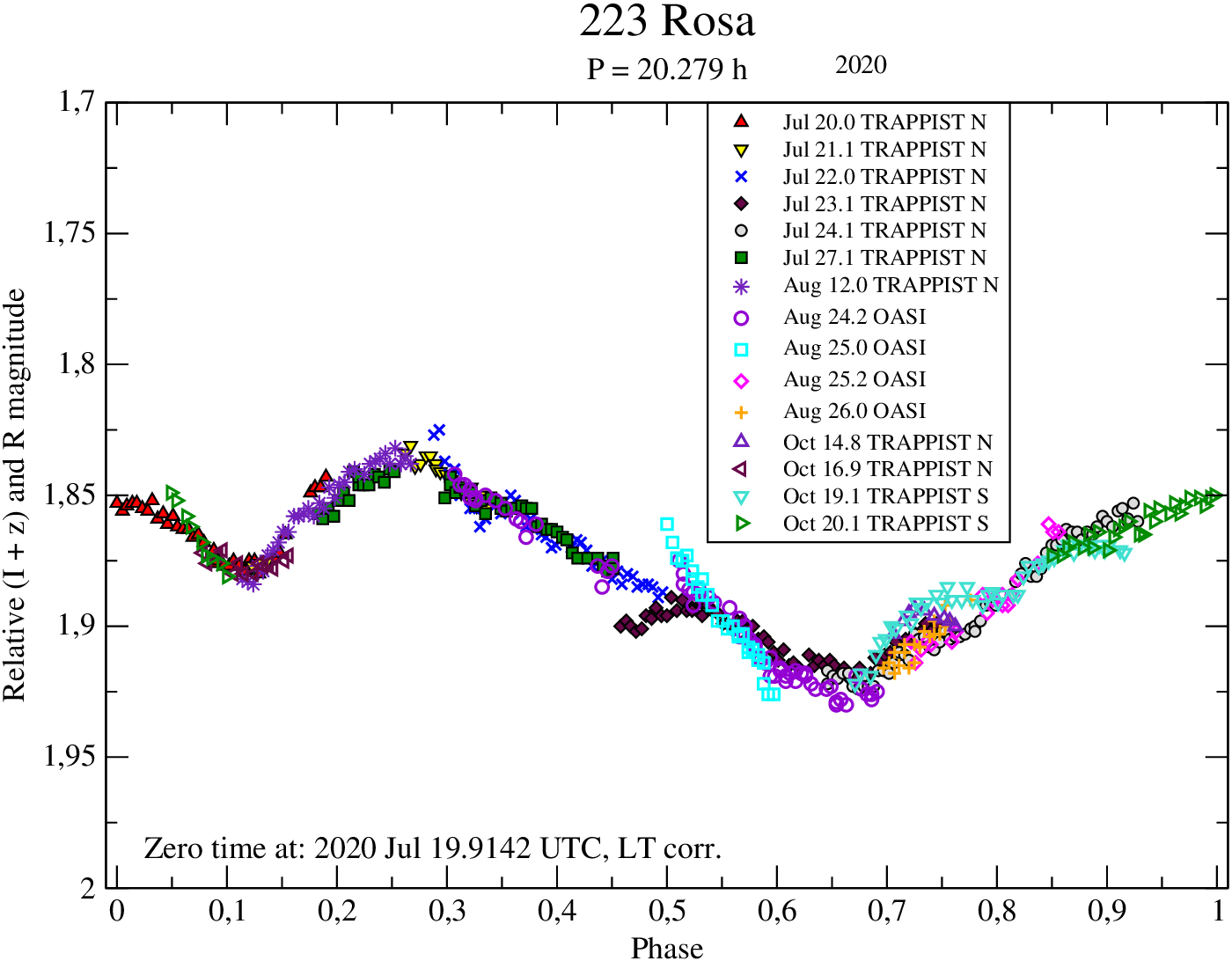} 
\captionof{figure}{Composite light curve of (223) Rosa from the year 2020.}
\label{223composit2020}
&
\includegraphics[width=0.44\textwidth]{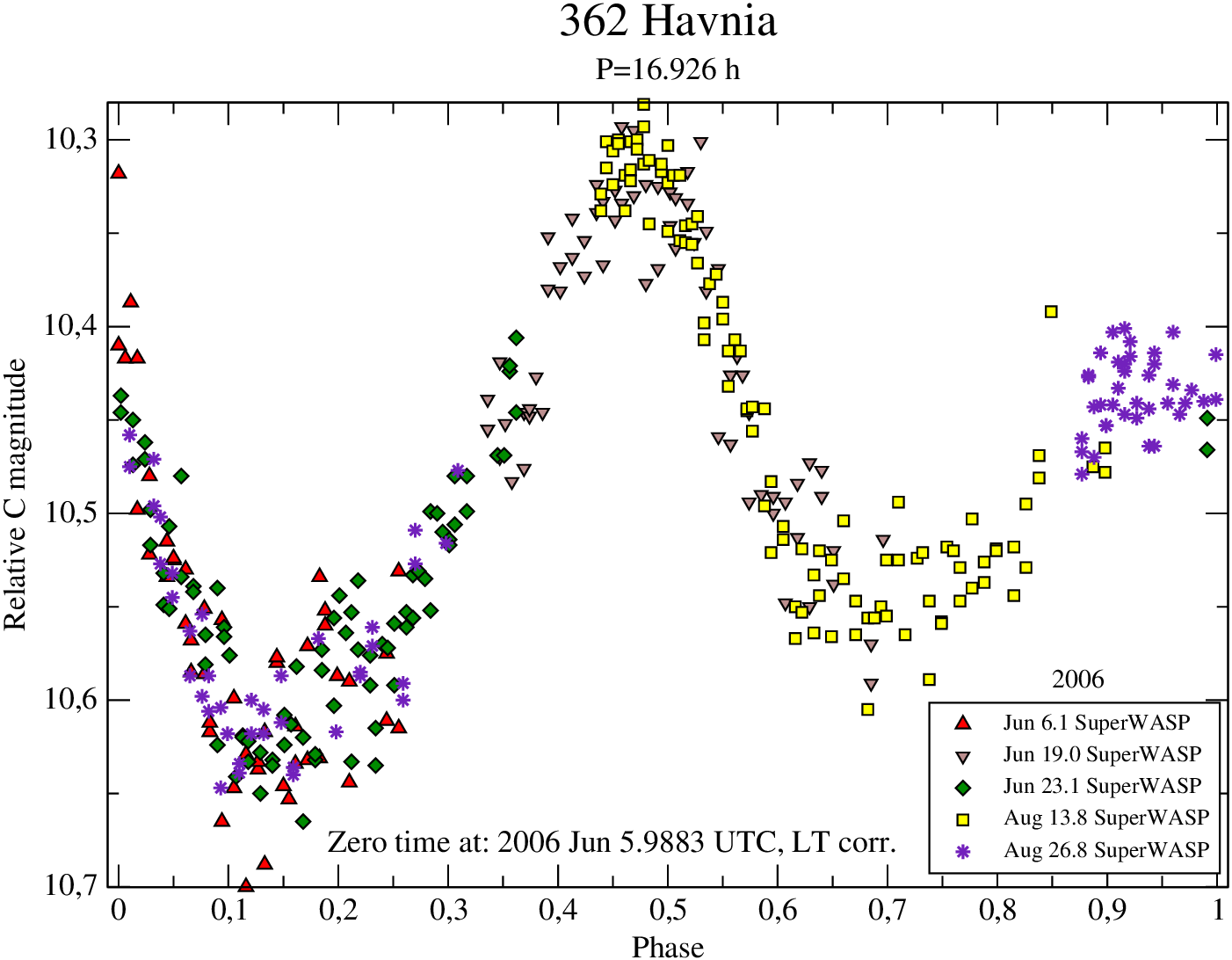} 
\captionof{figure}{Composite light curve of (362) Havnia from the year 2006.}
\label{362composit2006}
\\
\end{tabularx}
    \end{table*}

%Composits 4

    \begin{table*}[ht]
    \centering
\vspace{0.5cm}
\begin{tabularx}{\linewidth}{XX}
\includegraphics[width=0.46\textwidth]{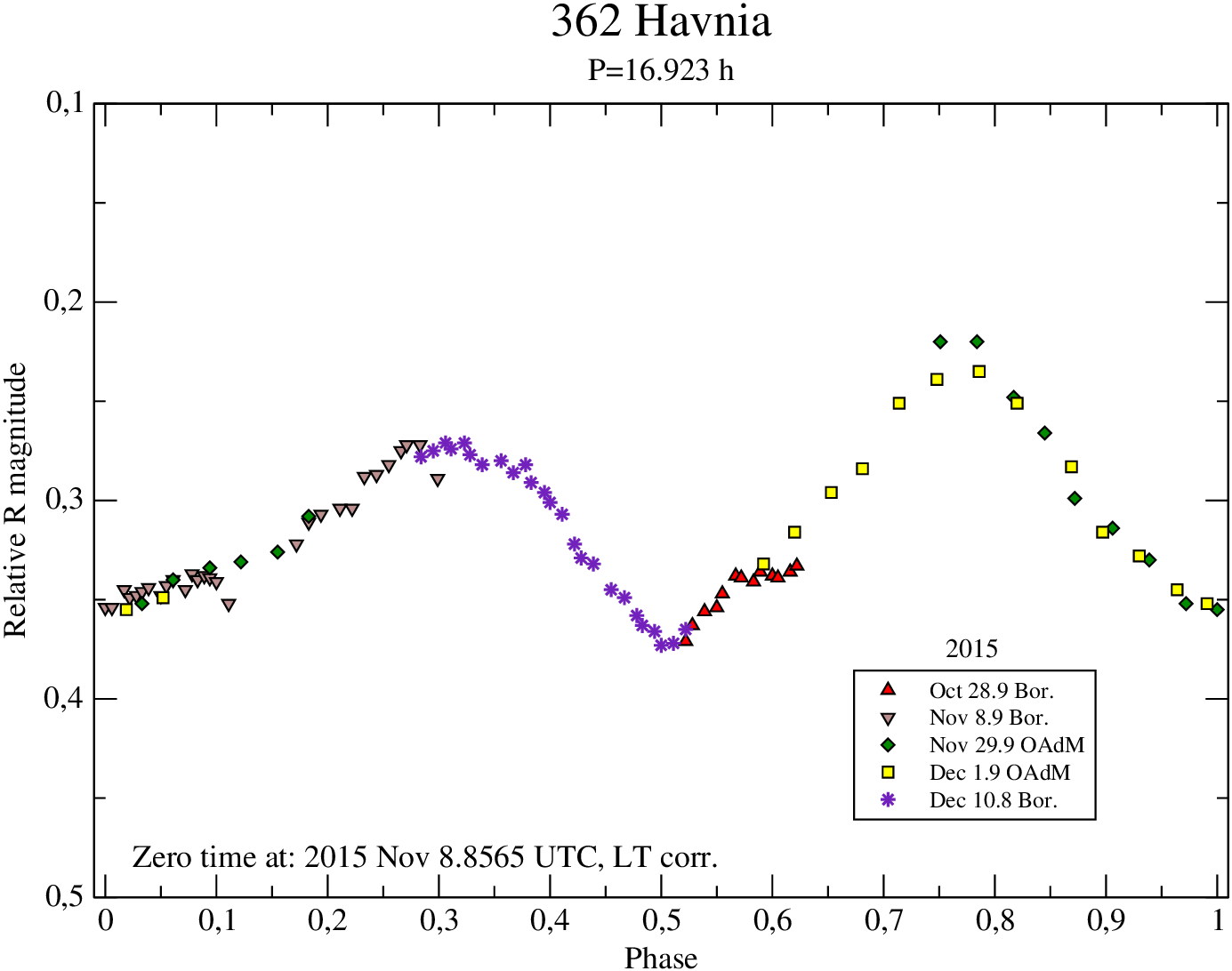} 
\captionof{figure}{Composite light curve of (362) Havnia from the year 2015.}
\label{362composit2015}
&
\includegraphics[width=0.46\textwidth]{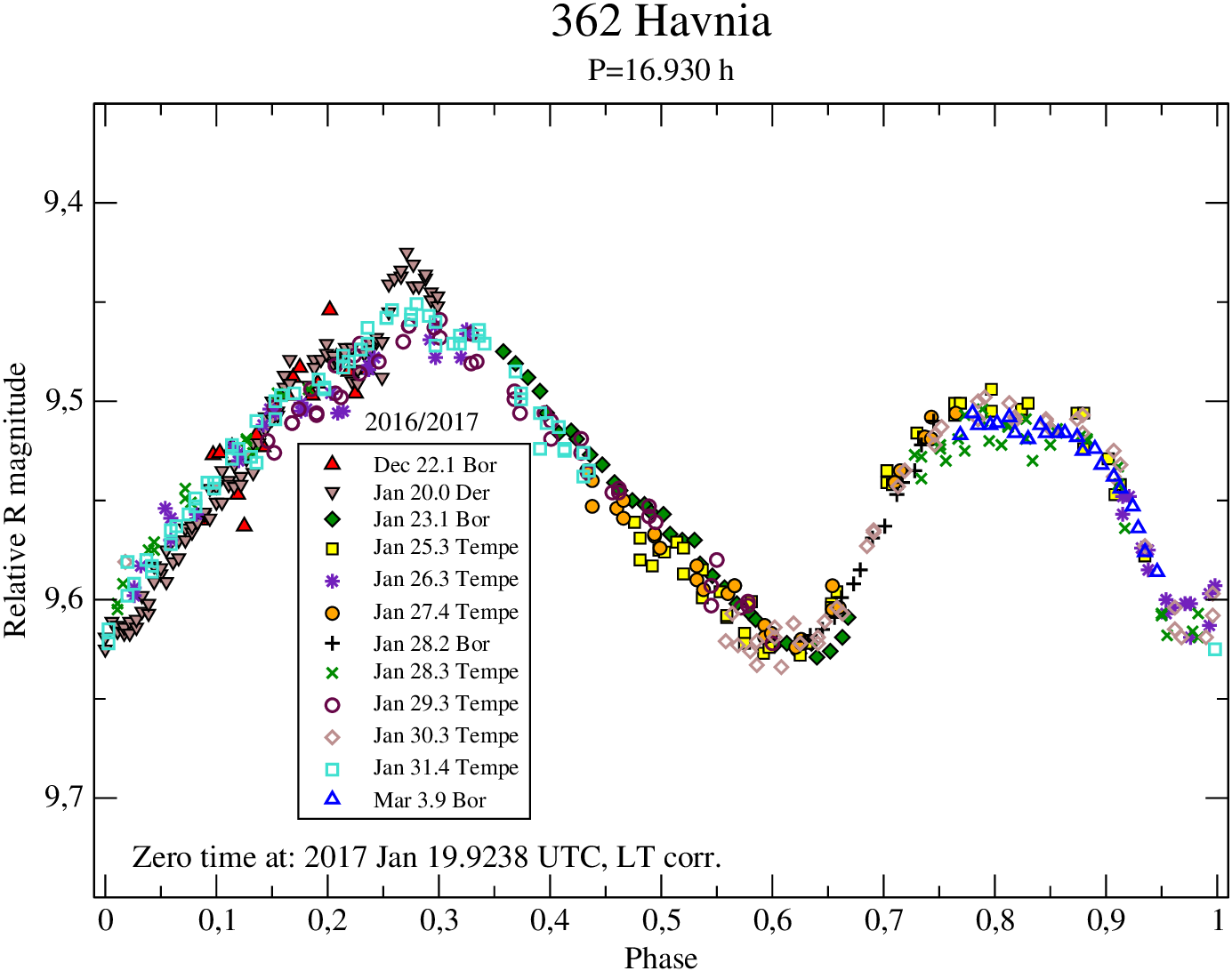} 
\captionof{figure}{Composite light curve of (362) Havnia from the years 2016-2017.}
\label{362composit2017}
\\
\includegraphics[width=0.46\textwidth]{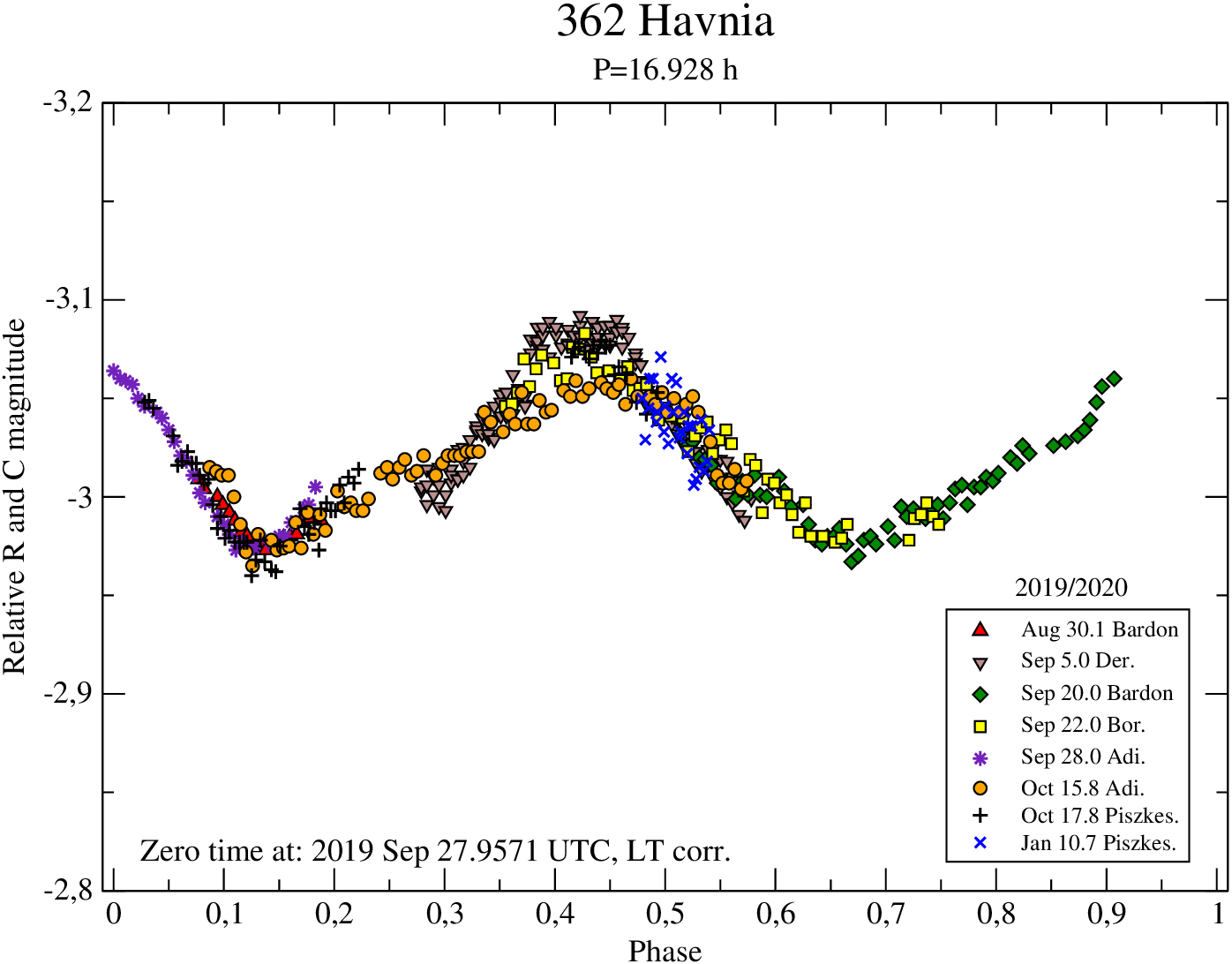} 
\captionof{figure}{Composite light curve of (362) Havnia from the years 2019-2020.}
\label{362composit2019}
&
\includegraphics[width=0.46\textwidth]{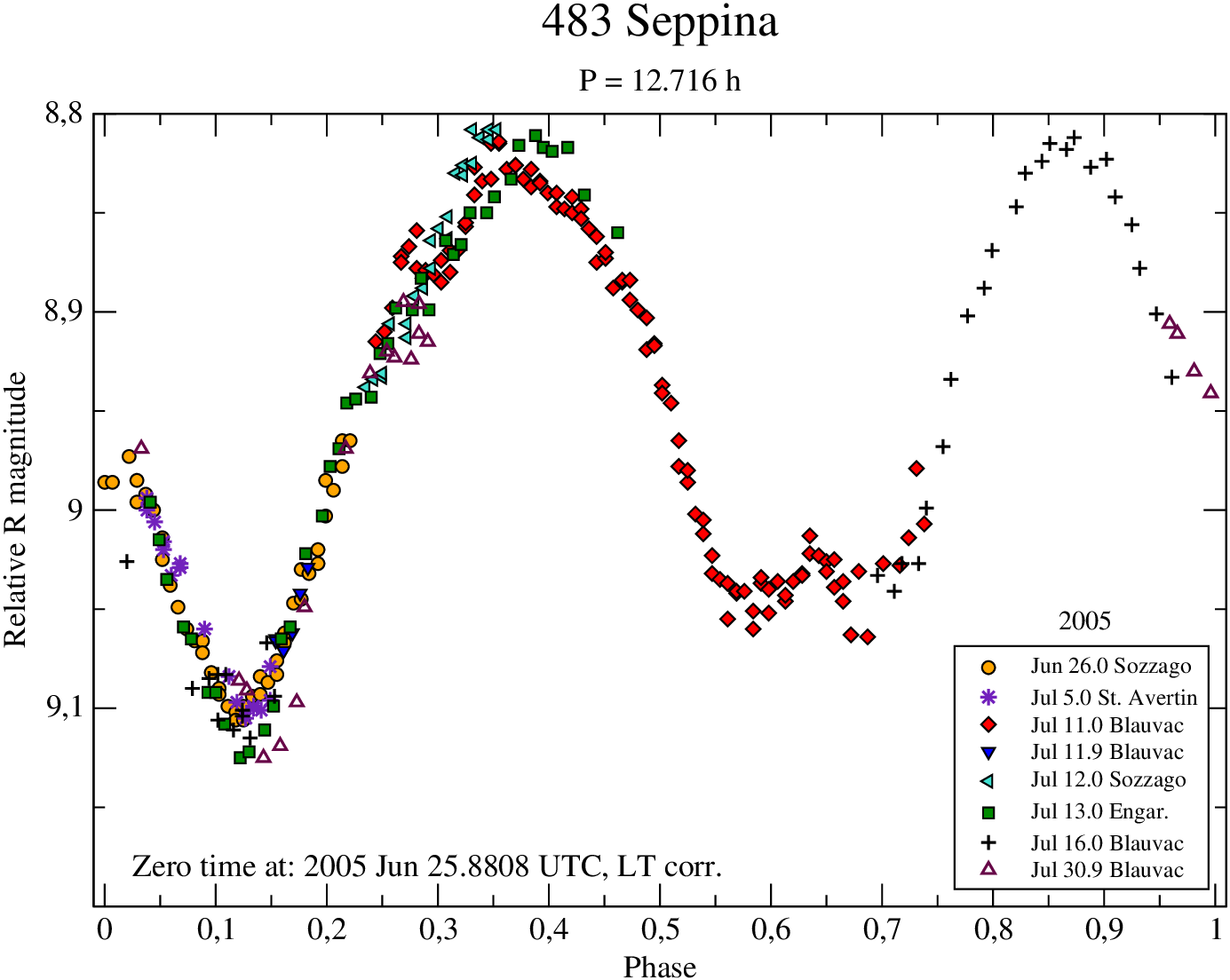} 
\captionof{figure}{Composite light curve of (483) Seppina from the year 2005.}
\label{483composit2005}
\\
\includegraphics[width=0.46\textwidth]{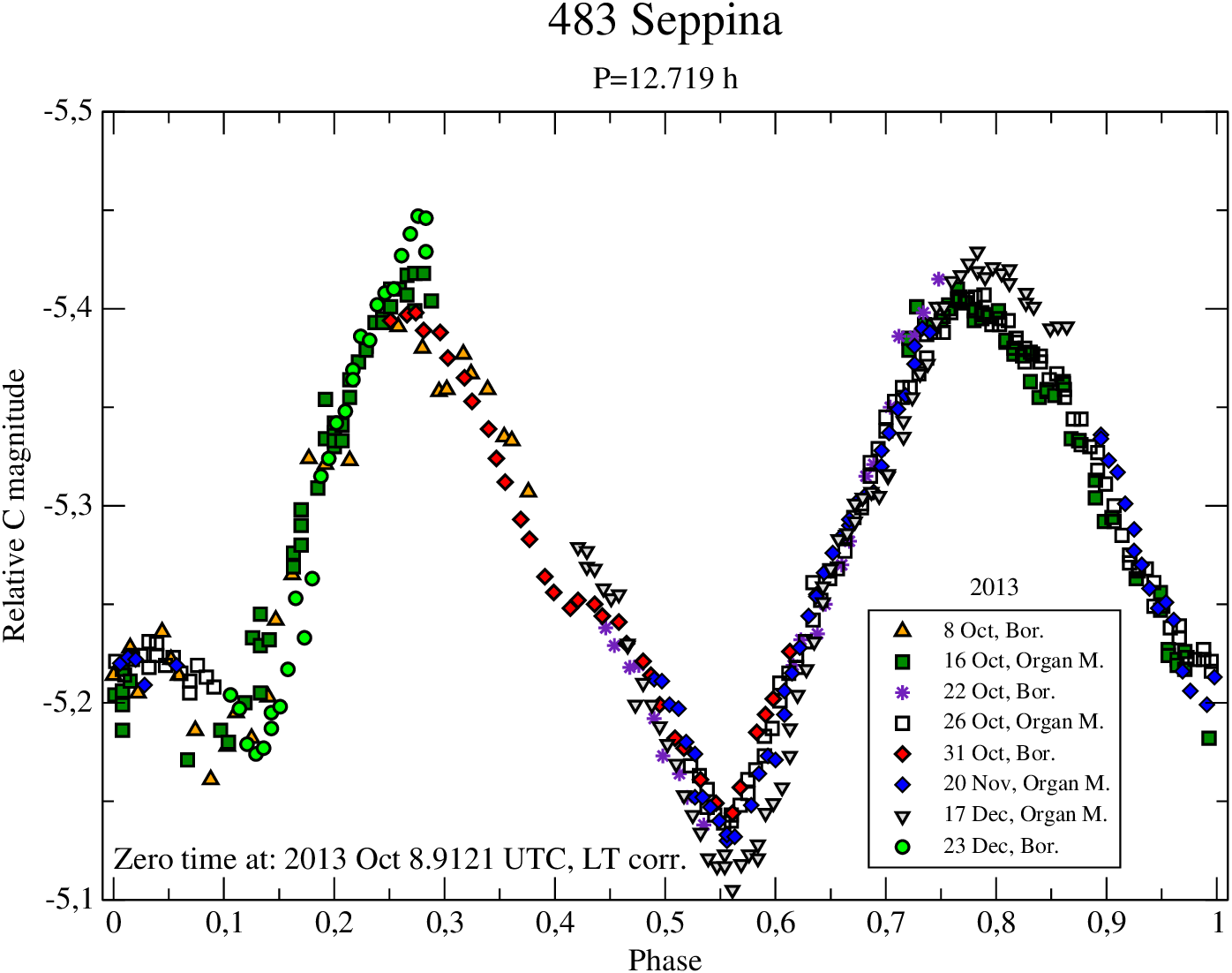} 
\captionof{figure}{Composite light curve of (483) Seppina from the year 2013.}
\label{483composit2013}
&
\includegraphics[width=0.46\textwidth]{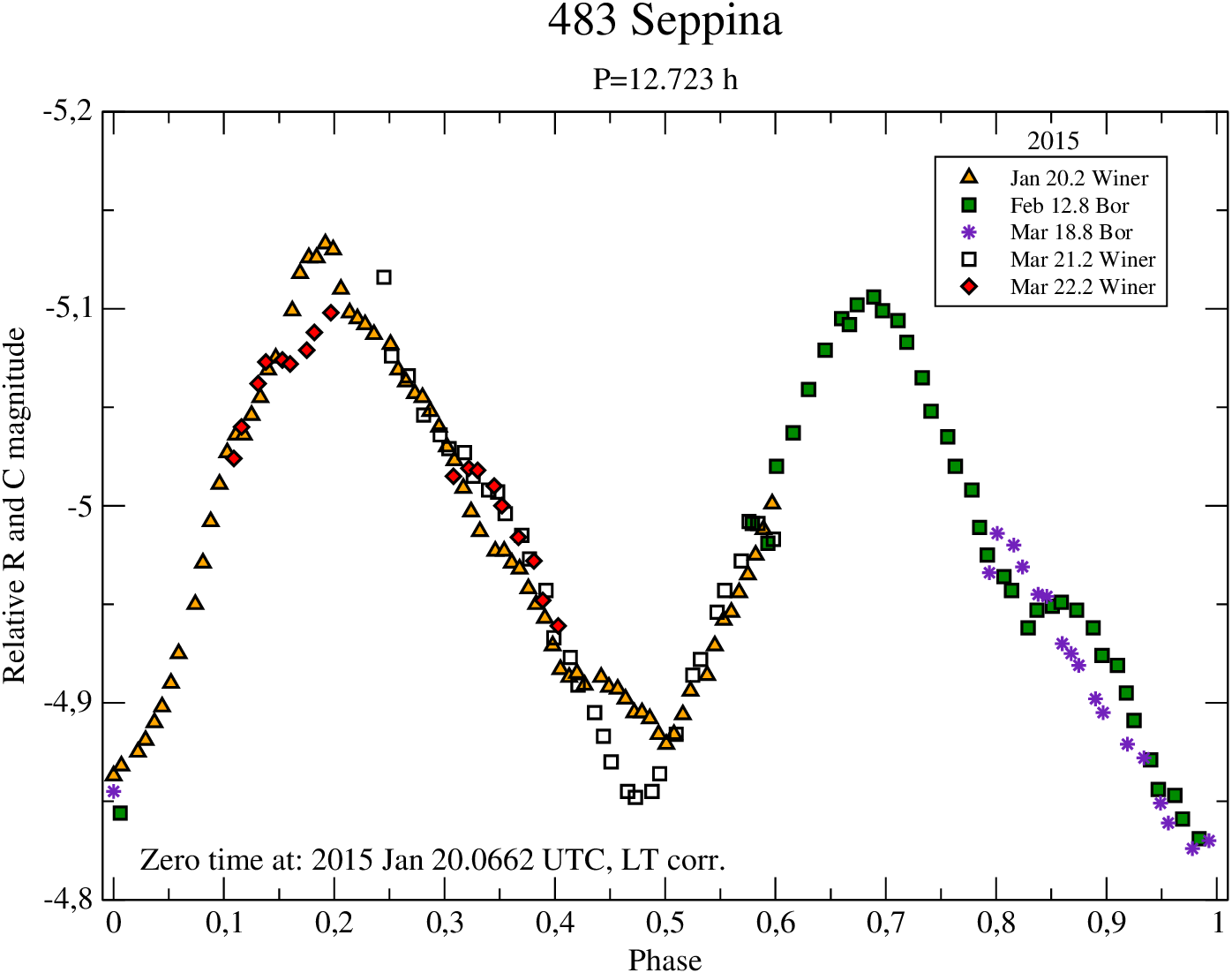} 
\captionof{figure}{Composite light curve of (483) Seppina from the year 2015.}
\label{483composit2015}
\\
\end{tabularx}
    \end{table*}

%\clearpage
%\vspace{0.5cm}

%Composits 5
    \begin{table*}[ht]
    \centering
\vspace{0.5cm}
\begin{tabularx}{\linewidth}{XX}
\includegraphics[width=0.46\textwidth]{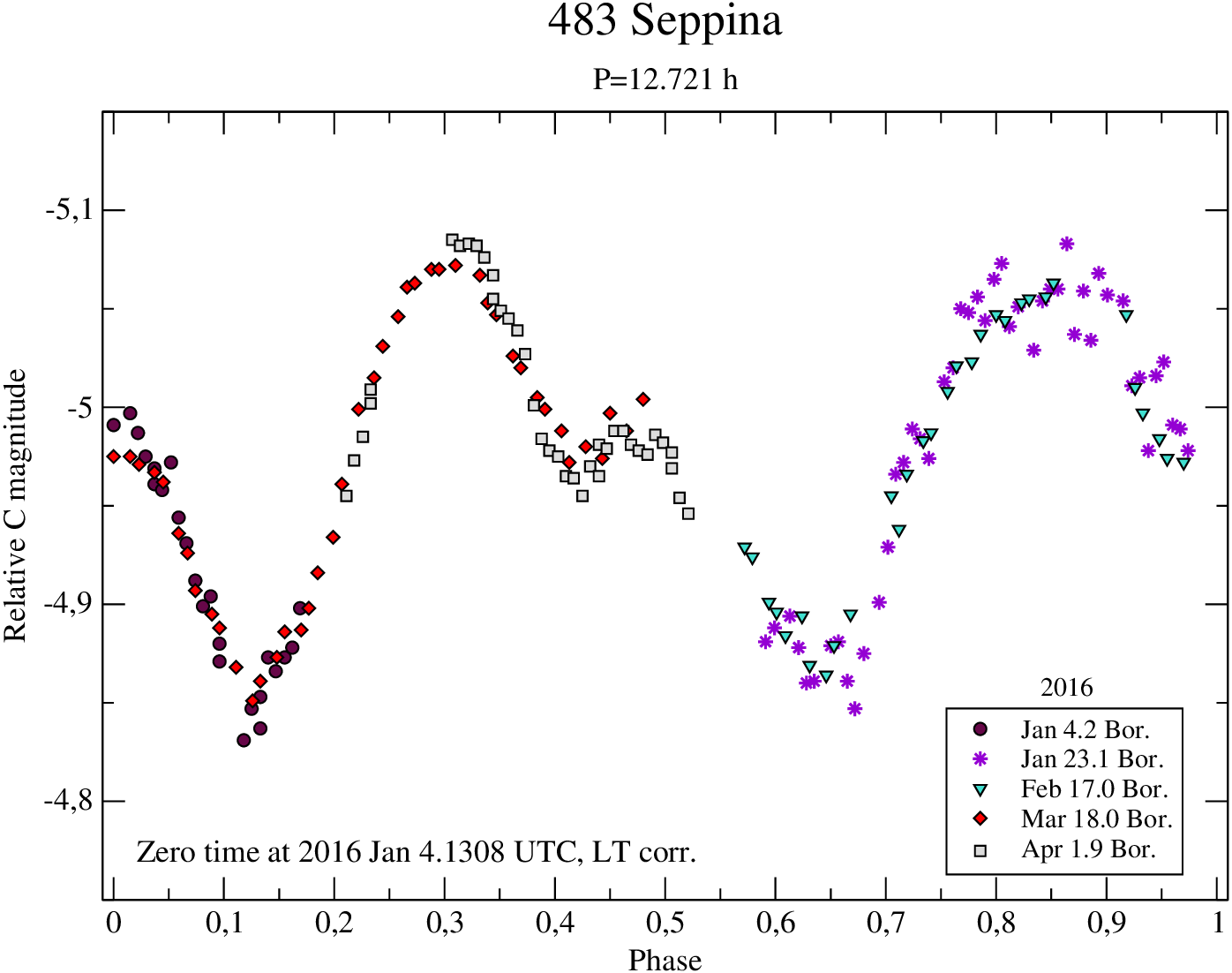} 
\captionof{figure}{Composite light curve of (483) Seppina from the year 2016.}
\label{483composit2016}
&
\includegraphics[width=0.46\textwidth]{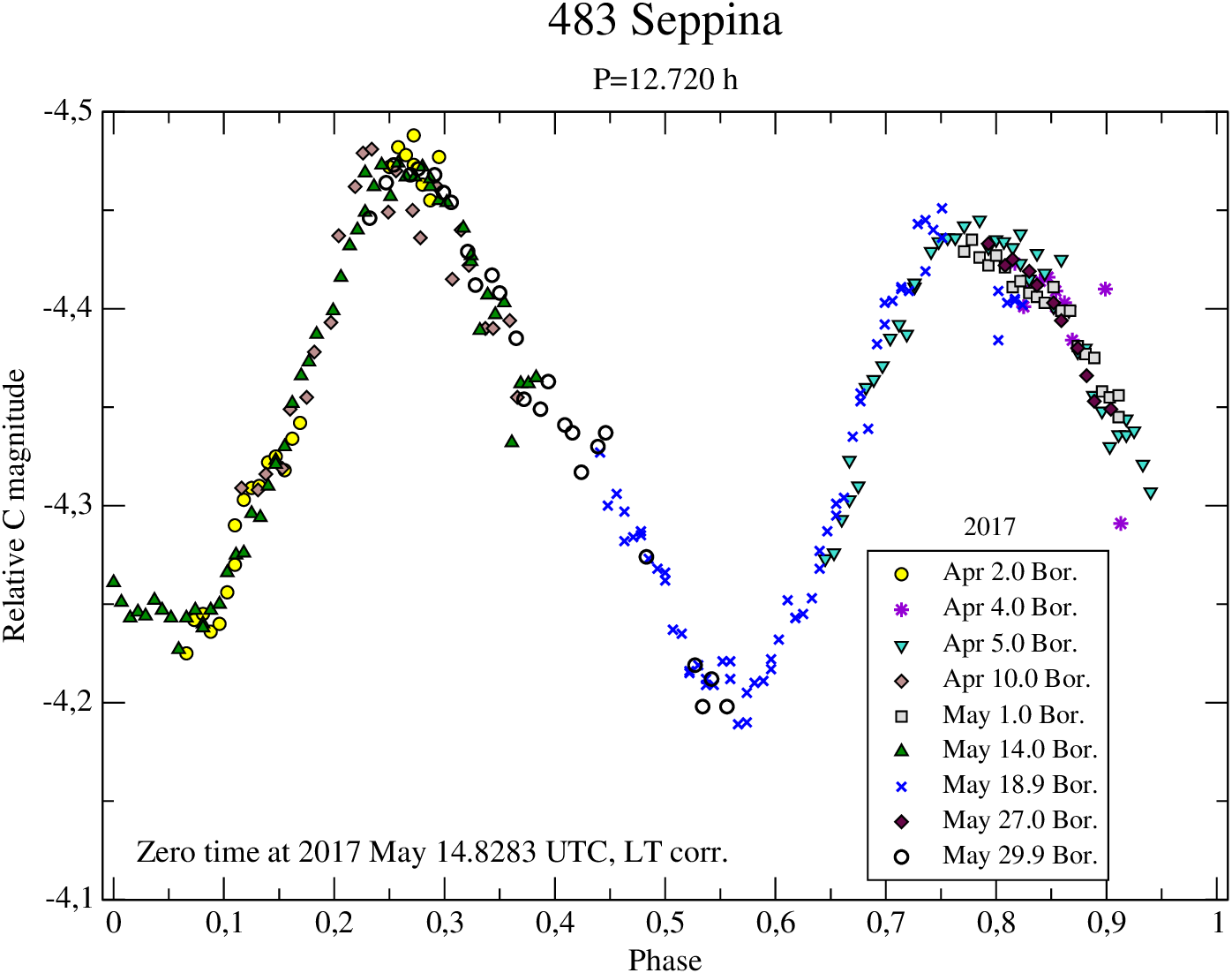} 
\captionof{figure}{Composite light curve of (483) Seppina from the year 2017.}
\label{483composit2017}
\\
\includegraphics[width=0.46\textwidth]{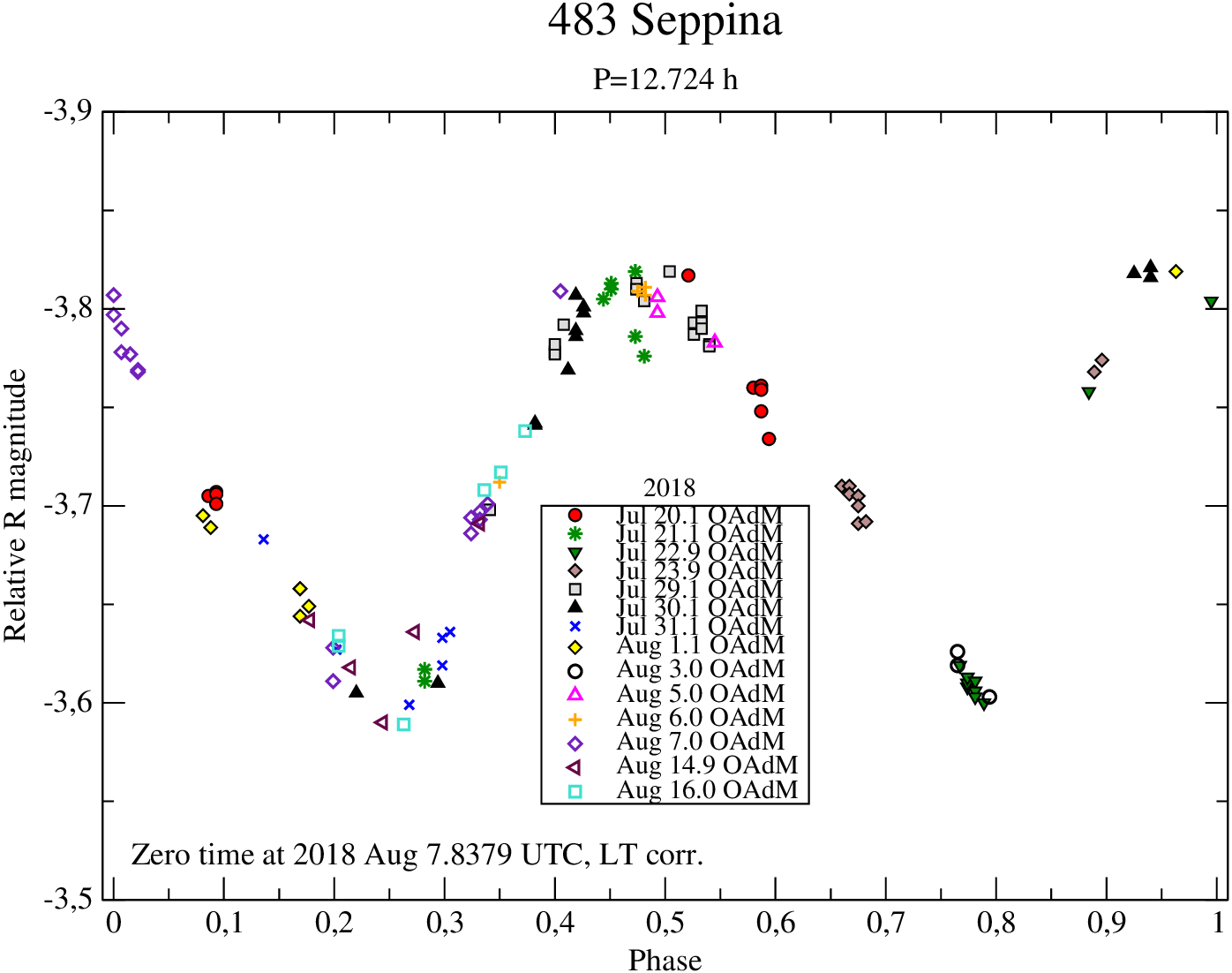} 
\captionof{figure}{Composite light curve of (483) Seppina from the year 2018.}
\label{483composit2018}
&
\includegraphics[width=0.46\textwidth]{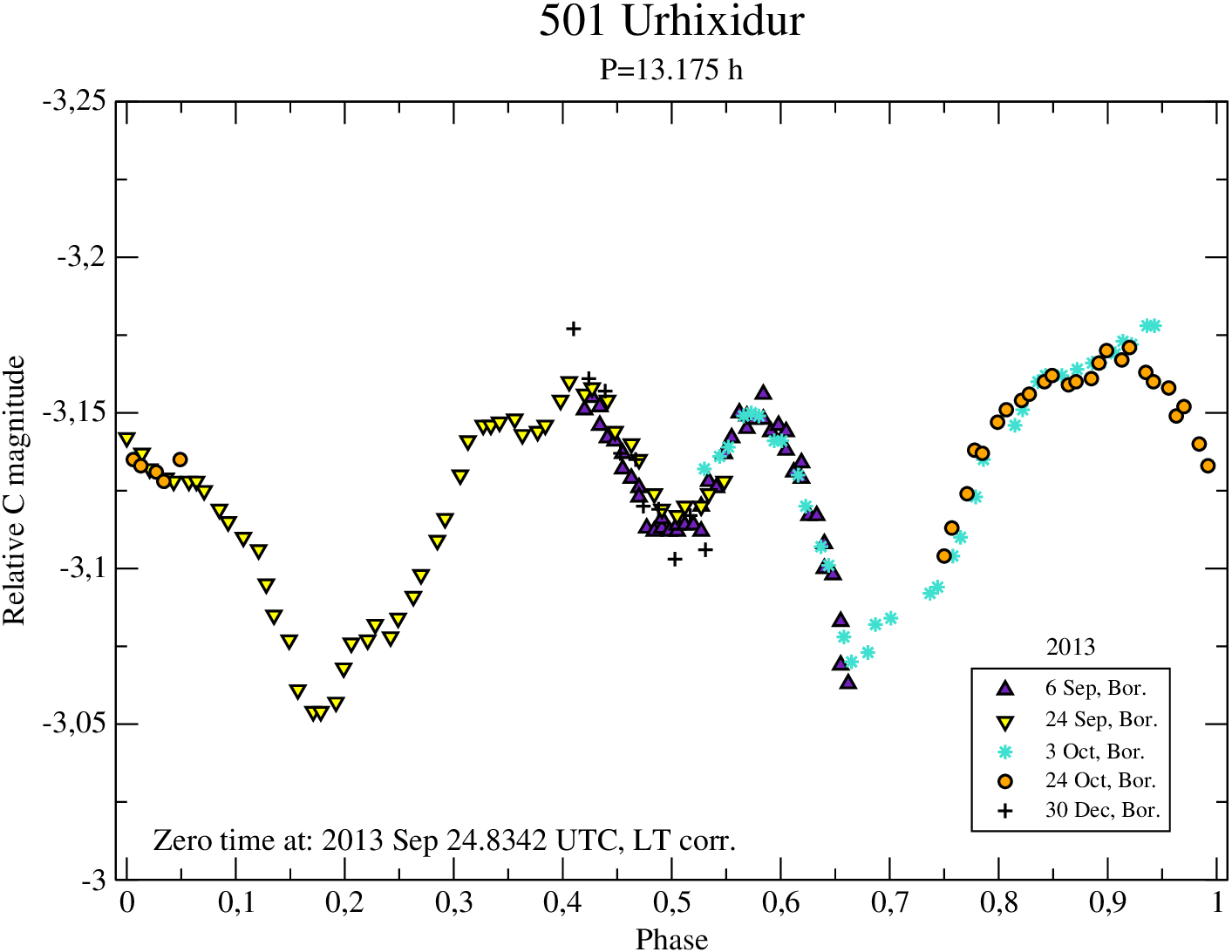} 
\captionof{figure}{Composite light curve of (501) Urhixidur from the year 2013.}
\label{501composit2013}
\\
\includegraphics[width=0.46\textwidth]{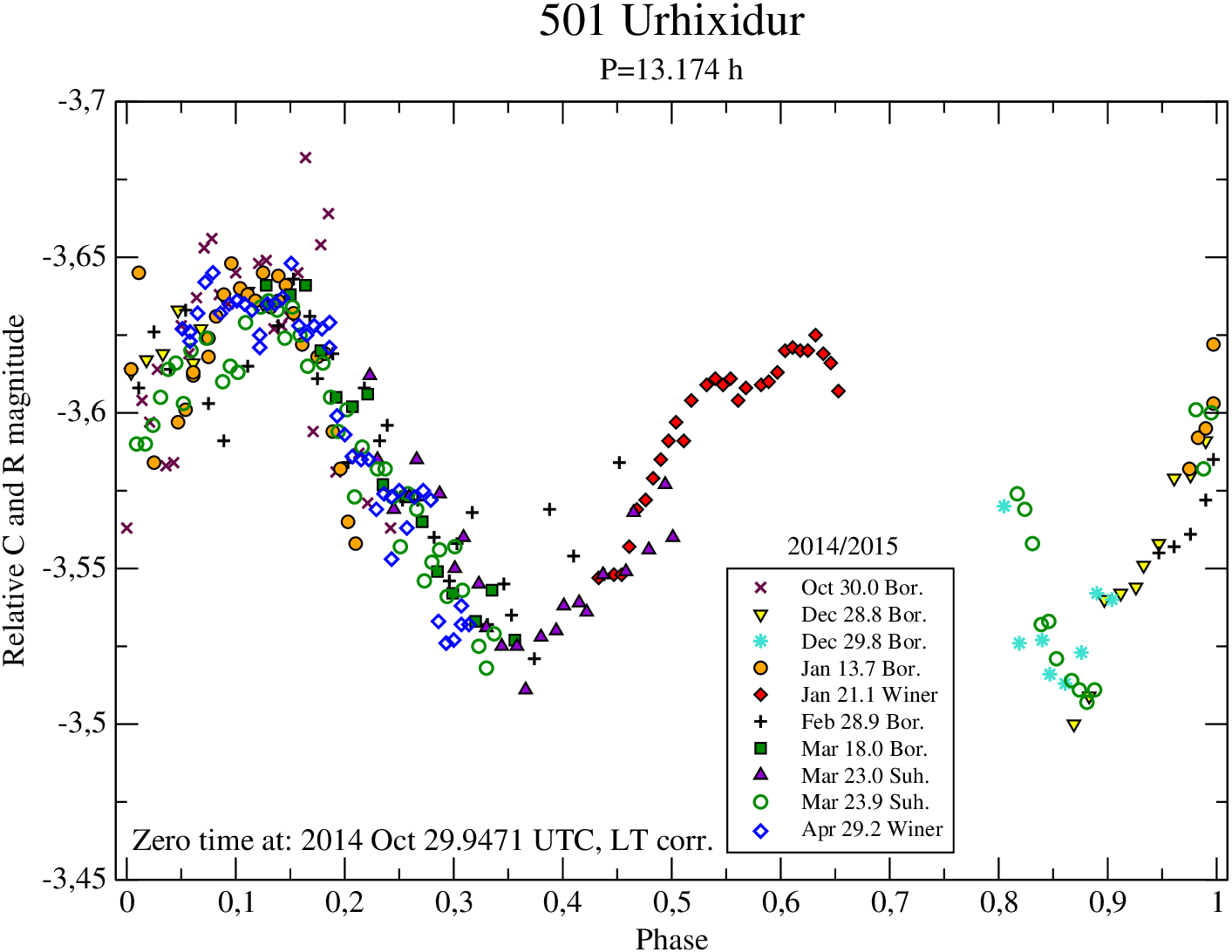} 
\captionof{figure}{Composite light curve of (501) Urhixidur from the years 2014-2015.}
\label{501composit2015}
&
\includegraphics[width=0.46\textwidth]{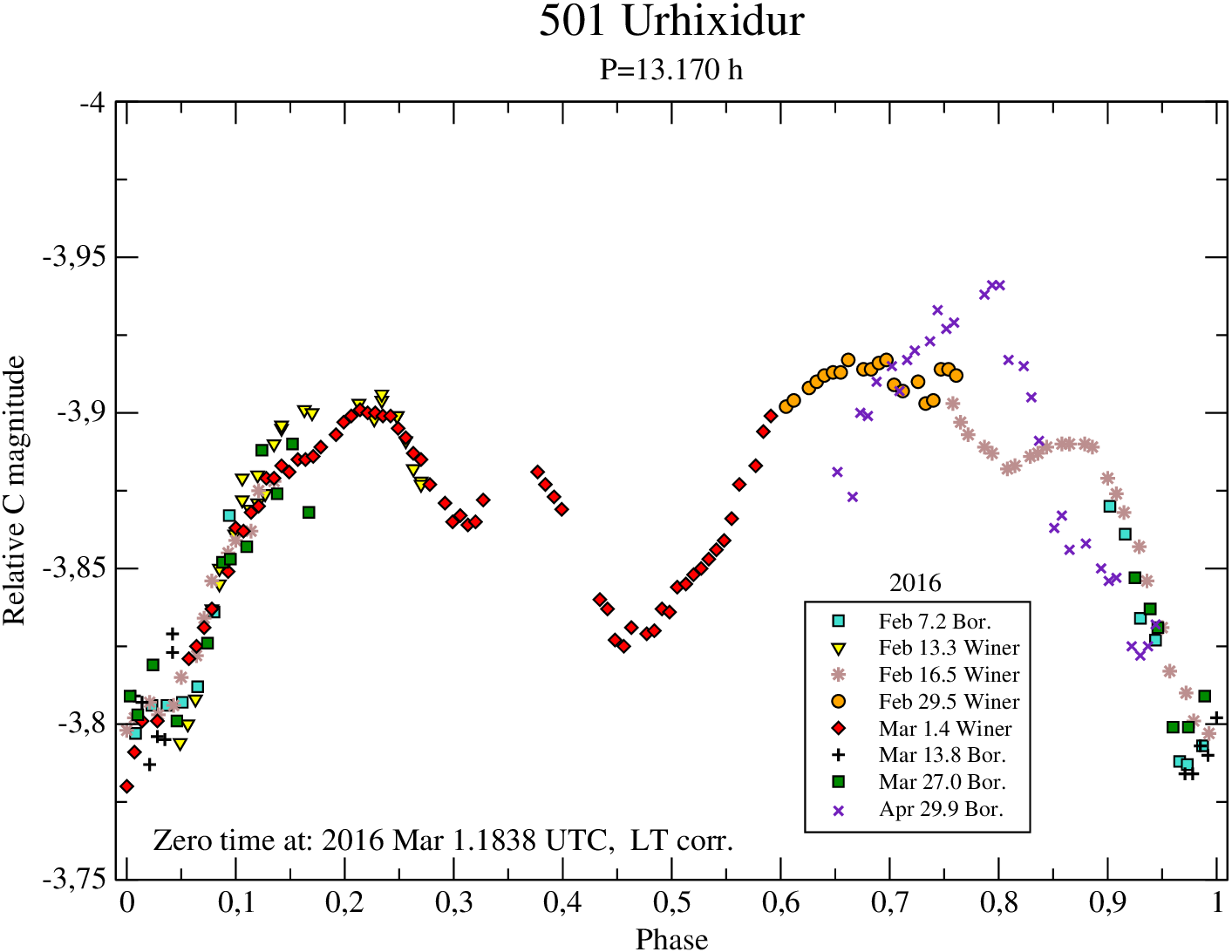} 
\captionof{figure}{Composite light curve of (501) Urhixidur from the year 2016.}
\label{501composit2016}
\\
\end{tabularx}
    \end{table*}

%Composits 6
    \begin{table*}[ht]
    \centering
\vspace{0.5cm}
\begin{tabularx}{\linewidth}{XX}
\includegraphics[width=0.46\textwidth]{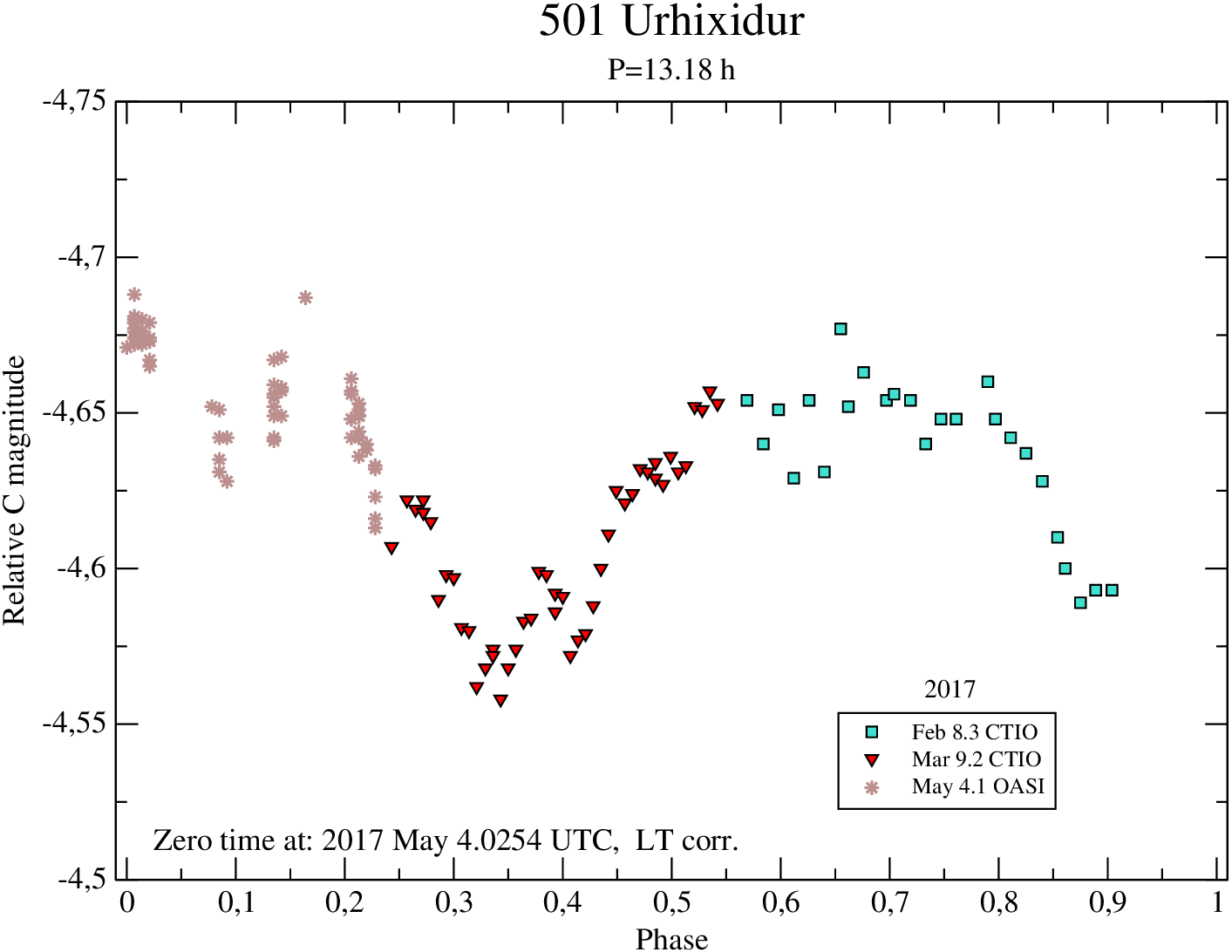} 
\captionof{figure}{Composite light curve of (501) Urhixidur from the year 2017.}
\label{501composit2017}
&
\includegraphics[width=0.46\textwidth]{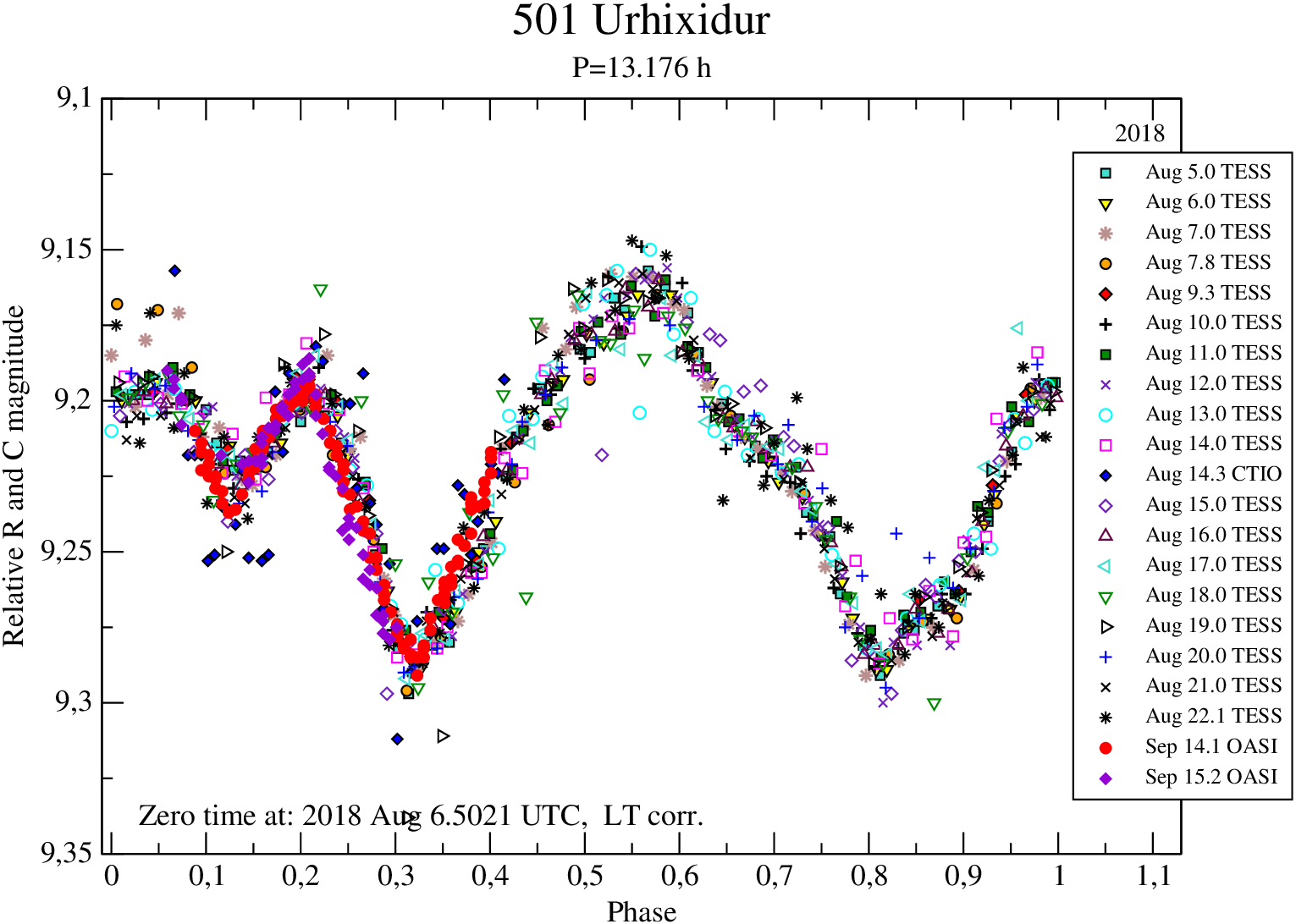} 
\captionof{figure}{Composite light curve of (501) Urhixidur from the year 2018.}
\label{501composit2018}
\\
\includegraphics[width=0.46\textwidth]{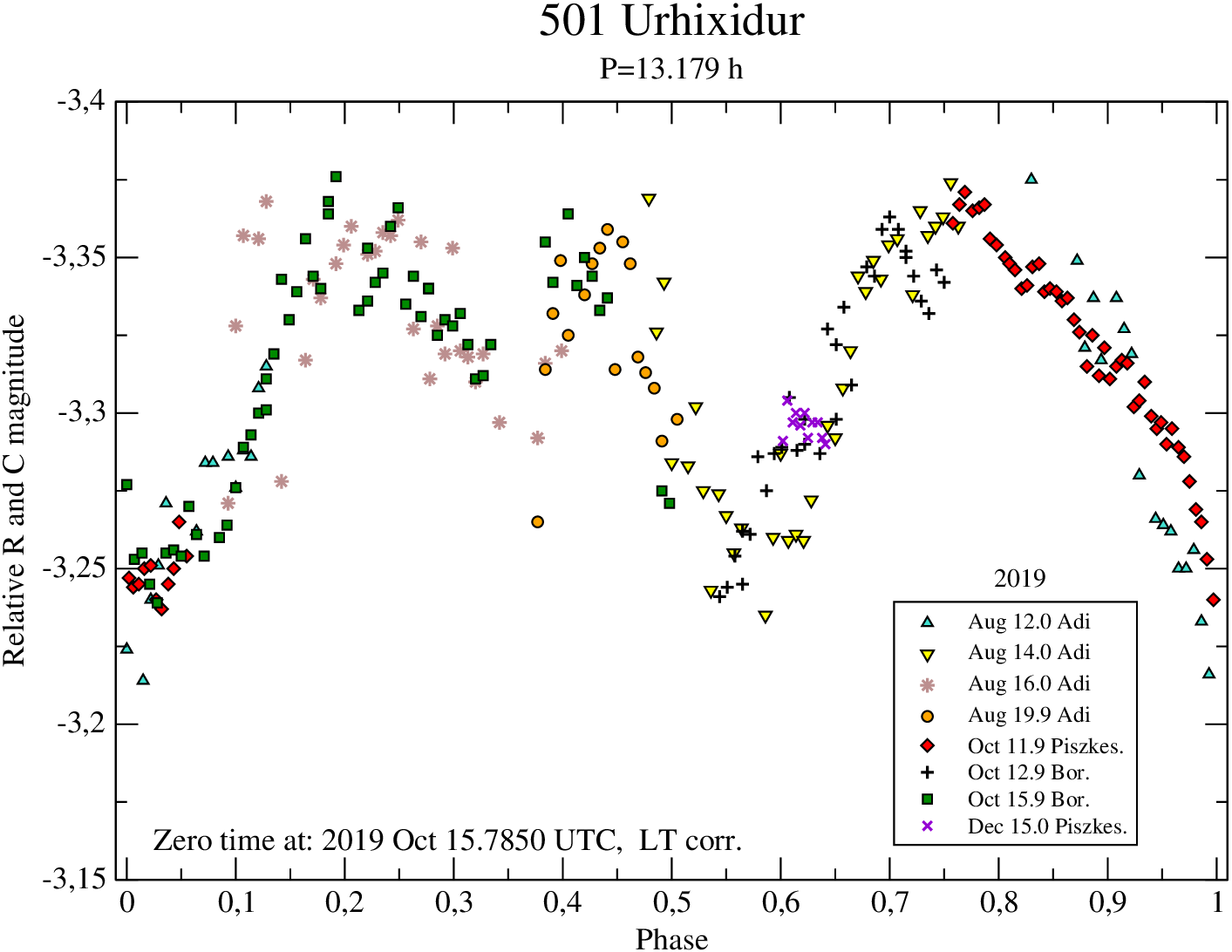} 
\captionof{figure}{Composite light curve of (501) Urhixidur from the year 2019.}
\label{501composit2019}
&
\includegraphics[width=0.46\textwidth]{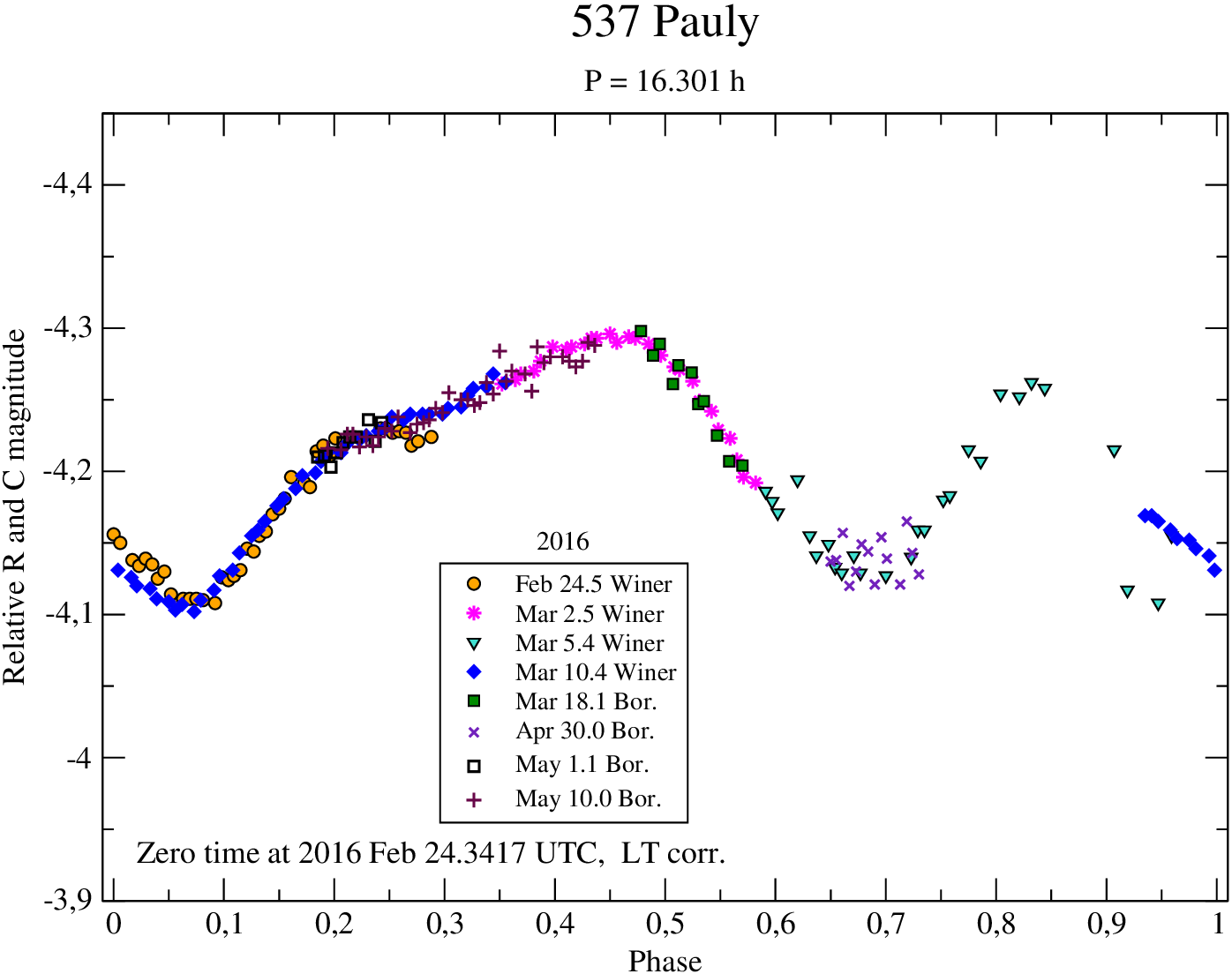} 
\captionof{figure}{Composite light curve of (537) Pauly from the year 2016.}
\label{537composit2016}
\\
\includegraphics[width=0.46\textwidth]{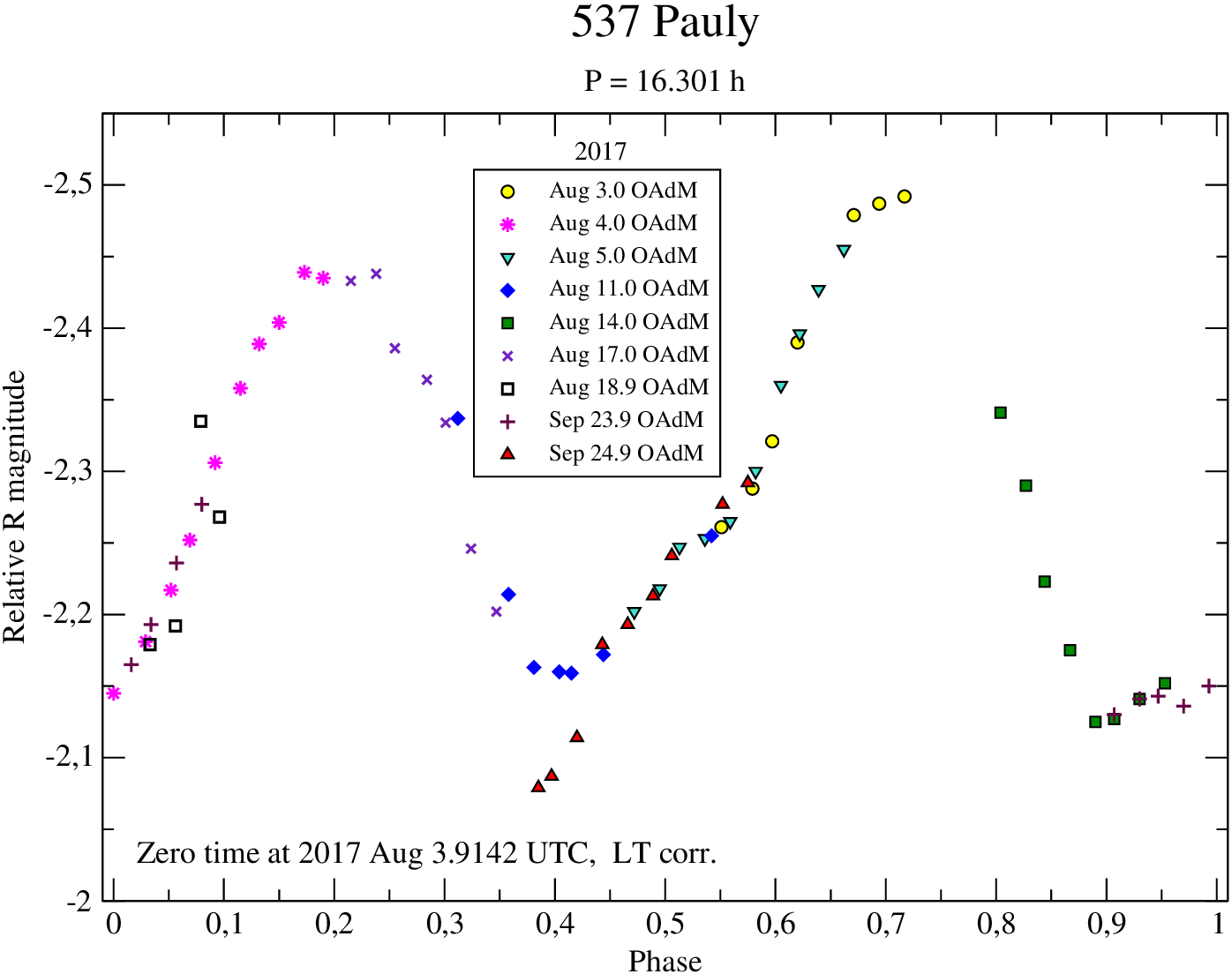} 
\captionof{figure}{Composite light curve of (537) Pauly from the years 2017.}
\label{537composit2017}
&
\includegraphics[width=0.46\textwidth]{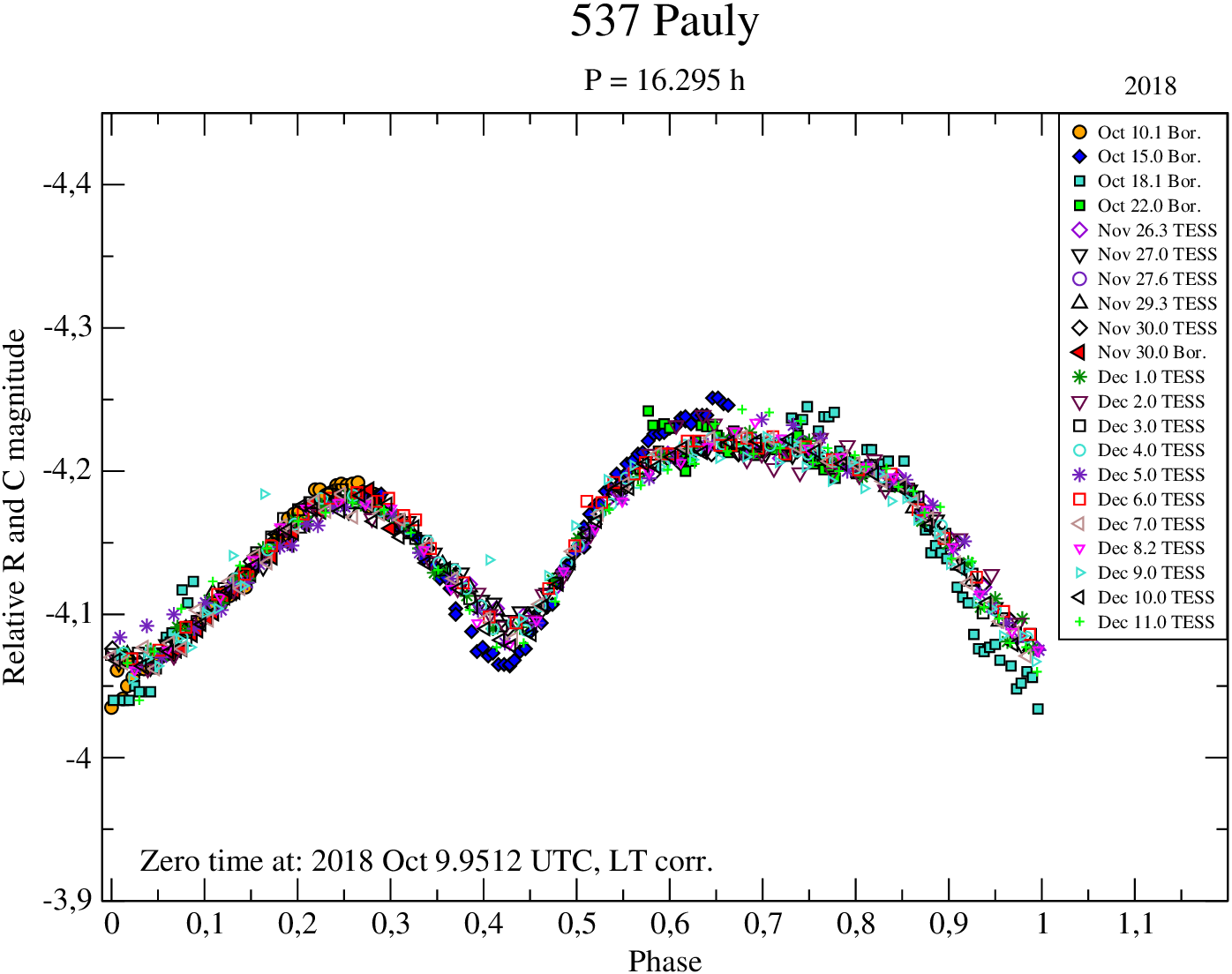} 
\captionof{figure}{Composite light curve of (537) Pauly from the year 2018.}
\label{537composit2018}
\\
\end{tabularx}
    \end{table*}
 
%Composits 7
    \begin{table*}[ht]
    \centering
\vspace{0.5cm}
\begin{tabularx}{\linewidth}{XX}
\includegraphics[width=0.46\textwidth]{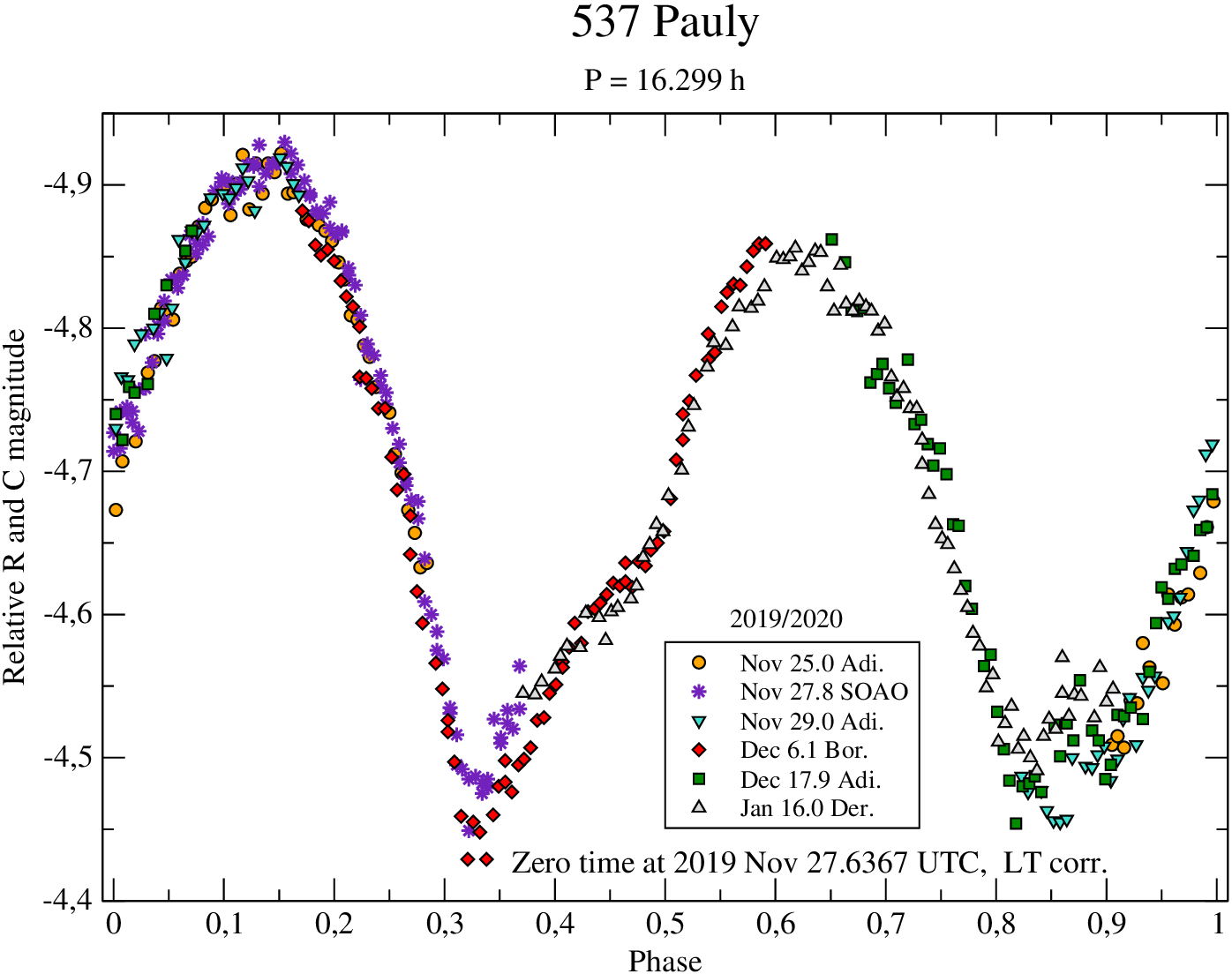} 
\captionof{figure}{Composite light curve of (537) Pauly from the years 2019-2020.}
\label{537composit2019}
&
\includegraphics[width=0.46\textwidth]{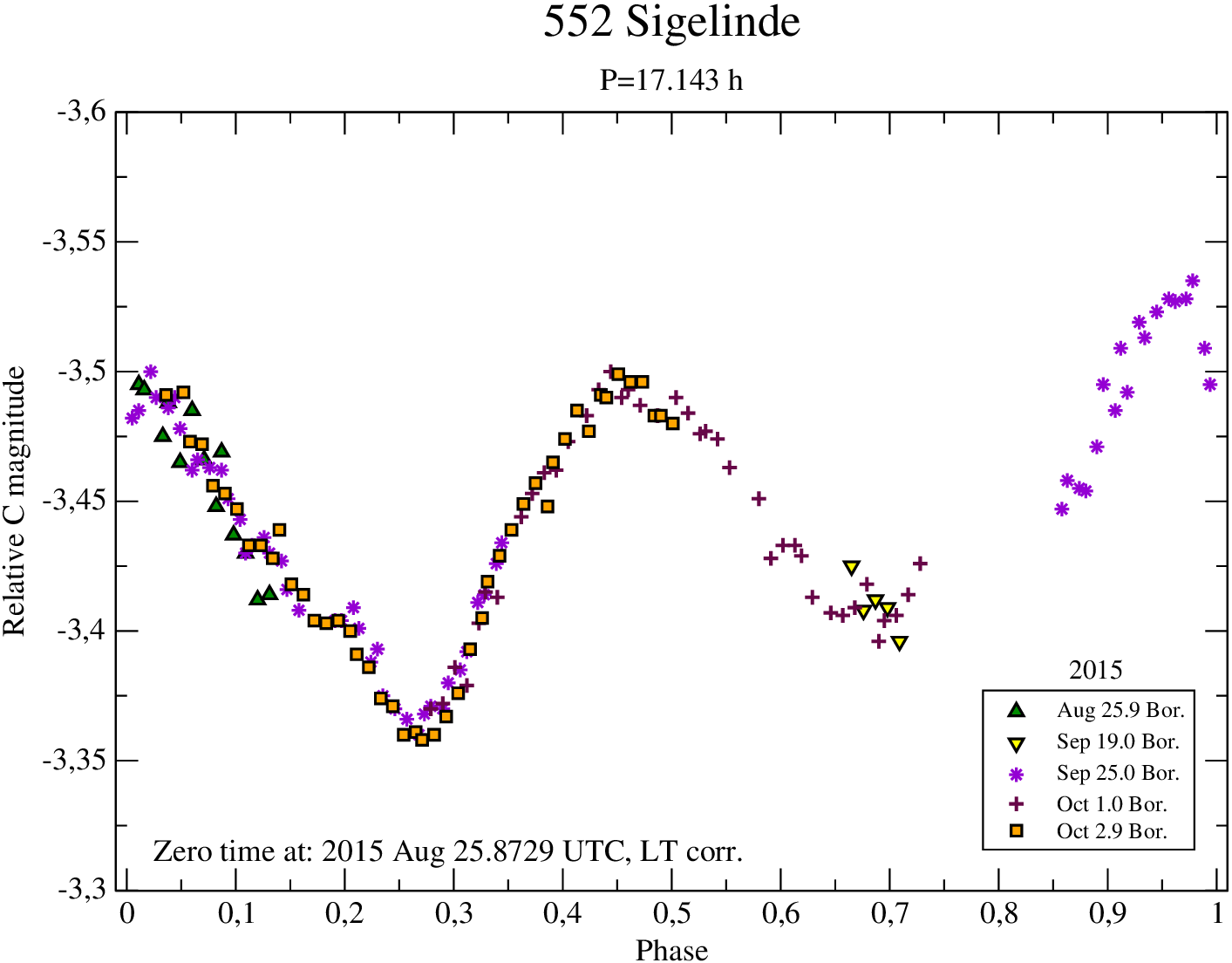} 
\captionof{figure}{Composite light curve of (552) Sigelinde from the year 2015.}
\label{552composit2015}
\\
\includegraphics[width=0.46\textwidth]{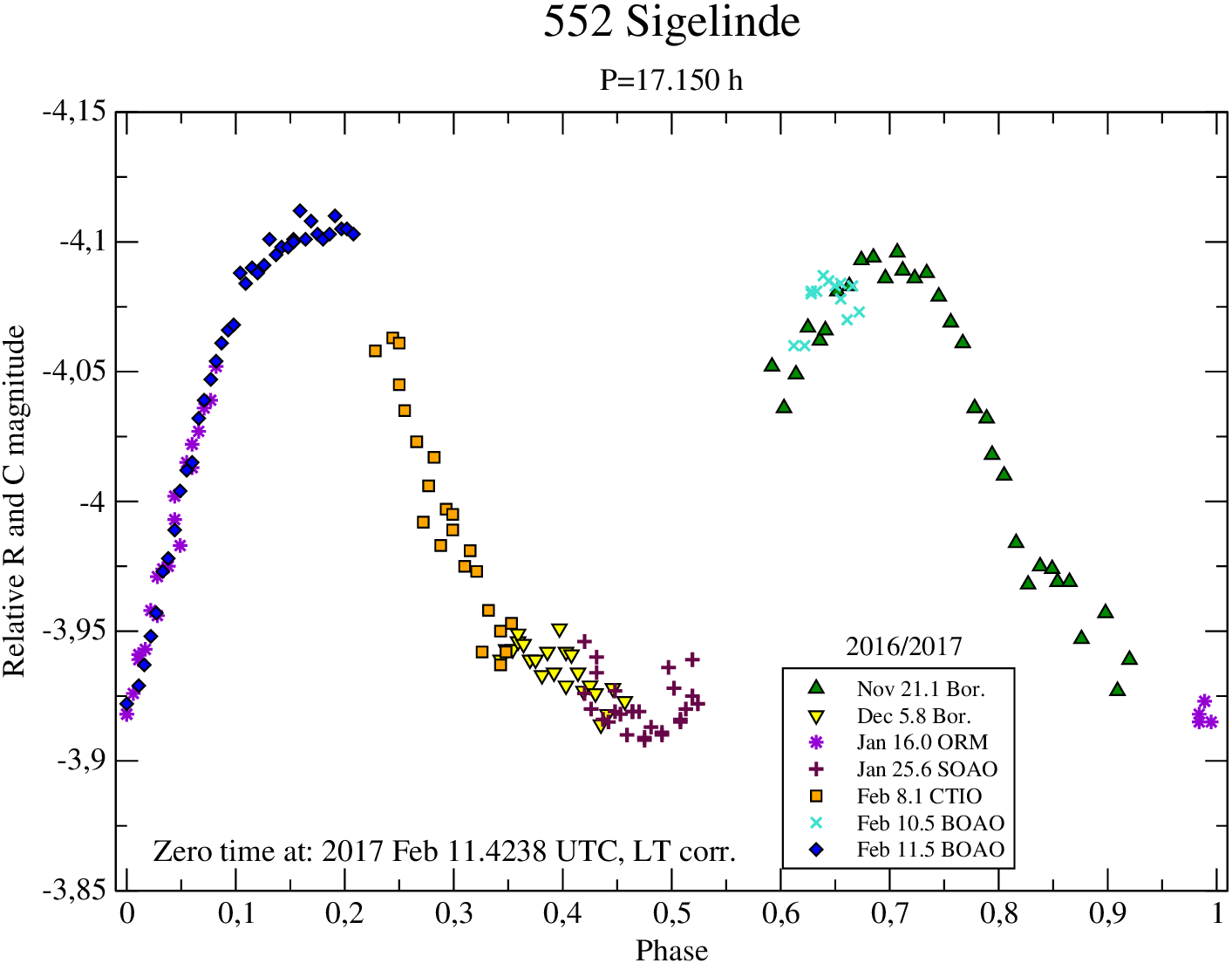} 
\captionof{figure}{Composite light curve of (552) Sigelinde from the years 2016-2017.}
\label{552composit2017}
&
\includegraphics[width=0.46\textwidth]{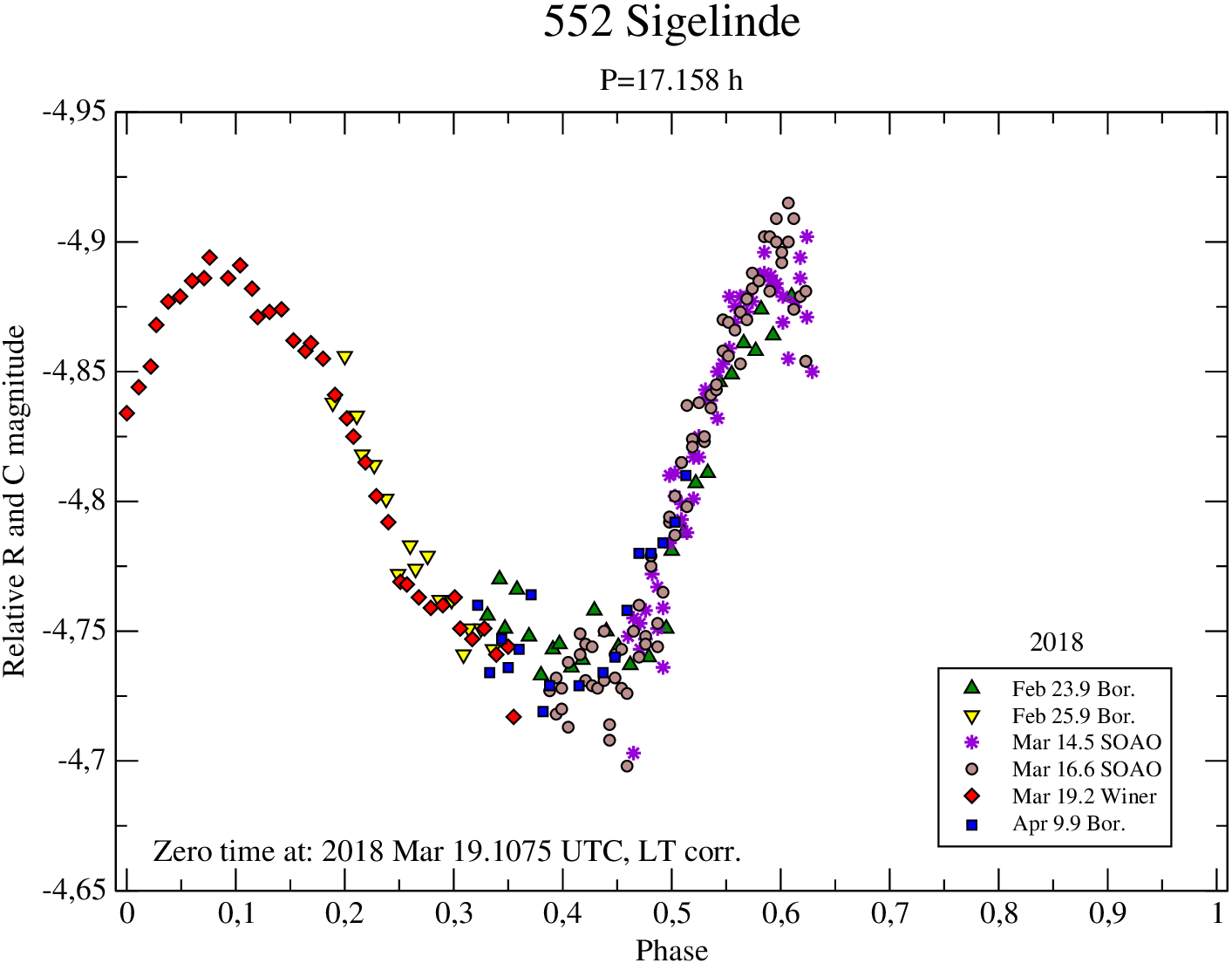} 
\captionof{figure}{Composite light curve of (552) Sigelinde from the year 2018.}
\label{552composit2018}
\\
\includegraphics[width=0.46\textwidth]{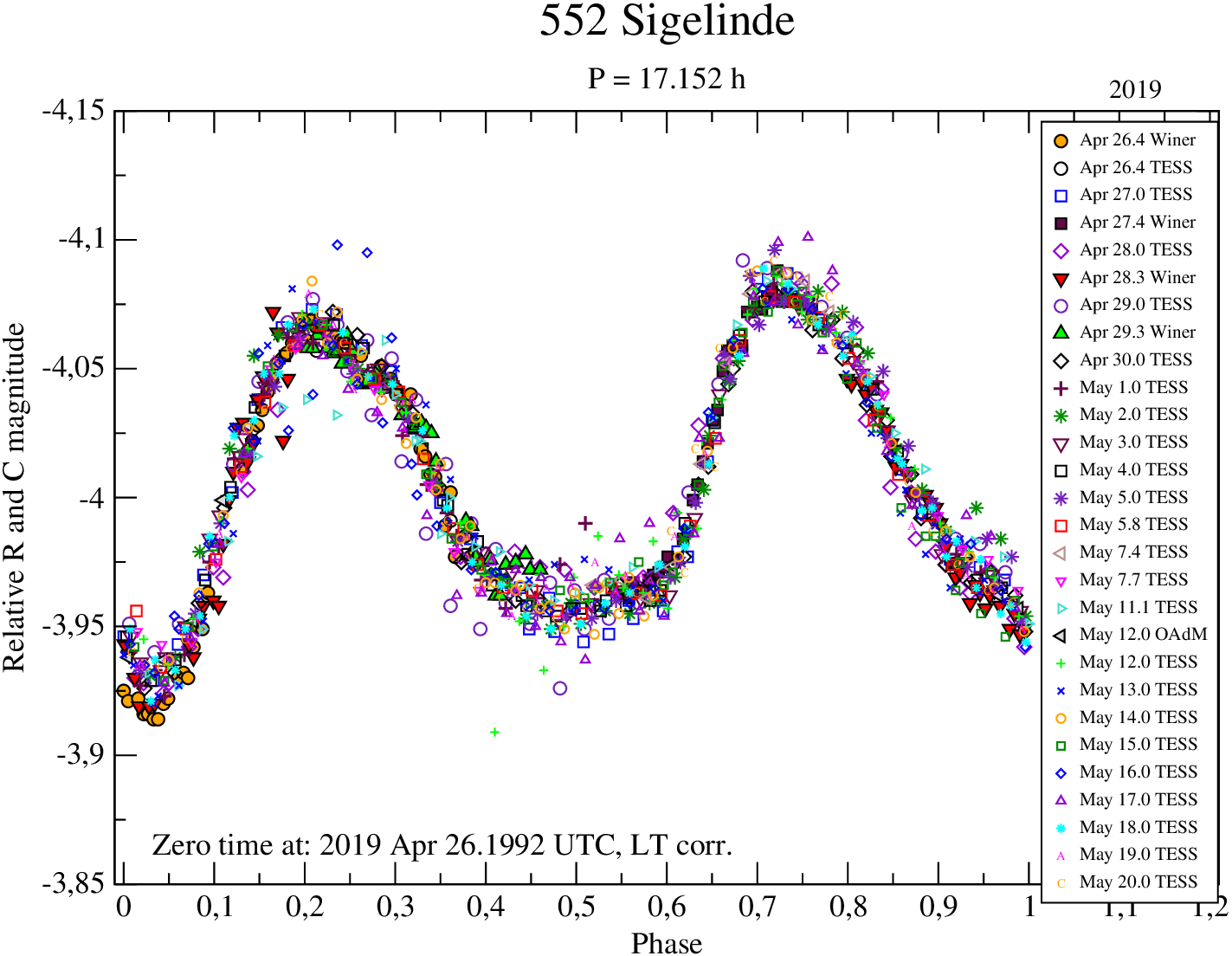} 
\captionof{figure}{Composite light curve of (552) Sigelinde from the year 2019.}
\label{552composit2019}
&
\includegraphics[width=0.46\textwidth]{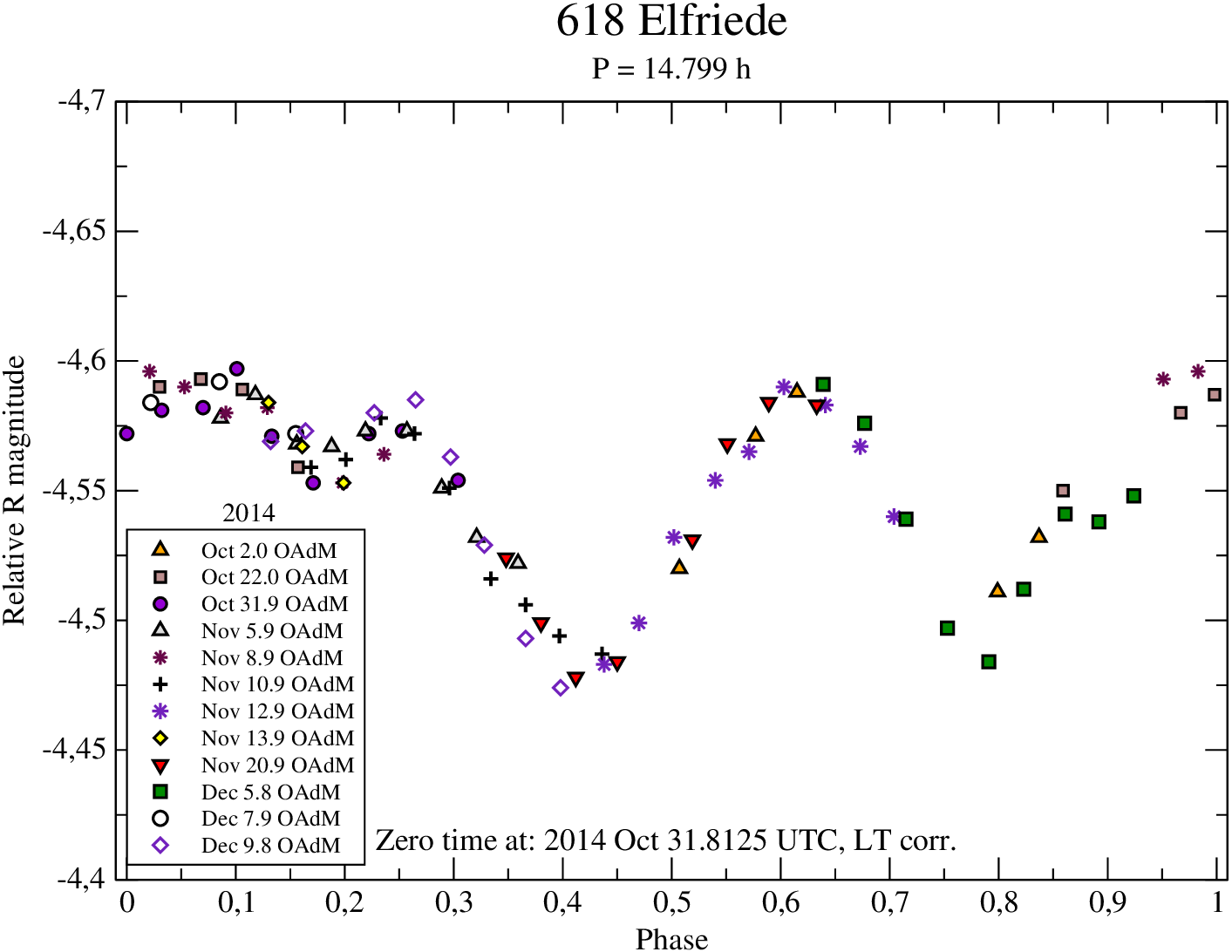} 
\captionof{figure}{Composite light curve of (618) Elfriede from the year 2014.}
\label{618composit2014}
\\
\end{tabularx}
    \end{table*}

%Composits 8
    \begin{table*}[ht]
    \centering
\vspace{0.5cm}
\begin{tabularx}{\linewidth}{XX}
\includegraphics[width=0.46\textwidth]{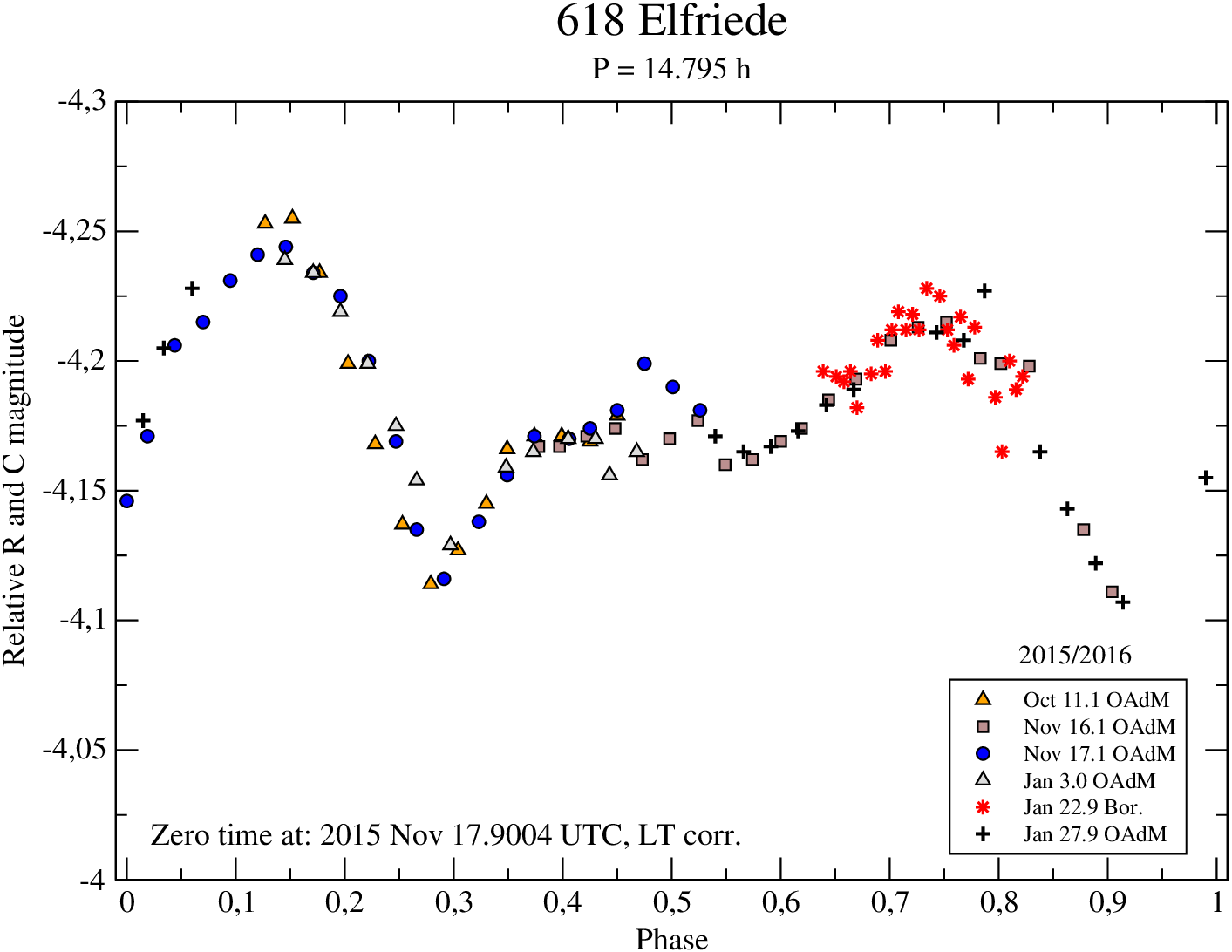} 
\captionof{figure}{Composite light curve of (618) Elfriede from the years 2015-2016.}
\label{618composit2015}
&
\includegraphics[width=0.46\textwidth]{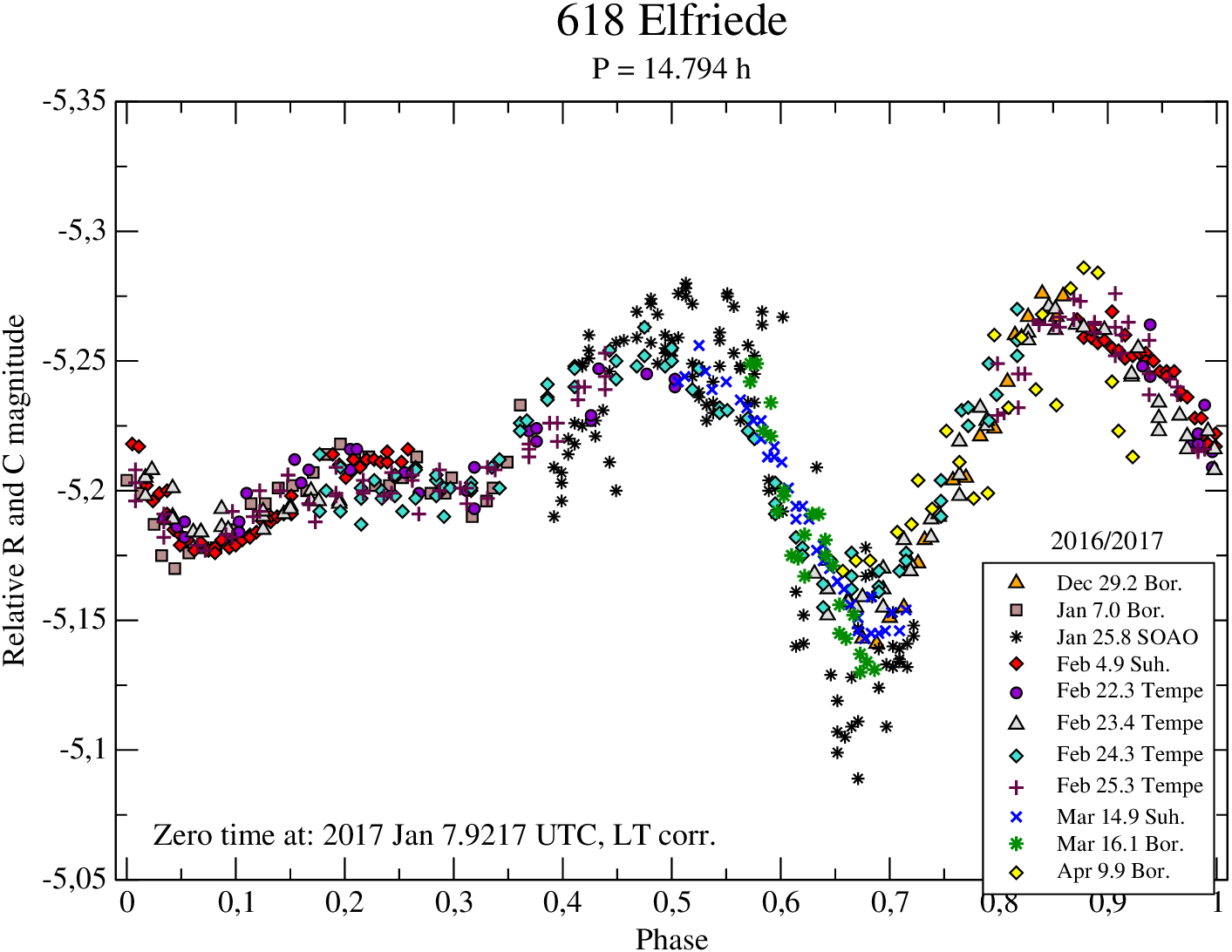} 
\captionof{figure}{Composite light curve of (618) Elfriede from the year 2017.}
\label{618composit2017}
\\
\includegraphics[width=0.46\textwidth]{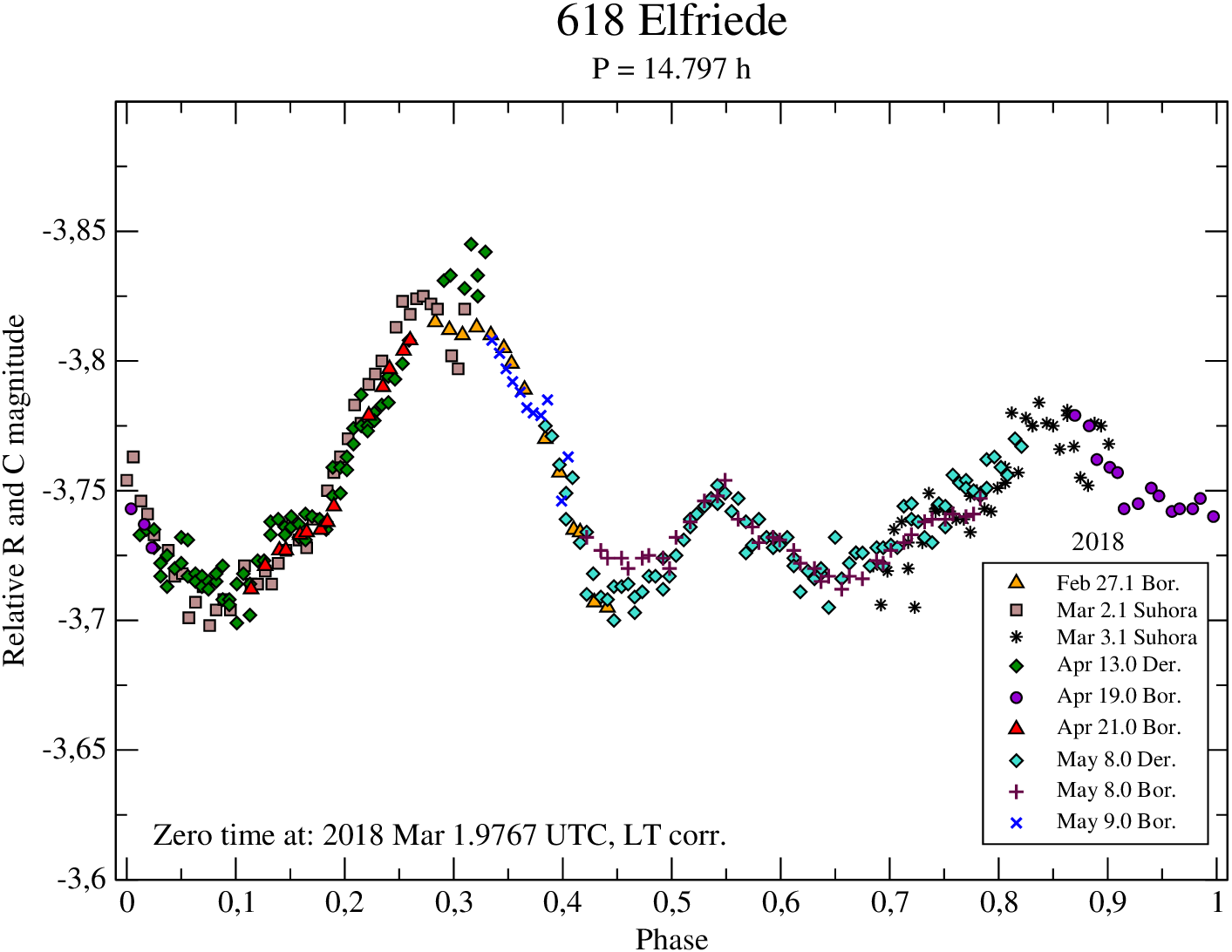} 
\captionof{figure}{Composite light curve of (618) Elfriede from the year 2018.}
\label{618composit2018}
&
\includegraphics[width=0.46\textwidth]{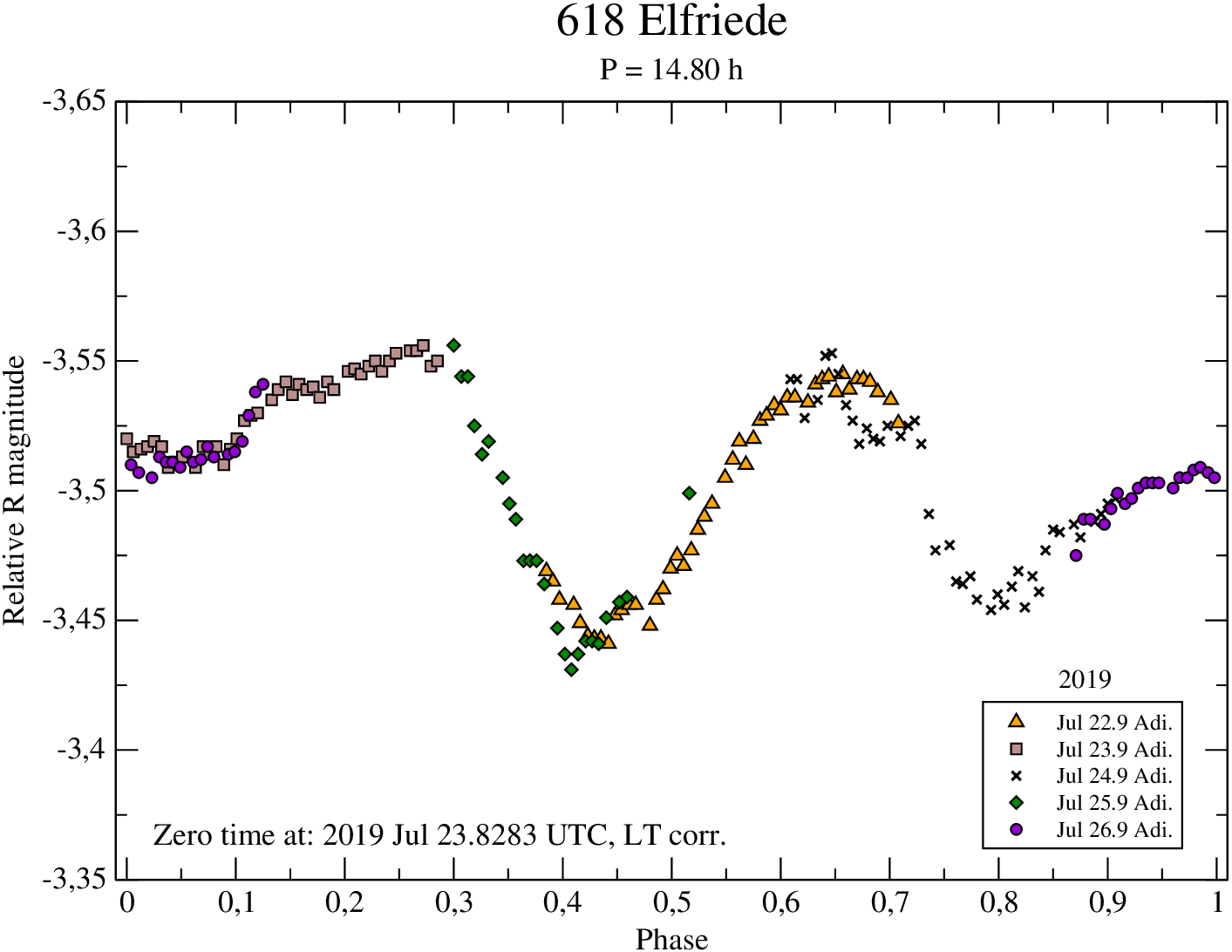} 
\captionof{figure}{Composite light curve of (618) Elfriede from the year 2019.}
\label{618composit2019}
\\
\includegraphics[width=0.46\textwidth]{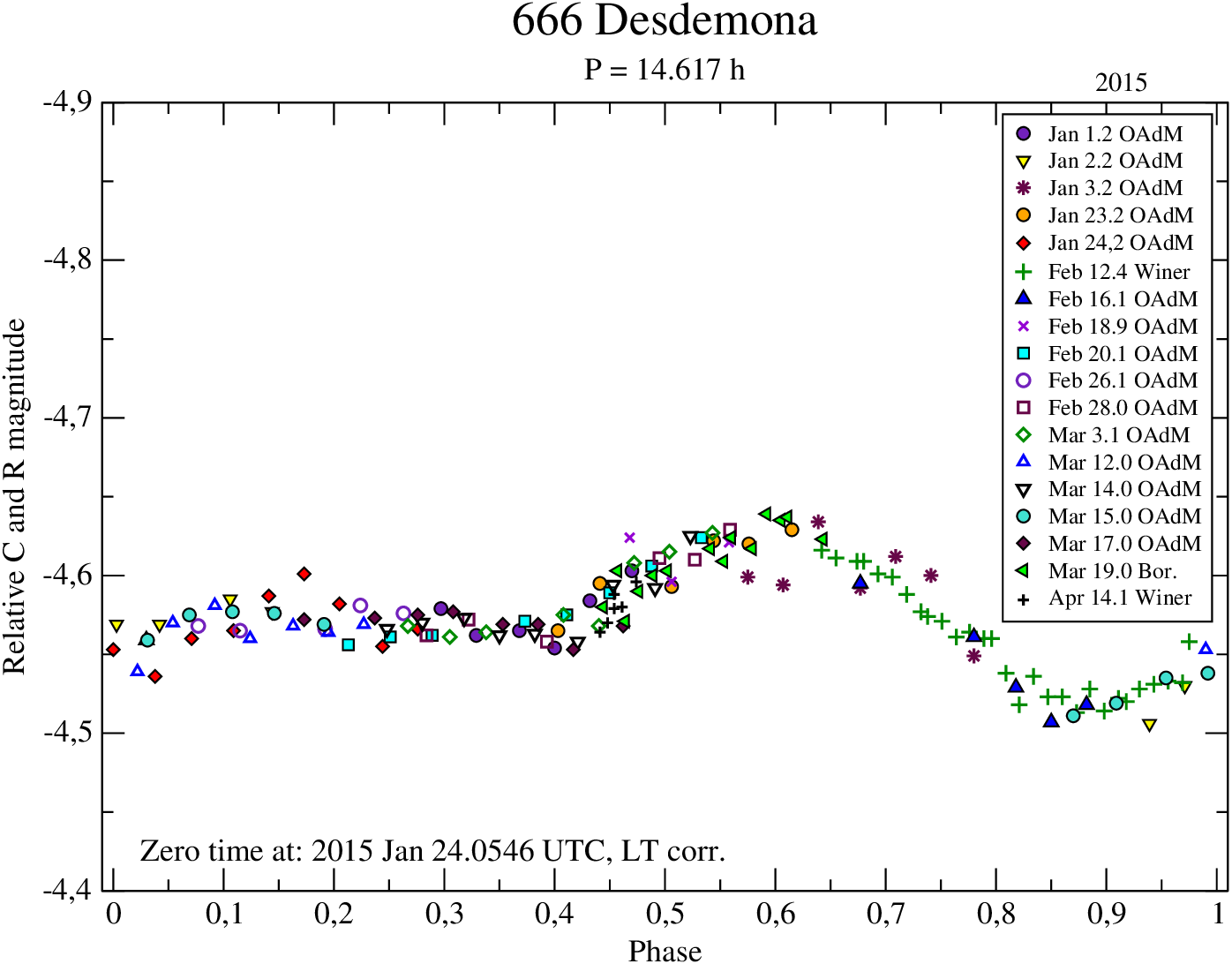} 
\captionof{figure}{Composite light curve of (666) Desdemona from the year 2015.}
\label{666composit2015}
&
\includegraphics[width=0.46\textwidth]{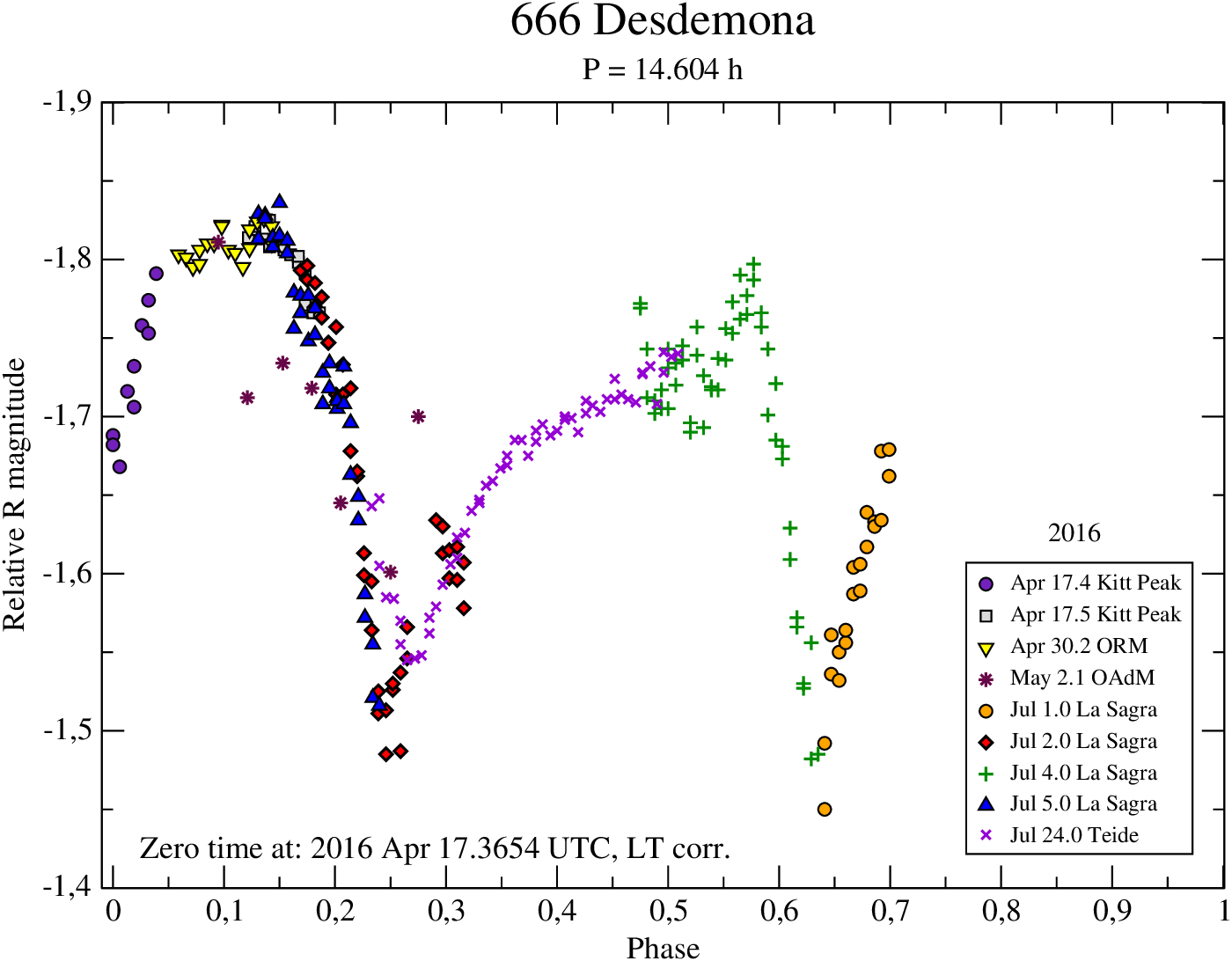} 
\captionof{figure}{Composite light curve of (666) Desdemona from the year 2016.}
\label{666composit2016}
\\
\end{tabularx}
    \end{table*}

%Composits 9
    \begin{table*}[ht]
    \centering
\vspace{0.5cm}
\begin{tabularx}{\linewidth}{XX}
\includegraphics[width=0.46\textwidth]{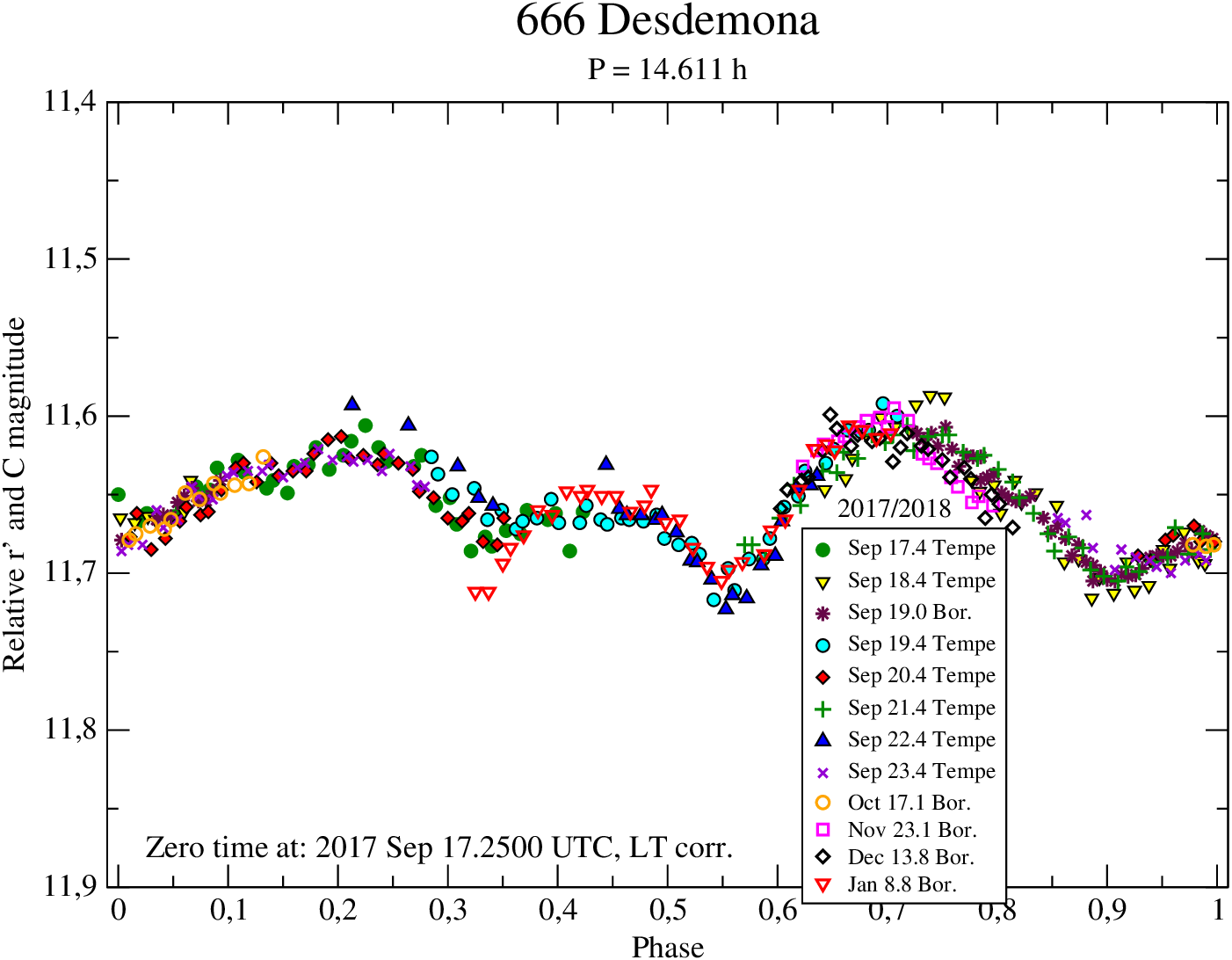} 
\captionof{figure}{Composite light curve of (666) Desdemona from the years 2017-2018.}
\label{666composit2017}
&
\includegraphics[width=0.46\textwidth]{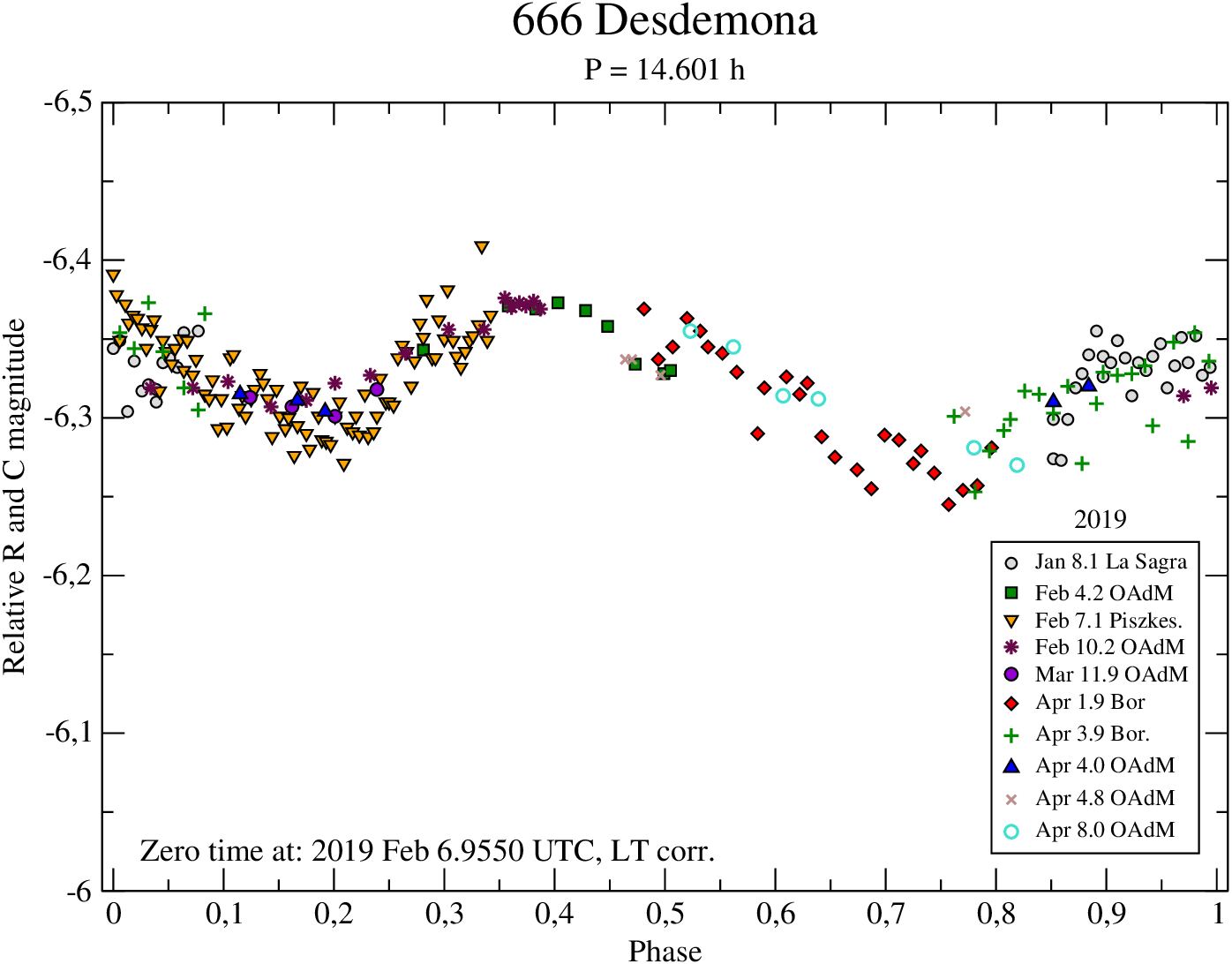} 
\captionof{figure}{Composite light curve of (666) Desdemona from the year 2019.}
\label{666composit2019}
\\
\includegraphics[width=0.46\textwidth]{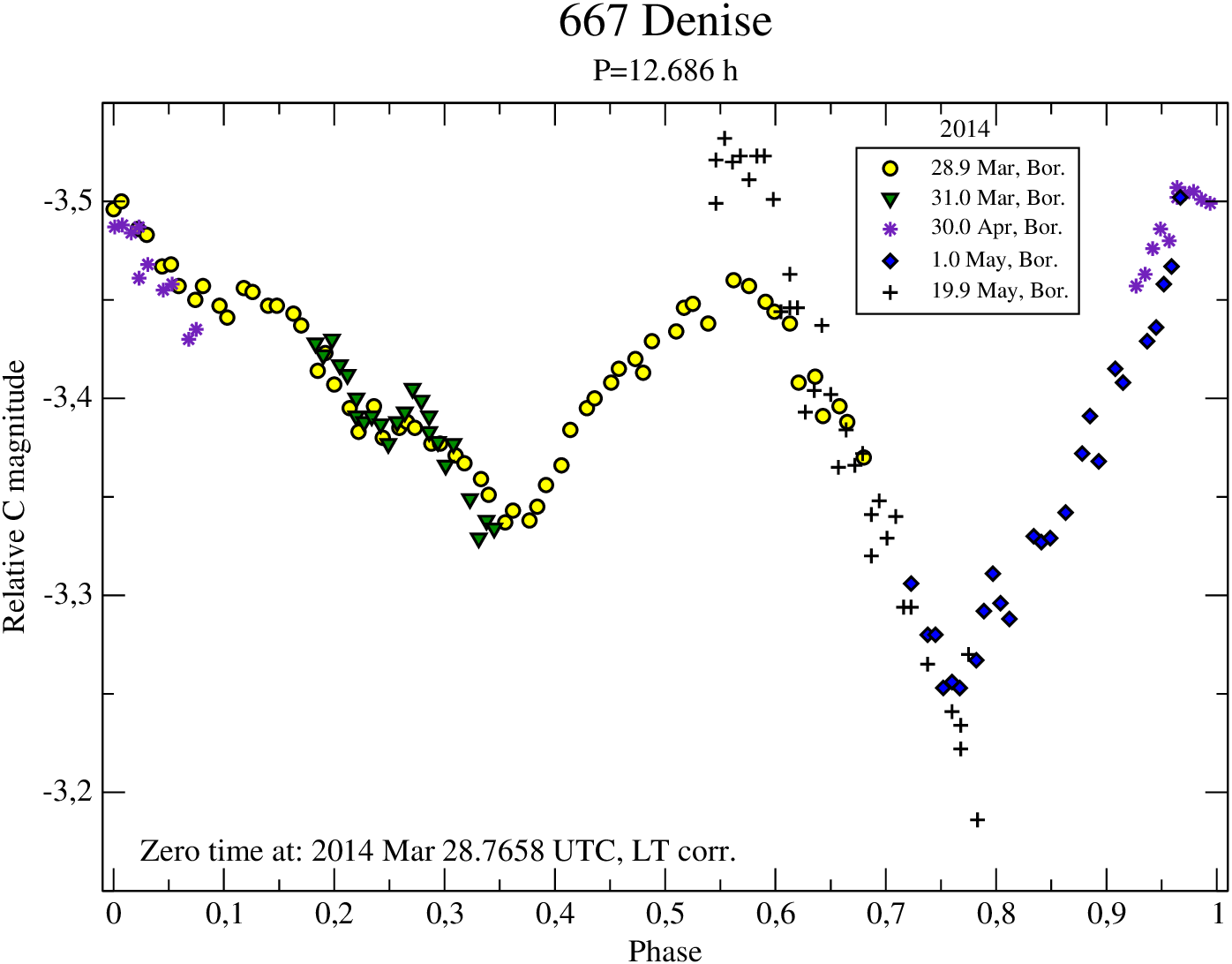} 
\captionof{figure}{Composite light curve of (667) Denise from the year 2014.}
\label{667composit2014}
&
\includegraphics[width=0.46\textwidth]{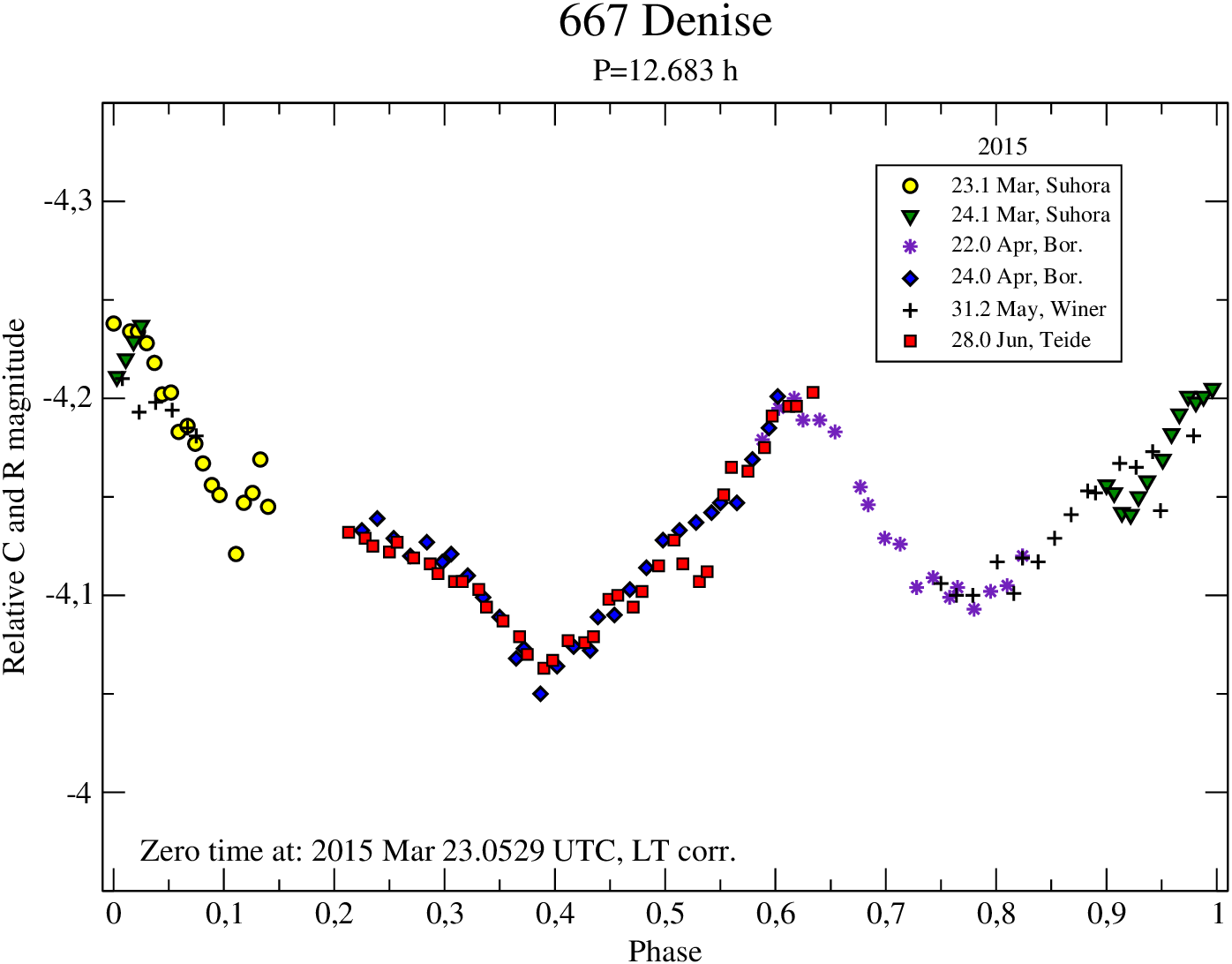} 
\captionof{figure}{Composite light curve of (667) Denise from the year 2015.}
\label{667composit2015}
\\
\includegraphics[width=0.46\textwidth]{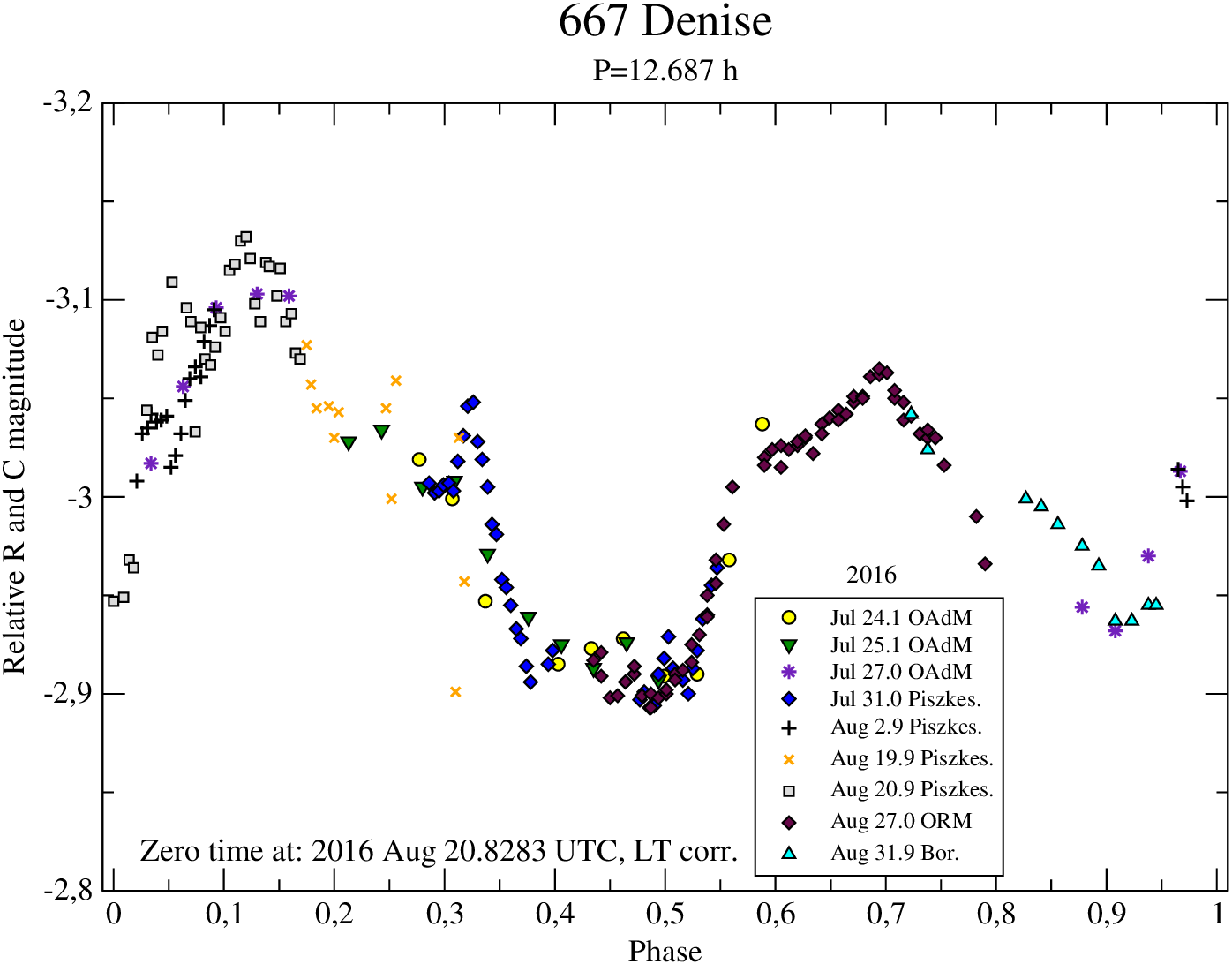} 
\captionof{figure}{Composite light curve of (667) Denise from the year 2016.}
\label{667composit2016}
&
\includegraphics[width=0.46\textwidth]{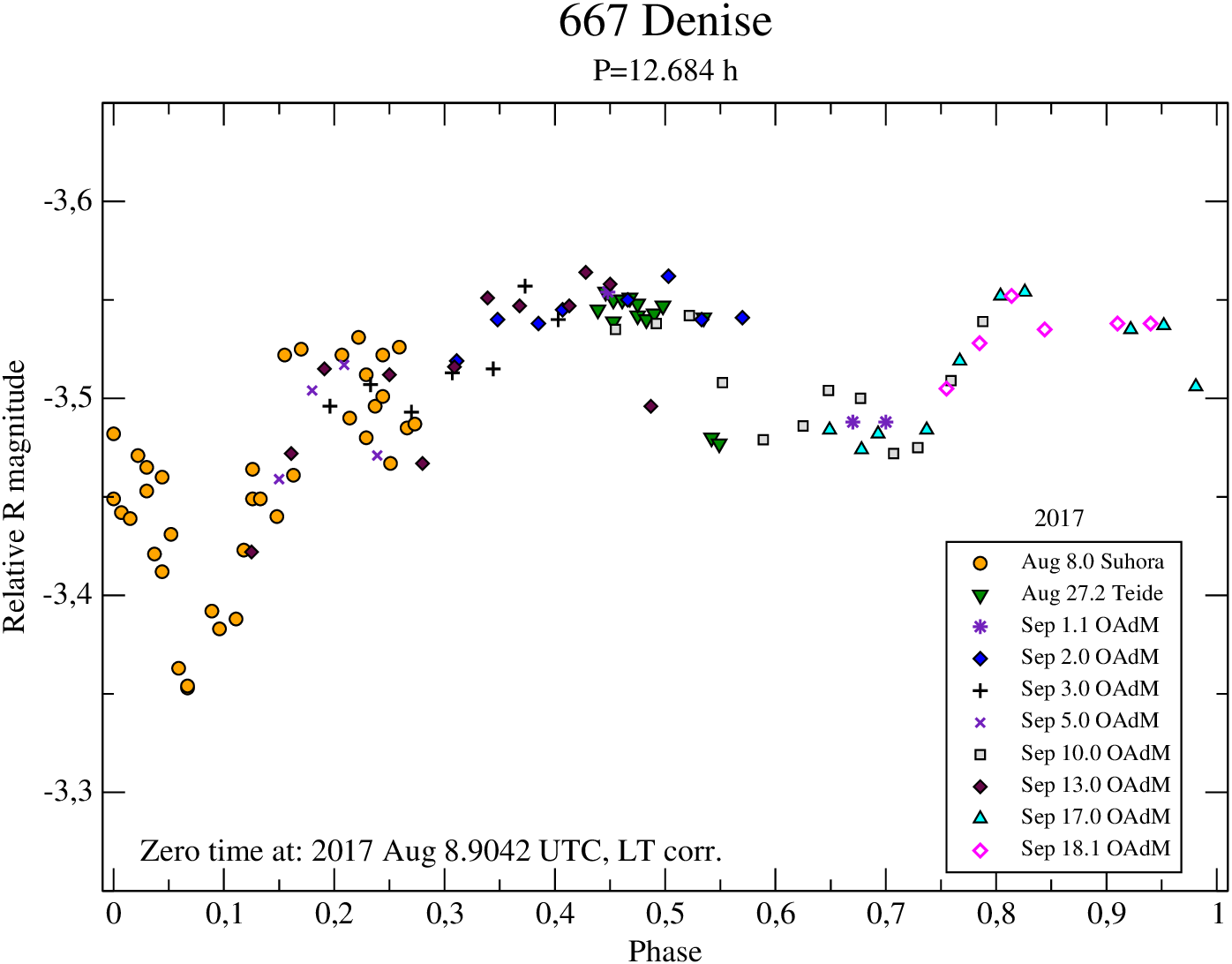} 
\captionof{figure}{Composite light curve of (667) Denise from the year 2017.}
\label{667composit2017}
\\
\end{tabularx}
    \end{table*}
 
%Composits 10
    \begin{table*}[ht]
    \centering
\vspace{0.5cm}
\begin{tabularx}{\linewidth}{XX}
\includegraphics[width=0.46\textwidth]{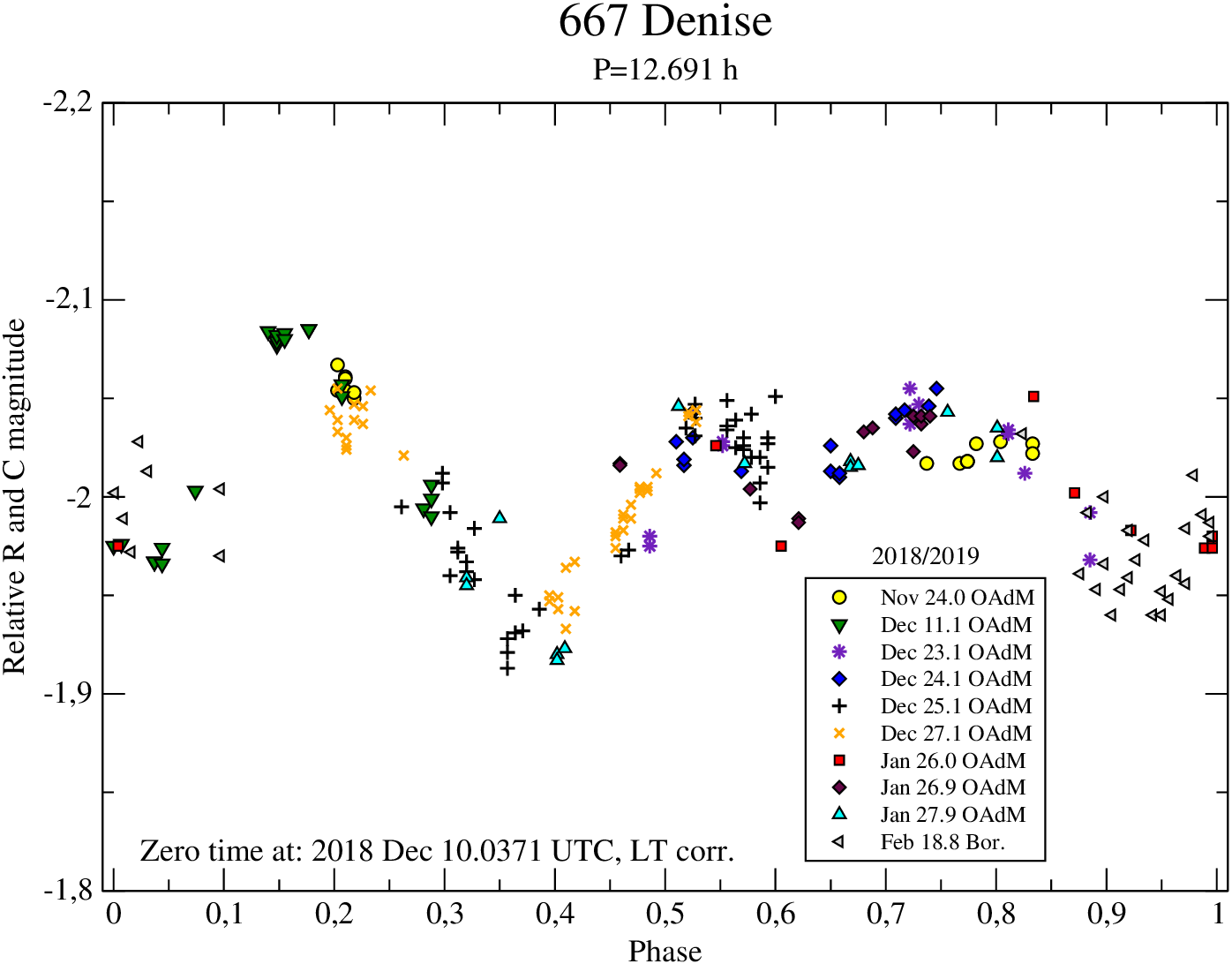} 
\captionof{figure}{Composite light curve of (667) Denise from the years 2018-2019.}
\label{667composit2018}
&
\includegraphics[width=0.46\textwidth]{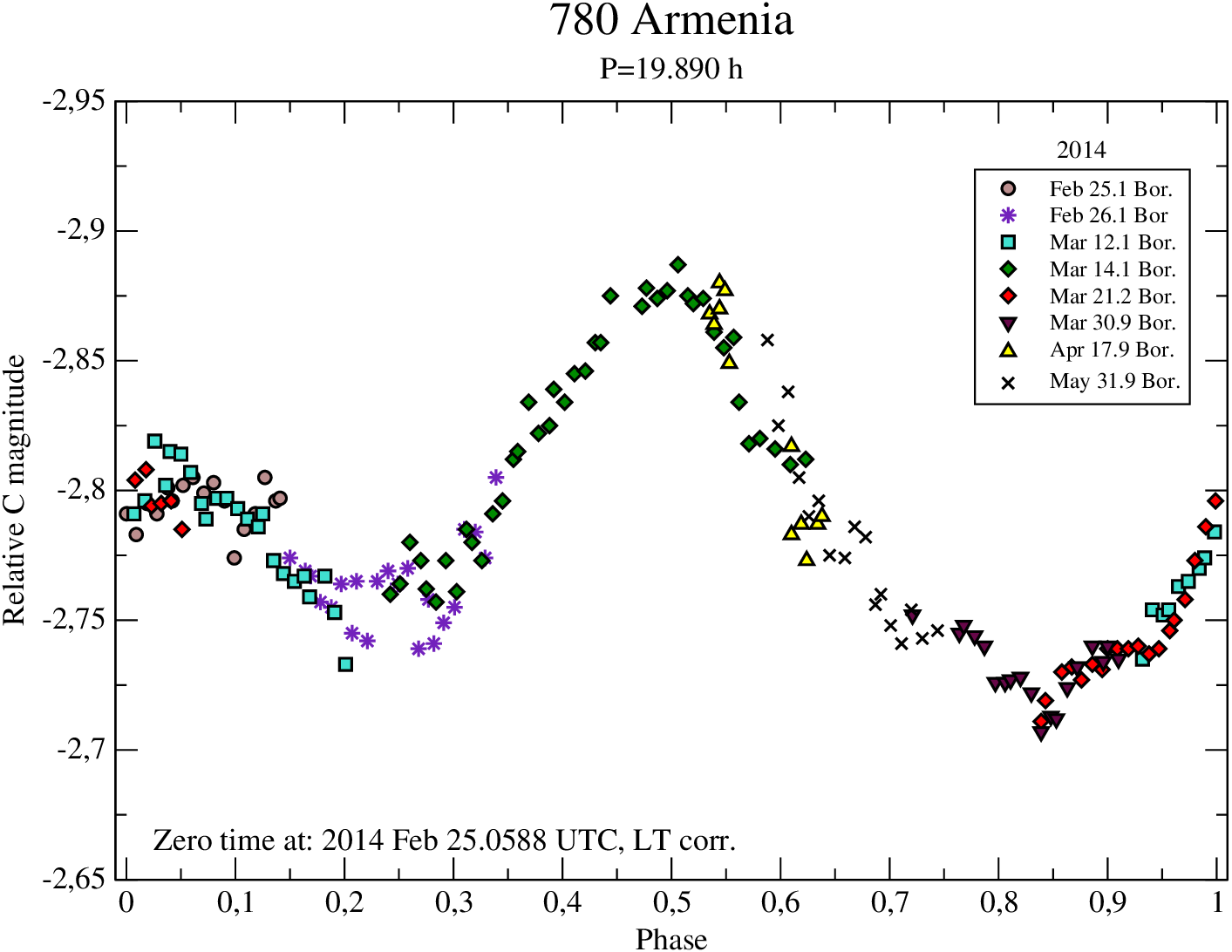} 
\captionof{figure}{Composite light curve of (780) Armenia from the year 2014.}
\label{780composit2014}
\\
\includegraphics[width=0.46\textwidth]{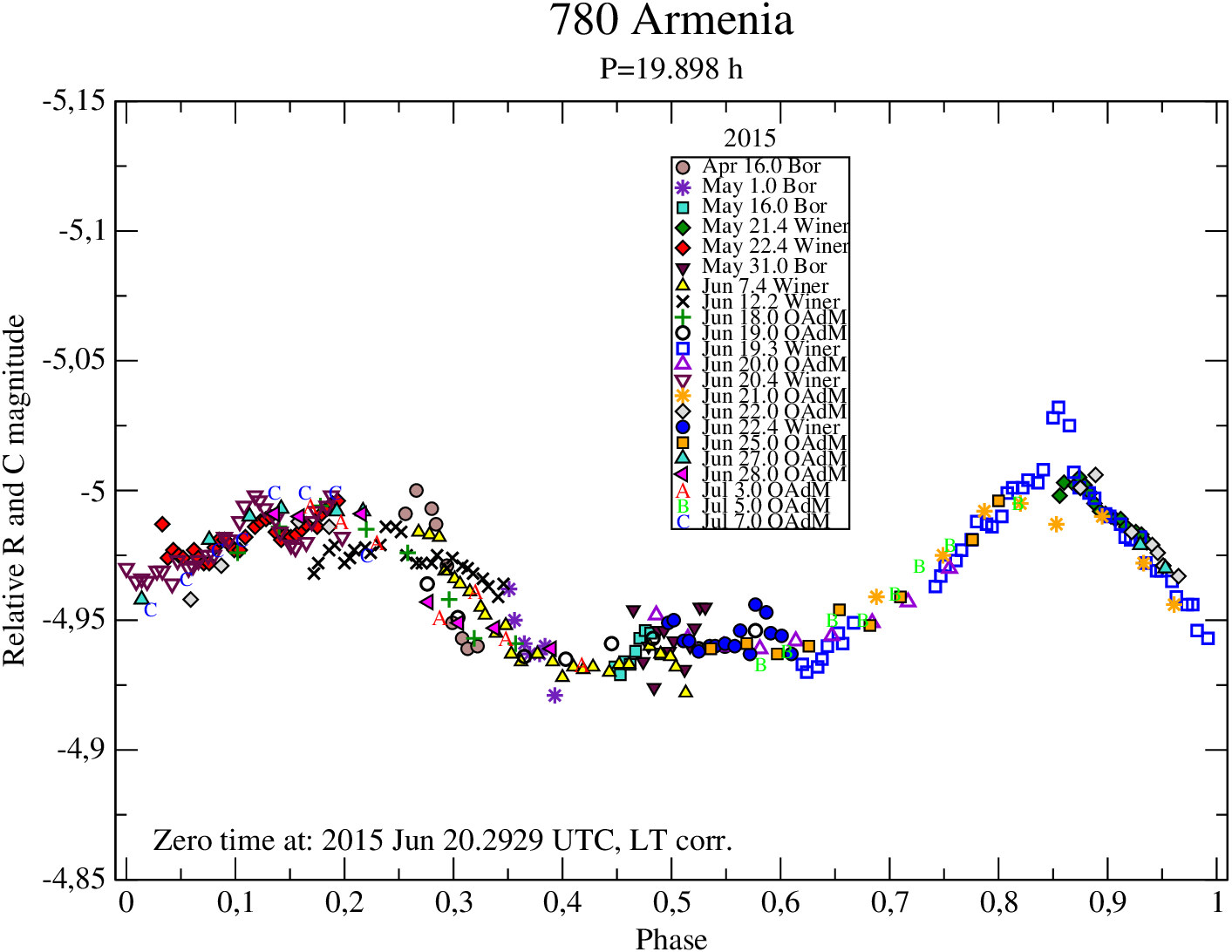} 
\captionof{figure}{Composite light curve of (780) Armenia from the year 2015.}
\label{780composit2015}
&
\includegraphics[width=0.46\textwidth]{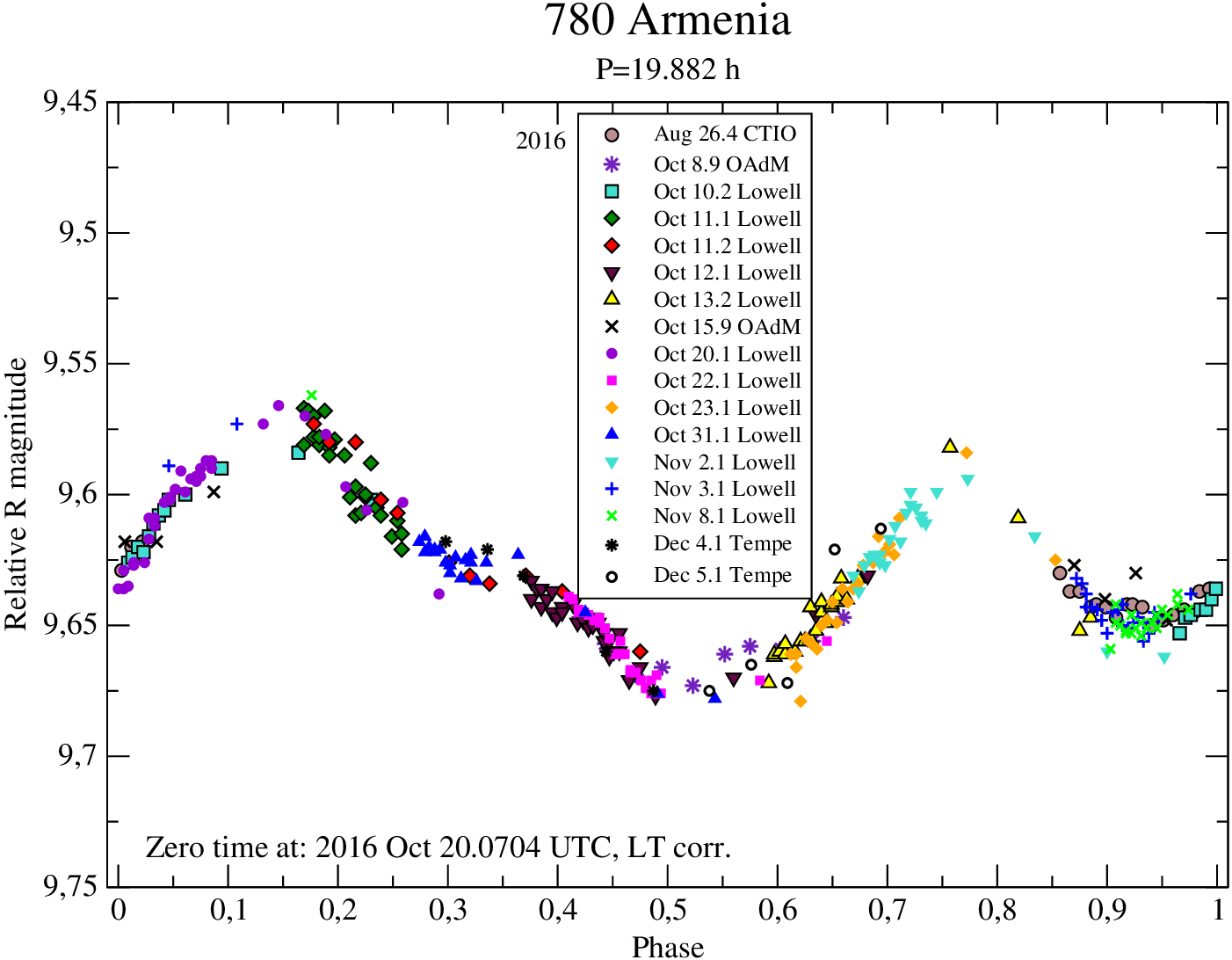} 
\captionof{figure}{Composite light curve of (780) Armenia from the year 2016.}
\label{780composit2016}
\\
\includegraphics[width=0.46\textwidth]{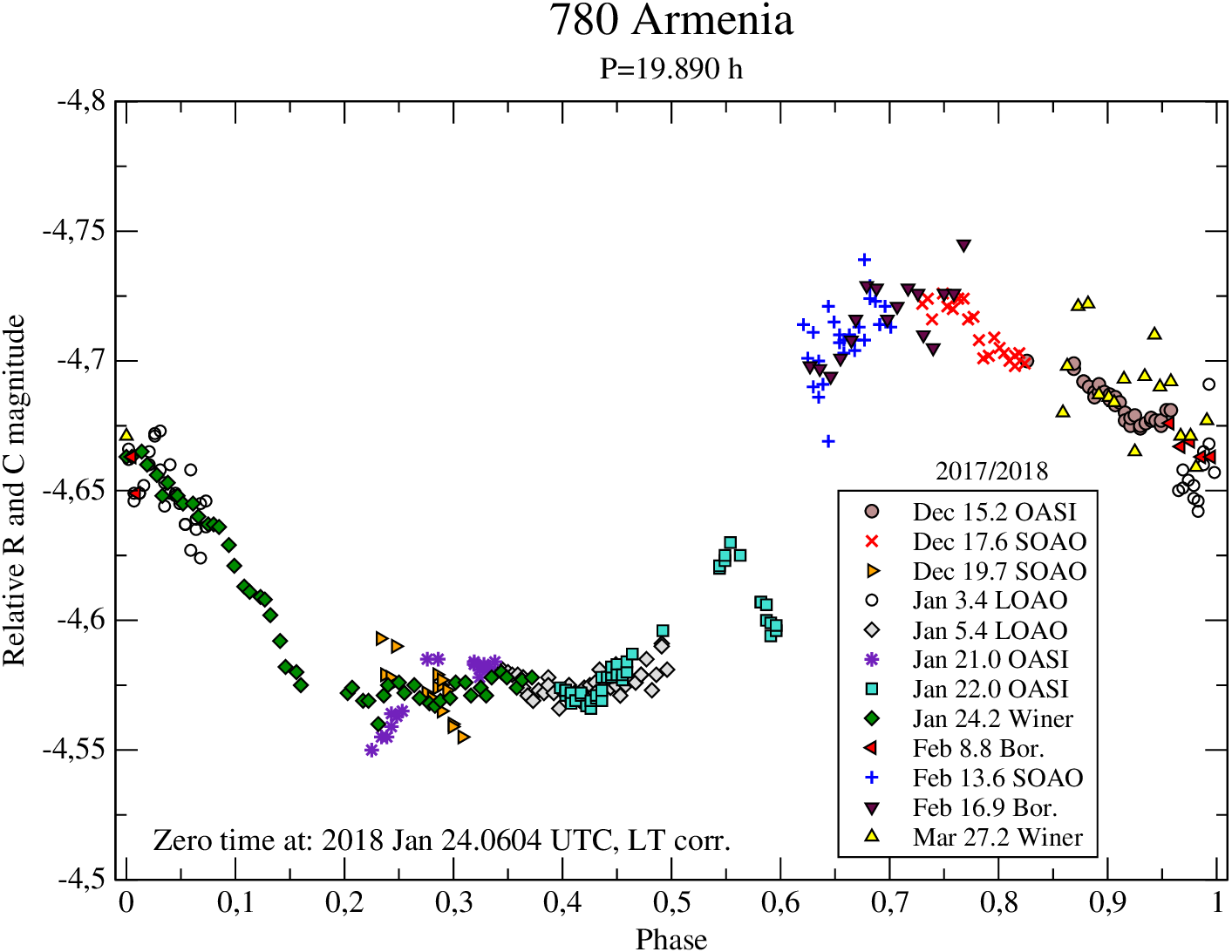} 
\captionof{figure}{Composite light curve of (780) Armenia from the years 2017-2018.}
\label{780composit2018}
&
\includegraphics[width=0.46\textwidth]{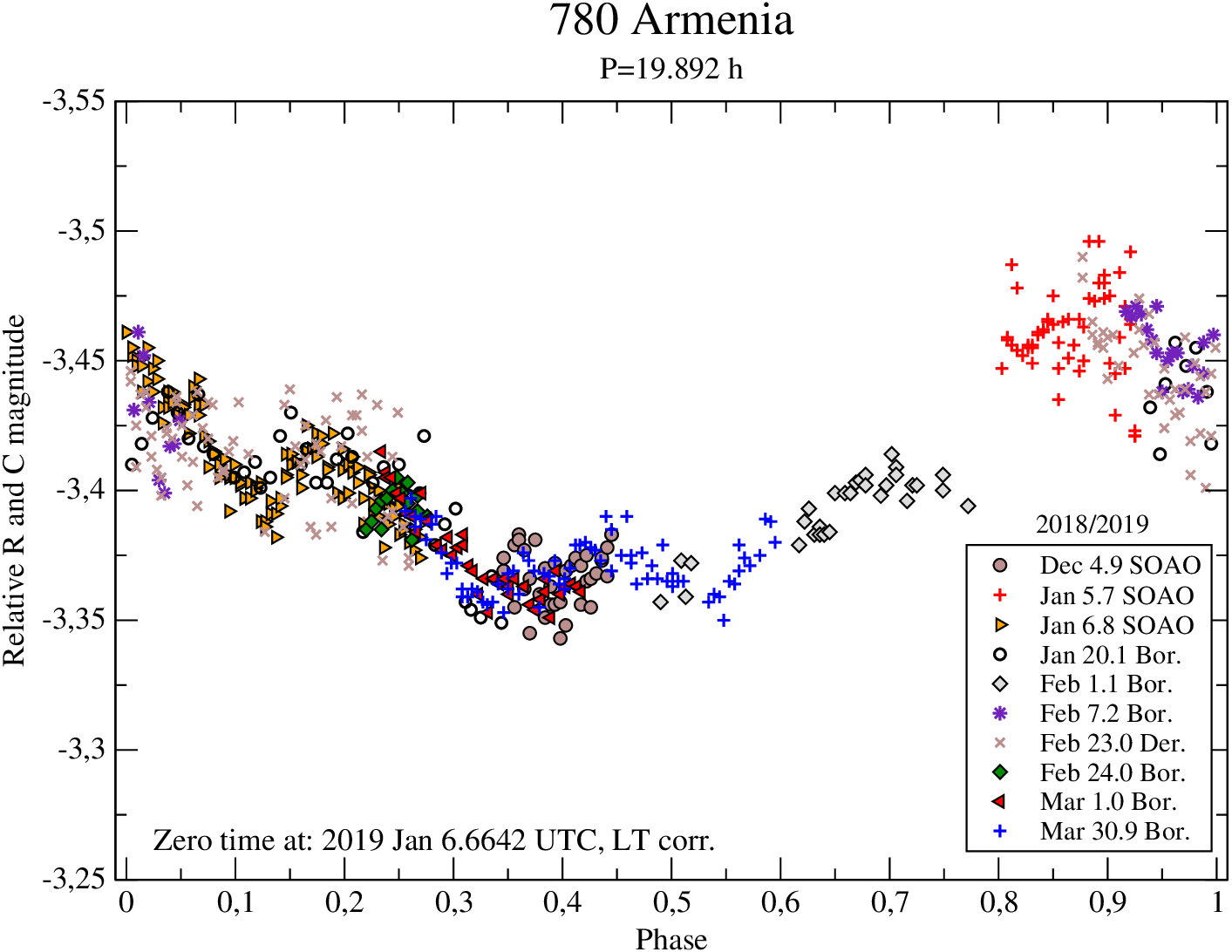} 
\captionof{figure}{Composite light curve of (780) Armenia from the years 2018-2019.}
\label{780composit2019}
\\
\end{tabularx}
    \end{table*}

%Composits 11
    \begin{table*}[ht]
    \centering
\vspace{0.5cm}
\begin{tabularx}{\linewidth}{XX}
\includegraphics[width=0.46\textwidth]{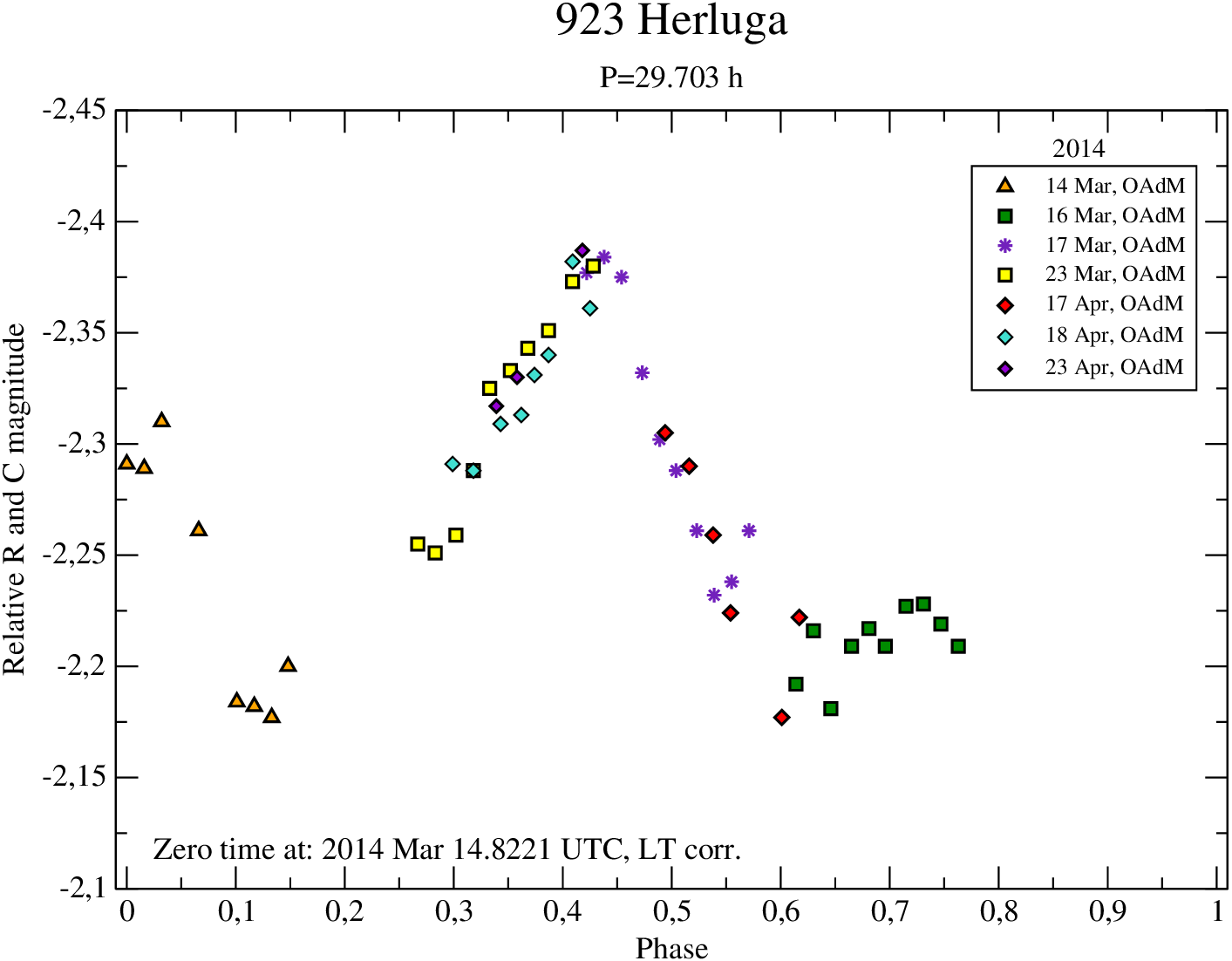} 
\captionof{figure}{Composite light curve of (923) Herluga from the year 2014.}
\label{923composit2014}
&
\includegraphics[width=0.46\textwidth]{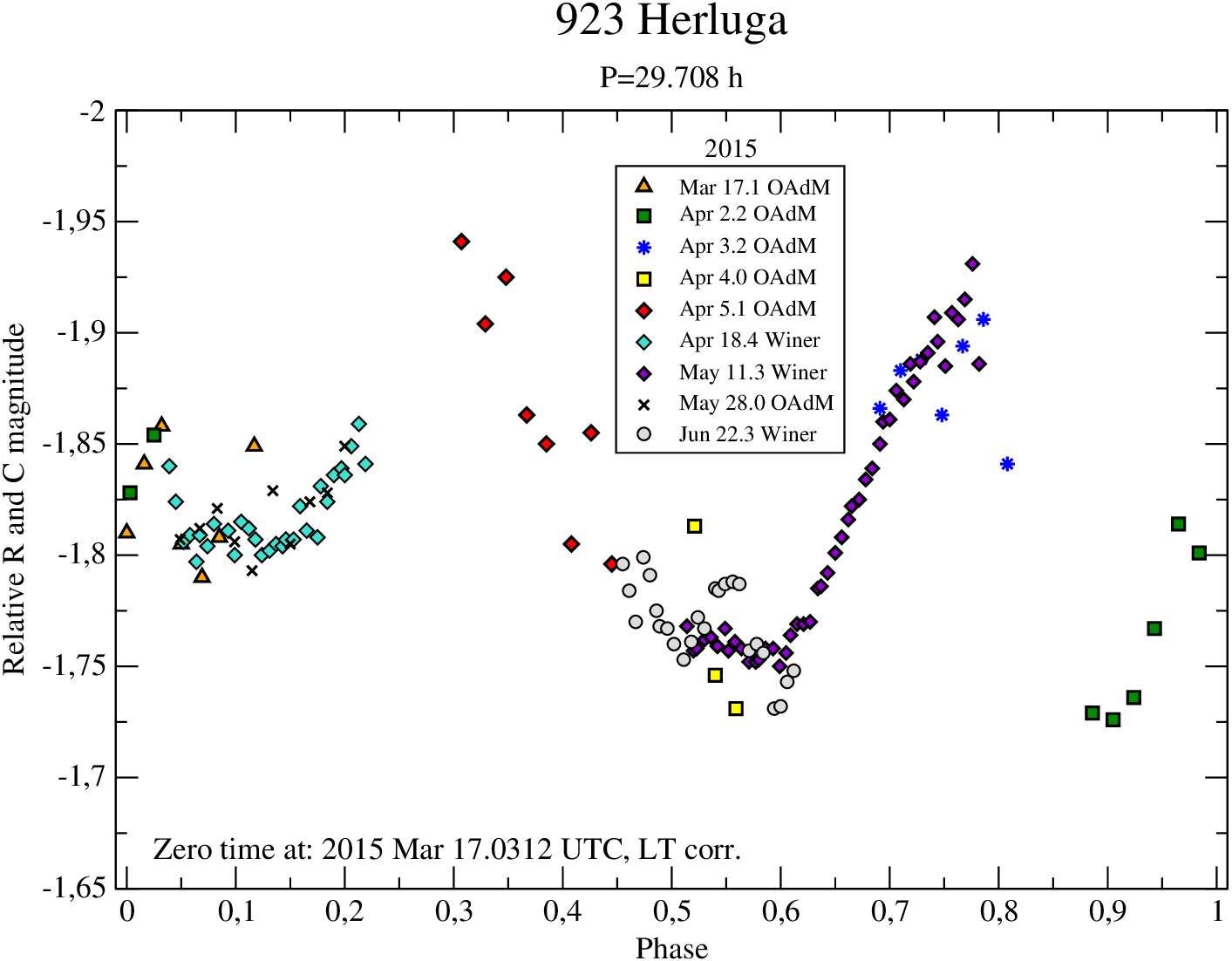} 
\captionof{figure}{Composite light curve of (923) Herluga from the year 2015.}
\label{923composit2015}
\\
\includegraphics[width=0.46\textwidth]{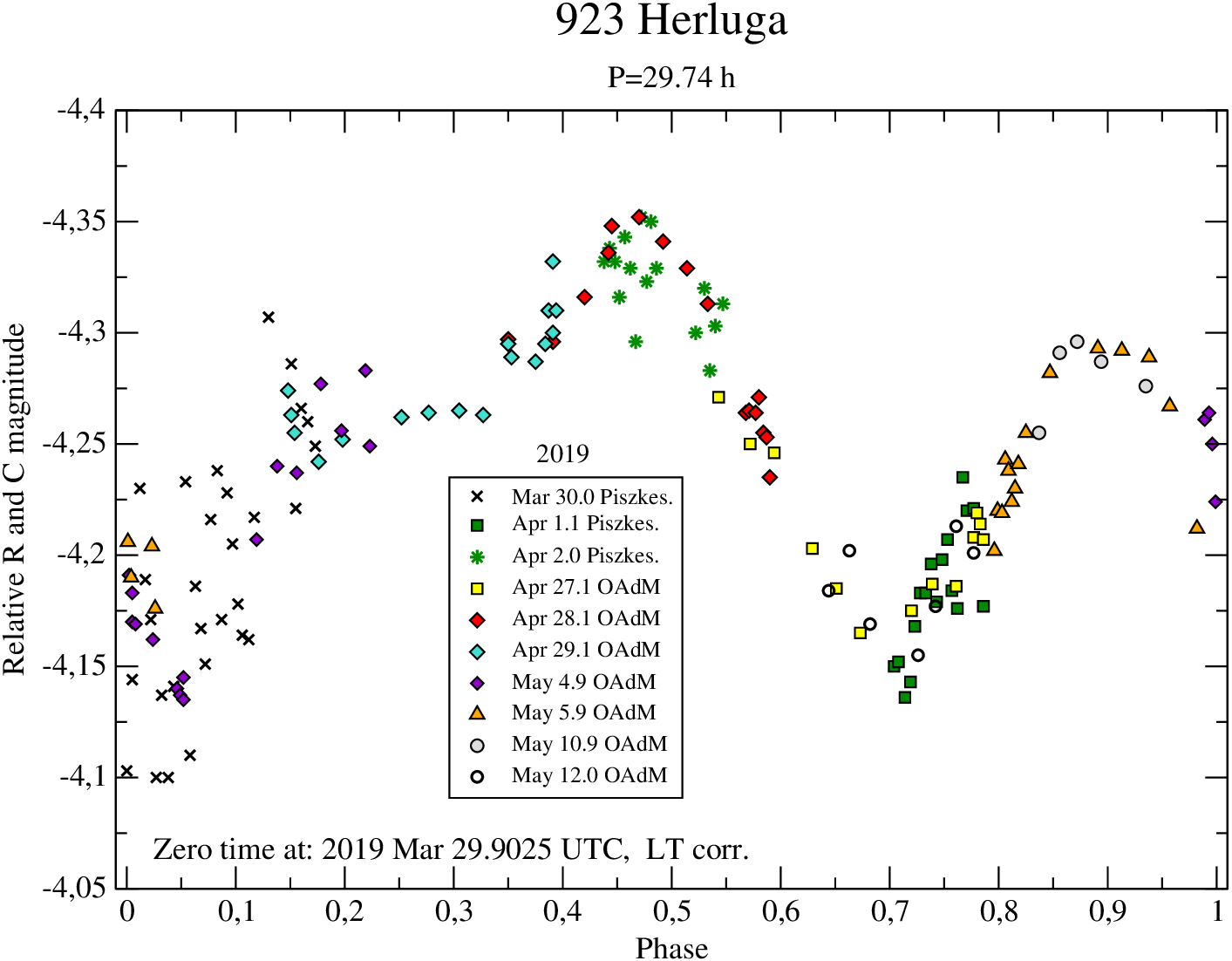} 
\captionof{figure}{Composite light curve of (923) Herluga from the year 2019.}
\label{923composit2019}
&
\includegraphics[width=0.46\textwidth]{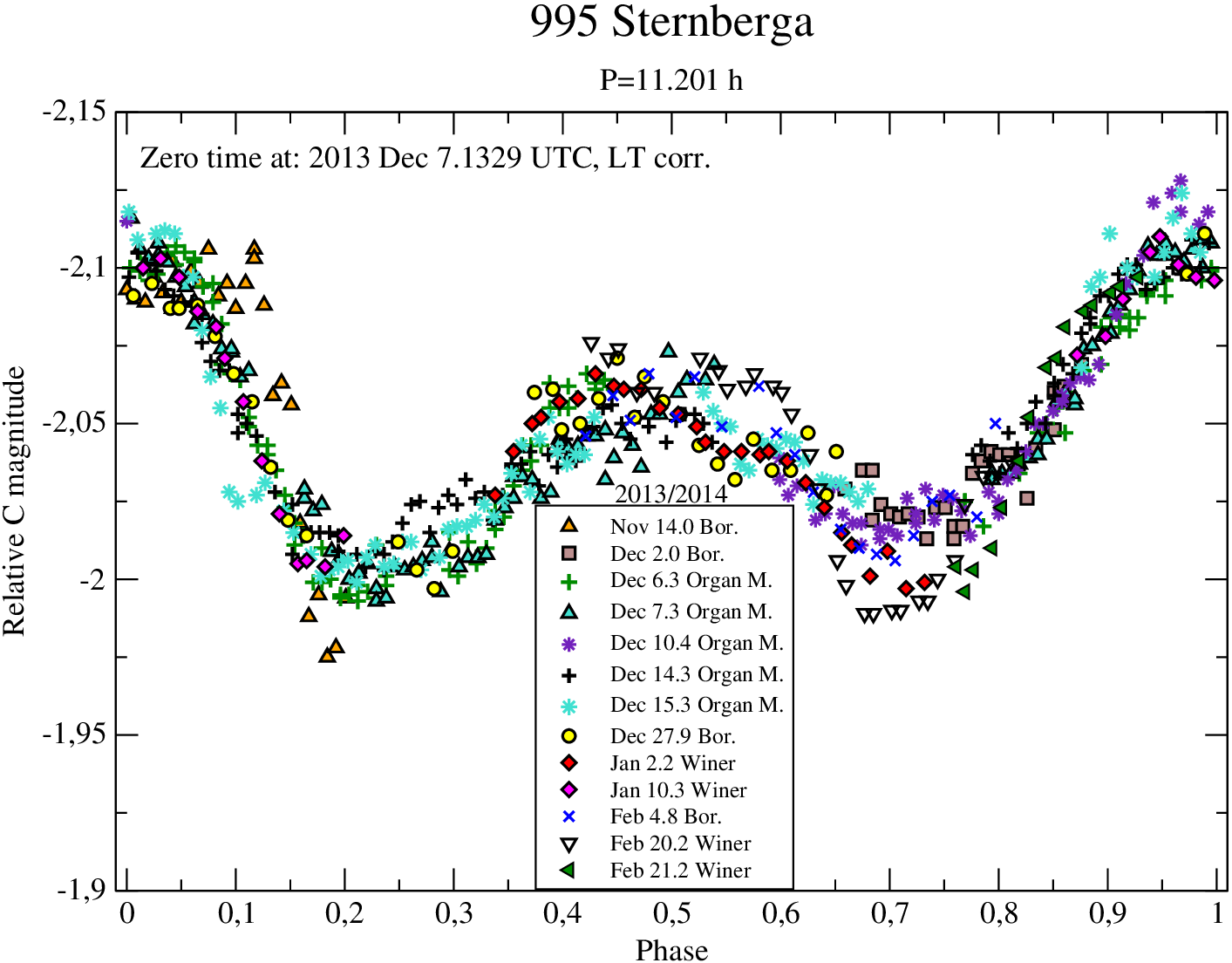} 
\captionof{figure}{Composite light curve of (995) Sternberga from the years 2013-2014.}
\label{995composit2013}
\\
\includegraphics[width=0.46\textwidth]{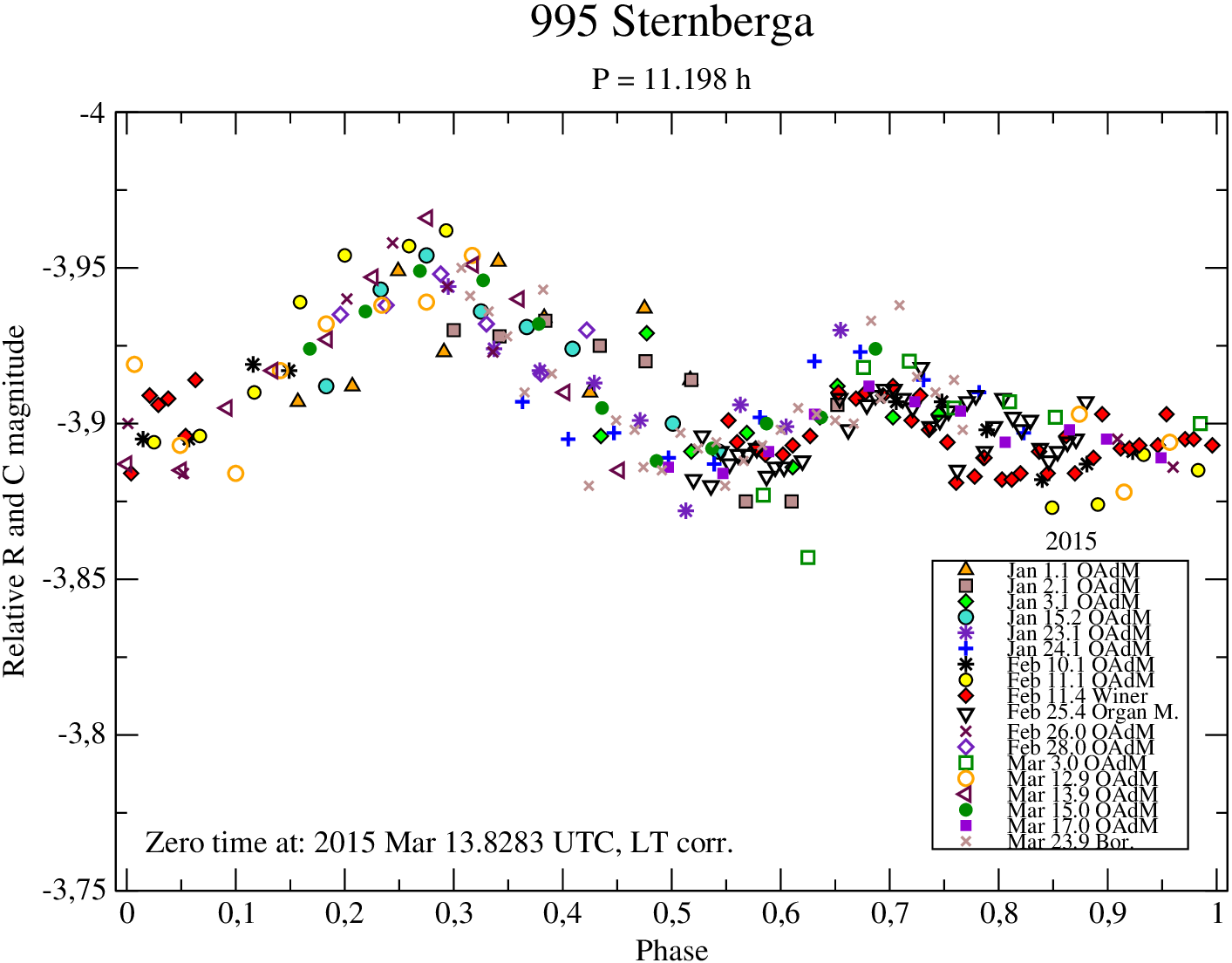} 
\captionof{figure}{Composite light curve of (995) Sternberga from the year 2015.}
\label{995composit2015}
&
\includegraphics[width=0.46\textwidth]{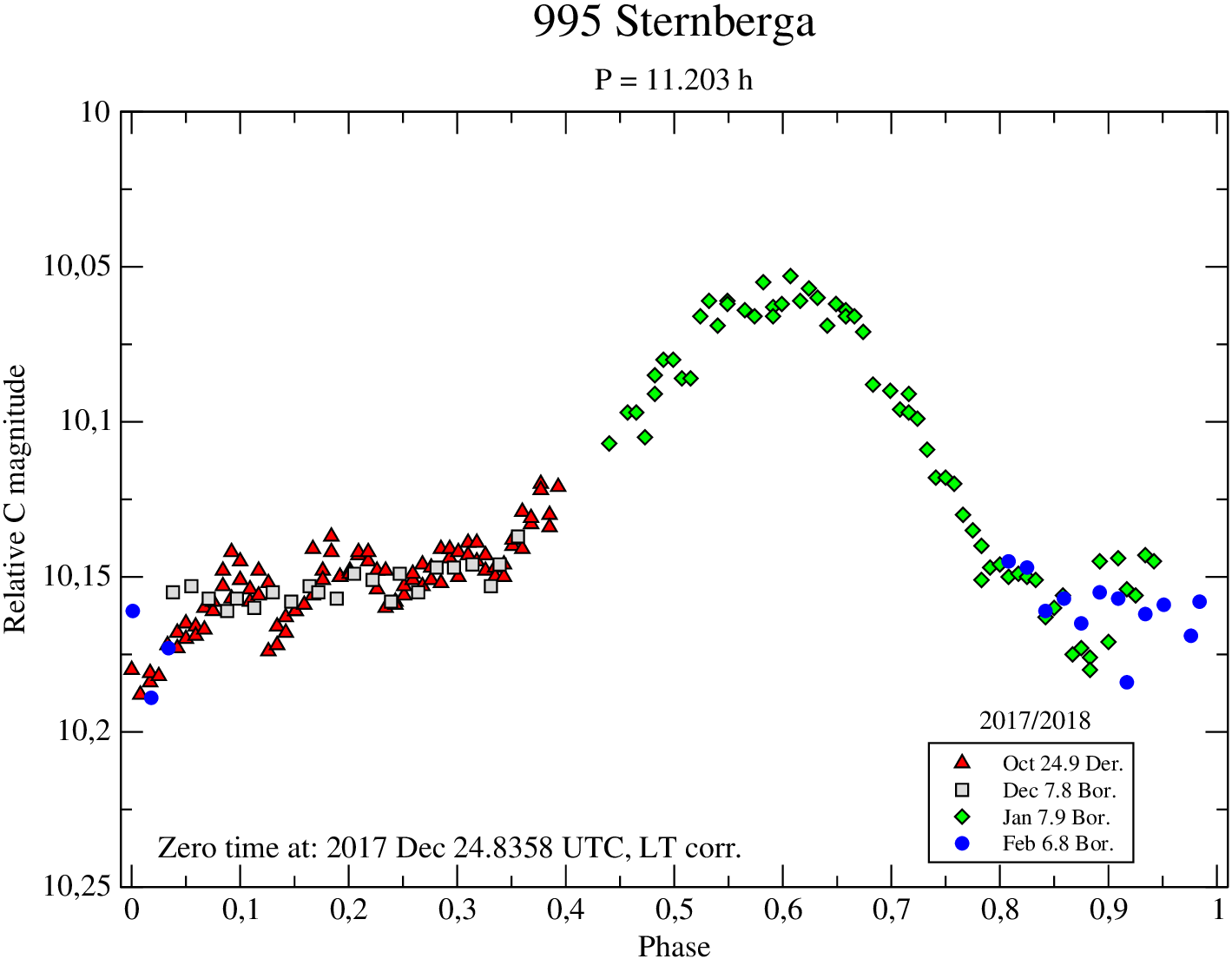} 
\captionof{figure}{Composite light curve of (995) Sternberga from the years 2017-2018.}
\label{995composit2017}
\end{tabularx}
    \end{table*}

\end{appendix}

\end{document}